\documentclass[sn-basic]{sn-jnl}

\usepackage{graphicx}%
\usepackage{multirow}%
\usepackage{amsmath,amssymb,amsfonts}%
\usepackage{amsthm}%
\usepackage{mathrsfs}%
\usepackage[title]{appendix}%
\usepackage{xcolor}%
\usepackage{textcomp}%
\usepackage{manyfoot}%
\usepackage{booktabs}%
\usepackage{algorithm}%
\usepackage{algorithmicx}%
\usepackage{algpseudocode}%
\usepackage{listings}%
\usepackage{upgreek}

\renewcommand{\mu}{\upmu}
\newcommand{\micron}{\,\ensuremath{\upmu\mathrm{m}}}

\newcommand{\rev}[1]{{#1}}

\begin{document}

\title[Is cosmic dust porous?]{Is cosmic dust porous?}

\author*[1]{\fnm{Alexey} \sur{Potapov}}\email{alexey.potapov@uni-jena.de}

\author[2]{\fnm{Martin R. S.} \sur{McCoustra}}\email{m.r.s.mccoustra@hw.ac.uk}

\author[3]{\fnm{Ryo} \sur{Tazaki}}\email{ryo.tazaki1205@gmail.com}
\author[4]{\fnm{Edwin A.} \sur{Bergin}}\email{ebergin@umich.edu}

\author[5,6]{\fnm{Stefan T.} \sur{Bromley}}\email{s.bromley@ub.edu}

\author[7]{\fnm{Robin T.} \sur{Garrod}}\email{rgarrod@virginia.edu}

\author[8]{\fnm{Albert} \sur{Rimola}}\email{albert.rimola@uab.cat}

\affil[1]{\orgdiv{Analytical Mineralogy Group, Institute of Geosciences}, \orgname{Friedrich Schiller University Jena}, \city{Jena}, \country{Germany}}

\affil[2]{\orgdiv{Institute of Chemical Sciences}, \orgname{Heriot-Watt University}, \city{Edinburgh}, \country{Scotland}}

\affil[3]{\orgdiv{Department of Earth Science and Astronomy, Graduate School of Arts and Sciences}, \orgname{The University of Tokyo}, \city{Tokyo}, \country{Japan}}

\affil[4]{\orgdiv{Department of Astronomy}, \orgname{The University of Michigan}, \city{Ann Arbor}, \state{MI}, \country{United States}}

\affil[5]{\orgdiv{Departament de Qu\'imica F\'isica, Institut de Qu\'imica Te\`orica i Computacional (IQTC)}, \orgname{Universitat de Barcelona}, \city{Barcelona}, \country{Spain}}

\affil[6]{\orgdiv{Instituci\'o Catalana de Recerca i Estudis Avançats (ICREA)}, \city{Barcelona}, \country{Spain}}

\affil[7]{\orgdiv{Departments of Astronomy and Chemistry}, \orgname{The University of Virginia}, \city{Charlottesville}, \state{VA}, \country{United States}}

\affil[8]{\orgdiv{Departament de Qu\'imica}, \orgname{Universitat Aut\`onoma de Barcelona}, \city{Barcelona}, \country{Spain}}

\abstract{There is a long-standing discussion in the astrophysical/astrochemical community as to the structure and morphology of dust grains in various astrophysical environments (e.g., interstellar clouds, protostellar envelopes, protoplanetary and debris disks, and the atmospheres of exoplanets). Typical grain models assume a compact dust core which becomes covered in a thick ice mantle in cold dense environments. In contrast, less compact cores are likely to exhibit porosity, leading to a pronounced increase in surface area with concomitant much thinner ice films and higher accessibility to the bare grain surface. Several laboratory experimental and theoretical studies have shown that this type of dust structure can have a marked effect on several physico-chemical processes, including adsorption, desorption, mobility, and reactivity of chemical species. Porous grains are thus thought to likely play a particularly important and wide-ranging astrochemical role. 

Herein, we clarify what is meant by porosity in relation to grains and grain agglomerates, assess the likely astrochemical effects of porosity and ask whether a fractal/porous structural/morphological description of dust grains is appropriate from an astronomical perspective. We provide evidence for high porosity from laboratory experiments and computational simulations of grains and their growth in various astrophysical environments. Finally, we assess the observational constraints and perspectives on cosmic dust porosity.

Overall, our paper discusses the effects of including porosity in dust models and the need to use such models for future astrophysical, astrochemical and astrobiological studies involving surface or solid-state processes.}

\keywords{Astrochemistry, Molecular processes, Solid state: refractory, (ISM:) dust, extinction, (ISM:) evolution}

\maketitle

\setcounter{tocdepth}{3}
\tableofcontents

\clearpage
\textbf{List of acronyms}\\

ADP - agglomerated debris particles 

AGB - asymptotic giant branch 

AFM - atomic force microscopy

ALMA - Atacama Large Millimeter/submillimeter Array

BCCA - Ballistic Cluster Cluster Aggregates 

BE - binding energy

CCA - cluster-cluster aggregation

CDA - Cassini’s Cosmic Dust Analyzer 

DDA - discrete dipole approximation

DFT - density functional theory

DHS - distribution of hollow spheres

DoLP - degree of linear polarization 

EDX - energy dispersive X-ray 

EMA - effective medium approximation

EPM - electrostatic potential map

GRF - Gaussian Random Field 

HST - Hubble Space Telescope 

IDP - interplanetary dust particles

IP - interatomic potential

IR - infrared 

ISM - interstellar medium

IUPAC - International Union of Pure and Applied Chemistry 

JWST - James Webb Space Telescope

MCMC - Markov-Chain Monte Carlo

MD - molecular dynamics

MLS - multilayered spheres

MM - molecular mechanics

MRN - Mathis, Rumpl, Nordsieck 

NASA - National Aeronautics and Space Administration

PAHs - polycyclic aromatic hydrocarbons 

PBC - periodic boundary conditions

PCA - particle-cluster aggregation

PDR - photon-dominated region

PSF - point spread function

pSPF - polarized intensity scattering phase function 

RGD - Rayleigh--Gans--Debye 

QM - quantum mechanical

SPF - scattering phase functions 

SED - spectral energy distribution

SEM - scanning electron microscopy 

TEM - transmission electron microscopy

tSPF - total intensity scattering phase function 

UV - ultraviolet 

VLT - Very Large Telescope

\section{Introduction}
\label{sec:intro}
\subsection{General remarks}
The properties of dust grains in various astrophysical environments, such as the envelopes of evolved stars, diffuse and dense interstellar clouds, protostellar envelopes, protoplanetary disks, debris disks, and the atmospheres of (exo)planets are the focus of much research for several obvious reasons. Dust grains: (i) are the building blocks of rocky planets, cores of gas giants, and minor bodies (comets, asteroids) in planetary systems; (ii) participate in the formation of planetary atmospheres; (iii) provide surfaces for key chemical reactions defining astrochemical networks; (iv) can play a catalytic role in those reactions; and (v) influence the thermodynamic properties of astrophysical media. 

A number of review papers discuss the role of dust grains in the aforementioned processes and we refer the reader to those papers for details \citep{RN381, Draine2003, RN1604, Henning2010, RN424, RN1508, RN1280, RN852, RN1361, RN1301, RN1533}. However, despite of all these reviews (based on much research), there is a parameter of dust grains, which may play a crucial role in the processes mentioned above but which has never been discussed in common from the observational, experimental, and theoretical points of view. This parameter is dust porosity.

There is considerable controversy over the porosity of cosmic dust grains and the astrophysical community is very far from reaching a consensus. Clarification of this question, particularly an affirmative answer to the question “whether dust grains in astrophysical environments are porous?” would lead to re-evaluation of many previous results and approaches and would open new doors for future experimental, theoretical, and observational studies. That discussion must rightly rest, in part, in the field of materials science where porosity is not a loosely defined concept but a multiscale property of solid materials.

\subsection{Definitions and key points} 

Porosity in a material is often defined as the ratio of the volume of voids/vacuum over the total volume. According to the International Union of Pure and Applied Chemistry (IUPAC): ``The structure is defined by the distribution in space of the atoms or ions in the material part of the catalyst and, in particular, by the distribution at the surface. The texture is defined by the detailed geometry of the void space in the particles of catalyst. Porosity is a concept related to texture and refers to the pore space in a material.'' Porous materials are classified into microporous $(d < 2\, \text{nm})$, mesoporous $(2\, \text{nm} \le d \le 50\, \text{nm})$, and macroporous $(d > 50\, \text{nm})$, where $d$ is the pore diameter \citep{RN1602}. \rev{This leads us to consider two sources of porosity; the material itself (which we might call intrinsic porosity) and the morphology of the dust grain agglomerate (which we might call extrinsic porosity).}

Thus, in speaking about cosmic dust porosity, we consider that dust grains in various astrophysical environments are not compact but consist of compact structural units and voids between these units and the fraction of the volume of voids over the total volume is high. Porous dust grains are also often described as fractal aggregates, which typically means that structural units of the aggregates are distributed in a 3D space with a dimension $\gamma < 3$, so that in a sphere of a radius $r$ around a fixed unit there are N $\sim$ $r$$^\gamma$ other units. There is no contradiction between these two definitions, however, porosity means the presence of “closed” voids between units (meaning the presence of an internal surface) and fractal structures can be completely “open” meaning absence of material in the lines of sight between the central unit and the edge of a grain aggregate (thus, only an external surface is present). 

There are three important points of the high porosity view of cosmic dust grains. Firstly, high porosity tends to lead to weak bonding between aggregated units and a corresponding fluffiness of the resultant grains, which should lead to more efficient growth/destruction of grain aggregates as compared to compact (strongly bound) grains. Secondly, high porosity implies the existence of a much larger surface area as compared to compact grains. A large surface area provides more sites for physico-chemical processes. Indeed, we should note that the catalytic formation of molecules on dust surfaces may be one of the most important astrochemical gas-grain interaction processes. For a definition of catalysis, we again refer to IUPAC: ``Catalysis is the phenomenon in which a relatively small amount of a foreign material, called a catalyst, augments the rate of a chemical reaction without itself being consumed'' \citep{RN1602}. Thirdly, “closed” voids between grain units may provide efficient confinement, storage and shielding of reactants and reaction products that adsorb and form on the internal dust surface within pores. 

\subsection{Evolution of dust grains with respect to porosity} 
Various types of solids are considered as components of cosmic dust, with carbon- and \rev{silicate-based} materials as major components and metallic oxides, carbides, sulfides, and metals themselves as minor components. In cold environments, refractory grains are mixed with molecular ices. Composition and origins of dust were recently discussed in a review paper and the reader can find details and relevant references there \citep{RN1301}. 

In brief, primary, nm-sized dust grains form in circumstellar envelopes of evolved stars \rev{and expanding shells of supernovae}, where they can reach sizes up to tens of nm \citep{RN823, RN820}. Here, the first stage of fractal aggregation may take place leading to the formation of micropores. After the formation, grains are pushed away from their parent stars, enter the ISM, travel through interstellar clouds, protostellar envelopes, and planet-forming disks, and finally act as building blocks of minor bodies, planets, and their atmospheres. Alternatively, star dust grains can be completely destroyed in the ISM and restart their evolution history there \citep{Draine2003}. The journey from the ISM to planetary systems is accompanied by the growth of grains from nm-sized particles to mm-sized aggregates \citep{RN903, RN902, Draine2003} and there is much evidence for the fractal growth and appearance of meso- and macropores as discussed in the following sections. 

Moreover, space weathering (by radiation fields, cosmic rays, shock waves) during the passage from source to sink has an impact on dust porosity, e.g., it might lead to pore deformation and collapse. While vice versa, the efficiency of space-weathering processes is dependent on the physical properties of grains, such as grain size, surface roughness, and porosity \citep{RN1603}.

\subsection{Dust grain models} 
There are several models describing the structure of dust grains in astrophysical environments. \rev{The still most popular model considers grains as a compact core of refractory materials, which in cold dense environments (dense interstellar clouds, protostellar envelopes and protoplanetary disks beyond the snowline) is covered by a thick ice layer, which in the onion model is assumed to comprise an inner polar layer and an outer apolar layer \citep{RN1601, RN714}.} In this model, dust and ice are physically separated due to the thick ice coverage and dust has no or very limited influence on the physico-chemical processes on and within ice mantles. The compact refractory core itself might also have an onion-like structure with silicate materials in the centre and a carbon-based accreted mantle \citep{RN1600, RN795}. Cold \rev{dense} astrophysical environments attract a lot of attention \rev{due to the rich chemistry, known from detections of many organic molecules, including direct precursors of prebiotic species.}

An alternative model considers porous fractal dust core and corresponding low ice coverage due to the dust/ice mixing and large dust surface area \citep{RN930}. This concept is based on the results of various laboratory experiments and grain growth models demonstrating that dust grains in the interstellar medium (ISM) and planet-forming disks should be highly porous. These experiments and models are discussed in the following sections. In this alternative model, the dust surface should be available for physico-chemical processes in cold \rev{dense} environments and may play an important (or in some cases crucial) role therein. The availability of the bare dust surface in cold environments was also shown in the experiments on H$_2$O agglomeration at low temperatures resulting in the presence of wet (water covered) and dry (bare dust) areas on the surface of dust \citep{RN955, RN954}.

In \rev{less dense (diffuse interstellar clouds) and} warmer \rev{(protostellar envelopes and protoplanetary disks inside the snowline and atmospheres of exoplanets)} environments, dust is perhaps a little simpler. There is unlikely to be a layer of adsorbed volatiles and hence in such environments the bare refractory grain surface is exposed. This surface may be inorganic in nature and comprised of silicate minerals or may, over time, have accreted a layer of carbonaceous materials. In both instances, the materials will have been exposed to \rev{stellar and} interstellar radiation fields for significant lengths of time.

\section{Optical properties of porous dust particles}\label{sec:optprop}

\rev{The problem of determining dust porosity in various astronomical environments relies on our understanding of the optical properties of porous dust particles, which describe how they absorb, scatter, and emit electromagnetic radiation. The purpose of this section is to provide a high-level summary of their optical properties that are relevant for later discussion. In the following, we highlight four key behaviors in the optical properties of porous dust particles. Quantitative behavior depends sensitively on the detailed morphology, size, and composition of the particles, and a complete discussion of the optical property is beyond the scope of this review. We limit our discussion to only qualitative behavior and aim to develop some basic intuition to the physical processes. }

\subsection{Qualitative behavior} \label{sec:quali}

\subsubsection{Emission and absorption: cross section} \label{sec:rgdabs}

\rev{Dust particles absorb the incoming electromagnetic wave and re-emit via thermal emission. The magnitude of thermal emission and absorption of dust particles is given by their absorption cross section, $C_{\mathrm{abs}}$, which has units of $\mathrm{cm}^2$. 
The incoming electric field induces polarization within a particle with polarizability $\alpha$, which depends on the particle size, shape, and complex refractive index. 
A simple piecewise expression for $C_{\mathrm{abs}}$ for a spherical grain with a radius of $a$ is \citep{Bohren83}:
\begin{equation}
C_{\mathrm{abs}} \simeq
\begin{cases}
k\, \mathrm{Im}[\alpha] & (ka \ll 1), \\
\pi a^2 & (ka \gg 1).
\end{cases} \label{eq:cabs}
\end{equation}
Here, $k=2\pi/\lambda$ is a wavenumber, $\lambda$ is a wavelength measured in vacuum, and $\mathrm{Im}[...]$ takes the imaginary part of a complex quantity $\alpha$. For small grains such that $ka\ll1$, $C_{\mathrm{abs}}$ is wavelength dependent via $k$ and $\alpha$ (Rayleigh limit), whereas for large grains it is wavelength independent as $C_\mathrm{abs}$ is given by its geometrical cross section (geometrical optics limit).}

\rev{First, we discuss the absorption property of porous dust particles.
As such, we consider a fractal aggregate consisting of $N$ spherical monomer grains, like Ballistic Cluster Cluster Aggregates (BCCA; $\gamma\sim2$) shown in Fig.~\ref{fig:spf}.
We assume that each monomer grain is optically soft ($|m-1|\ll1$) and has a small phase shift ($ka_0|m-1|\ll 1$), where $m$ is a complex refractive index and $a_0$ is the monomer radius. 
For $\gamma\lesssim2$, geometrical overlapping of monomers along sight lines is negligible, and all monomers in the aggregate would be uniformly illuminated/excited by the external field so that absorption by each monomer may be treated as if it would be isolated. In other words, the aggregate is ``transparent'' under such conditions. Hence, the absorption cross section of each monomer would be approximated by $k\mathrm{Im}[\alpha_0]$, where $\alpha_0$ is the polarizability of each monomer grain without the presence of neighboring monomers. 
As a result, the total amount of absorption by the aggregate would be the sum of that by all monomers:
\begin{equation}
C_{\mathrm{abs}}(\lambda) \simeq N C_{\mathrm{abs},0}(\lambda) = N k\, \mathrm{Im}[\alpha_0]  \label{eq:rgdcabs}
\end{equation}
where $C_{\mathrm{abs},0}$ is the absorption cross section of a single monomer. 
The wavelength dependence of absorption/emission, or spectroscopic properties, of the aggregate is determined by that of each monomer grain. In other words, in a spectroscopic quantity, large porous aggregates may appear as small particles \citep[e.g.,][]{Min06}. As long as the transparent conditions are met, this equation may apply to whatever the aggregate size is.}

\rev{The above approximation is called the Rayleigh--Gans--Debye (RGD) approximation (or equivalently 1st-order Born approximation), which only holds in the limit of a higher porosity and a lower refractive index \citep[see][for more detailed conditions]{Berry86, Sorensen01, Tazaki16, Tazaki18}.
If those conditions are not met, each monomer would feel an electric field different from the external one due to shadowing by other monomers in the aggregate and/or electromagnetic interaction between neighboring monomers \citep{Wright87,Bazell90,Kozasa92,Henning95,Stognienko95,Mackowski95,Mackowski06,Kohler11,Tazaki18,Lodge24}. Despite its limited applicability, such a view is still useful to gain intuition on why large porous aggregates often exhibit spectroscopic properties similar to small particles.}

\rev{This idea could also be applied to qualitatively assess the influence of dust porosity on solid state features. The information of the absorption features (i.e., enhanced absorption at a resonant wavelength of a solid material) is imprinted in the refractive index (and then polarizability). In the Rayleigh limit ($ka\ll1$), the feature is clearly observable because $C_\mathrm{abs}$ directly reflects $\alpha$, but would be weakened for large grains ($ka\gg1$) because $C_\mathrm{abs}$ is independent of the wavelength (Eq.~\eqref{eq:cabs}).
For large porous aggregates, we may still observe an absorption feature because their absorption property is govern by monomers' (or small particles') property (Eq.~\eqref{eq:rgdcabs}). Qualitatively similar behavior has been reported by a number of numerical studies \citep{Min06,Vosh08,Kolokolova07,Vaidya11,Kataoka14,Gupta16}.}

\subsubsection{Emission and absorption: polarization} \label{sec:abspol}

\rev{In the ISM, dust particles are often aligned with the interstellar magnetic field \citep{Andersson2015}. Aligned particles produce polarized light via dichroic extinction or dichroic emission. This occurs because they preferentially absorb or emit light polarized along specific directions, effectively acting as polarization filters or emitters.}

\begin{figure}[ht]
   \centering
   \includegraphics[width=0.75\textwidth]{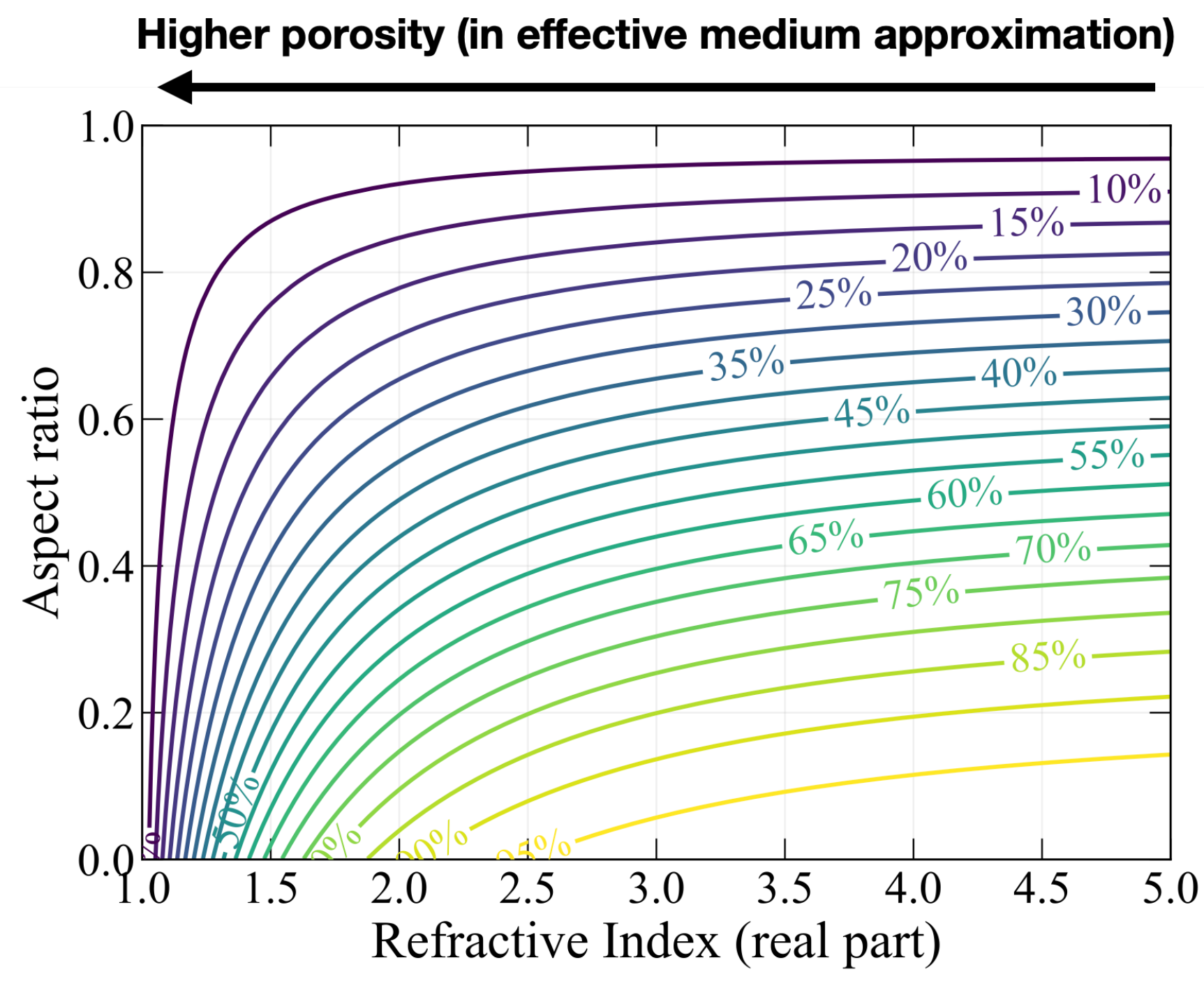}
      \caption{Degree of linear polarization calculated using Eq.~(\ref{eq:alignpol}, $ka\ll1$) for oblate spheroids with various axis ratios and the real part ($n$) of refractive index. The imaginary part ($k$) is assumed to be small such that $k\ll n-1$.}
         \label{fig:aligndop}
\end{figure}

\rev{A simple piecewise expression for the degree of linear polarization (DoLP) produced by an aligned spheroid is
\begin{eqnarray}
\mathrm{DoLP} &=& \frac{C_\mathrm{abs,\perp}-C_\mathrm{abs,||}}{C_\mathrm{abs,\perp}+C_\mathrm{abs,||}},\\
\frac{C_\mathrm{abs,\perp}}{C_\mathrm{abs,||}}&\simeq&
\begin{cases}
\frac{\mathrm{Im}[\alpha_\perp]}{\mathrm{Im}[\alpha_{||}]} & (ka \ll 1), \\
1 & (ka \gg 1),
\end{cases} \label{eq:alignpol}
\end{eqnarray}
where we have used $C_{\mathrm{abs},j} = k\, \mathrm{Im}[\alpha_j]$ for $ka\ll1$ and $j=\perp, \parallel$ refers to the incident electric field polarized perpendicular or parallel to the grain’s short axis, respectively. In this case, $a$ is the radius for the longest axis of the spheroid. For spherical grains ($C_{\mathrm{abs},\perp} = C_{\mathrm{abs},\parallel}$), $\mathrm{DoLP}=0$. For spheroidal grains with $C_{\mathrm{abs},\perp} \neq C_{\mathrm{abs},\parallel}$, we observe a polarized light. 
Equation~\eqref{eq:alignpol} captures an overall behavior of numerical results for DoLP reported in \citet{Cho07},
although there may be a systematic error particularly for $ka\gg1$ \citep{Guillet20}. 
Figure~\ref{fig:aligndop} shows a DoLP in the $ka \ll 1$ regime for various sets of an aspect ratio and the real part of the refractive index. The real part of the refractive index plays a crucial role in determining DoLP; it essentially controls the extent how much of the geometrical asymmetry of spheroid (long/short axis ratio) is converted into the polarization signal.}

\begin{figure}[ht]
   \centering
   \includegraphics[width=0.75\textwidth]{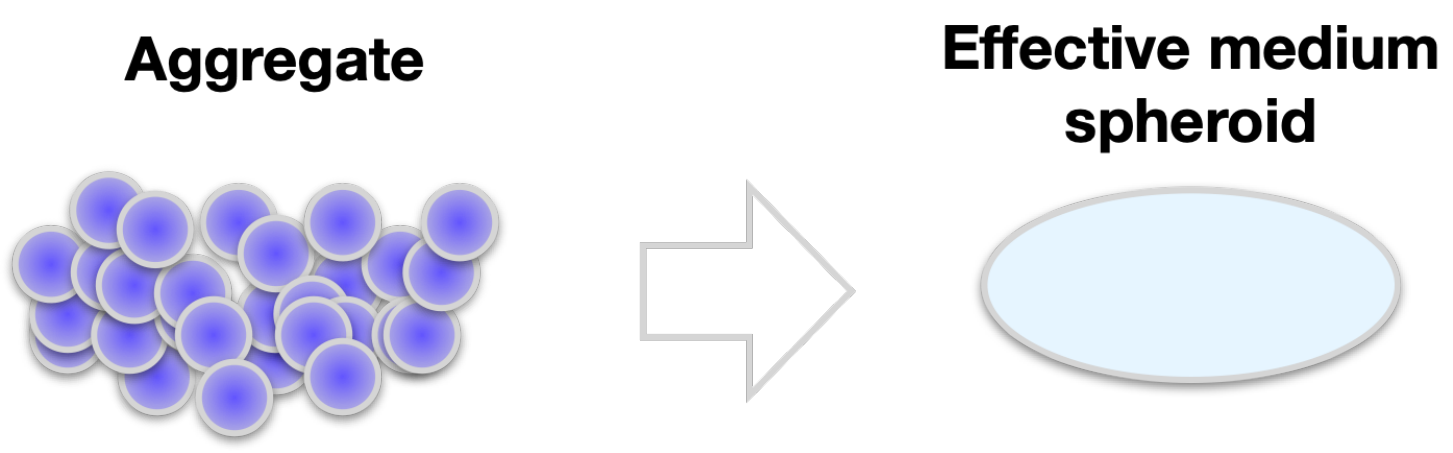}
      \caption{A cartoon showing the concept of the spheroidal analog approximation of a porous aggregate.}
         \label{fig:saa}
\end{figure}

\rev{How does the porosity influence the DoLP by aligned particles? As with the previous case, let's consider porous dust aggregates. 
Porous aggregates may be approximated by a spheroid with an effective medium refractive index (Fig.~\ref{fig:saa}). In fact, \citet{Draine2025} recently demonstrated that the optical properties of irregular porous aggregates can be well approximated by spheroids using the EMT with the Bruggeman mixing rule. When the volume filling factor $f$ (with porosity $= 1 - f$) is small ($f\ll1$), the Bruggeman mixing rule coincides with the Maxwell--Garnett (MG) mixing rule. Under the MG approximation, \citet{Kataoka14} showed that $n_\mathrm{eff} - 1 \propto f$, where $n_\mathrm{eff}$ is the real part of the effective refractive index of porous aggregates. If $f\to0$, $n_\mathrm{eff}\to1$. This means that a higher porosity (lower $f$) leads to a lower effective real refractive index, which in turn leads to a lower degree of linear polarization (Fig.~\ref{fig:aligndop}). Thus, porous grains are less efficient polarizers. The above trend has been confirmed by several numerical studies \citep{Kirch19, Draine2021, Draine2024_aggregates}. }

\rev{Figure~\ref{fig:aligndop} also highlights a parameter degeneracy between assumptions about intrinsic dust composition (i.e., the refractive index $n$ before introducing porosity) and the grain aspect ratio. Also, the observed polarization fraction depends on an efficiency of grain alignment, which should also be dependent on dust porosity \citep{Herranen21}.}

\begin{figure}[ht]
   \centering
   \includegraphics[width=\textwidth]{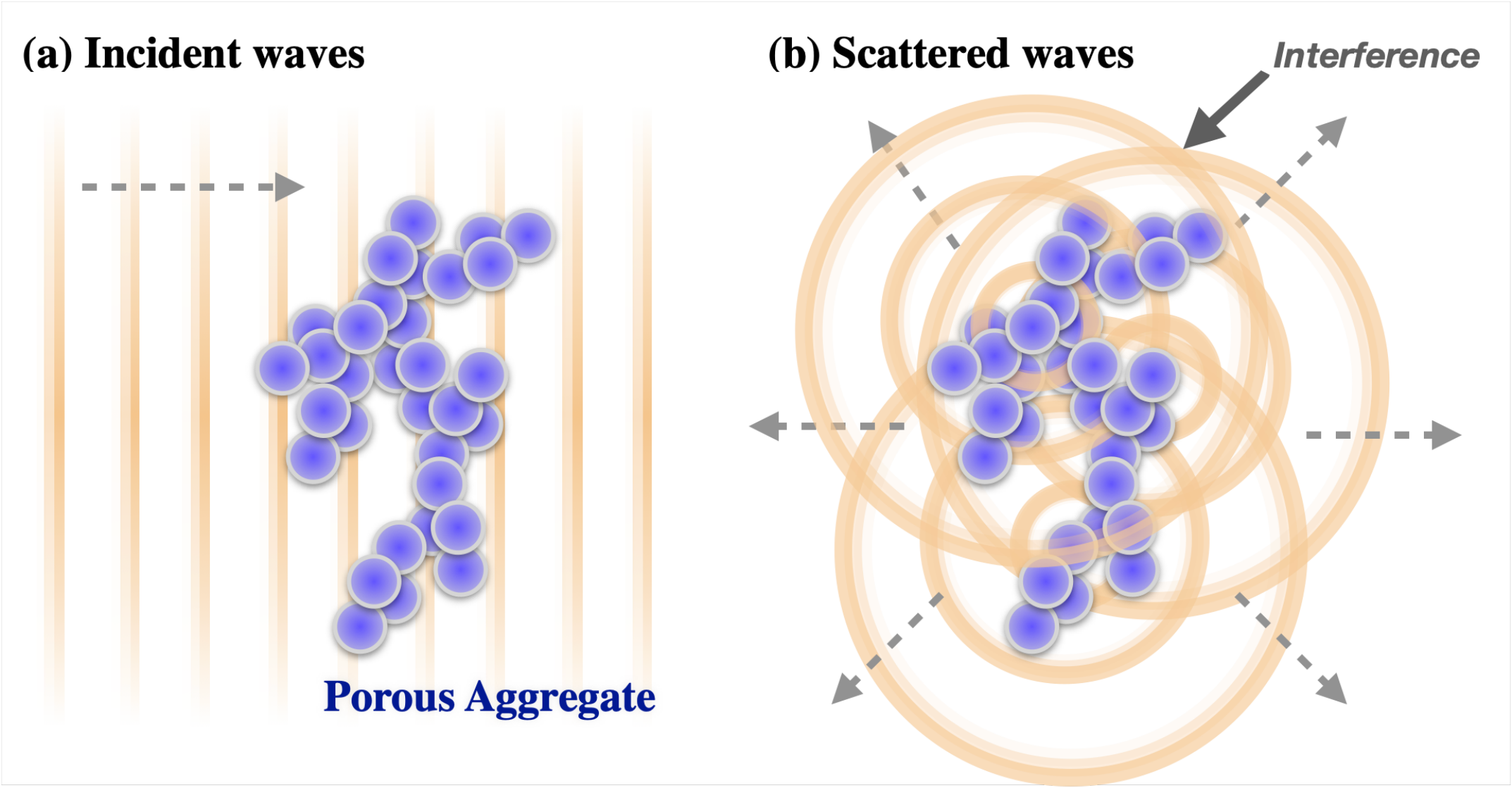}
      \caption{Plane wave incidents on porous aggregates of spherical monomer grains (a) and scattered waves from them (b).}
         \label{fig:rgbcartoon}
\end{figure}

\subsubsection{Scattering: cross section and phase functions} \label{sec:scacs}

\rev{Light scattering has received considerable attention in the context of dust porosity because it directly encodes morphological information about porous dust aggregates, such as their size and internal structure (e.g., porosity).}

\rev{We begin with the simplest case, that is the scattering cross section of a spherical particle, which may be expressed by:
\begin{equation}
C_{\mathrm{sca}}\simeq  
\begin{cases}
\frac{k^4}{6\pi} |\alpha|^2 & (ka \ll 1), \\
\pi a^2 & (ka \gg 1),
\end{cases}\label{eq:csca}
\end{equation}
where the first case corresponds to the Rayleigh scattering limit and the second to the geometrical optics limit \citep{Bohren83}.}

\rev{As in Sect.~\ref{sec:rgdabs}, we follow the RGD approximation to qualitatively explore how porosity affects light scattering. In the RGD framework, the scattering cross section is given by \citep{Berry86, Sorensen01}:
\begin{equation}
C_{\mathrm{sca}}(\lambda) = N^2 C_{\mathrm{sca},0}(\lambda) \cdot F_s(\lambda),\label{eq:rgbcsca}
\end{equation}
where $C_{\mathrm{sca},0}$ is the scattering cross section of an individual monomer, and $F_s$ is a function factor that encapsulates the spatial configuration of the monomers (i.e., porosity). Physically, interference of scattered waves play important role in determining scattering properties (see Fig.~\ref{fig:rgbcartoon}), and $F_s$ describes the degree of coherence among scattered waves. When all scattered waves interfere constructively, we have $F_s = 1$. When part of scattered waves undergoes destructive interference, we have $F_s<1$.
For a fractal aggregate, we have a simple scaling law \citep{Berry86,Sorensen01,Tazaki18}
\begin{equation}
F_s \sim
\begin{cases}
1 & (ka_\mathrm{agg} \ll 1), \\
(ka_\mathrm{agg})^{-\gamma} & (ka_0 \ll 1 \ll ka_\mathrm{agg}), \\
1/N & (ka_0 \gg 1),
\end{cases}\label{eq:rgbfs}
\end{equation}
where $a_0$ and $a_\mathrm{agg}$ are the radii of the monomer and aggregate, respectively, and $\gamma$ is the fractal dimension. The progression from the first to the third regime may be considered as wavelength dependence: the first case ($ka_\mathrm{agg} \ll 1$) corresponds to the long-wavelength limit, while the third case ($ka_0 \gg 1$) corresponds to the short-wavelength limit. Therefore, by measuring wavelength dependence of scattered light, such as colors, one may constrain $a_\mathrm{agg}$, $\gamma$, and $a_0$.}

\rev{The wavelength dependence of scattered light (color) is one of important diagnostics of dust properties, and it is often explained by the single scattering albedo of dust particles, defined by
\begin{equation}
\omega(\lambda) = \frac{C_{\mathrm{sca}}(\lambda)}{C_{\mathrm{sca}}(\lambda)+C_{\mathrm{abs}}(\lambda)}
\end{equation}
Using Eqs.~\eqref{eq:cabs} and \eqref{eq:csca}, one can anticipate $\omega\propto k^3 \propto \lambda ^{-3}$ for small grains (Rayleigh scattering) and $\omega\sim 0.5$ for large grains (geometrical optics). Note that in Rayleigh scattering we have used $C_\mathrm{sca}\ll C_\mathrm{abs}$ and assumed that the wavelength dependence of polarizability is negligible. Rayleigh scattering gives a bluish color, whereas geometrical optics gives a gray color.} 

\begin{figure}[ht]
   \centering
   \includegraphics[width=0.75\textwidth]{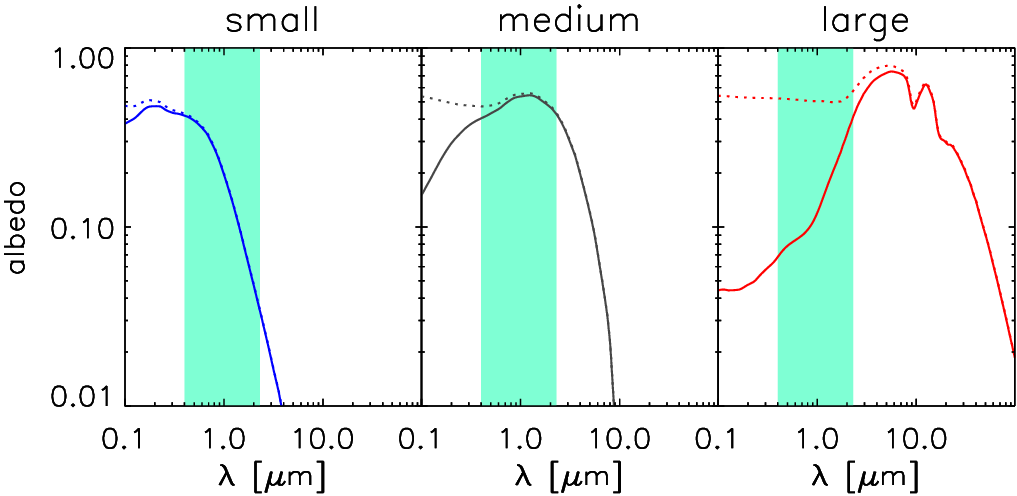}
      \caption{The \rev{effective albedo $\omega_\mathrm{eff}$} (solid lines) and \rev{the single scattering albedo $\omega$} (dashed lines). In the \rev{former}, scattered light flux is integrated \rev{from $34^\circ$ to $126^\circ$}, while it is integrated for all scattering direction \rev{(i.e., from $0^\circ$ to $180^\circ$)} in \rev{the latter}. Left, middle, and right panels represent the results for a grain size of 0.08\,$\mu$m, 0.25\,$\mu$m, and 2.5\,$\mu$m, respectively. The green area represents the range of wavelengths covered by the HST scattered light images of the protoplanetary disk around a young star HD 100546. Image reproduced with permission from \citet{Mulders13}, copyright by ESO.}
         \label{fig:colors}
\end{figure}

\rev{The single scattering albedo may not accurately represent the color of observed scattered light, particularly when a grain scatters light anisotropically. This is because the single scattering albedo, $\omega$, is defined based on $C_\mathrm{sca}$, which represents the total energy scattered into \emph{all} directions. In contrast, observations often sample scattered light at specific scattering angles. As a result, when scattering is anisotropic, the wavelength dependence of the observed scattered light does not necessarily match that of $\omega$. To address this, \citet{Mulders13} introduced the concept of the effective scattering albedo, $\omega_\mathrm{eff}$, which is defined by integrating scattered light only over the observable range of scattering angles. This quantity is illustrated in Fig.~\ref{fig:colors}. The difference between $\omega$ and $\omega_\mathrm{eff}$ increases as scattering becomes more anisotropic. For small grains ($ka \ll 1$), since Rayleigh scattering is nearly isotropic, we have $\omega \approx \omega_\mathrm{eff}$. In contrast, for large grains ($ka \gg 1$), anisotropic scattering causes $\omega_\mathrm{eff}$ to be significantly lower than $\omega$. In such cases, contrary to the gray scattered light implied by $\omega$, the color of scattered light may appear reddish, as shown in Fig.~\ref{fig:colors} \citep[see also][]{Min16}.
}

\rev{What about scattered light colors of porous aggregates? 
We can discuss the single scattering albedo of the aggregate using Eqs.~\eqref{eq:rgdcabs}, \eqref{eq:rgbcsca}, and \eqref{eq:rgbfs} by ignoring the wavelength dependence of polarizability. When the aggregates are small ($ka_\mathrm{agg} \ll 1$), scattering obeys Rayleigh scattering, producing blue scattered light. If aggregates are large, but each monomer is still smaller than the wavelength ($ka_0 \ll 1 \ll ka_\mathrm{agg}$), the albedo is 
\begin{equation}
\omega=\frac{C_\mathrm{sca}}{C_\mathrm{sca}+C_\mathrm{abs}}\simeq NF_s(\lambda) \frac{C_\mathrm{sca,0}}{C_\mathrm{abs,0}}\propto \lambda^{\gamma-3},\label{eq:rgbalbedo}
\end{equation}
where we have used $C_\mathrm{sca} \ll C_\mathrm{abs}$ and $C_\mathrm{sca,0}/C_\mathrm{abs,0}\propto\lambda^{-3}$ as each monomer obeys Rayleigh scattering. For $\gamma\sim2$ (i.e., BCCAs), the albedo follows $\propto \lambda^{-1}$. This means that even if aggregates are large compared to the wavelength, low-dimension fractal aggregates may appear bluish in scattered light. 
Finally, if the wavelength is shorter than the size of the monomers ($ka_0 \gg 1$), we obtain $\omega\sim\omega_0$, meaning that aggregates would exhibit a similar scattered light color to that of each monomer. }

\rev{Another important diagnostic of dust porosity is the scattering phase function (SPF), which describes angular dependence of scattering. Within the RGD approximation, the elements of the scattering matrix are expressed as \citep{Sorensen01,Tazaki16}:
\begin{equation}
S_{ij}(\theta) \simeq N^2 S_{ij}^{\mathrm{mono}}(\theta) \cdot \mathcal{S}(q),\label{eq:sij}
\end{equation}
where $q = 2k \sin(\theta/2)$, $\theta$ is the scattering angle, and $\mathcal{S}(q)$ is the structure factor, characterizing the interference pattern of scattered waves due to the internal monomer arrangement. For a fractal aggregate, we have simple relations:
\begin{equation}
\mathcal{S}(q) \sim
\begin{cases}
1 & (qa_\mathrm{agg} \ll 1), \\
(qa_\mathrm{agg})^{-\gamma} & (qa_0 \ll 1 \ll qa_\mathrm{agg}), \\
1/N & (qa_0 \gg 1).
\end{cases}
\end{equation}
Since $q$ depends on both the wavenumber $k$ and the scattering angle $\theta$, the transition from the first to the third regime (for fixed $k$) may be considered as the angular dependence of scattering. The first case corresponds to forward scattering, while the third case corresponds to large-angle (e.g., backward) scattering. Therefore, by measuring the intensity of scattered light as a function of scattering angle, one may constrain $a_\mathrm{agg}$, $\gamma$, and $a_0$.}

\rev{Figure~\ref{fig:spf} shows some examples of total intensity scattering phase function (tSPF; proportional to $S_{11}$) and/or polarized intensity scattering phase function (pSPF; proportional to $-S_{12}$) computed by the discrete dipole approximation (for irregular grains) and the $T$ matrix method (for aggregates) taken from the {AggScatVIR} database \citep{Tazaki23-AggScat}. Scattering becomes progressively anisotropic with increasing particle radius regardless of particle morphology; however, the detailed angular dependence of the SPF differs for different particle morphology.}

\subsubsection{Scattering: polarization} \label{sec:scapol}

\rev{Finally, we discuss scattering polarization by dust grains. We start again from the simplest case, that is, a single spherical grain. A simple piecewise expression for the DoLP of scattered light by spherical grains is \citep[e.g.,][]{Tazaki21ice}:
\begin{equation}
\mathrm{DoLP} \simeq
\begin{cases}
\frac{(1-\cos^2\theta)}{(1+\cos^2\theta)} & (ka \ll 1), \\
\mathrm{often~low}& (ka \gg 1, ka\mathrm{Im}[m-1]\ll1), \\
\frac{R_\perp-R_{||}}{R_\perp+R_{||}}& (ka \gg 1, ka\mathrm{Im}[m-1]\gg1), \\
\end{cases} \label{eq:dop}
\end{equation}
where $R_\perp$ and $R_{||}$ are the Fresnel reflectances for the electric field vectors perpendicular and parallel to the plane of incidence. 
The first case corresponds to Rayleigh scattering. Since the cosine factor vanishes at $\theta=90^\circ$, Rayleigh scattering produces a 100\% DoLP at this angle.
The second and third cases ($ka\gg$1) corresponds to geometrical optics, and former and latter are transparent and opaque particle cases, respectively.
For the transparent case, DoLP is often (but not always\footnote{One such an example is light scattering by a water droplet in the atmosphere causing rainbow. A DoLP of rainbow is larger than that of Rayleigh scattering. This is because internal reflection of light within a droplet occurs close to Brewster angle.}) lower than that of Rayleigh scattering. For the opaque case, $R_{||}$ vanishes at $\theta=\pi - 2 \Theta_\mathrm{B}$, where $\Theta_\mathrm{B}=\tan^{-1}(n)$ is the so-called Brewster angle and scattering at this angle is referred to as Brewster scattering. Thus, the Brewster angle gives a 100\% DoLP. Compared to Rayleigh scattering which is peaking at $90^\circ$, Brewster scattering peaks at scattering angles smaller than $90^\circ$ for $n\ge1$.} 

\begin{figure}[htbp]
   \centering
   \includegraphics[width=\textwidth]{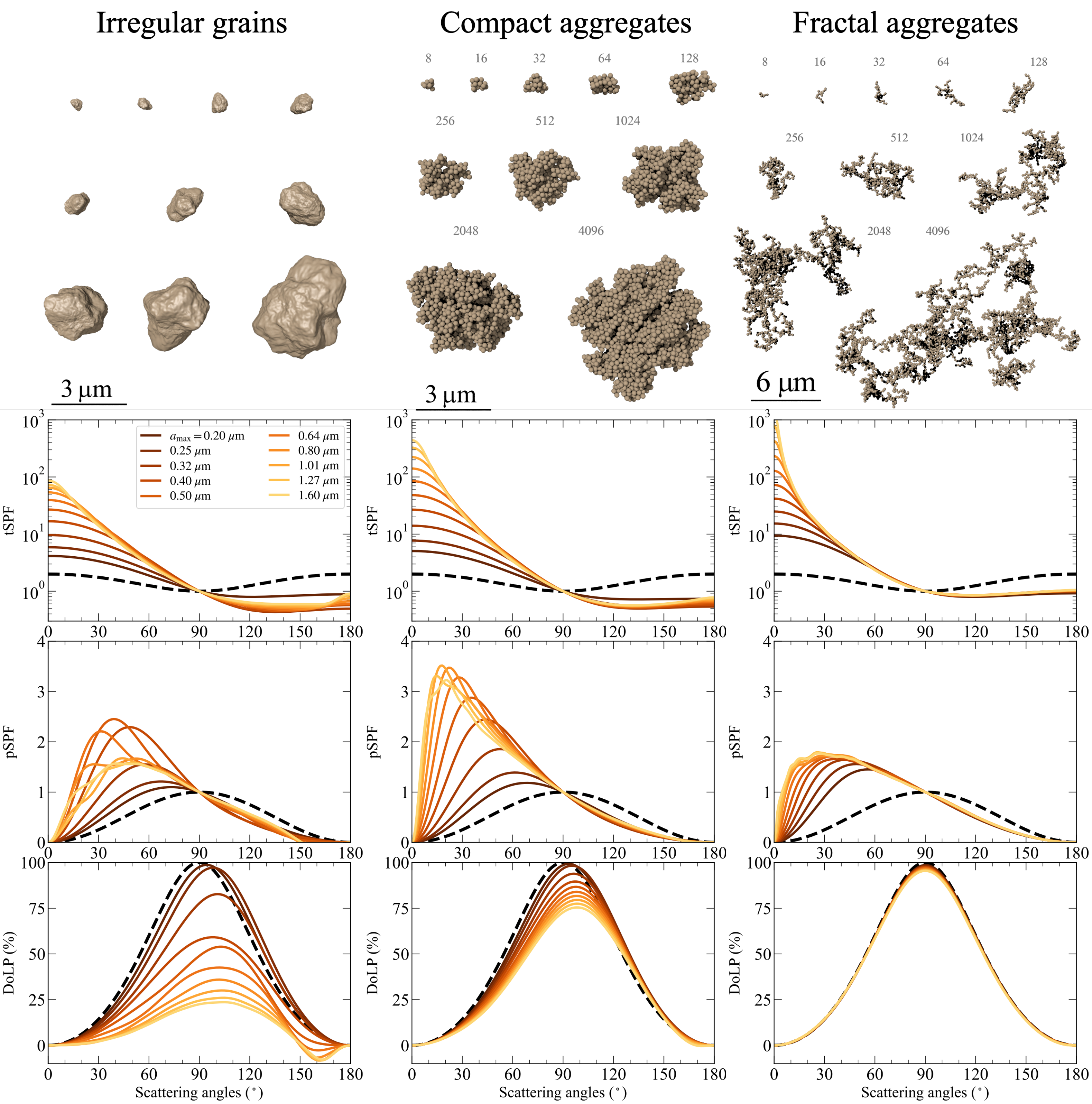}
      \caption{Numerical models of non-porous irregular grains (left column), low-porosity compact aggregates (a Ballistic Aggregate model called BAM2 originally proposed in \citealt{Shen08}) (middle column), high-porosity fractal aggregates with a fractal dimension of 1.9 (or Ballistic Cluster Cluster Aggregates) (right column). The panels show tSPF (top), pSPF (middle), and DoLP (bottom) at a wavelength of 1.04\micron. All models have a refractive index of $1.49+0.0108i$. 
      Numbers shown next to each aggregate represents the number of monomers, where each monomer has a radius of 0.1\micron.
      Note that tSPF is shown in logarithmic scale, whereas pPSF is in linear scale. Assuming particle size distribution obeying $a^{-3.5}$ (0.2\micron$\le a \le a_{\max}$) with $a$ being the mass-equivalent sphere radius, the scattering properties averaged over the size distribution are shown. Note that the mass-equivalent sphere radius represents a radius of a sphere that has the same mass of each non-spherical particle, and therefore, an apparent size of each particle could be much larger than this value. Dashed lines represent the Rayleigh scattering solution, which is similar to the scattering properties of each individual monomer of aggregates.These calculation results are taken from the {AggScatVIR} database  \citep{Tazaki22,Tazaki23}. }
         \label{fig:spf}
\end{figure}

\rev{Figure~\ref{fig:spf} shows the DoLP of numerical dust particles. The DoLP of irregular grains (the lower left panel) shows the increasing grain radius lowers DoLPs. This is due to a transition from Rayleigh scattering ($ka\ll1$) to the second case ($ka \gg 1, ka\mathrm{Im}[m-1]\ll1$). Those grains are not opaque enough to enter the third regime  ($ka \gg 1, ka\mathrm{Im}[m-1]\gg1$) and therefore we do not see a high polarization due to Brewster scattering. }

\rev{What about scattering polarization of porous aggregates?
In light of the RGD approximation, we can calculate the DoLP of scattered light by Eq.~\eqref{eq:sij}. The DoLP for unpolarized incident light is given by the ratio of the two scattering matrix elements, and we obtain
\begin{equation}
\mathrm{DoLP} \equiv -\frac{S_{12}}{S_{11}} = -\frac{S_{12,\mathrm{mono}}}{S_{11, \mathrm{mono}}} \label{eq:dop2}
\end{equation}
indicating that the polarization properties of large aggregates are governed by the monomer characteristics. If monomers are single-sized and spherical, their polarization characteristics follows Eq.~\eqref{eq:dop}. 
For example, if monomers are small ($ka\ll1$), porous aggregates will show a polarization similar to $(1-\cos^2\theta)/(1+\cos^2\theta)$. In contrast, if monomers are large and transparent ($ka \gg 1, ka\mathrm{Im}[m-1]\ll1$), DoLP from porous aggregates becomes low because monomers are not efficient polarizer anymore.
These behavior has been well confirmed by numerical studies \citep{West91, Kozasa93,Lumme97,Petrova00,Petrova04,Kimura01,Kimura06,Bertini07,Shen09,Kolokolova10,Tazaki16,Halder18,Tazaki22} and laboratory studies \citep{Zerull93,Gustafson99,Volten07,TobonValencia22,Gomez24}. 
Figure~\ref{fig:spf} shows the DoLP of fractal aggregates made of Rayleigh-scattering monomers (rightmost column). The results show that DoLPs remain essentially similar to Rayleigh scattering despite the fact that aggregate sizes are much larger than the wavelength.}

\subsubsection{Short summary}

\rev{The qualitative behavior discussed above can be summarized as follows.
\begin{enumerate}
\item Emission and absorption by large porous aggregates behave similarly to those of their constituent monomers.
\item Porous aggregates are less efficient polarizers for dichroic extinction and emission.
\item Scattered light from large porous aggregates differs from that of compact spherical grains of the same size, and carries information about the aggregate size and internal structure.
\item Scattering polarization from large porous aggregates resembles that of their constituent monomers.
\end{enumerate}
}

\rev{The above concepts have been applied to constrain dust porosity in various astrophysical environments, as we will discuss in Sect.~\ref{sec:porobs}.
For example, the detection of a certain level of dichroic polarization can place an upper limit on dust porosity knowing grain composition, since higher porosity tends to reduce the DoLP. This approach has been applied to polarization arising from grain alignment in both the interstellar medium (Sect.~\ref{sec:ism}) and protoplanetary disks (Sect.~\ref{sec:obs}). Analyses of SPFs have also been used to constrain size and morphology of fractal aggregates in planet-forming disks (Sect.~\ref{sec:obs}). 
}

\subsection{Toward further quantification: approximations, numerical and experimental approaches}

\rev{In Sect.~\ref{sec:quali}, we summarized only qualitative behavior. The discussion was incomplete in the sense that we limited our discussion mostly to porous dust aggregates that the RGD approximation is valid. We summarize some major approaches to find quantitative optical properties of porous dust grains.}

\rev{Calculating the optical properties of non-spherical grains is not an easy task as numerical approach is usually required, i.e., solving Maxwell's equations given non-spherical particle boundary conditions. There are various numerical techniques that can accurately calculate the optical properties of non-spherical grains, such as the $T$-matrix method and the discrete dipole approximation \citep[DDA; see, e.g., for a comprehensive textbook by][]{Mishchenko00}. A number of numerical simulations has been performed to understand the optical properties of porous dust grains in astronomical context \citep[e.g.,][]{Wright87, Kozasa92, Kozasa93, Stognienko95, Kimura01, Kimura06, RN812, Shen08, Shen09, Kolokolova10, Kohler11, Kirchschlager14, Min16, Tazaki16, Tazaki18,Tazaki22, Tazaki23, Halder18, Lodge24a, Draine2024_convex,Draine2024_aggregates,Draine2025}. For readers interested in light scattering by cometary dust, we refer to \citet{Kimura2016review}.}

\rev{Although the above numerical methods provide accurate results, they are still computationally demanding. To ease the computational cost, various approximate methods have been adopted in observational modeling of cosmic dust. The most common way to calculate the optical properties of porous dust grains is to use the Mie theory \citep{VdH57, Bohren83} together with the effective medium approximation},  where one defines an effective refractive index of a mixture of different materials (including vacuum) assuming a mixing rule prescription.
There are two popular prescriptions for the mixing rule: Maxwell--Garnett and Bruggeman mixing rules \citep{Bohren83,Chylek00}, although other prescriptions are also used \citep{Ossenkopf91, Ossenkopf1994, Henning96, Vosh99,Min08}. 
The Mie theory with the EMA may reproduce optical properties of some types of porous grains, such as spherical grains with small vacuum inclusions \citep{Wolff98, Voschchinnikov05,Vosh07}; however, its validity for arbitrary-shaped porous dust particles is by no means trivial. For example, \citet{Tazaki16} highlighted that the EMA leads to significant errors in scattering properties when applied to fractal dust aggregates with a fractal dimension of $\gamma\sim2$. \rev{There have been proposed alternative methods that offer better accuracy than the Mie theory with EMA, while remaining computationally tractable}, such as the Distribution of Hollow Spheres (DHS) \citep{Min03, Min05, Min16}, the multilayered spheres (MLS) \citep{Voschchinnikov05}, the mean-field approximation \citep{Berry86, Botet97, Rannou97, Tazaki18, Tazaki21,Lodge24a,Rannou24}, \rev{the spheroid analog approximation for compact dust aggregates \citep{Draine2025}.}

Laboratory experiments are \rev{another important approach to study optical properties of non-spherical grains} \citep{Greenberg61, Zerull93, Gustafson99, Kolokolova01,Hadamcik02, Hadamcik06, Hadamcik07, Volten07, Munoz12, Munoz20, Munoz21, TobonValencia22, TobonValencia24,Gomez24, Renard24}. \rev{These studies not only provide benchmarks of numerical simulations, but also useful to study the optical properties of larger particles that are often computationally demanding.}

\section{Observations} \label{sec:porobs}

\subsection{Dust porosity in the Solar System} \label{sec:dustsolar}
\begin{figure}[ht]
   \centering
   \includegraphics[width=\textwidth]{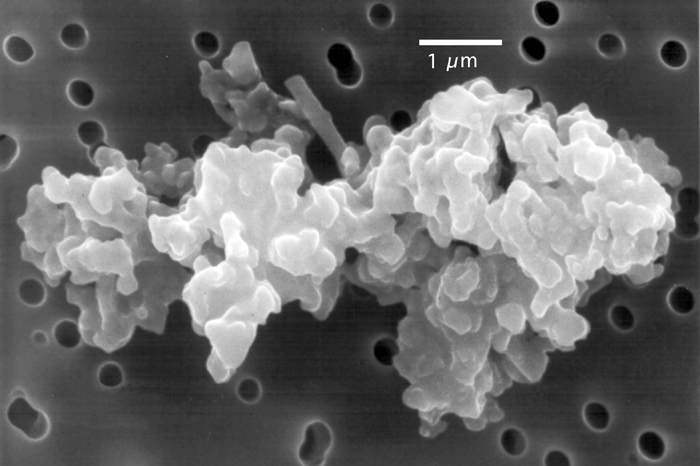}
      \caption{Scanning electron microscope image of an interplanetary dust particle with nearly chondritic composition.  Credit: D. Brownlee and E. Jessberger, adapted from \citet{Jessberger2001}. Available with a Creative Commons Attribution 2.5 License.
              }
         \label{fig:idp}
   \end{figure}

In the following sub-sections we focus on the properties of dust particles in the ISM and planet-forming disks as inferred via remote observation. However, materials in the Solar System provide an opportunity to explore the properties of solids more directly, and in some instances we can even touch extraterrestrial particles by hand. An example of a highly-porous \rev{interplanetary dust particle (IDP)} is shown in Fig.~\ref{fig:idp} and the reader can find many similar examples in the NASA catalog \citep{RN1225}. The image shows a non-spherical aggregate with several clear areas of vacuum. The size of this particular IDP is of order 10~$\mu$m; IDP sizes span from from 1~$\mu$m to 1~mm \citep{Jessberger2001}.  Analysis of dust scattering at mid-infrared wavelengths with Spitzer \citep{Pagani2010} and high precision JWST spectroscopic observations of ice absorption profiles \citep{Dartois2024} demonstrate that grains grow to $\sim$1~$\mu$m in the dense (n $>$ 10$^4$~cm$^{-3}$) interstellar medium.   As such the typical IDPs (with size $>$ 1~$\mu$m) clearly are representative of modest aggregate growth in the ISM and more substantive growth within the solar nebular disk.

Another example is the cometary particles
from comet 81P/Wild 2 captured by the Stardust mission. The sizes and morphologies of these particles are diverse, varying from tens of nanometers to hundreds of micrometers and from dense grains of about 3 g
cm$^{-3}$ to fluffy aggregates with densities of 0.3 g cm$^{-3}$ \citep{RN811}. Seven candidate interstellar particles are identified with compact particles having smaller sizes, of a few hundred nm, in contrast to the low density particles having sizes of $>$ 1 $\mu$m \citep{RN1677}. The easiest way to understand these structures is to assume that interstellar dust particles in this size range consist of materials with low-density and high porosity \citep{RN1677}.  \rev{We note that \citet{2016ApJ...818..133S} suggests that at least 2 of these particles are likely not interstellar in origin.}

Analysis of radar observations of comet Hyakutake \citep{RN813} are consistent with these results as they indicate a backscatter from porous grains (0.3 g cm$^-$$^3$) with the grain size distribution ranging from 1 $\mu$m to 1 cm in the coma of the comet. Simulations of the Ulysses interstellar dust measurements in the Solar System has led to a similar result, a decrease of the density of dust particles with an increase of their radius \citep{RN1678}. Fluffy grain aggregates with the fractal dimensions of 1.5 and 2.9 were considered by the modeling of the light scattering properties of the cometary dust \citep{RN812}. The conclusion based on these results is that cometary dust constitutes of fluffy particles that originate from fractal aggregation of sub-micron sized grains \citep[see also][]{Kimura2016review}. Analysis of in situ data from Cassini's Cosmic Dust Analyzer (CDA) suggests the detection of interstellar dust particles in the Saturnian system \citep{RN1909}. There particles are mainly compact small particles in the size regime of less than a few hundred nanometers observed by the CDA.

Analysis of the results from the Rosetta mission reveal the detailed properties of the comet 67P/Churyumov–Gerasimenko and its dust particles \citep[see e.g.,][]{RN1533}. Some of dust particles collected by the Rosetta are reminiscent of the CP (Chondritic porous)-IDP \citep[see][for a summary of particle morphology]{Guttler19}. It was found that dust particles collected in the coma of 67P can be divided in two families: compact and fractal particles. \rev{For the compact family,} \citet{Fulle16} estimated the porosity of the compact dust particles detected by the GIADA instrument on board the Rosetta spacecraft and found it to be $\sim48\%\pm8\%$. 
\rev{For the fractal family, particles are found to have (extrinsic)}porosity of more than 99\% \citep{Fulle15, Bentley16, Mannel16, Mannel19}. Fractal aggregates are likely to be formed in the early phase of dust coagulation (thus, they are directly linked to interstellar dust) \citep[][and references therein]{Fulle17}, as seen in near-IR disk observations \citep{Tazaki23}. 

\subsection{Dust porosity in the diffuse and dense ISM} \label{sec:ism}

\begin{figure}[htbp]
   \centering
   \includegraphics[width=0.6\textwidth]{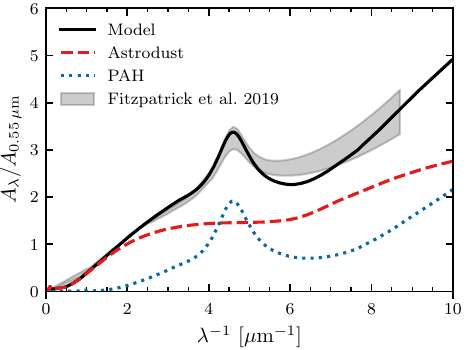}
   \includegraphics[width=\textwidth]{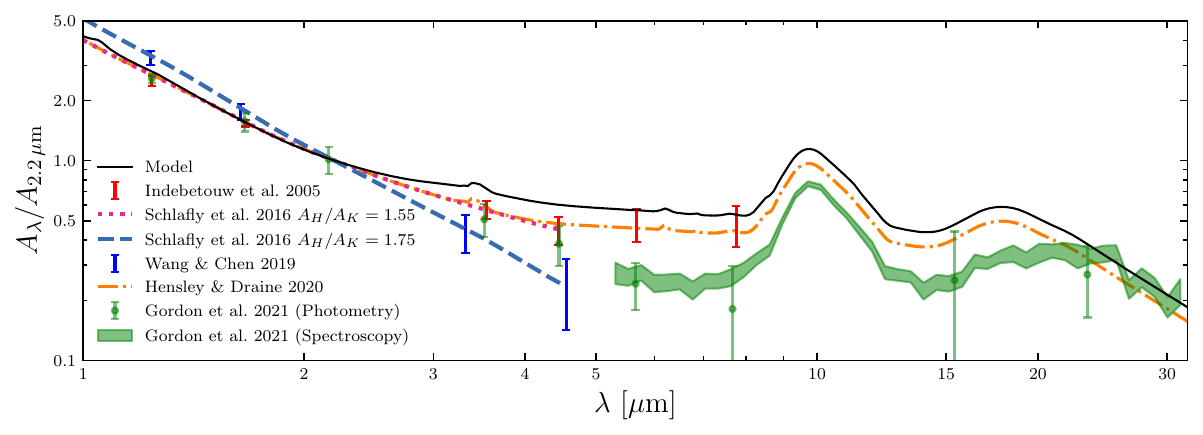}
      \caption{Plot of the interstellar grain extinction curve as compiled and modeled by \citet{Hensley2023}.  (Top Panel): Extinction ($A$ in magnitudes) normalized to 0.55~$\mu$m plotted as a funcition of inverse wavelength.  The gray shaded area is the average galactic extinction curve (from observations) taken from \citet{Fitzpatrick2019}.  The black line is the model fit which is broken down into contributions from Astrodust (red; composite grains containing silicates and other materials including hydrocarbons) and polycyclic aromatic hydrocarbons (blue; PAHs).  (Bottom panel): Extinction curve in the infrared normalized to 2.2~$\mu$m.   Observed values of the extinction are as noted in the figure with references noted here \citep{Indebetouw2005, Schlafly2016, Schlafly2017, Wang2019, Hensley2020, Gordon2021} and the model fitis shown as the black curve. Images reproduced with permission from \citet{Hensley2023}, copyright by the author(s).
              }
         \label{fig:extinction}
   \end{figure}

Dust has been known to be present in the diffuse interstellar medium since the pioneering work of \citet{Trumpler1930}.  Much of our understanding of dust extinction and properties arose as new windows with the spectrum of light were opened up within the ultraviolet (UV) through the infrared (IR) and into the far-infrared/submillimeter.  The most prominent effect of dust is the wavelength dependent extinction of starlight which is known as the extinction curve.  A sample of the extinction curve is shown in Fig.~\ref{fig:extinction}.
Notable aspects in the optical/UV (top panel) are the significantly stronger effects of extinction towards shorter UV wavelengths and the prominent ``bump'' at 0.2175$~\mu$m that is believed to arise from a carbonaceous component within the population \citep{Strecher1965, Fitzpatrick1986}. In Fig.~\ref{fig:extinction} this is associated with polycyclic aromatic hydrocarbons or PAHs \citep{Joblin1992, Hensley2023}; but alternate suggestions exist \citep{Jones2013}. A central determination is that interstellar dust has a size distribution that follows a size$^{-3.5}$ power law distribution with maximum sizes of $\simeq$0.2~$\mu$m (or larger) with most of the mass found in large grains and the surface area carried by small grains \citep{MRN1977, Draine1984}.   Sight lines towards dense regions show a shift in the extinction curve \citep{Fitzpatrick2019} that relates to greater extinction per H atom beyond 0.3~$\mu$m \citep{Draine2003}; this is believed to be the result of grain growth by coagulation and accretion in the denser regions of the ISM.

In the infrared the overall trend is a decrease in extinction with increasing wavelength and towards even longer wavelengths the energy absorbed by grains is released via a continuous emission spectrum peaking near 100--200~$\mu$m \citep{Dwek1997}.  Prominent features in this regime are the excess absorption seen in bands associated with silicates near 10 and 18~$\mu$m.   Dust emission is widely used as a tracer of both the dust and the gas mass within dense ($n$ $>$ 1000~cm$^{-3}$) regions of the galaxy \citep{Hildebrand1983}.

Another important observational constraint associated with interstellar grains is that both the absorption and emission is \rev{partially} polarized \citep{Levertt1951, Hildebrand2000, Andersson2015}.   In the UV/optical the peak polarization (linear) is of order a few percent near 0.55~$\mu$m \citep{Serkowski1975}.  A defining characteristic of the polarization is that while extinction rises from the visible to the UV, the polarization percentage peaks in the visible with a monotonic decline towards both shorter and longer wavelengths \citep{Serkowski1975, 1996AJ....112.2726A}.  The presence of polarization peaking in the visible part of the spectrum suggests that interstellar grains are not spherical, with typical sizes of $\sim$0.1~$\mu$m, and that they are aligned with the galactic/local magnetic field  \citep{Levertt1951, Andersson2015, Draine_ismbook}.  A final point is that dust grains are also responsible for the scattering of starlight which is a major contributor to the overall extinction \citep{Whittet_dustbook}.  

Given these observational constraints a wide range of detailed models with different assumptions with respect to grain composition and population(s), optical properties/methodology, presence/absence of ices have been constructed \citep[][to name just a few]{Draine1984,Greenberg1990,Ossenkopf1994, Weingarnter2001, Jones2013, Hensley2023}.   These contributions, and the full import of dust within astrophysics are nicely summarized in the formative reviews of interstellar dust by \citet{Mathis1990, Draine2003, Whittet_dustbook, Henning2010}.
Here we will focus on the question of porosity and the potential evidence of its presence.

Many of the early papers within the field explored the question of dust porosity either as a central aspect \citep{Mathis1989, Hage90} or as an important open question \citep{Draine1984}. There were several aspects that drove this exploration.  
First and foremost imaging studies of chondritic interplanetary dust particles (IDP), which are believed to originate from comets \citep{MacKinnon1987}, find structures that are highly porous and fragile \citep{Bradley_IDP}. These aggregate-like structures would be consistent with \rev{low velocity coagulation of lower mass particles (v~$\lesssim$~1~cm/s for mass $<$~10$^{-6}$ g)} as seen in laboratory experiments and might be expected to form inside denser regions of the interstellar medium \citep{Tielens1989, Guttler19}.  An example of one such image is shown in Fig.~\ref{fig:idp}. 
   
Cosmic elemental abundances and the interstellar extinction curve present another possible driver towards a need for porous grains; this issue is particularly acute for carbon.  Models of interstellar extinction adopt an assumed size distribution of grains with an assumed composition.  The composition of interstellar grains is estimated through observations of ultraviolet gas absorption lines of atoms and molecules towards stellar sources that lie behind low to modest extinction diffuse gas clouds; one such example is $\zeta$~Oph \citep{Savage1996}.  These measurements give an estimate of the total column and in many instances absolute abundances relative to hydrogen.  Using a standard for cosmic abundances, such as the Sun, the amount missing from the gas is attributed to the solids or the interstellar grains. Initial models of grains adopted the solar standard, but the local ISM cosmic carbon abundance appear to be sub-solar within young B stars \citep{Nieva2012}.  We note that this is not uniformly the case and measurements towards older populations (F, G, K) find abundances closer to solar \citep[see][]{Hensley2021}.   Regardless,  the observed fraction of carbon gas in particular provides a tension between some models which require more material than cosmically available to be present in the grain population to match extinction \citep{Mathis1996}.  
Dust grains dominate the extinction at wavelengths where their size is nearly equivalent ($\lambda \sim 2\pi \times$ size).  With strong extinction near $\sim$0.1~$\mu$m (see Fig.~\ref{fig:extinction}) this implies sizes near 0.1~$\mu$m, close to the maximum size of the MRN distribution.
Large fluffy grains with areas of vacuum offer a way to retain the observed extinction with less material.
However, areas of vacuum can also affect the optical and scattering  properties of materials as they interact with radiation \citep[][]{Jones1998,Voshchinnikov2006, Draine2003}.  This interaction relates to the composite polarizability of individual atoms within the substrate and hence polarizability voids within a given volume can affect both scattering and absorption at some modest level \citep{Voshchinnikov2006}.
An additional motivation for large fluffy grains was argued to be needed to account for the X-ray scattering halos surrounding point sources such as Nova V1974 Cycgi \citep{Mathis1995}.
Overall, the issue of cosmic abundances and scattering within X-ray halos as a motivation for porous grains, appears to be of less important in more modern simulations \citep{Smith1998, DraineTan2003,Li2005,Hensley2021}.
\rev{For an X-ray scattering halo for GX13+1, dust porosity has been constrained to be less than 55\% \citep{Heng09}.}
At present, models for the diffuse ISM with an assumed porosity of 20\% can match the observational constraints in terms of the overall extinction as a function of wavelength, polarization, scattering and dust emission \citep{Hensley2023}.

Looking at the denser regions of the ISM there is a generic expectation of grain growth to be consistent with changes in the extinction curve at optical/UV \citep{Fitzpatrick2019} and infrared wavelengths \citep{McClure2009}. Detailed models of grain evolution suggest the formation of aggregates which can have porosities $\sim$30--80\% \citep{RN903,Ossenkopf1994, Ormel2009}.  However, it should be noted that models exist that match the observational constraints yet have little to no porosity \citep{Ysard2019}.  More generically dust grain aggregates in the ISM (with vacuums and without) have an increased far-infrared/mm wavelength opacity coefficient \citep{Ossenkopf1994, Ormel2011,Kohler2012} relative to the models of the diffuse ISM. Observations of dust extinction at near infrared wavelengths and the mm/sub-mm continuum emission along the same line of sight towards molecular clouds find agreement with this prediction \citep{Kramer2003}. Thus, grain growth in the dense ISM is present, as is clear from the JWST spectoscopic observations \citep{Dartois2024}, and this growth may foster an increase in grain porosity. 

\begin{figure*}[ht]
    \centering
    \includegraphics[width=\textwidth]{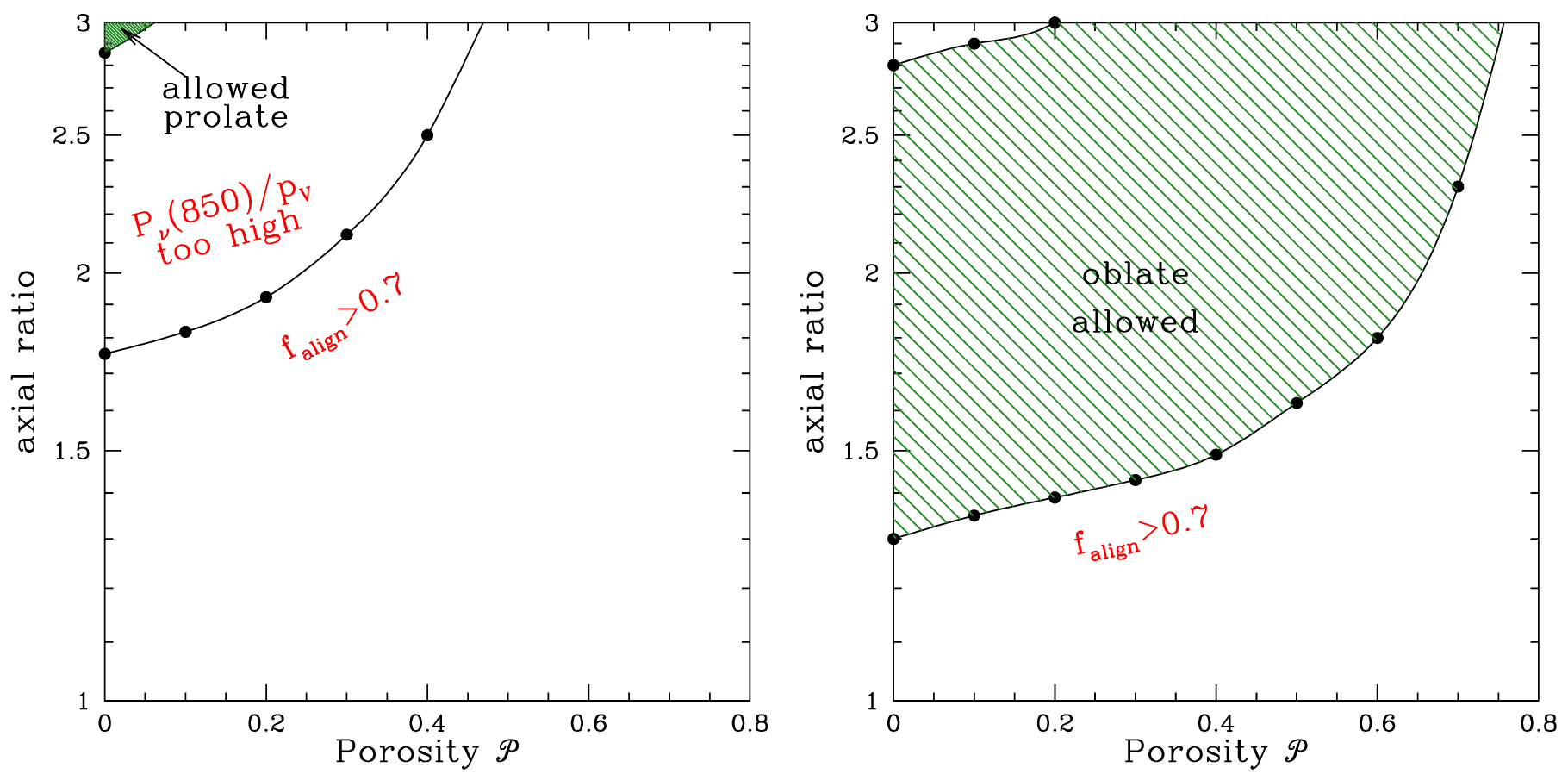}
    \caption{Summary figure of the allowed solutions of ISM grain models with variable porosity when compared to the monochromatic polarized 850~$\mu$m intensity normalized per unit V-band polarization in the diffuse ISM.  Models are shown for both prolate and oblate spheroids. The green shaded regions show the allowed axial ratios which are limited for prolate \rev{spheroids} to the upper left corner. Image reproduced with permission \rev{from} \citet{Draine2021}, copyright by AAS.}
    \label{fig:polarization}
\end{figure*}

The strongest limits on porosity at present is dust polarization and, to a lesser extent, the overall dust temperature.   In the latter case models of fluffy grains with high porosity ($>$80\%) appear to have increased opacities and subsequently become too cold to match the equilibrium temperature of large grains in the diffuse ISM \citep{Mathis1990, Dwek1997fluffy}.  Further these grains may be difficult to construct \citep{Jones1998}.
In terms of dust polarization in Fig.~\ref{fig:polarization} we show the allowed solutions for ISM grain models by \citet{Draine2021}, whether prolate or oblate, as constrained by starlight polarization and the polarized emission at submillimeter wavelengths.  Based on these models oblate shapes are preferred and porosity is limited to $\le$0.75.  Additional work \citep{Draine2024_aggregates} discusses the formation of aggregates and their observational properties.  These models also suggest 
porous grains form with either extreme elongation or flattening; if this is not the case then they must have reduced values of porosity.

Overall growth models suggest that porosity is likely fostered as grains grow.  However, if porous grains are present in the diffuse ISM then some external process may need to be present to reduce their porosity, unless the process of growth itself (i.e. collisions) leads to compactness.  This sets the stage for the next phase -- the disk -- where significant grain growth is fostered inside the dust-rich midplanes.  Here the question of porosity powered by this growth becomes even more central.

\subsection{Dust porosity in planet-forming disks} \label{sec:obs}

In planet-forming disks, dust grains undergo coagulation by mutual collisions and then form a cluster of grains called dust aggregates. 
The intergrain voids are the source of (\rev{extrinsic})porosity. 
\rev{Porosity has been a major subject in this field because it directly affects how dust grains grow in the disks, eventually affecting formation of planetesimals \citep[see reviews by][]{Johansen14,Blum18}. Once planetesimals are formed, they collide and produce smaller fragments in disks, like IDPs in the Solar System.}

\rev{Recent disk observations start to show some clues on dust porosity in disks. However, no consensus on observational constraints on dust porosity has been reached. This is partly because of incoherent modeling methods/techniques among different studies  (e.g., light scattering calculations) used to infer dust porosity. Also, the number of protoplanetary disks analyzed in this context is still limited. In this section, we aim to summarize the current status --rather inconclusive though-- of dust porosity in planet-forming disks.}
\rev{Interpretation of multiwavelength observations of planet-forming disks depends if the disk is optically thin or not. Hence, we discuss} two evolutionary stages of planet-forming disks \rev{separately}: protoplanetary disks (Sect.~\ref{sec:ppdobs}) and debris disks (Sect.~\ref{sec:ddobs}), \rev{where the latter is older and is optically thin at all wavelengths, and the former is not.} 

\subsubsection{Protoplanetary disks} \label{sec:ppdobs}

Dust porosity in protoplanetary disks has been inferred from modeling of the following observational results:
\begin{itemize}
\item Optical/near-infrared (IR) scattered light observations (e.g., HST, VLT, Subaru, Gemini),
\item Mid-IR spectroscopy (e.g., Spitzer, JWST),
\item Millimeter(mm)-wave thermal and scattered radiation (e.g., ALMA).
\end{itemize}
\rev{One of the major challenges here is that the disk is optically thick at most of the wavelengths at which we observe. The immediate consequence is that different wavelengths trace dust particles in different regions of the disk (disk surface for IR vs. closer to the midplane for mm). 
Since dust particles in different regions may have different sizes and porosity  (see Sect.~\ref{sec:collexp}), we cannot directly combine multiwavelength observables to determine a single dust model. Dust properties (e.g., porosity) should be modeled each observing wavelength separately. }

\paragraph*{Optical/near-IR scattered light observations.} 
In this wavelength region, we probe relatively small grains (typically around $\sim1$\micron) floating in atmospheric regions of a protoplanetary disk. Those grains are directly illuminated by the central radiation source, such as a star or thermal radiation from the innermost regions, and grains scatter the incoming light toward the observer. 

\begin{figure}[ht]
   \centering
   \includegraphics[width=\textwidth]{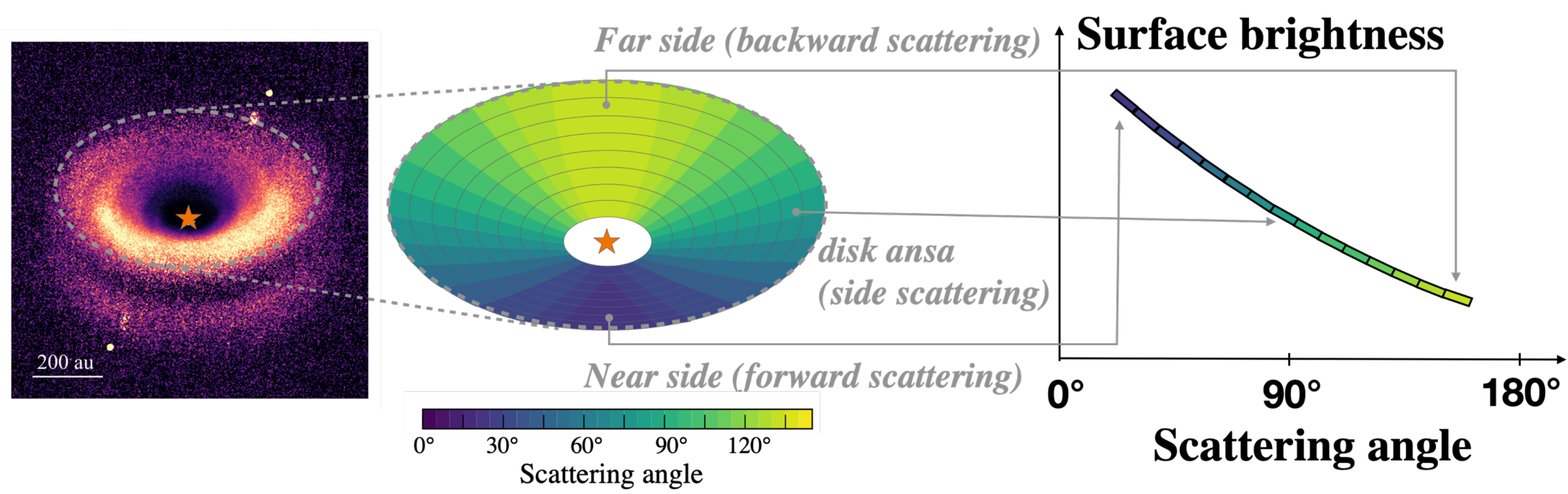}
      \caption{Basic concept of scattered light mapping to extract an SPF from an observed scattered-light image.}
         \label{fig:mapping}
\end{figure}

\rev{As explained in Sects.~\ref{sec:scacs} and \ref{sec:scapol}, the information of dust properties (including porosity) is imprinted in the SPF, the DoLP, and scattered light colors. By conducting observations, we obtain a scattered light image of a protoplanetary disk with each pixel in the image containing total intensity, polarized intensity, or DoLP. By comparing disk images taken at different wavelengths, we can also derive scattered light colors. How do we measure the SPF? The most common way is to use a technique called scattered light mapping (Fig.~\ref{fig:mapping}), where we compute scattering angles at each location of the disk surface and then map the observed surface brightness as a function of the scattering angle \citep[e.g.,][]{Stolker16,Roumesy25}. Although scattered light mapping is only applicable to axisymmetric disks, there have been measurements for several protoplanetary disks so far \citep{McCabe02,Avenhaus14HD100546,Takami14,Ginski16, Stolker16, Ginski23, Chen24, Roumesy25}.}

\rev{
\citet{Pinte08} modeled disk-scattered light of IM Lup using the Mie theory with EMA and found that it is either explained by a dust porosity of 80\% or ice-rich non-porous grains based on the asymmetry parameter (the first cosine moment of a tSPFs). 
\citet{Perrin09} and \citet{Dykes24} studied the disk-scattered light around AB Aur using the Mie theory with EMA and found a dust porosity of 55\% and 60-80\%, respectively, based on the observed DoLP. \citet{Chen24} modeled SPFs, DoLP and color information toward the disk around HD 34700 A using the Mie theory/DHS with EMA and obtained a dust porosity of 40\%. Taken at face value, those studies favor dust grains with some amount of porosity; however, it is worth keeping in mind that those values may be affected by what grain size distribution is assumed \citep[e.g.,][]{Chen24}. 
\citet{Canovas13} reported VLT/NACO observations for HD 142527 in the H band and found that the DoLP was relatively low. Based on this result, they argued that the dust grains in the disk surface are likely compact and not highly porous. However, this interpretation needs to be taken with care because a low DoLP could be due to the size effect of monomers within highly porous aggregates \citep{Tazaki22} (Section~\ref{sec:scapol}). \citet{Mulders13} focused on disk-scattered light color around HD 100546. The color is found to be reddish, requiring grains a few microns in size (see also Sect.~\ref{sec:scacs}). 
Along with a dynamical argument, the authors pointed out the possibility that the grains may be porous to the order of 99\% with sub-micron-sized monomers, though the scattering properties of the aggregates were not considered. According to the RGD light scattering theory (see Sect.~\ref{sec:scacs}, Eq.~\eqref{eq:rgbalbedo}), aggregates with a porosity $>99\%$ (likely having a low fractal dimension; $\gamma<3$) will produce bluish scattered light \citep{Tazaki2019a}. Therefore, it remains unclear whether fluffy aggregates of sub-micron-sized monomers explain the observed reddish colors.}

\begin{figure}[ht]
   \centering
   \includegraphics[width=\textwidth]{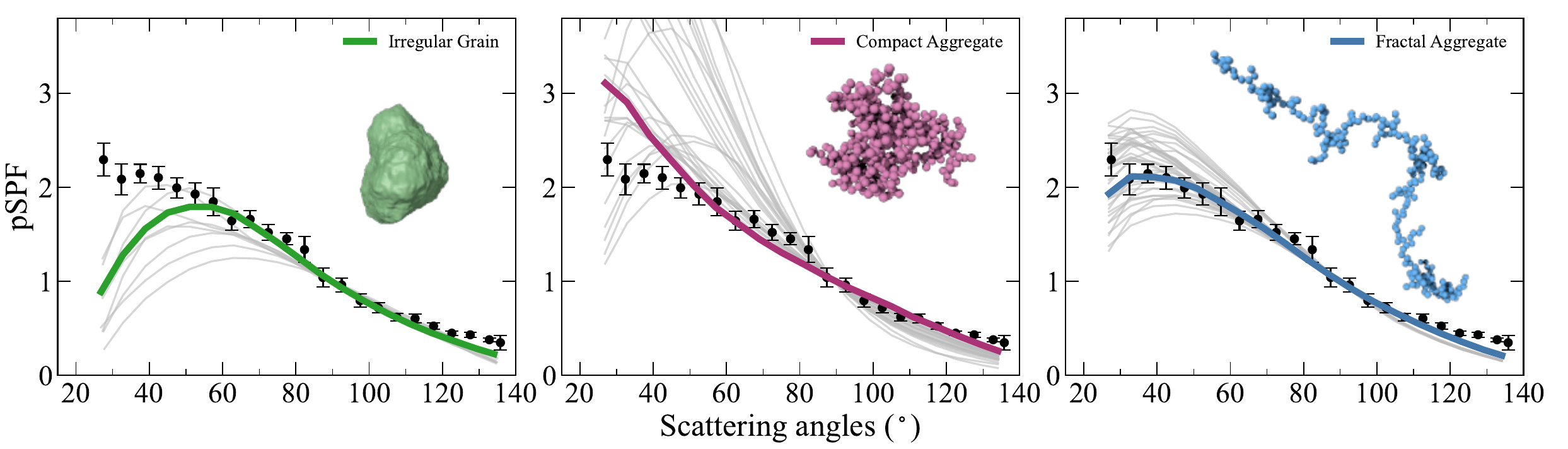}
      \caption{Polarized scattering phase function (pSPF) extracted for the disk around IM Lup (black points with error bars) and radiative transfer models for irregular grains (left), compact aggregates (middle), and fractal aggregates (right). Thin gray lines in each panel correspond the results obtained for changing aggregate/monomer sizes, porosity, and composition, and the thick solid lines represent the best fit among them for each class of particle morphology. Particle images shown in each panel is the dust model used in the best-fitting model. See \cite{Tazaki23} for more detailed information.} 
         \label{fig:rt23}
\end{figure}

\rev{The above studies have used the Mie theory or DHS with EMA to constrain dust porosity. Another line of research is to model disk-scattered light using a more sophisticated light scattering models. 
\citet{Tazaki23} studied disk-scattered light around IM Lup with $T$-matrix light scattering simulations of fractal dust aggregates ($1.1\le\gamma\le3$). They found that fractal aggregates can naturally explain the observed pSPF, as shown in Fig.~\ref{fig:rt23}. Fractal aggregates suggested in their modeling has a characteristic radius greater than $\sim2$\micron\  and $\gamma\lesssim2$ and with a monomer radius of $0.2$\micron. The inferred aggregate structure suggests that they form via Brownian motion, as suggested from laboratory and numerical experiments (see Sect.~\ref{sec:collexp}). Fractal dust aggregates in the IM Lup disk implies that cometary fractal aggregates might have formed in the early-stage planet-forming disks (Sect.~\ref{sec:dustsolar}).}

\rev{Looking ahead, two major challenges remain in constraining dust porosity from disk-scattered light observations.
First, the systematic uncertainty in inferred porosity is still poorly understood due to the diversity of modeling approaches and the lack of a robust diagnostic signature. A key limitation is perhaps the non-uniqueness of model solutions: extensive exploration of parameter space is needed to assess degeneracies, but such surveys are often computationally prohibitive. This is because accurate modeling requires not only detailed light-scattering calculations, but also radiative transfer simulations, as SPFs and DoLPs are significantly affected by multiple scattering and limb brightening effects \citep{Tazaki23, Takami13, Ma22}. These challenges underscore the need for more accurate and computationally efficient modeling techniques. Second, it is essential to collect as much observational information as possible, including SPFs, DoLPs, and color indices. Fortunately, recent advances in observational facilities, instrumentation, and data reduction techniques have made it increasingly feasible to obtain accurate and quantitative measurements of scattered light. For instance, pSPFs have been obtained for about 10 protoplanetary disks, revealing two distinct behavioral categories that may correspond to different dust populations \citep{Ginski23, Roumesy25}. Similarly, measurements of the DoLP are rapidly increasing, owing to the improved stability and quality of reference differential imaging (RDI) observations \citep{Hunziker21, Tschudi21, Ren23, Chen24}. Color measurements are also becoming more common \citep{Fukagawa10, Mulders13, Zhong24, Avenhaus18, Ginski24, Ma24a, Ma24b}. However, several caveats must be considered when interpreting these observations. Surface brightness and color analyses can be affected by temporal variability if multi-wavelength images are taken at different epochs \citep{Watson07, Rich19, Ma24b}. Additionally, different wavelengths may trace different surface heights in the disk, introducing further complexity in interpreting color trends \citep{Duchene04, Pinte07}. In some cases, disk surface brightness may be influenced by shadowing from the inner disk \citep{Acke09, Garufi22}, meaning that observed faintness does not necessarily reflect the scattering properties or reflectivity of the dust grains if this is the case.}

\paragraph*{Mid-IR spectroscopy.} 
In this wavelength region, we can observe thermal emission \rev{most dominantly} from lukewarm grains ($\sim300$ K), which are located at a radial distance of $\lesssim10$ au.
In particular, the mid-IR domain is rich in solid-state features of various minerals, providing another way to constrain dust properties, including porosity. One of the most well-studied mid-IR features would be the 10\micron\ silicate feature \citep[see e.g., a review by][]{Henning2010}. 

\citet{Vosh08} studied the effect of porosity on the 10\micron\ silicate feature using the MLS \citep[e.g.,][]{Voschchinnikov05}. They fitted the observed silicate feature from $\lambda=$ 8\micron\ to 12\micron\ for 47 sources (30 T-Tauri stars and 17 Herbig Ae/Be stars) observed with the Spitzer Space Telescope \citep{Schegerer06,vanBoekel05,Sargent06,SiciliaAguilar07} by assuming small grains that are in the Rayleigh domain. The derived porosity for 47 sources varies from $\approx39\%$ to $\approx98\%$ with an average value of $64\pm15\%$. 
Recently, \citet{Jiang24} studied mid-IR spectra for PDS 70 with JWST using a porous irregular grain model generated via the Gaussian Random Field (GRF) algorithm \citep{Min07}. They computed the optical properties of those grains with DDA and found that the porous-GRF model reproduces the observed mid-IR spectra better than DHS and laboratory spectra by \citet{Tamanai06a,Tamanai09}.

Although there are studies favoring the presence of porous grains based on mid-IR spectra, the detailed structure of dust grains, such as fractal dimension and the monomer size, still remains unclear. \rev{Furthermore,} it is worth keeping in mind that the absorption property of grains in the Rayleigh regime is sensitive to the particle shape \citep{Bohren83,Min06JQSRT,Mutscheke09}. \rev{In other words, the failure of spherical-grain models (like the Mie theory) should not be interpreted as (indirect) evidence for porous grains; the failure may simply be attributable to asphericity of grains.}

\paragraph*{Millimeter-wave thermal and scattered radiation.} 
In this wavelength \rev{region} we observe thermal emission from cold (sub)millimeter-sized grains \rev{in a region close to the disk midplane}. There are two main approaches to constrain dust porosity. One is to use mm-wave polarization and the other one is to perform multi-band analysis (i.e., the spectral index), or the combination of both. The strongest constraint on porosity at present appears to be from mm-wave polarization, as described below.

\rev{Thanks to its sensitivity and spatial resolution,} ALMA opened up a new window to observe mm-wave polarization of disks.
It is suggested at least two polarization mechanisms are needed to explain the observed polarization patterns \citep{Stephens17}. 
One is self-scattering \citep{Kataoka15,HaifengYang16}, which can naturally explain \rev{polarization patterns uniformly oriented parallel to the disk minor axis} \citep[e.g,][]{Hull18,Dent19}. Another one is polarized thermal emission from aligned grains \citep{Cho07,Bertrang17,Tazaki17,Guillet20,Thang24,Kataoka19,DanielLin24Align}, although the detailed alignment process/direction remains being debated \citep{Andersson2015, HaifengHang19}. 
These two mechanisms are not exclusive each other and may work simultaneously \citep{Mori21,Stephens23,DanielLin24HLTau}.

\begin{figure}[ht]
   \centering
   \includegraphics[width=0.75\textwidth]{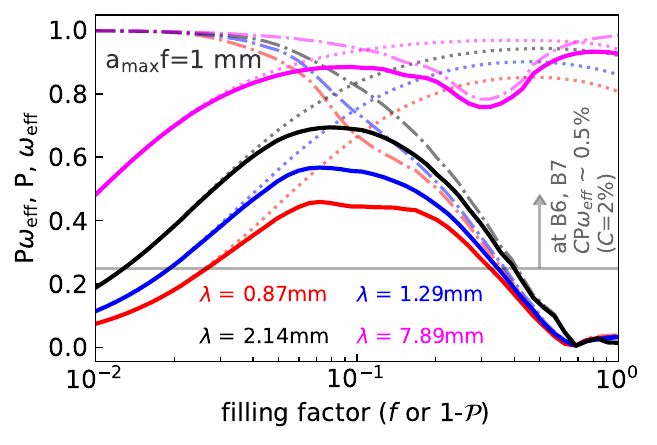}
      \caption{The scattering polarization efficiency at millimeter wavelength. $P$ is the degree of polarization (dot-dashed), and $\omega_\mathrm{eff}$ is the (effective) scattering albedo (dotted). Their product $P\omega_\mathrm{eff}$ (solid lines) gives an efficiency of scattering polarization, which only exceeds the required level to explain the observations at a porosity of 70--97\% in the HL Tau disk. Image reproduced with permission from \citet{SZhang23}, copyright by the author(s).}
         \label{fig:zhang}
\end{figure}

\rev{Theoretical studies have suggested that a less-porous dust origin is favorable for the mm-wave polarizations, regardless of whether it is due to scattering \citep{Tazaki2019b,Brunngraber21} or alignment \citep{Kirch19}. This has been supported by subsequent, more detailed observational modeling \citep{SZhang23,Ueda24}. }\citet{SZhang23} studied the multi-band fluxes and polarization observations of the HL Tau disk by using the Mie theory with EMA. They estimated the viable porosity range to be 70--97\% (Fig.~\ref{fig:zhang}), and a porosity $>99\%$ is unlikely. \citet{Ueda24} constructed a dust evolution model in a self-gravitating disk using a dust evolution code {DustPy} \citep{Stammler22} \rev{and applied it to the mm-wave observations for the IM Lup disk}. The model favors grains with 80\% porosity for the IM Lup disk. \rev{\citet{Kirch19} studied mm-wave polarized thermal radiation from aligned grains for the disk around HD 142527 and found that a grain porosity of $\le 70\%$ is needed unless the grain's axis ratio is extremely large. \citet{Ohashi23} studied the dust porosity in the DG Tau disk; however, the observations are found to be explained by both non-porous and porous (80\%) models, and hence, the porosity for this source remains not well constrained.}

Another approach for constraining porosity is multi-band mm-wave flux analysis. 
\rev{If disks are optically thin, we may constrain dust porosity from spectral index of mm-wave flux \citep{Miyake93, Natta04, Draine06, Testi14, Cuzzi14, Kataoka14}. However,} there is growing evidence for disks being (at least marginally) optically thick at mm wavelengths \citep[e.g.,][]{Chung24}, making the porosity constraint not straightforward as anticipated for the optically thin case. For a higher optical thickness, the effect of scattering also affects the observed emission \citep{Birnstiel18, Zhu19, Liu19, Sierra20, Ueda20}.
To find a dust property, simultaneous fitting of multi-band data from mm to cm-wave by taking into account scattering has been performed\citep[e.g.,][]{CarrascoGonzalez19,Sierra21,Macias21,Ueda22,Guidi22,Ohashi23,SZhang23}.
\citet{Guidi22} performed multi-band analysis for HD 163296 with five wavelengths and found that low-porosity grains (with 25\% porosity) better reproduce the data compared to high-porosity grains (with 80\% porosity). 
\rev{More recently, \citet{Yoshida25} proposed a novel approach to breaking the degeneracy between dust temperature and dust scattering albedo in the TW Hya disk using CO line emission. They derived a dust porosity of $<96\%$ and a power-law index of the size distribution $>-4.1$. These porosity range is consistent with what has been inferred from mm-wave polarization measurements.}

\rev{Overall, observations of mm-wave polarization and multi-band analysis thus far point to a dust porosity of approximately 90\% or less, ruling out extreme porosity greater than 99\%. The less-porous particles in disks might be link to cometary particles found in the comet 67P/Churyumov-Gerasimenko with GIADA instrument, which have a porosity of $\sim48\%\pm8\%$ (see Sect.~\ref{sec:dustsolar}), although dust evolution models in disks still do not fully explain how these less-porous particles were formed (see Sect.~\ref{sec:collexp}).}

It is worth keeping in mind in above modeling studies that grain's optical properties are still mostly modeled by using the Mie theory or DHS with the EMA. 
\rev{There are some studies claiming potential issues of using the Mie theory.} For example, \citet{Kirchschlager20} demonstrated that non-porous spheroid grains show scattering polarization behavior different from the mass-equivalent counterpart of the Mie theory. \citet{DanielLin22} also pointed out that the scattering properties of laboratory non-porous grains show qualitatively different polarization properties from the Mie theory's prediction. Further careful studies on light scattering properties and its application to the actual disk observations are mandatory to push forward our understanding of dust porosity in disks.

\subsubsection{Debris disks} \label{sec:ddobs}

Debris disks are circumstellar disks found around (pre)main sequence stars. 
Although these disks are placed in a later evolutionary stage of protoplanetary disks, they are not simply the leftover of protoplanetary disks. Dust grains (and sometimes gas) in debris disks are likely secondary origin (i.e., those replenished by collisions of planetesimals), as primordial grains are short-lived due to orbital decay via Poynting Robertson drag or blown-out by radiation pressure \citep{Burns79}. Therefore, dust properties in those disks are likely different from those found in protoplanetary disks and are perhaps more similar to those seen in the Solar System, e.g., IDP (Fig.~\ref{fig:idp}). Hence, studies of debris disks offer a unique opportunity to investigate how the Solar System is different from other extrasolar systems. 

\rev{Determining the dust porosity of debris disks poses fewer challenges than for protoplanetary disks, because they are optically thin at all wavelengths and disk geometry is constrained relatively well compared to protoplanetary disks. Nevertheless, there have been several challenges in observational modeling of debris disks.}
\rev{For example, d}ust properties inferred from SED fitting are often in conflict with those inferred from scattered light observations [HD 181327: \citet{Lebreton12}, HD 207129: \citet{Krist10}, HD 32297: \citet{Rodigas14}, HR 4796A: \citet{Rodigas15}], although a more flexible assumption in the refractive index (and then dust composition) may reconcile this issue \citep{Rodigas15, Ballering16}. Furthermore, even if we only aim to model scattered light observations, \rev{different fitting outcomes are sometimes obtained} depending on which of total or polarimetric observational data is used \citep{Arriaga20, Duchene20}. The issue \rev{may imply} limitations of simple light scattering models used, such as the Mie theory. \rev{For example,} \citet{Olofsson22} found pSPF extracted for HD 32297 could be well explained by a light scattering model of fractal aggregates (MMF; \citealt{Tazaki18}), whereas not when the Mie theory or DHS is used. In this way, observations have been challenging our understanding of the optical properties of dust particles in debris disks. In the following, we will highlight studies on dust porosity for a few well-studied debris disks: \rev{HR 4796 A (A0 star, $10\pm3$ Myr old; \citealt{Bell15}), AU Mic (M star, $22\pm3$ Myr; \citealt{Mamajek14}), and $\beta$ Pic (A6 star, $\sim20$ Myr; \citealt{MiretRoig20}). }

\begin{figure}[ht]
   \centering
   \includegraphics[width=0.75\textwidth]{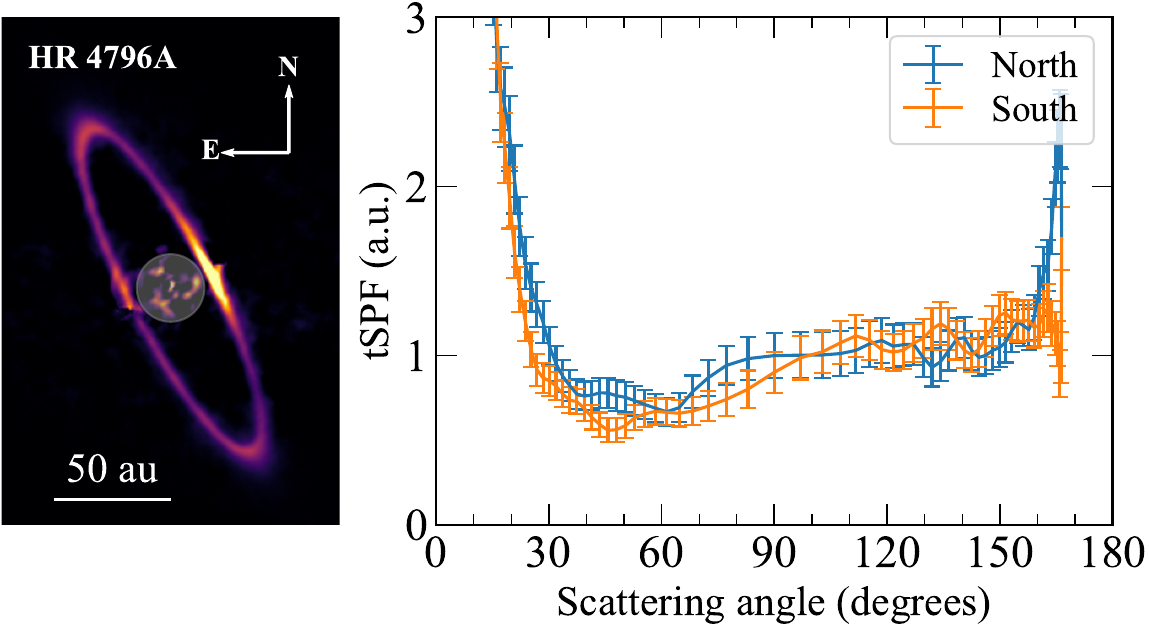}
      \caption{Total intensity image of HR 4796A (left) and the tSPFs extracted for the northern and southern sides (right) reduced by \citet{Milli17}. The image was reduced with the masked classical angular differential imaging \citep[mcADI:][]{Milli12}. The tSPFs were corrected for the mcADI flux losses and the convolution effect by the instrumental PSF, but not for the limb brightening, which results in a small hump at $\sim90^\circ$ \citep[see][]{Olofsson20}. The peak beyond $160^\circ$ seen in the northern SPF is likely an artifact.}
         \label{fig:milli17}
\end{figure}

\paragraph*{HR 4796A:} This is the one of the most well-studied debris disks in the context of dust characterization. HR 4796A harbors a large and narrow ring with a semi-major axis of $\sim1''\approx77$ au that is inclined by $\sim76^\circ$ degrees from pole-on (Fig.~\ref{fig:milli17}). 

\rev{Dust properties for this source has been inferred from SED \citep{Augereau99, Li03_HR4796,Rodigas15}, reflectance spectra \citep{Debes08, Kohler08, Rodigas15,Milli17,Ren23debriscolor}, SPFs \citep{Schneider09,Perrin15,Milli15,Milli17,Milli19, Olofsson20,Ren20,Chen20,Arriaga20,Schmid21}.}
\citet{Rodigas15} performed simultaneous modeling of reflectance colors and SED using the Mie theory with the EMA (Bruggeman). They found that 
\rev{the best fit porosity to SED and reflectance spectra are different: the former favors a higher porosity ($\sim70\%$) and the latter does a lower porosity ($\sim30\%$). }
When the two \rev{constraints} (reflectance spectra+SED) were combined, the model only provides a mediocre fit with a slight preference for a lower porosity ($\sim10\%$).
\rev{The SPF for this source has a unique trend: a strong forward scattering peak with a gradual enhancement of intensity toward larger scattering angles (Fig.~\ref{fig:milli17}). 
Such behavior can be either due to iron-rich grain composition \citep{Chen20} or grain morphology, i.e., large particles with rough surface \citep{Milli19}.}
In the presence of surface roughness, the incident light onto the dust surface will be diffusely reflected, rather than specularly reflected  (e.g., Fresnel reflection), and diffuse reflection can make back scattering rather bright, as with the case of lunar phase \citep{Hapke81, Mukai82, Min10}. 
\citet{Min16} demonstrated using DDA calculations that light scattering by compact dust aggregates \rev{(a fractal dimension $\gamma\sim3$)} shows the enhanced back scattering \citep[see also][]{Grynko20,Penttila21} \rev{similar to what is seen in the HR 4796 disk}, whereas such a trend is absent for fluffy dust aggregates \rev{($\gamma\sim2$)} with monomers smaller than the wavelength \citep{Volten07, Tazaki16}. From laboratory measurements, \citet{Munoz17} studied SPF of mm-sized non-spherical compact grains at optical wavelength and found a similar backscattering enhancement. \rev{These studies suggest that, if the enhanced backward scattering is due to grain morphology, less porous grains are favorable for explaining SPFs, although quantitative porosity value remains unknown.}

Overall, dust properties inferred from various observations of HR 4796A seem to be in favor of the presence of less porous grains, based on SED analysis \citep{Rodigas15} and the observed enhanced backscattering behavior \citep{Milli17, Milli19}. However, the narrowing down more detailed properties are still to be done. 

\paragraph*{AU Mic:}
AU Mic is one of a few debris disks around M-type stars.
The strongest constraint on dust porosity in AU Mic seems to be coming from polarization measurements with HST/ACS at 0.6\micron\ \citep{Graham07}. 
They \rev{found} a DoLP rising steeply from 5\% to 35\% between 20 and 50 au and reaching a maximum of $\sim40\%$ within 80 au. 
Based on the radial dependence of observed surface brightness and DoLP, \citet{Graham07} found grains that show a strong forward scattering and a high degree of polarization are compatible with the data. By using the Mie theory with the real part of refractive index being treated as a free parameter, they obtained $n=1.03\pm0.03$. This values correspond to a porosity of 91--94\% depending on if the matrix is either ice or rock assuming the Maxwell--Garnett mixing rule.
Qualitatively similar conclusions have been obtained in more recent studies by \citet{Fitzgerald07_aumic,Shen09,Arnold22}. In \citet{Fitzgerald07_aumic}, they constructed grain models based on multiwavelength surface brightness data and SED assuming the refractive index of \citet{Mathis1989} (a porous mixture of silicate, carbon, and ice) or astronomical silicate. They found a compatible fit quality to surface brightness and SED regardless of porosity, but the compact model is ruled out to explain the observed DoLP reported in \citet{Graham07}. \citet{Shen09} performed light scattering simulations of Ballistic Aggregates using DDA and compared the results of \citet{Graham07} and found aggregates with a porosity of $\sim60\%$ (so-called BAM2 model, see Fig.~\ref{fig:spf} for exemplary morphology of BAM2). 
\citet{Arnold22} carried out MCMC fitting to estimate grain properties based on spatially resolved scattered light spectra obtained with HST/STIS \citep{Lomax18}. Porous spheres or agglomerated debris particles (ADP)\citep{Zubko15}, where both of them has a porosity of $\sim76\%$, fit equally well to the wavelength and radial dependence of surface brightness, whereas non-porous spherical models only show a poor fit. They also confirm that their porous models are compatible with the DoLP reported in \citet{Graham07}. 
\rev{Using a Herschel 70 \micron\ image}, \citet{Matthews15} pointed out that a halo component is spanning from 40 to 140 au, which is beyond the outer edge of the planetesimal \rev{belt} ($\sim40$ au). Due to low luminosity of AU Mic, radiation pressure is insufficient to blow out small grains, and hence the stellar winds might blow them out \citep{Augereau06,Matthews15,Arnold19}. \citet{Matthews15} pointed out that high porosity grains are favorable for an efficient transport to the halo region via the stellar winds due to their larger cross sections. \rev{Note that their argument is for a halo region, and hence, dust porosity around the planetesimal belt at $\sim40$ au might not necessarily have the same porosity as the halo region.}
To summarize, all modeling efforts for the AU Mic disk (particularly for outer disk regions) so far point to the presence of porous grains with a porosity of $\sim60$--$94\%$ based on the Mie theory with the EMA, ADP models, or BAM2 models.

\paragraph*{$\beta$ Pic:} 
\rev{Earlier modeling of $\beta$~Pic's dust properties revealed an apparent conflict: \citet{Li98}, based on SED modeling, favored highly porous aggregates, whereas \citet{Golimowski06}, using scattered light images and color, favored less porous grains. \citet{Ballering16} found that this tension can be resolved when a wider parameter space in grain composition is explored. They modeled SED and scattered light with the Mie theory with EMA, and their best-fit dust model is made of predominantly astronomical silicate and refractory organics with little amount of vacuum and ice}. Although they do not discuss if the model shows a compatible level of DoP ($\sim10\%$ at K band) \citep{Tamura06}, it seems not challenging as non-porous grains readily show a low DoP (see, e.g., Fig.~\ref{fig:spf}). 

Recently, \citet{Rebollido24} observed the disk with JWST/NIRCam and MIRI and found that the secondary disk has a bluish color in the mid-IR, whereas it is reddish in the main disk. Moreover, they found a unique structure called `cat's tail', and such a structure could be explained by the presence of dust particles characterized with a high radiation pressure efficiency, i.e., fluffy aggregates made of organics. Based on these results, they proposed that the main disk and the secondary disk (including the cat's tail) are composed of different types of dust grains: the former is more non-porous grains and the latter is more fluffy aggregates. 

To summarize this subsection, debris disk observations have so far suggested various porosity: less porous grains \rev{in the ring region in the HR 4796A disk}, porous grains \rev{at least in the halo region of the AU Mic disk}, and possible a mixture of high and low porosity grains \rev{for the main disk and secondary disk} in $\beta$-Pic, respectively. 
\rev{Dust in a single system is liekly not characterized by a single porosity value, but instead exhibits a range of porosities within the disk. This idea is supported by observations of comet 67P, where both compact and fractal dust families have been identified (Sect.~\ref{sec:dustsolar}). In the three debris disks discussed above, the inferred porosity values differ, which may reflect variations in dust properties depending on the region observed, specifically, how close the observations probe to the planetesimal belt.}

\section{Laboratory and modeling studies}
In the first two subsections, we describe results of laboratory experiments and dust growth models relevant to the ISM and protoplanetary disks. Regarding the experiments, we consider two cases: (i) formation of grains in envelopes of evolved stars (mainly of asymptotic giant branch (AGB) stars) and their subsequent growth in the ISM and planet-forming disks and (ii) recondensation of grains in the diffuse ISM. The latter is related to the idea of a practically complete destruction of “stardust” in the ISM by supernova shocks \citep{Draine2003, RN1393, RN821}. Then we briefly discuss cosmic dust materials. The last subsection is devoted to atomistic modeling of cosmic dust. 

\subsection{Dust growth in the interstellar medium}
\rev{In the experiments simulating the formation of cosmic dust in envelopes of evolved stars and their subsequent growth in the ISM, gas-phase condensation techniques, such as laser ablation and condensation as first demonstrated by Smalley et al. in their foundational work on fullerenes \citep{RN425}, laser- and plasma-enhanced pyrolysis as commonly used in chemical vapour deposition (see \citealt{RN2003} for a review), and electron/ion beam sputtering as widely used in deposition of coatings and thin films (see \citealt{RN2004} for a review), combined with deposition of condensed particles on substrates are used. The typical low pressure (a few mbar) regime of quenching atmospheres applied for condensation zones is comparable to the pressure conditions for dust condensation in AGB stars \citep{RN1077}. Deposition of grains simulates the dust growth in the ISM and leads to the formation of porous fractal grain aggregates.}

For example, researchers of the Laboratory Astrophysics Group of the Max Planck Institute for Astronomy have for many years performed laser ablation of graphite targets for carbon dust and silicon-based targets (e.g., MgSi, FeSi, MgFeSi) for silicate dust in their work. Evaporated species condense in a quenching atmosphere of a few mbar He/H$_2$ (for carbon grains) or He/O$_2$ (for silicate grains) forming nm-sized particles.  Condensed particles are extracted adiabatically from the ablation chamber decoupling the particles from the carrier gas and generating a particle beam, which is directed onto a substrate. To characterize the condensed materials and to determine their composition, structure and morphology, sophisticated ex situ analytical methods such as high-resolution transmission electron microscopy (TEM), scanning electron microscopy (SEM), atomic force microscopy (AFM), and energy dispersive X-ray (EDX) spectroscopy have been used by the group. TEM and EDX spectroscopy deliver a direct view inside the grains and the information on the composition, structure, sizes, and shapes of individual grains. SEM and AFM provide information on the morphology of the condensed grain layers. 

The analysis of dust grain samples shows that the dust layers consist of individual spherical particles of a few nm in size and porous random-shaped aggregates of up to several tens of nm in size. Thus, the morphology of dust particles on a substrate can be understood as a layer of dust aggregates having a fractal structure (fractal distribution of grain monomers) and a very high (up to 90\%) porosity. Such grain aggregates present analogous of dust grains in the ISM, and in protostellar envelopes and planet-forming disks, for which they represent building blocks of larger aggregates having the similar structure and morphology (considering the fractal nature of the grain growth). 

\begin{figure}[ht]
   \centering
   \includegraphics[width=\textwidth]{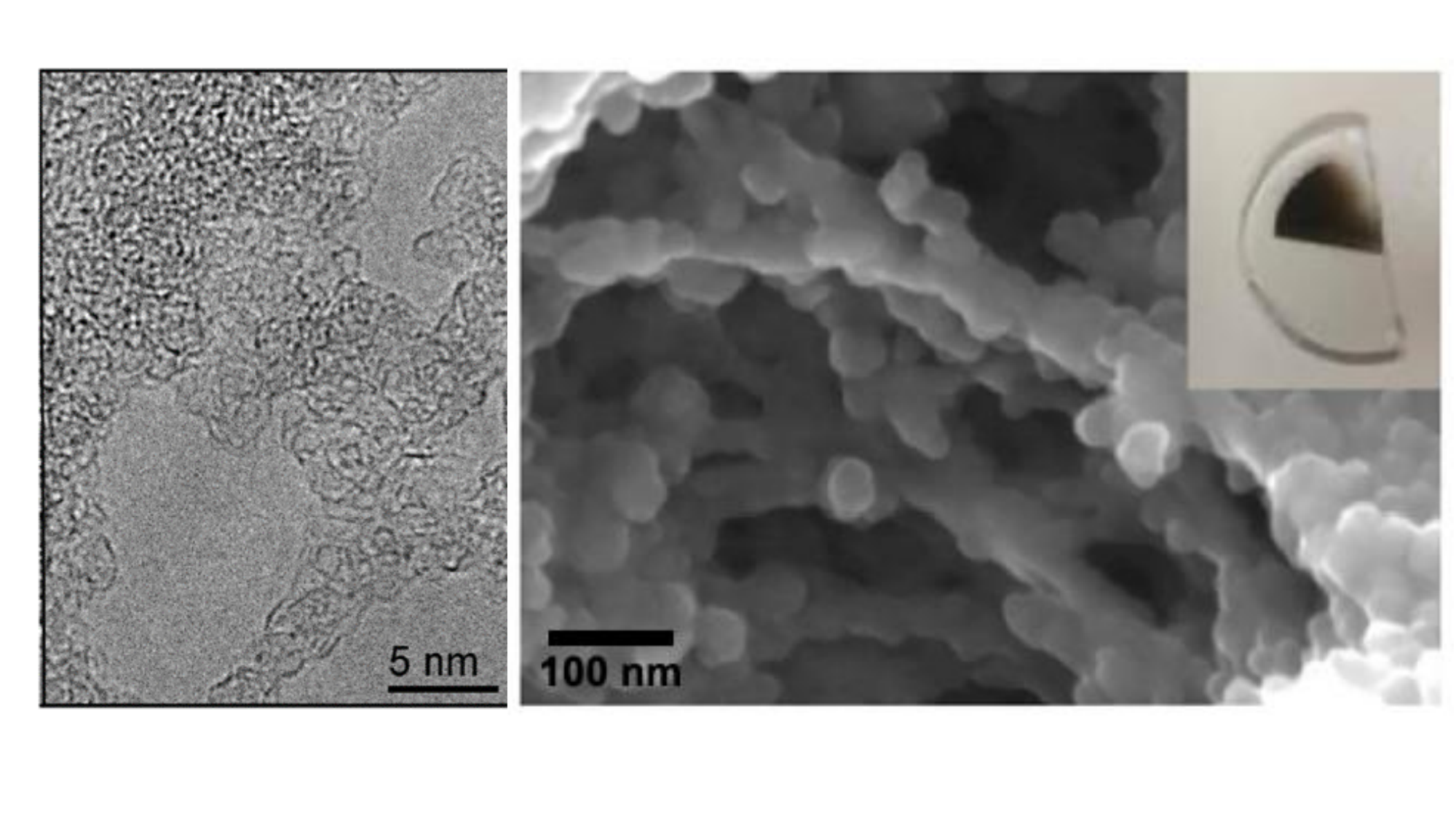}
      \caption{TEM (left) and SEM (right) images of amorphous carbon (left) and amorphous silicate (right) grains produced by gas-phase condensation of nm-sized grains and their subsequent deposition onto a substrate. Inset: a laboratory sample -- a layer of carbon grains on a KBr substrate. Images reproduced with permission from [left] \citet{RN930}, copyright by the author(s); and [right] from \citet{RN767}, copyright by APS.}
         \label{fig:figure_dust_analogues}
\end{figure}

Microscopy images of dust grains produced by laser ablation and subsequent deposition on a substrate are presented in Fig.~\ref{fig:figure_dust_analogues}. These are typical images, and the reader can find many similar examples in the literature (e.g., \citealt{RN1650, RN407, RN408, RN784, RN353, RN767, RN1715}). What is clearly seen in the images is a very high porosity of dust grain aggregates on the level of nanoparticles as well as large aggregates. 

\begin{figure}[ht]
   \centering
   \includegraphics[width=\textwidth]{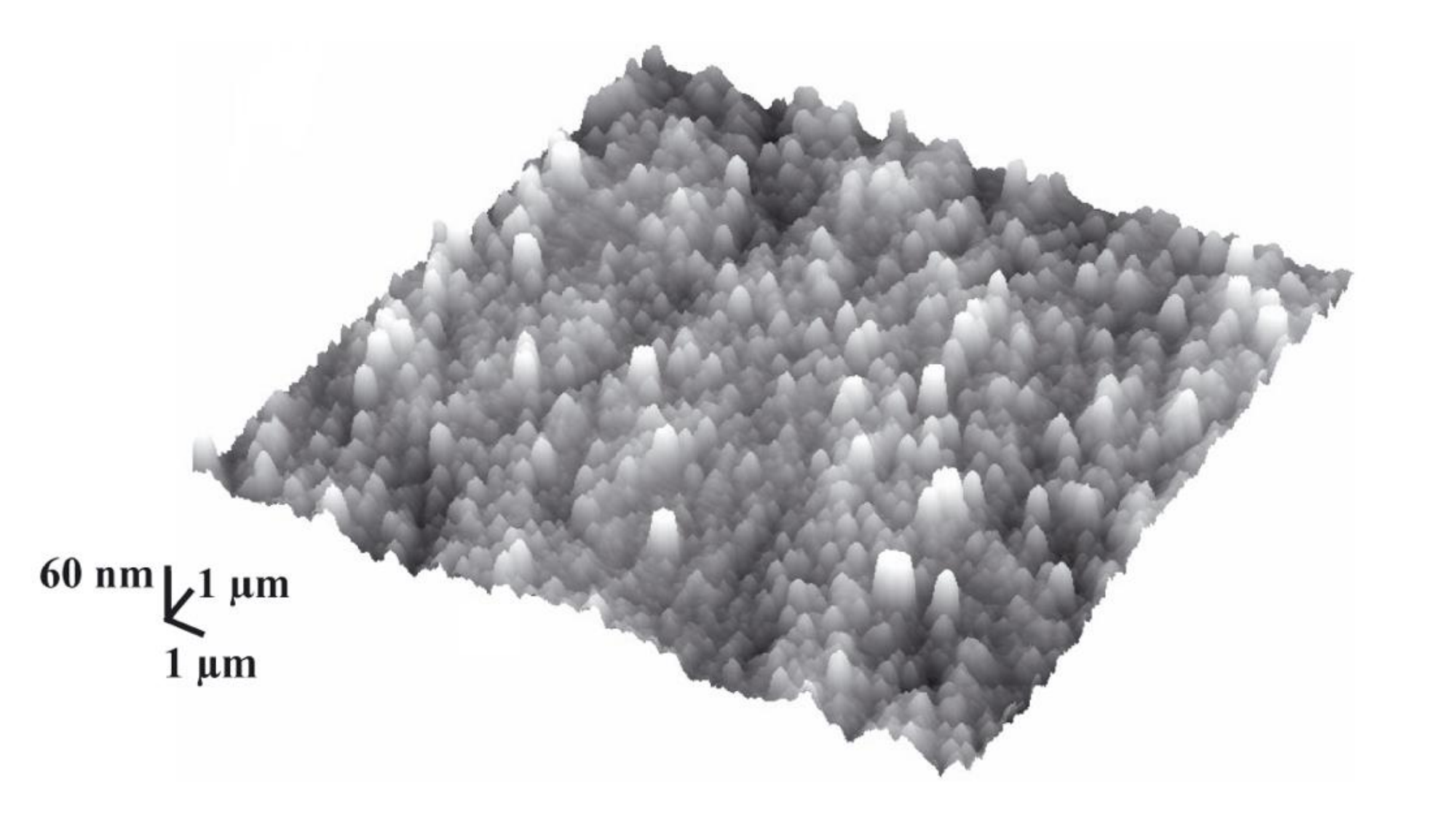}
      \caption{AFM image of a 200 nm thick amorphous silica thin film physical vapour deposited on a polished metal substrate. Image reproduced with permission from \citet{RN1137}, copyright by the author(s).}
         \label{fig:figure_afm}
\end{figure}

A much less sophisticated experimental approach can be taken using standard thin film deposition technologies. For example, McCoustra and co-workers have for over 15 years employed amorphous silica deposited on polished copper or stainless-steel substrates in their work (see, e.g., \citealt{RN712, RN1136}). Simple electron-beam evaporation is used to promote physical vapour deposition on room temperature substrates. The resulting thin films, illustrated by the AFM image in Fig.~\ref{fig:figure_afm}, are sufficiently thick that there is no access to the underlying substrate from the gas phase and there is no likelihood of metal substrate-mediated hot electron chemistry. High porosity of the dust layer is clearly seen.

An alternative to electron beam evaporation is ion sputtering in which ions mechanically remove material from a target for deposition on the substrate is being explored in some laboratories \citep{RN1698}. The presence of reactive gases can ensure deposition of oxides, nitrides, and other materials. Specific designs of sputter source, (e.g., \citealt{RN1699, RN1701, RN1700, RN1702}) permit growth of films and powders with size-selected particles. Either approach can be extended to produce thin films of virtually any of the materials in Table~\ref{tab:1} below (see Sect.~3.3). 

In the experiments simulating the re-formation of cosmic dust in the ISM \citep{RN825, RN824, RN982}, laser vaporization of SiO, silicate, or graphite targets was carried out followed by matrix isolation of laser-vaporized atomic and molecular species, e.g., Si, Mg, SiO, and carbon chains in neon or argon. After annealing and complete evaporation of the Ne/Ar matrix, the formation of silicate and carbon grains was detected. The condensed grains had the structures and porosity like the grains produced by laser evaporation and subsequent deposition (exemplified by Fig.~\ref{fig:figure_dust_analogues}).

\begin{figure}[ht]
   \centering
   \includegraphics[width=\textwidth]{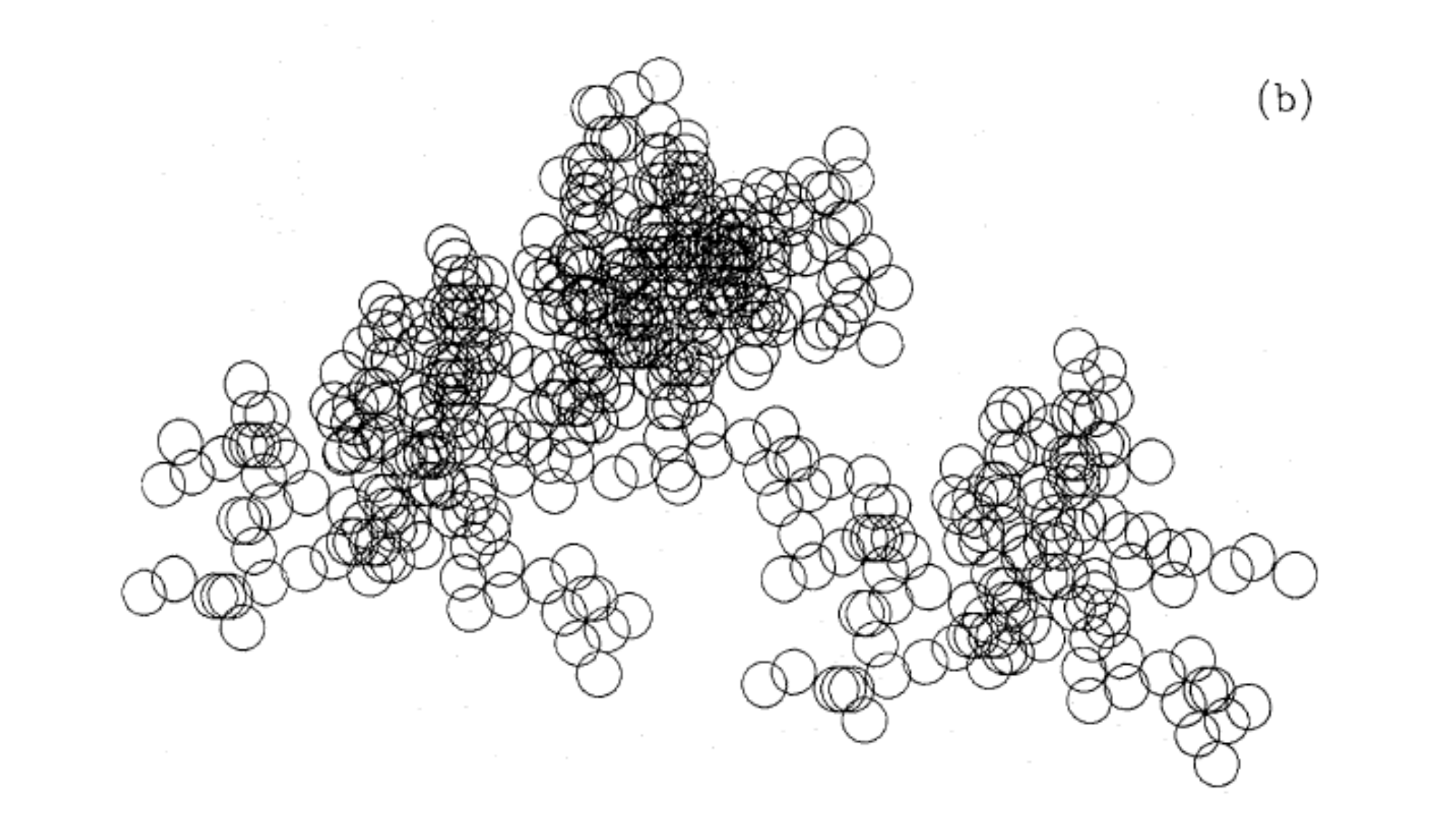}
      \caption{A result of numerical simulations of dust growth. Clusters consisting of 512 single grains of identical size grown by ballistic cluster--cluster agglomeration. Image reproduced with permission from \citet{RN903}, copyright by ESO.}
         \label{fig:figure_dust_ossenkopf}
\end{figure}

The aforementioned experimental results are in agreement with the results of modeling of grain growth in the ISM \citep{RN903, Ossenkopf1994, Ormel2009, RN1910} showing porosity of dust aggregates in the range of $\sim$30--80\%. Two main approaches were used in these models to simulate growth of spherical grains of a constant internal density: (i) particle-cluster aggregation (PCA), where an aggregate is created by adding each constituent particle from a random direction and (ii) cluster-cluster aggregation (CCA), where an aggregate is created by sticking with the same-sized cluster from a random direction \citep{Meakin91, Mukai92}. The CCA approach leads to more fluffy and porous grain aggregates. Figure~\ref{fig:figure_dust_ossenkopf} shows an example of the results of CCA simulations of dust growth in the dense ISM from \citep{RN903} where the porosity of dust aggregates of $\sim$80\% was demonstrated. We note here that if we compare Fig.~\ref{fig:figure_dust_analogues} (grain aggregates from experiments), Fig.~\ref{fig:figure_dust_ossenkopf} (grain aggregates from simulations), and Fig.~\ref{fig:idp} (a real extraterrestrial particle), we will find a great similarity.   

\rev{Space weathering may efficiently change the composition and structural, morphological (porosity), optical and chemical properties of the dust. Interactions of heavy ions and X-rays, analogues of cosmic rays, with analogues of cosmic dust grains have been studied \citep{RN1405, RN353, RN1925, RN1856}. Structural and compositional changes of dust grains have been demonstrated, however, no attention has been paid to the porosity and surface reactivity of dust grains, as has been done, for example, for thermal annealing. Thermal annealing of silicate grains causes the coalescence of grains leading to an increase of the sizes of individual grains and to a consistent decrease of the porosity of grain aggregates. Moreover, the grain surface becomes chemically less active. It was shown that the average size of silicate grains increases from about 6 to 30 nm and their porosity decreases from about 90\% to 40\% with annealing from room temperature to 900$^\circ$C \citep{RN1715}. Also, not much attention has been paid to changes in optical properties of analogues of cosmic dust grains due to space weathering. There is only a handful of studies of the influence of UV and ion irradiation \citep{RN1924, RN1923, RN1405, RN1925}. The influence of temperature has been investigated a bit more intensely, see \citep{RN1534} for a review.}

\rev{Absence of interest to changes in porosity and surface reactivity of dust grains due to space weathering was, probably, due to the most popular model (compact dust) described in Sect.~\ref{sec:intro}, considering that dust has no or very limited influence on the physico-chemical processes on its surface. However, in the alternative model (fractal dust) these changes may influence and, in some cases, define the efficiency of such processes as well as optical properties of dust grains. The latter may dramatically influence interpretation of astronomical spectra. Thus, new results on the influence of space weathering on the optical, morphological and chemical properties of analogues of cosmic dust grains would provide an important piece of
puzzle of the evolution of cosmic dust grains in astrophysical environments.}

The results on thermal annealing of dust grains are relevant to warm/hot environments, such as inner parts of protostellar envelopes and planet-forming disks and the atmospheres of exoplanets. As for exoplanet atmospheres, simulations have shown that the porosity of particles in atmospheric clouds may affect the structure and composition of clouds and has to be taken into account \citep{RN1433}. \rev{Influence of high temperatures on the growth of grains and on the porosity, chemical activity and optical properties of grain aggregates are open questions representing doorways to future research.}

\subsection{Dust growth in protoplanetary disks} 
\label{sec:collexp}
\begin{figure}[ht]
   \centering
   \includegraphics[width=\textwidth]{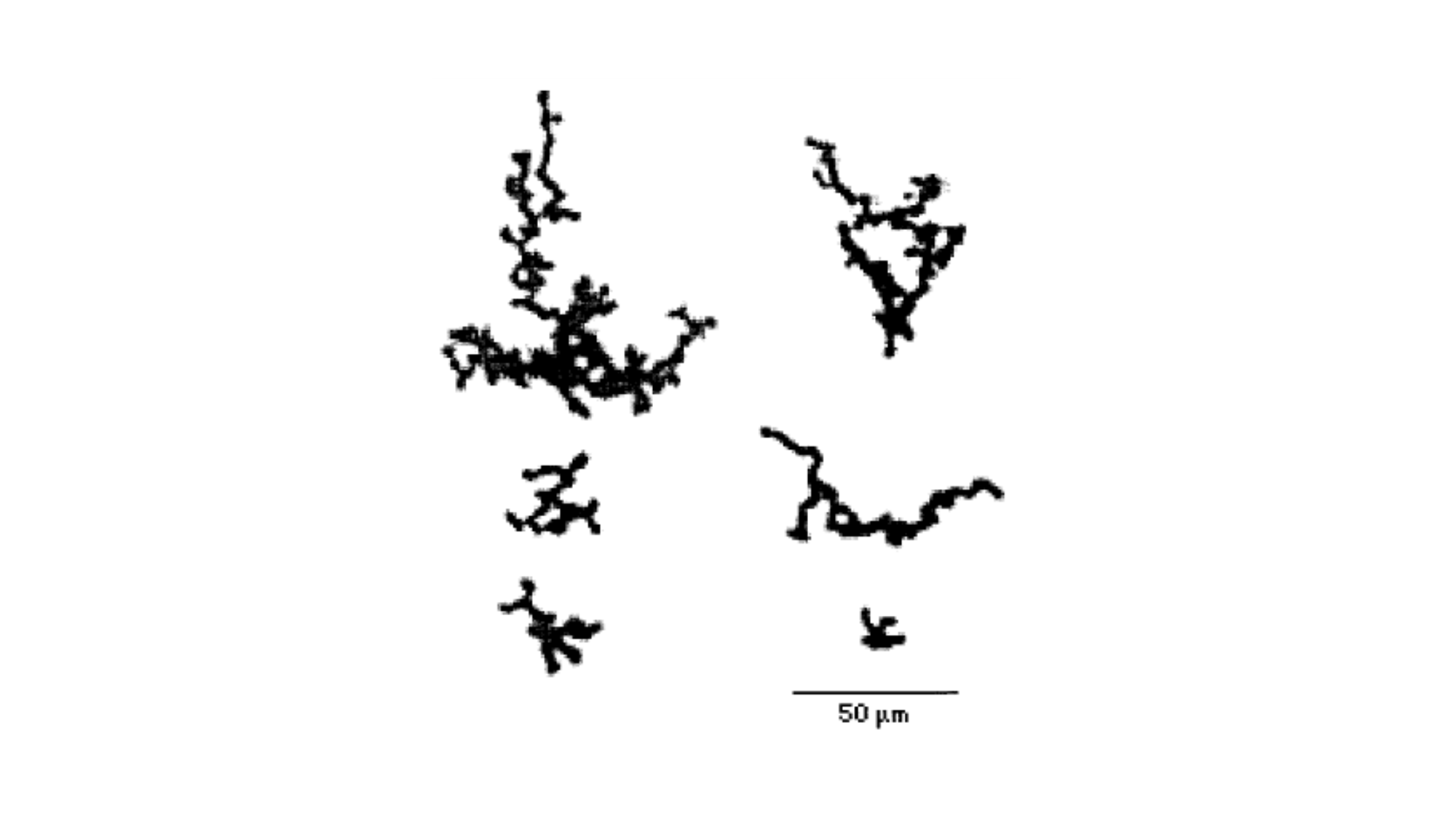}
      \caption{Examples of aggregates observed in the dust jet extracted from the turbomolecular pump. The monomers are monodisperse SiO$_2$ spheres with 1.9 $\mu$m diameter. Image reproduced with permission from \citet{RN785}, copyright by Elsevier.}
         \label{fig:figure_dust_collisions}
\end{figure}

In protostellar envelopes and planet-forming disks, grains continue to grow. The first stages of dust coagulation in disks lead to the formation of fractal aggregates of up to at least several $\mu$m sizes. Laboratory collisional studies \citep{RN785, RN1652, RN786, RN1604, RN1653} and \citep{Blum18} for a review corresponding to the initial stage of dust aggregation showed that aggregation of micrometer-sized grain monomers leads to the formation of fluffy particles with a fractal dimension of 1 – 2 and the porosity up to more than 90\%. An example of porous grain aggregates formed through collisions is shown in Fig.~\ref{fig:figure_dust_collisions}. In those collisional experiments \citep{RN785}, SiO$_2$ grains sedimented with constant velocity in a tube-shaped vacuum chamber forming a confined jet of high number density of grains.
A similar result, highly-porous aggregates (volume filling factor 0.11 ± 0.01) was obtained in collisional experiments on $\mu$m-sized water ice particles \citep{RN1651}. In those experiments, the particles and their aggregates were produced by spraying dispersed water droplets into a dry, cold gas (nitrogen) environment with subsequent sedimentation of the particles. 
The results of experimental collisional studies are mainly in agreement with the results of modeling of grain growth in protoplanetary disks also showing high porosity of dust aggregates at the first stage of dust coagulation \citep{RN1900, RN900, RN901, RN1865, RN1898, RN1901, RN1906, RN1907, RN1482}. 


\begin{figure}[ht]
   \centering
   \includegraphics[width=\textwidth]{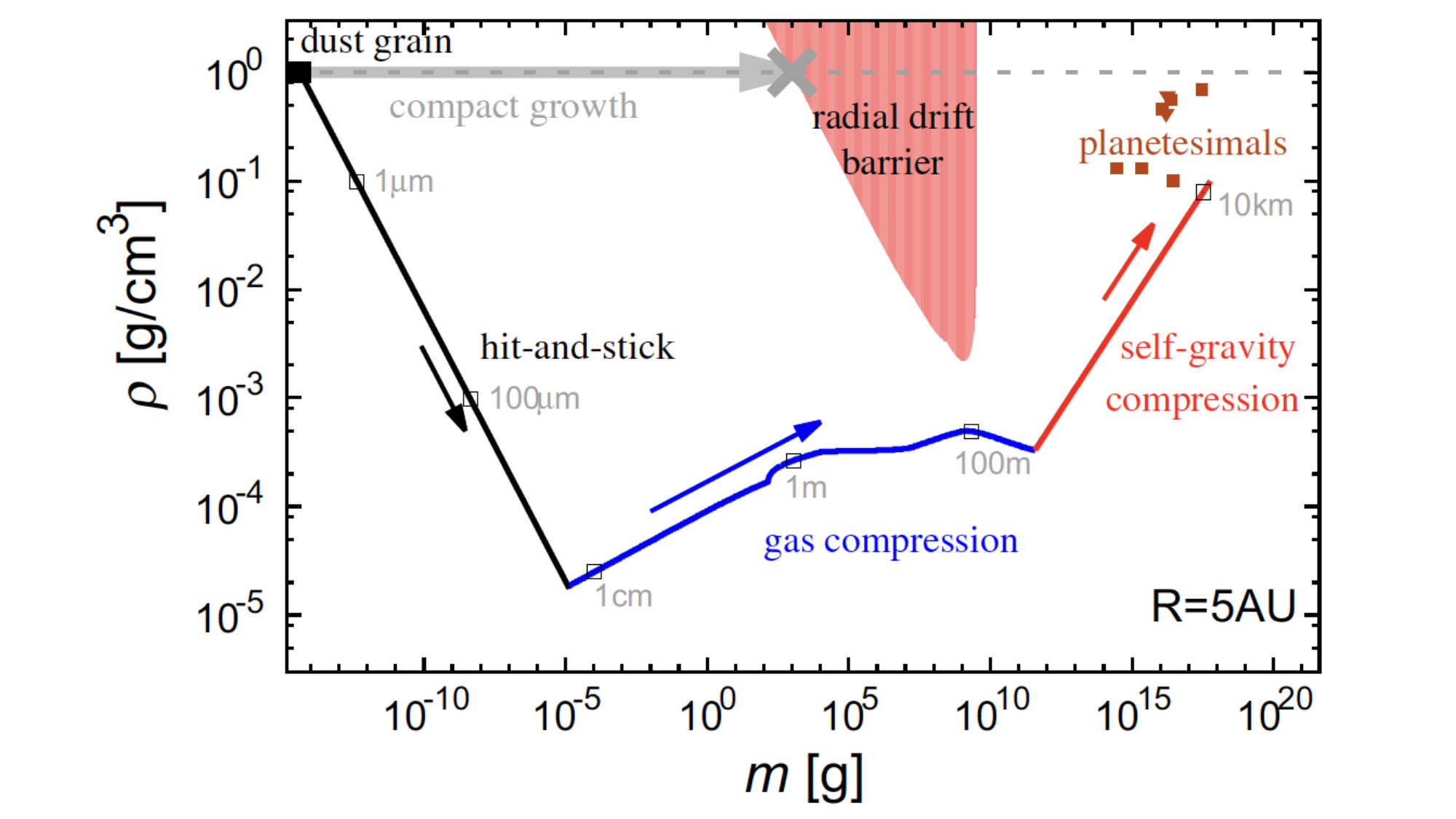}
      \caption{Pathways in the planetesimal formation in the minimum mass solar nebula model. The gray line shows the constant density evolutional track, which corresponds to the compact growth. The black, green, blue, and red lines are the evolutional track through dust coagulation via fluffy aggregates. Each line represents different mechanisms of dust coagulation, which are hit-and-stick, collisional compression, gas compression, and self-gravity compression. The red shaded region represents where the radial drift timescale is less than the growth timescale, which is equivalent to radial-drift region. The brown squares indicate the properties of comets, and the triangles represent their upper limit. The radii of dust aggregates for 1 $\mu$m, 1 cm, 1 m, 100 m, and 10 km are also written. Image reproduced with permission from \citet{RN788},  copyright by ESO.}
         \label{fig:figure_dust_disks}
\end{figure}

Although the initial growth leads to the formation of fluffy, open-structured aggregates, their later porosity evolution is rather controversial if such a structure is maintained as they grow to (sub)millimeter sizes. 
This is mainly because collisional strength of aggregates, which depends on the monomer size and composition as well as porosity of aggregates, and the impact velocity are still highly uncertain in a disk environment \citep{Blum18}.
One of the crucial factors for the porosity of grain aggregates created by collisions can be the impact speed. 
Depending on the impact speed, collisions result in sticking without deformation (hit-and-stick) for low velocities, sticking with deformation/compaction or bouncing of grain aggregates for moderate velocities, and catastrophic disruption (fragmentation) for higher velocities \citep{RN1653}. Usually, the collision speed increases with the aggregate size, and the hit-and-stick tend to happen in the first stage, whereas the latter two happen in later stages of dust aggregation.
Figure~\ref{fig:figure_dust_disks} taken from \citet{RN788} shows evolution of the internal density of dust aggregates during their growth, assuming perfect sticking (no bouncing and no fragmentation).
The results confirm that dust particles can evolve into highly porous aggregates at the first hit-and-stick stage (porosity reaching up to $\sim99.99\%$) and stay fractal even in the collisional compression stage, as suggested from MD simulations \citep{RN901, RN900}. In this scenario, dust aggregation can directly lead to the formation of planetesimals without invoking any hydrodynamic clumping of solid particles \citep{RN1865, RN788, Arakawa16, Homma18, Kobayashi21}, although gravitational instability could be triggered during the fluffy growth \citep{Michikoshi17}.
There are mechanisms that are thought to inhibit fluffy growth, such as bouncing/efficient compaction/fragmentation/erosion upon collision  \citep[e.g.,][]{Zsom10,Krijt15,Krijt16,Tanaka23,Lorek18,Estrada22,Estrada23,Michoulier24}. When bouncing collision occurs, the porosity is reduced to $\sim60\%$ after sequential bouncing collisions \citep{Weidling09, Zsom10}. Note that there is a debate if fluffy aggregates are subject to bouncing collisions between numerical and laboratory studies \citep{RN1863, Seizinger13,Kothe13,Brisset17}, although more recently, \citet{Arakawa23} pointed out that not only porosity but also aggregate size affects the probability of bouncing, which might explain why numerical and experimental studies show apparently different conclusions about bouncing. Another effect that could potentially be important is collisional fragmentation \citep{Dominik97,Paszun09,RN900, RN1898,Hasegawa21}. If collisional fragmentation happens, we would not be able to complete the proposed journey to icy planetesimals.

In summary, although it is widely accepted that growing aggregates have some porosity, but it remains rather inconclusive how much porosity they will have due to uncertainties in a number of conditions such as adhesion energy (surface composition) of the monomers, size of the monomers, and the impact velocity. With that said, disk observations now start to tell us about porosity, as we discussed in Sect.~\ref{sec:ppdobs}. Those observational studies suggest that aggregates may have a fluffy structure at least when they are micron sizes (see Fig.~\ref{fig:rt23}), but (sub)millimeter-sized aggregates seem to have a porosity less than $\sim99\%$. The occurrence of collisional fragmentation has been favored from disk observations as well \citep{RN1858, RN1862}. However, it is worth pointing out that those (sub)millimeter-sized aggregates might be the secondary origin, that is, they are not primordial aggregates of interstellar grains, but those produced by collisions of massive bodies like planetesimals, as proposed in \citet{Tatsuuma24}.

\subsection{Cosmic dust materials}
The choice of materials for laboratory experiments is guided by knowledge from estimates of the elemental inventory in space derived from a combination of Big Bang and stellar nucleosynthesis and tempered by observations of materials in a range of spectroscopic windows. \citet{RN1696} in \citet{RN1695} nicely describes the historical development of our understanding of dust mineralogy in linking spectroscopic observations (e.g., the 217 nm bump in the interstellar extinction curve commonly attributed to presence of graphitic grains, and the observations of silicate vibrations in the infrared consistent with both amorphous and crystalline materials) with the composition of meteorites as they represent an end point for processing of carbonaceous and silicaceous materials through the many phases of star formation and ending in cometary environments \citep{RN1694}. \citep{RN1697}, in the same volume, highlights the discovery of fossil grains in primitive meteorites as representative of the parent dust cloud from which the Solar System formed. 

From this discussion, we can establish an inventory of likely materials on which to base further laboratory exploration. This is presented in Table~\ref{tab:1}.
It is interesting to note that none of these materials is known naturally to be micro- or mesoporous. Porosity in these materials is most likely to be associated with amorphisation of the material and surface heterogeneity. However, physico-chemical alteration is space environments can produce materials that exhibit such porosity, e.g., carbon nanotubes and related materials, or aluminosilicates (zeolites). The latter are well-known terrestrially as size and shape selective industrial catalysts.   

\begin{table}[htbp]
\caption {Materials inventory of the major minerals for dust models in laboratory astrophysics experiments based on Table~1 from \citet{RN1697}. This list ignores the metallic elements also likely to be present.} 
\label{tab:1}
\centering
\begin{tabular}{c} 
 \toprule
 \textbf{Carbonaceous Materials} \\ 
 \midrule
 Amorphous carbon \\ 
 Graphite \\
 Diamond \\
 PAHs \\
 Fullerences \\
 \midrule
 \textbf{Silicon Derivatives} \\
 \midrule
 Silicon Carbide (SiC; carborundum) \\
 Silicon Nitride (Si$_3$N$_4$) \\
 \midrule
 \textbf{Oxides} \\
 \midrule
 Silicon Dioxide (SiO$_2$, silica) \\
 Aluminium Oxide (Al$_2$O$_3$; sapphire) \\
 Magnesium Aluminium Oxide (MgAl$_2$O$_4$, spinel) \\
 \midrule
 \textbf{Silicates} \\ 
 \midrule
 (Fe,Mg)$_2$SiO$_4$, Olivine \\
 (Fe,Mg)$_2$Si$_2$O$_6$, Pyroxene \\
 \bottomrule
\end{tabular}
\end{table}

\clearpage
\subsection{Atomistic modeling of nanoporous materials}

Attempts to model the effects of dust grain porosity typically assume that a general physical understanding of such systems can be obtained without consideration of the atomistic level structure and properties. Following this assumption, porous dust models are typically created using aggregates of monomers with sizes ranging between a few nanometers to a few microns. Such studies have provided estimates of how porosity could affect dust extinction (Sect.~\ref{sec:optprop}) and have simulated the evolution of porosity in dust aggregates in various astrophysical environments (Sect.~\ref{sec:porobs}). Here, we focus on models of porous systems in which atoms are the fundamental building blocks. Such atomistic level modeling is typically applied to systems possessing pore sizes smaller than 10 nm, where it can provide a uniquely detailed perspective.  


Although there are currently very few atomistic modeling studies that have been directly applied to astronomical dust, such methods have been used for decades to understand technologically important porous materials. \rev{As astronomically relevant examples, we refer to studies on nanoporous silica-based materials to illustrate how atomistic modeling can provide otherwise unobtainable physico-chemical insights.} Atomistic modeling initially involves the bottom-up atom-by-atom construction of realistic porous structures. Subsequently, these modeling methods can be used to theoretically characterize the physico-chemical properties of such systems, such as: atomic-scale structural features (e.g., detailed morphology of pores and surfaces), spectroscopic characteristics (e.g., IR and UV spectra), and phenomena occurring within pores (e.g., adsorption and diffusion of chemical species). 

As previously mentioned in this review, the porosity of cosmic dust remains largely unexplored, and its atomistic level characterization is even less understood. Therefore, we provide a brief background to atomistic modeling methods applied to porous silica-based materials, which could potentially be applied to the structural modeling of porous cosmic dust. We note that, although the typical pore sizes in cosmic dust are still unknown, most reported atomistic modeling studies of porous silicas have focused on materials with pore diameters $<$ 10 nm due to their industrial importance.

\subsubsection{Atomistic modeling methods}

Firstly, we briefly introduce the most commonly employed methods to model materials at an atomistic level. The most computationally efficient methods are based on classical molecular mechanics (MM), for which current state-of the-art simulations can treat systems possessing up to a few million atoms. MM methods use interatomic potentials (IPs), also known as force fields, to calculate the energy, structure, and properties of the system, neglecting the electrons of the system. Typically, IPs consist of a set of analytical functions that have been empirically parameterized to represent different atomic-level interactions (e.g., chemical bonds, electrostatics, van der Waals). Some IPs also incorporate extended parameter sets to describe the energetics of chemical reactions (e.g., ReaxFF [\citealt{ReaxFF-SiO,ReaxFF-Zeolite}]). Well-parameterized IPs are available for describing the interatomic interactions in some astronomically relevant dust materials such as silicates \citep{Pedone2022} and water ice \citep{TIP4P2005,CO-H2O_2014,CO2-H2O_2014}. However, for more chemically complex systems (e.g., numerous different molecular species interacting within a porous material) accurate parameterization is more difficult and suitable IPs are often lacking. As such, the applicability of MM modeling is limited to systems for which IP parameters have been developed. We note that, as an alternative to using empirically determined parameters, machine learning is now being widely used to develop more accurate and general IPs which are starting to be used for modeling astronomical dust \citep{Tang2023}. Even with such advances, the fundamental limitation of MM modeling is that it does not explicitly describe electronic degrees of freedom. As such, MM modeling cannot be reliably applied in scenarios where a detailed understanding of electronic effects is important (e.g., astrochemical reactions, UV adsorption spectra).

To overcome the electronic limitations of MM methods, one can use quantum mechanical (QM) methods. QM methods are often referred to as first principles methods, as they account for the electronic structure of the system from solving the electronic Schr\"odinger equation. Among the various available QM modeling methodologies, density functional theory (DFT) is the most popular and widely used due to its relatively high computational efficiency. This efficiency derives from computing the properties of the system via a function of the electronic density (i.e., a density functional) rather than explicitly from the wavefunction. As an accurate general description of this functional for arbitrary systems of atoms is still unknown, hundreds of DFT functionals have been developed. In practice, only a few DFT functionals are commonly used (e.g., B3LYP [\citealt{B3-1993,LYP-1988}], PBE [\citealt{PBE1996}]\dots), which generally provide a reasonably accurate description of the electronic structure and related properties for a wide range of molecular and solid-state systems. Generally, DFT methods provide a more accurate and fuller description of a system than IP-based MM methods. 

Despite the computational effectiveness of DFT methods, such simulations are limited by the size of the systems they can handle. Currently, with the aid of high-performance computing and massively parallel calculations, such calculations are still limited to models containing up to a few thousand atoms. This constraint places much stronger size limits on the applicability of DFT modeling to treat porous dust systems compared to MM methods. As a compromise between DFT and MM methods, semiempirical methods offer a relatively computationally inexpensive QM-based approach to overcome the size limitations of DFT. Semiempirical methods are derived from pure QM methods but incorporate various approximations (e.g., omission of bi-electronic integrals) and empirical parameters to avoid costly computations. Due to these simplifications and parameterizations, semiempirical methods are faster than their QM counterparts, enabling the treatment of large chemical systems that are otherwise untenable with pure QM methods. However, similar to MM methodologies, the accuracy of semiempirical methods heavily depends on the suitability of the parameterization for the specific case under study. There are various semiempirical methods, but the most frequently used are those based on tight binding approaches, which derive from DFT such as SCC-DFTB \citep{SCC-DFTB}. Recently, for example, Grimme and coworkers developed the GFN-xTB family \citep{GFN-xTB}, a set of tight binding-based methods that model covalent H-bond, and dispersion interactions with improved accuracy, and are parameterized for almost the entire periodic table of elements (up to Z = 86).

\begin{figure}[ht]
   \centering
   \includegraphics[width=\textwidth]{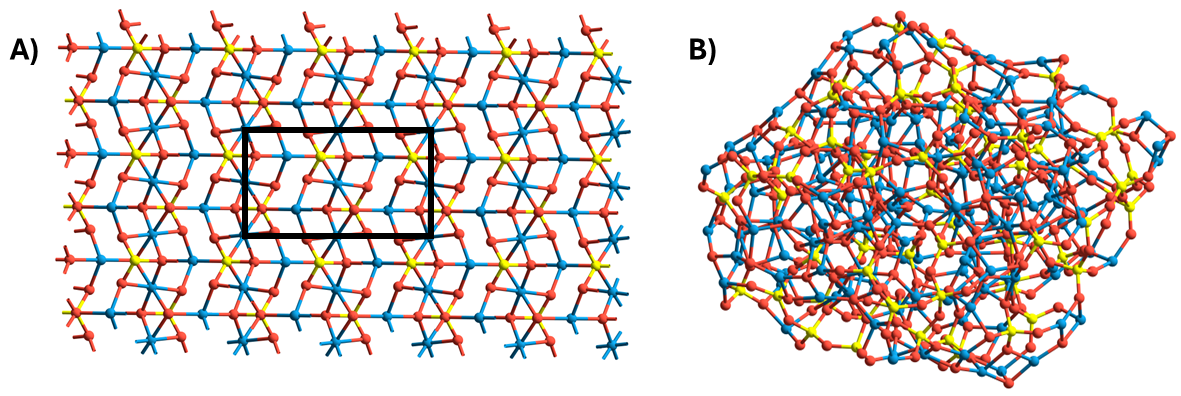}
      \caption{Atomistically detailed structural models for forsterite (Mg$_2$SiO$_4$). A) a periodic approach, modeling crystalline forsterite (repeat unit cell outlined in black), and B) a cluster approach, modeling a finite amorphous system. Atom color key: Si – yellow, Mg – blue, O – red.}
         \label{fig:periodic_vs_cluster}
\end{figure}

Apart from the choice of calculation method, there are two main system-based approaches to atomistically modeling materials: i) the periodic system approach and ii) the cluster approach \citep{Minerals-Rimola2021}. Periodic systems involve a repeating unit cell defined by lattice parameters (i.e., the cell lengths along the three spatial directions and the angles between them) to create a periodic system that extends infinitely by applying periodic boundary conditions (PBC). For bulk materials, this repetition is three dimensional (3D), while for surfaces, it is two dimensional (2D). In contrast, the cluster approach models the material as a discrete finite system without considering PBC. Figure~\ref{fig:periodic_vs_cluster} shows examples of these two approaches as applied to an extended silicate crystal and an amorphous silicate nanoparticle \citep{Zamirri2019}. 

The simplest way to approximately model a porous material at the atomic scale is by using an extended (“open”) surface model. Here, for suitably large pores, we can assume that the local properties of a porous wall are similar to those of an isolated extended surface in vacuum. For example, both extended silica surfaces and the interior surfaces of porous silica have exposed silanol (Si-OH) groups, which largely determine their adsorptive properties \citep{Rimola2013}. Therefore, an extended silica surface model with an appropriate silanol density can also serve as a model for the Si-OH covered interior pore surface of a mesoporous silica material \citep{Ugliengo2008,DellePiane2014-JPCA}. Within this framework, open surfaces can be modeled using either a cluster or a periodic approach. Clearly, however, open surfaces and real surfaces inside a porous material are not the same, and this approximation ignores several potentially important factors. For example, modeling a porous material with an extended surface neglects the long-range influence of the surrounding pores and the local effects of an enclosed pore surface. Such issues are particularly important in phenomena like adsorption and diffusion of chemical species, which become increasingly confined as the pore size becomes smaller. To address the approximations of an open surface approach, more accurate atomistic models explicitly account for the pore structure. Below we divide our discussion of such nanoporous models into two parts, following the conventional classification of porous silica-based materials based on pore diameter ($d$): i) mesoporous materials with $d = 2 - 50$ nm, and ii) microporous materials with $d < 2$ nm. In both cases, we briefly discuss possible physico-chemical properties of these systems that could be astronomically relevant.   

\subsubsection{Mesoporous models}

Mesopores (with $d$ $\geq$ 2 nm) are large compared to the dimensions of most small molecules likely to be encountered in astrophysical environments (e.g., H$_2$, CO, H$_2$O, CH$_3$OH). Such species can thus diffuse through such materials relatively unimpeded, at least as far as pore size is concerned. During diffusion, molecules will repeatedly collide with the interior pore walls, likely leading to eventual adsorption at interior pore wall sites where they interact the most strongly. On Earth, water molecules tend to strongly interact with the surfaces of amorphous silica where they dissociate and form surface bound hydroxyl groups (i.e., silanol groups: Si-OH). As such, the interior pore walls of laboratory-synthesized mesoporous silicas are covered with silanol groups. For mesoporous silicate dust particles, we may also expect interior silanols, but from arising from reactions with hydrogen and atomic oxygen \citep{Jones1984}, rather than from water molecules. The presence of such strongly bound Si-OH groups on extraterrestrial silicates has been confirmed in cometary material \citep{Mennella_2020} and in lunar silicates \citep{Zeller1966}. Chemisorbed hydroxyl groups are susceptible to desorption via interaction with strong UV radiation \citep{Imai1996}. The coverage of silanol groups on the exterior surface of silicate dust grains in the diffuse ISM is thus likely to be low. However, silanol groups within a mesoporous silicate material would be more protected from external radiation and would be more likely to persist at higher coverages.

Experimentally synthesized stable mesoporous silica materials usually possess relatively dense amorphous pore walls. Models of such amorphous porous materials can be derived from an initially non-porous model (typically a dense crystalline phase) which is amorphized using molecular dynamics (MD) simulations through a melt/quench scheme \citep{Ugliengo2008,Williams2016,DellePiane2018}. MD solves Newton's equations of motion for atoms, thus providing a time-dependent description of the system's atomistic dynamic evolution. Performing MD simulations at high temperatures on an initially crystalline system induces the atoms to disorder (i.e. induces melting). Subsequent rapid decrease of the simulation temperature quenches the system while maintaining the disordered atomic positions.  

\begin{figure}[ht]
   \centering
   \includegraphics[width=\textwidth]{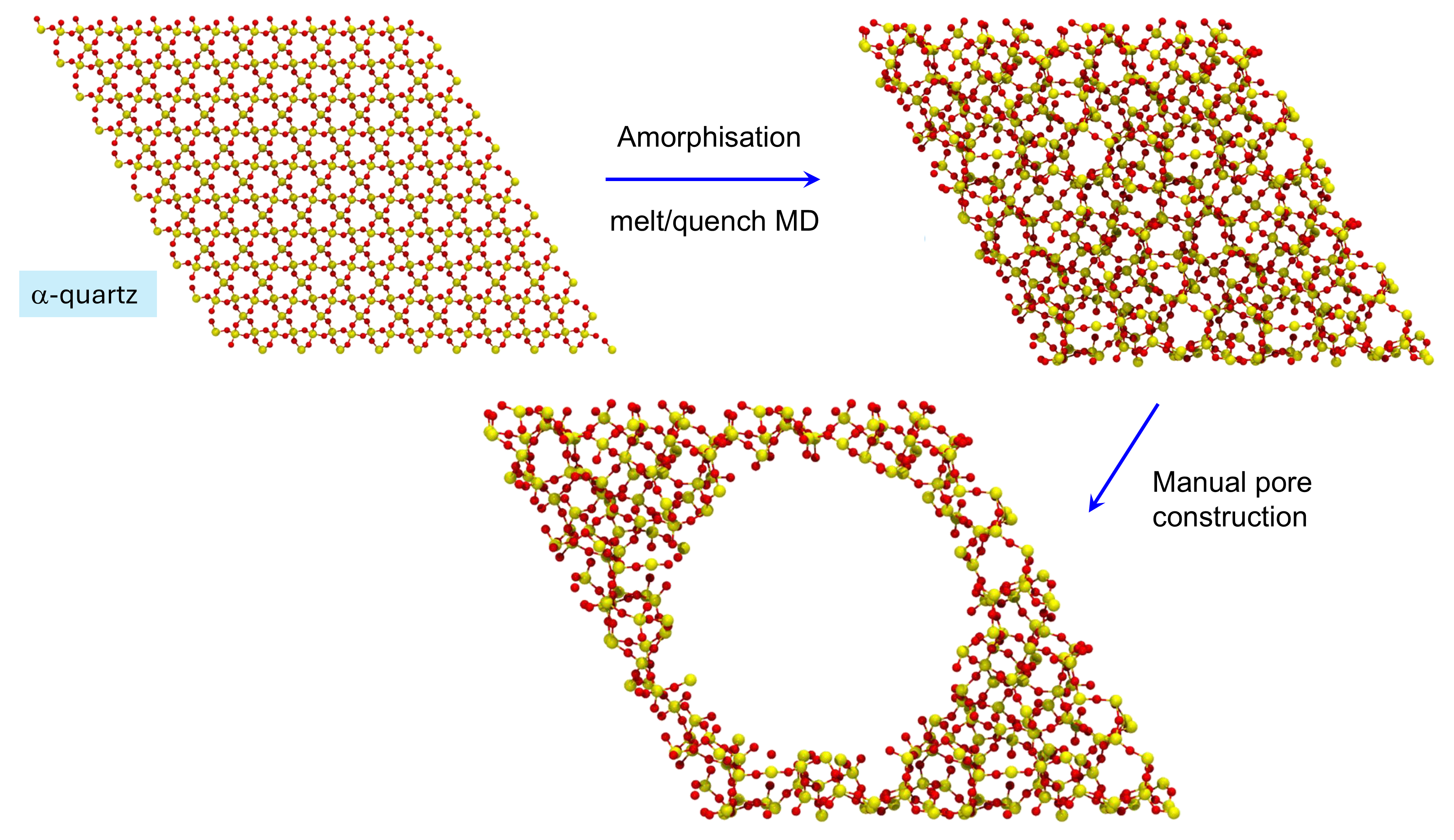}
      \caption{Atomistic modeling of a porous system, using the transformation from a periodic cell of crystalline $\alpha$-quartz to a cell of an amorphous porous silica containing one mesopore. The first step consists in the amorphization of the initially crystalline structure by molecular dynamics (MD) simulations following a high temperature melt and low temperature quench process. In the second step atoms are removed to construct a pore. Finally, the system is structurally relaxed. Atom color key: Si – yellow, O – red.}
         \label{fig:mesoporous-modelling-procedure}
\end{figure}

\begin{figure}[ht]
   \centering
   \includegraphics[width=\textwidth]{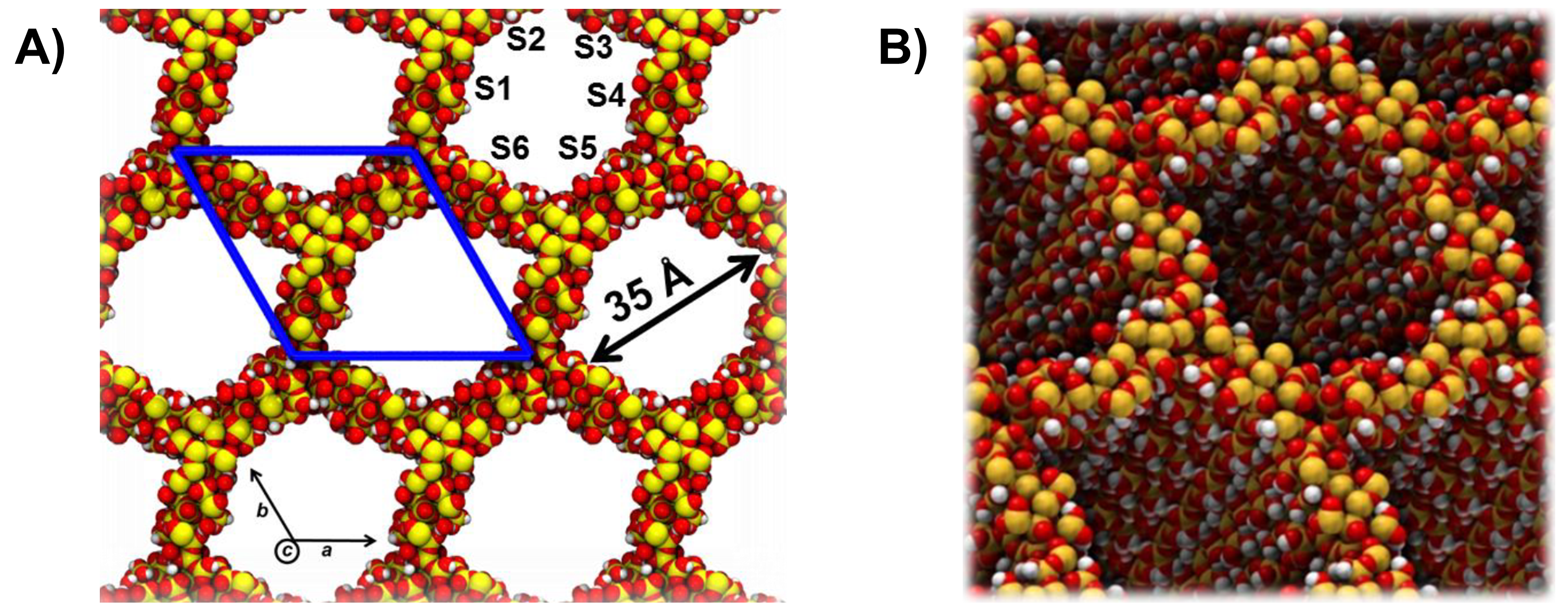}
      \caption{A) Atomistic model of a mesoporous silica after addition of Si-OH groups and application of PBC. The unit cell is outlined in blue. The pore diameter is also highlighted (35 Å), along with potential binding sites (S1-–S6). B) Three-dimensional view of the atomistic mesoporous model. Atom color key: Si – yellow, O – red, H - white.}
         \label{fig:mesoporous-silica}
\end{figure}

To generate atomistic models of mesoporous materials one can then remove atoms from the non-porous dense analogue to create pore spaces with a desired size and shape. Here, it is crucial to proceed cautiously during atom removal to ensure structurally stable and chemically sound structures. For example, for ionic materials (e.g., silicates) stoichiometry and electroneutrality must be maintained \citep{Minerals-Rimola2018}, while for molecular materials (e.g., water), the chemical bonds of the molecular moieties should not be broken \citep{Zamirri2018,Ferrero2023}. For example, in the case of mesoporous silica this atom removal process often results in under-coordinated oxygen atoms that can be saturated with hydrogen atoms to form Si-OH groups. Note, that one can also choose to add extra silanol groups to the interior pore surfaces to create mesoporous models with desired Si-OH surface coverages. Finally, the as-generated porous system is structurally relaxed by minimizing the forces on all atoms and the stress on the unit cell. Figure~\ref{fig:mesoporous-modelling-procedure} illustrates the computational procedure to generate a realistic atomistic model for a mesoporous amorphous silica from periodic cell extracted from crystalline $\alpha$-quartz following: (i) amorphization of the initial non-porous crystalline structure through melt/quench MD simulations, (ii) removal of atoms to create pore spaces, and (iii) optimization of the system to achieve a structurally relaxed (stable) structure. The advantage of using a periodic approach when applying this procedure is that application of PBC to a suitable of the final cell with a single pore intrinsically generates a perfectly regular array of pores. Figure~\ref{fig:mesoporous-silica} shows the result of applying PBC to a porous unit cell (outlined in blue) to generate an array of pores to represent an extended mesoporous silica material \citep{Ugliengo2008}.

Use of such models (see Fig.~\ref{fig:mesoporous-silica}) can help in providing detailed atomic scale insights into the characteristic IR response of Si-OH groups within mesoporous silicates \citep{Ugliengo2008} and how small molecules (e.g., H$_2$O) diffuse through such a Si-OH covered pore systems \citep{Coasne2012}. Much experimental work has focused on the behavior of water in mesoporous silicas, where the combined roles of nanoporosity and surface Si-OH groups can be significant \citep{Pore_size_vs_hydroxylation_MMM2016}. In such systems, nanoscale confinement leads to a decrease in the melting/freezing point temperature, density, and surface tension of water \citep{Knight_water_confinement_2019} which could have implications for astrochemical ice formation/stability in mesoporous grains. These physico-chemical changes are fundamentally linked to the underlying changes in hydrogen bonding in these systems \citep{H-bond_confined_water_ACIE2008, Trapped_water_network_PCCP2011}. Atomistic modelling of water-filled mesoporous silicas with selected pore sizes and Si-OH coverages and can provide highly detailed insights into the IR spectroscopic shifts associated with such changes \citep{Simulated_IR_confined_water_JCP2021, Simulated_IR_confined_water_JCP2016}, which may be observationally detectable.        

\subsubsection{Microporous models}

\begin{figure}[ht]
   \centering
   \includegraphics[width=\textwidth]{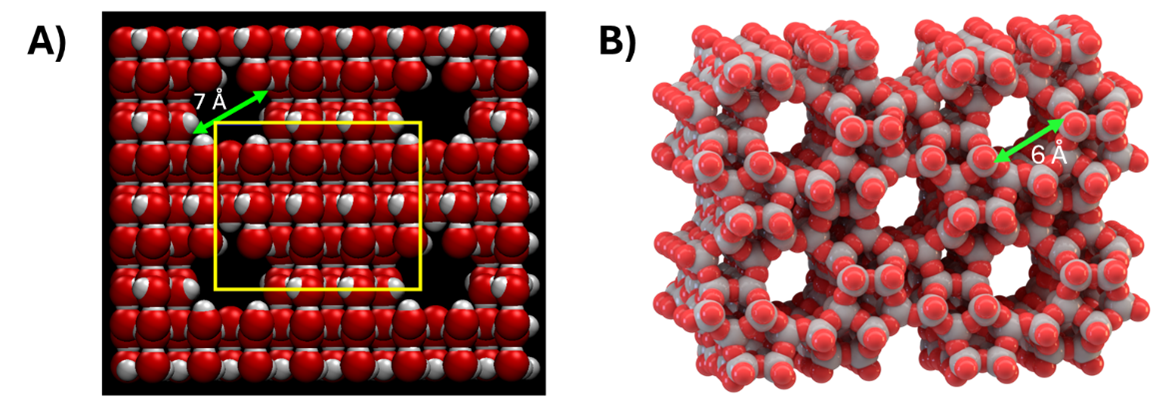}
      \caption{A) Microporous water ice (oxygen – red, hydrogen – white) with pores generated by removing water molecules from the crystalline P-ice phase. The pore diameter is also highlighted (7 Å) and the unit cell is shown in yellow. B) A crystalline microporous silica zeolite (oxygen – red, silicon - gray) with pore diameter indicated.}
         \label{fig:microporous-modelling}
\end{figure}

Microporosity refers to smallest pore sizes ($d < 2$ nm), which can be of the order of the sizes of small molecules. In principle, one can generate microporous models by following the strategy outlined for mesoporous models, by removing atoms from a dense precursor material. Figure~\ref{fig:microporous-modelling}A illustrates this process for microporous water ice, in which small pores are generated by removing water molecules from a compact crystalline ice phase. For silicates, many microporous solids are known which have complex crystalline structures that are distinct from any dense phase. These materials, known as zeolites (see Fig.~\ref{fig:microporous-modelling}B), have been widely studied by atomistic modeling methods due to their technological applications involving adsorption, diffusion and separation of small molecules \citep{Smit2008, Diffusion_in_zeolites_Koppens2000}. Different zeolites have distinct degrees of porosity made up from micropores and channels with characteristic sizes and topologies. These, often subtle, differences can have a relatively large  influence on the ease with which different small molecules diffuse through these materials. 

The smallest micropore sizes ($d <$ 0.5 nm) can be similar in size to numerous astronomically abundant small molecules such as H$_2$, CO and H$_2$O. Here, we are in the highly confined regime where even a single molecules can significantly interact with the pore walls in multiple directions. Simple atomistic scale models of such highly confined molecules can rationalise how the size of the micropores with respect to the confined molecules can contribute to differences in diffusion and adsorption. Such models, for example, predict that the physisorption of small molecules inside small micropores can be up to eight times larger than that on the corresponding flat surface \citep{Derouane1988}. For such situations, detailed atomistic modeling of confined H$_2$ molecules has also shown that the flexibility of the silicate structure is essential for facilitating interpore diffusion \citep{vandenBerg-2004a}. As such, the structural response of the microporous system needs to taken into account for understanding diffusion rates of highly confined molecular species \citep{vandenBerg-2004b}.

Although studies of zeolites can provide some insights of astronomical relevance, these crystalline silica-based materials with well-ordered micropore systems are unlike the disordered multiscale porosity likely exhibited by porous silicate dust (e.g. as in IDPs). An alternative method for constructing atomistic models of porous materials is through aggregation. In principle, aggregation can bring together particles of many different sizes to form porous fractal aggregates in which a hierarchy of pore sizes can exist. Here, we consider the formation of a microporous silicate from the aggregation of ultrasmall nanosilicate grains \citep{Bromley_nanosilicate_review}. Nanosilicates with diameter less than 3 nm are likely to be produced in high numbers in the ISM due to processing of larger grains by SNe shockwaves \citep{Jones1996}. Although it has been estimated that nanosilicates account for a relatively small mass fraction of available silicon (approximately 5-10 $\%$ \citep{Hensley_2017, Li_2001} these dust species are so small that they would be numerically much more abundant than larger dust species. 

Nanosilicates could eventually potentially aggregate in more protected translucent or molecular cloud environments \citep{RN825}. Fractal nanosilicate aggregates may also be formed during circumstellar dust nucleation and growth \citep{Kimura2022}. We also note that laboratory-made fractal aggregates of nano-sized silicate particles can reproduce observed IR spectra of silicate dust \citep{Ishizuka2015}. We expect that the astrochemical and astrophysical impact of microporosity in nanosilicate aggregates to have some similarities with those observed for microporous silicate zeolites.

\begin{figure}[ht]
   \centering
   \includegraphics[width=\textwidth]{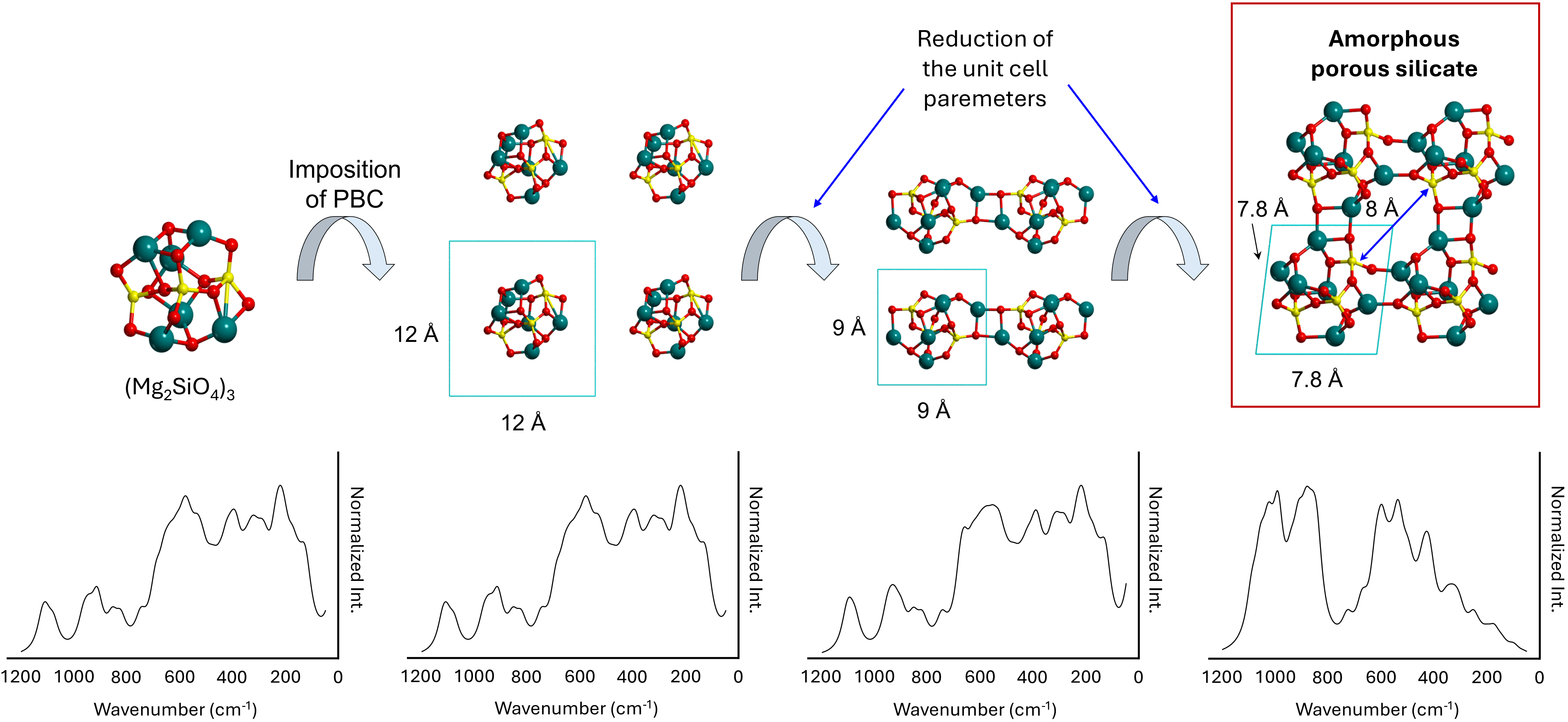}
      \caption{Top: Atomistic modeling of an aggregation-induced microporous silicate. From left to right: A (Mg$_2$SiO$_4$)$_3$ nanosilicate is selected. PBC are imposed by introducing the isolated nanosilicate into a 12 $\times$ 12 $\times$ 12 Å cubic unit cell. Reducing the unit cell parameters to 9 $\times$ 9 $\times$ 9 Å and optimising the structure induces an initial interaction between clusters of adjacent unit cells. The process of cell size reduction and structural optimisation is stopped once a minimum energy is reached. In this case a microporous silicate structure with 8 Å diameter micropores is obtained. The unit cell at each step is shown in blue. Atom color key: Si – yellow, Mg – blue, O – red. Bottom: Simulated IR spectrum corresponding to each structure throughout the aggregation process. For further details see \cite{Rimola-Bromley2021}.}
         \label{fig:amorphous-porous-silicate}
\end{figure}

The process of aggregation can be approximately modelled by placing an isolated nanosilicate in a repeat simulation cell using PBC, and then gradually reducing the unit cell parameters. As the unit cell size is reduced, nanosilicate structures in adjacent cells start to interact with each other and form bonds. Continuation of this process, with structural optimization at each cell size reduction step, eventually creates a three-dimensional microporous structure. The upper part of Fig.~\ref{fig:amorphous-porous-silicate} illustrates this procedure in which an olivinic (Mg$_2$SiO$_4$)$_3$ nanosilicate is used to produce a periodic microporous silicate model \citep{Rimola-Bromley2021}. In the lower part of Fig.~\ref{fig:amorphous-porous-silicate} the calculated IR spectra are shown for each respective step of the aggregation process. Observed IR spectra of silicate grain populations are typically dominated by two broad fetaures centred at around 9.8 $\mu$m (i.e. 1020 cm$^{-1}$) and 18 $\mu$m (i.e. 556 cm$^{-1}$) corresponding to Si-O stretching and O-Si-O bending vibrational modes, respectively. Due to their ultrasmall size and non-bulklike atomstic structures, nanosilicate grains are predicted to have characteristically different IR spectra. Typically, the IR spectra of nanosilicates exhibit more peaks with frequencies that are shifted with respect to the IR spectra of larger amorphous grains \citep{Macia_nanosilicate_2019, Zeegers_2023}. Such a spectrum can be seen in the lower right of Fig.~\ref{fig:amorphous-porous-silicate}. Upon aggregation into a microporous silicate the IR spectra evolve towards a more typical silicate spectrum. However, even when fully aggregated, the IR spectrum still retains some fine structure which may indicate that that nanoporous silicates could be observationally distinguished from denser silicate grains. Clearly, further atomistic modeling and laboratory experiments are required to consolidate these preliminary findings.      

\subsection{Astrochemical modeling}
Much of the interest in porosity in astrochemical models has been driven by laboratory studies of ice composition and structure, but there are shared features between the ice and grain porosity scenarios, as well as between the models used to simulate them. \cite{Awad05} developed a model for chemistry on an ice structure with crack-like porous features; the model was devised to replicate laboratory experiments, such that CO molecules in a CO--H$_2$O ice mixture would be partially protected from reactive H atoms adsorbing onto the surface. The competition between efficient production of H$_2$ on the outer ice surface via H recombination, and the more limited penetration of H into the cracks, allowed some of the CO to avoid conversion to H$_2$CO or CH$_3$OH.

\cite{PB06} constructed the first rate equation-based chemical model to simulate grain porosity explicitly, working under the hypothesis that porosity could provide the higher rate of grain-surface H$_2$ production deemed necessary to explain H$_2$ abundances in photon-dominated regions (PDRs). The model considered only the grain-surface chemistry, and was limited to the H+H$\rightarrow$H$_2$ system under steady-state conditions. Diffusive hydrogen was allowed to exist in two locations; the external grain surface that was coupled to the gas phase via adsorption and desorption; and an internal system of pores accessible via diffusion through ``edge'' sites (making up 5\% of external binding sites) that linked the internal and external surfaces. While both H and H$_2$ could pass through the edge sites, desorption within the pores was not allowed, on the assumption that such species would be re-adsorbed on an inner wall. Thus, H atoms in particular could be retained within the pores at greater abundance than would be otherwise expected, increasing their conversion to H$_2$. Different levels of porosity, expressed as the ratio of internal to external binding sites were tested, ranging from 10--1000. The effect of the inclusion of porosity in the model was to extend the temperatures at which efficient H$_2$ production could be produced (although not to high enough temperatures to explain efficient H$_2$ production in PDRs). 

The {GRAINOBLE} model by \cite{Taquet12} is the only rate equation-based treatment of a full gas-grain interstellar chemical network to consider porous structures. Technically, this model apparently considered porosity in the grains as well as in the ice mantles, but was more focused on the latter. Their treatment of grain-surface chemistry employed a so-called multilayer model, of a type similar to the one first proposed by \cite{HH93}, allowing active chemistry only in the upper layer of the ice (or initially on the bare grain surface), with material beneath being stored as layers of bulk ice. As with \cite{PB06}, the porous structures could be entered from the surface via edge sites; however, whereas \cite{PB06} considered an expansive porous structure much larger than the outer surface, \cite{Taquet12} envisioned a system of pores whose available reactive surface was defined to be less than that of the outer surface layer. While the model prohibited pore-bound species from desorbing, the generically small size of each individual pore surface (9 binding sites) would make the trapping effect much weaker, in comparison with that seen in Perets \& Biham's model. Furthermore, the Taquet model rendered trapped pore material chemically inert once it was incorporated into the bulk ice. This model therefore likely says little about the effects of \emph{grain} porosity in particular. The recent Monte Carlo kinetics model by \cite{Song24} also concentrates exclusively on the effects of ice porosity.

\begin{figure}[ht]
   \centering
   \includegraphics[width=\textwidth]{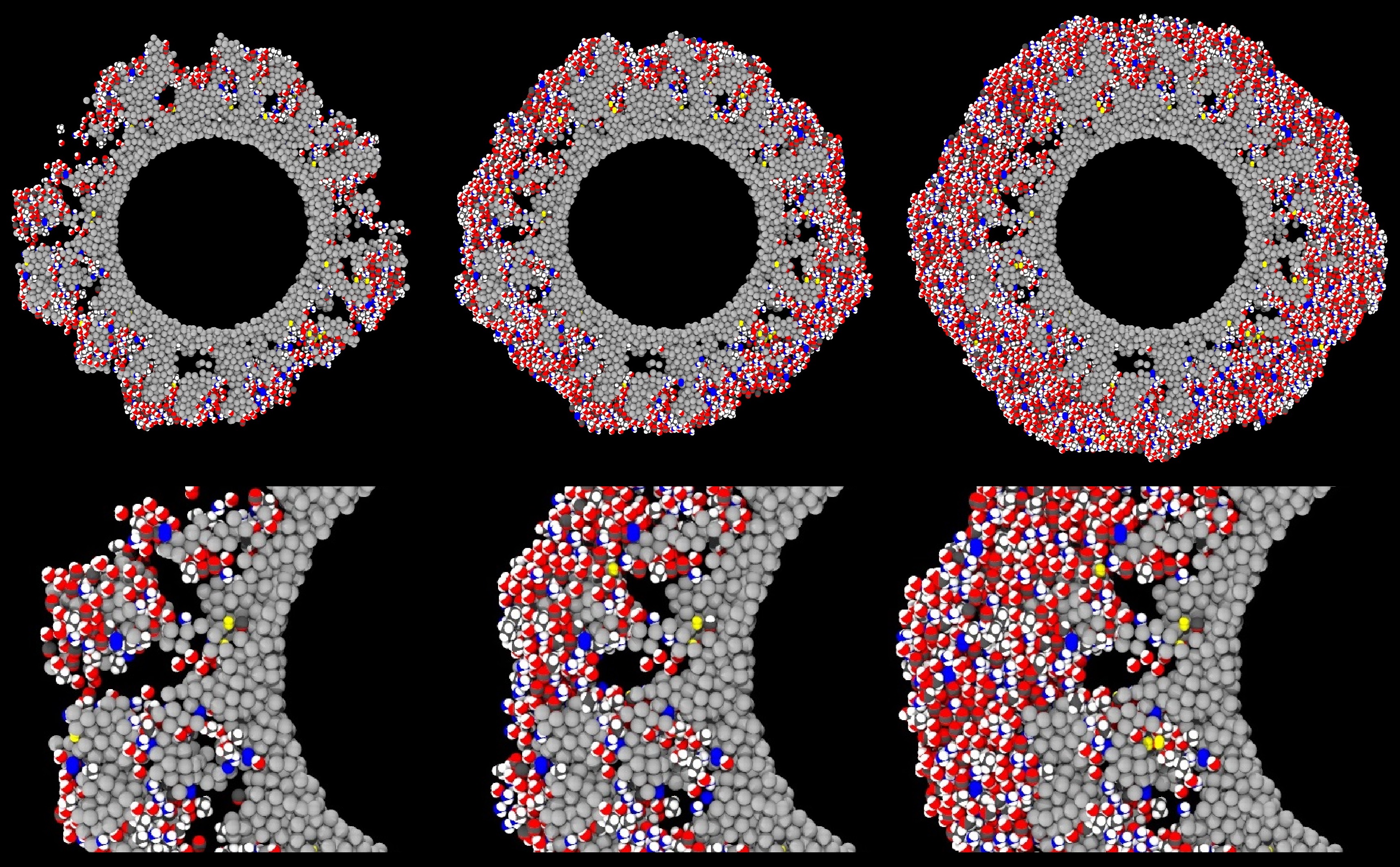}
      \caption{Cross-sectional images produced from chemical simulations on a 100~\AA~radius porous grain, using the off-lattice Monte Carlo kinetics model {MIMICK}. Results at three moments in the early stages of ice build-up are shown; lower panels represent a zoomed-in view of a section of each upper panel. Gray spheres represent the carbon atoms composing the dust grain. The solid inner region of the grain was removed to simplify the calculations; this inner void plays no part in the model. Grain-surface atoms are represented by white (H), red (O), black (C) and blue (N) spheres. Molecules are similarly represented as collections of such atoms in appropriate structures. H$_2$ molecules are highlighted in yellow. Image reproduced from \cite{CG21}, copyright by the author(s).}
         \label{fig:cg21}
\end{figure}

The models by \cite{CG21} currently represent the only attempt to simulate explicitly the grain-surface chemistry that would occur on a porous interstellar grain. Their {MIMICK} model simulates surface chemical activity using an off-lattice microscopic Monte Carlo kinetics technique in which the positions of atoms and molecules are calculated at run-time, based on the local surface potentials, which also govern the mobilities of those species on the surface. The model can simulate chemistry on any pre-defined 3-D surface structure. \cite{CG21} created a selection of porous and non-porous grains by first producing a spherical, carbonaceous grain (with amorphous surface structure), and then randomly depositing further grain atoms using a hit-and-stick method. The result was a broadly spherical grain with an open porous structure. Gas-phase CO molecules and H, C, O and N atoms (with fixed total budget) were then deposited onto the surface over periods on the order of 10$^4$--10$^5$~yr. These species and their chemical products were allowed to diffuse, react and thermally desorb.  The desorption treatment included the explicit calculation of the trajectory of the ejected species, allowing it to re-adsorb onto any surface obstructing its path, and thus allowing the more volatile species in the model to be retained more easily within the pores than on the outer surface.

Cross-sections through one of the porous grains at three periods during the build-up of ice are illustrated in Fig.~\ref{fig:cg21}. In some areas, the grain-pore structures are seen not to become completely filled with molecules, but are nevertheless closed over by the growing ice mantle, leaving voids within the deeper regions. With the formation of a thick ice mantle, the porous structure was lost and the grain took on a regular and roughly spherical morphology.

The overall chemical composition of the ice was found not to vary substantially between the porous and non-porous grain cases; nor indeed was the production of the volatile H$_2$ molecule much altered. The main effect of the porous structure was found instead to be the trapping of H$_2$. This was caused by a combination of (i) porous structures that could be enclosed by the growing ice, and (ii) the enhanced binding energies experienced by H$_2$ molecules when encountering larger species (such as water) or highly irregular microscopic surface structures in the pores. The trapped H$_2$ did not need to be formed inside the pores; the porous structures acted rather as a trap for otherwise mobile, volatile species formed elsewhere. The authors posited that such an effect would be larger if gas-phase H$_2$ adsorption were considered (which is found to be a substantial computational hurdle in models of this kind). It was also speculated that other volatile, and largely non-reactive, species such as atomic He could become trapped deep within the ices in a similar fashion.

\section{Impact and potential implications of porosity}

\subsection{Atomistic and astrochemical modeling}

\subsubsection{Calculating properties of nanoporous systems}

Once an atomistic structural model has been generated, its physicochemical properties can be theoretically characterized. Regarding observational astronomy, spectroscopic properties are probably most relevant. Here, atomistic modeling can simulate a range of spectroscopies (e.g., IR, UV, microwave) with varying degrees of accuracy which can be compared with observations. The significance of using atomistic modeling in this way is that it directly links the spectral properties to an atomic-level (and/or quantum-level) description of the system in question. For example, using the atomistic structure of a silicate dust grain we can calculate the atomic displacements associated with all vibrational normal modes and their associated dipolar variations, and thereby simulate IR spectra. An example of this type of calculation is shown in Fig.~\ref{fig:amorphous-porous-silicate}. By analysing the detailed structures associated with the simulated spectra and comparing these spectra with the observed IR spectra, we can infer possible characteristics of the astronomical dust under study \citep{Zamirri2019, Zeegers_2023}. We note that other non-observational spectroscopic properties, such as Raman and solid-state NMR spectra, can also be simulated using atomistic modeling \citep{DellePiane2014-JPCC, DellePiane2015-JPCC}. Here, simulated data can be compared with data from laboratory-based experimental characterization (e.g., of porous dust analogues, \citep{Dib}).

\begin{figure}[ht]
   \centering
   \includegraphics[width=0.4\textwidth]{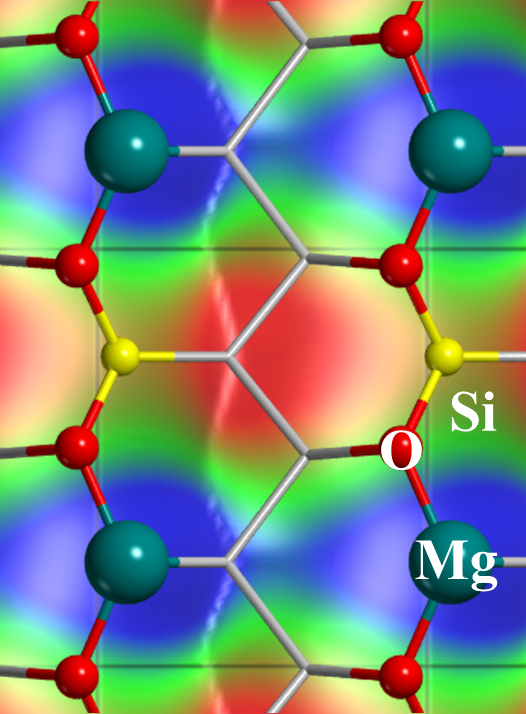}
   \caption{Electrostatic potential map corresponding to a repeat fragment of the crystalline forsterite Mg$_2$SiO$_4$ (010) surface. Blue regions represent positive electrostatic potentials, mainly centered on the Mg$^{2+}$ cations. Red regions represent negative electrostatic potentials, mainly centered on the O$^{2-}$ anions.}
         \label{fig:EPM-silicate}
\end{figure}

Besides spectroscopy, atomistic modeling is also able to simulate the adsorption and diffusion of chemical species inside porous materials. One fundamental property of great interest in astrochemistry is how strongly molecules interact with dust grain surfaces (i.e., molecular binding energies – BEs, \citep{Minissale2022}). The first step in calculating the BE of an adsorbed molecule is to consider the adsorption of a single molecule. The high surface area of interior pores walls possess several potential binding sites with varying energetic stabilities and, consequently, different BEs. Therefore, detailed sampling of the pore walls to cover all possible binding sites is of paramount importance to obtain good statistical averages. Identification of likely strong adsorption sites on pore walls can make use of electrostatic potential maps (EPMs). EPMs highlight regions of positive and negative electrostatic potentials, indicating where interactions with negative and positive charges would be strongest, respectively. Figure~\ref{fig:EPM-silicate} shows the EPM of a crystalline forsterite (010) surface \citep{Navarro-Ruiz2014}, in which the Mg$^{2+}$ cations are associated with regions of positive electrostatic potential, while O$^{2-}$  anions are associated with a negative electrostatic potential. These maps are useful for identifying binding sites that follow the electrostatic complementarity principle, such as ion/dipole interactions (e.g., between a Mg$^{2+}$ cation of the silicate surface and an O atom of a molecular adsorbate).

\begin{figure}[ht]
   \centering
   \includegraphics[width=\textwidth]{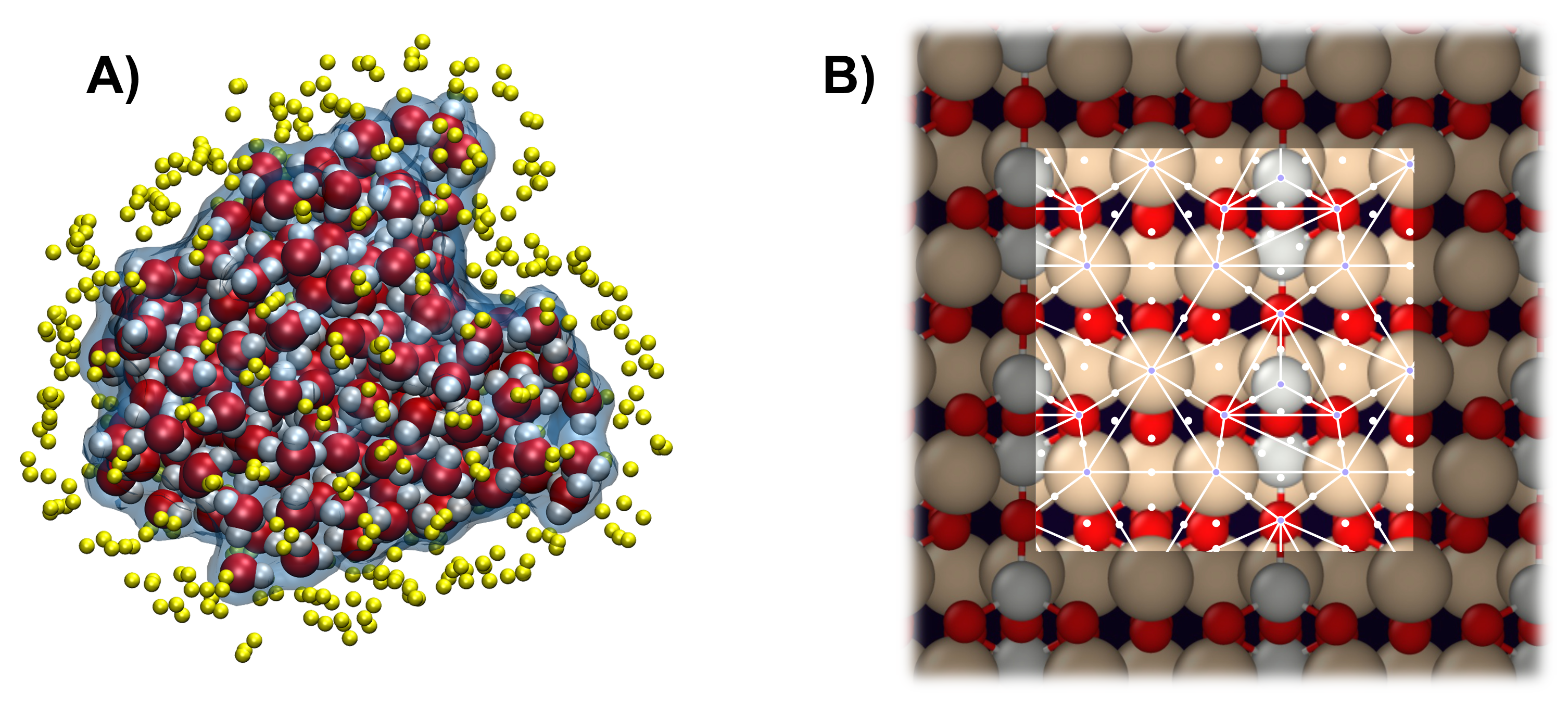}
      \caption{Automated procedures for an exhaustive sampling of binding sites: A) through the ACO-FROST program \citep{Germain2022} for large H$_2$O ice clusters, where the yellow grid points represent potential binding sites, B) through the Polycleaver program \citep{PolyCleaver} for periodic ionic surfaces (in this case the crystalline forsterite (010) surface), where the network point represent potential surface binding sites.}
         \label{fig:binding-sites}
\end{figure}

For small pore sizes, sampling of binding sites can be done manually since the number is expected to be limited. However, for large pore sizes, manual sampling can be overwhelming, and automated procedures can facilitate the task. Such automated procedures are currently implemented in programs for sampling binding sites of external surfaces of interstellar interest, such as the ACO-FROST program \citep{Germain2022} for sampling large H$_2$O ice cluster models and the Polycleaver program \citep{PolyCleaver} for sampling extended periodic ionic surfaces (see Fig.~\ref{fig:binding-sites}A and B, respectively). Applying these algorithms to porous materials can be beneficial for accurately sampling the space of possible binding sites for a single molecule.

Once a set of binding sites has been identified, the adsorbed molecule together with the porous system can be optimized using a suitable modeling method to find the most energetically stable adsorption configuration. Comparing the energy of such a configuration with those of the separate porous system and the isolated relaxed molecule allows one to calculate BEs for each adsorbed configuration. From a sampling of many such configurations a statistically reliable distribution of the BEs can be obtained. Additionally, calculating the IR spectra of the adsorption complexes can help identify any adsorption-induced vibrational shifts, which can be compared with observational and experimental data.

Exhaustive sampling of binding sites can also be useful for increasing the loading of adsorbed species, eventually forming monolayer structures of adsorbates covering the pore walls \citep{Gierada2016}. This building-up process allows for the derivation of BE variations as adsorbate loading increases. For high loading scenarios, it is recommended that these “static” optimization evaluations of BEs (effectively at 0 K) be complemented with finite temperature MD simulations. The use of MD can capture dynamic effects, which can be significant at the temperatures found in warmer astrophysical environments. For instance, for adsorption based on H-bond interactions, MD simulations can capture the dynamics of H-bonds, showing how they continuously break and reform. For adsorption based on weak dispersive forces, MD simulations allow visualization of the dynamic behavior of adsorbates on the pore wall, providing a more comprehensive atomic-scale picture beyond the static snapshot given by geometry optimizations \citep{DellePiane2016-TCA,DellePiane2013-JCTC}.

Within the context of porous dust, using atomistic modeling to sample the BEs of water molecules inside a porous silicate model would be interesting due to the likely deposition of water ice inside the pores of cosmic dust \cite{RN930}. Such modeling could be performed for a range of water loadings and the corresponding simulated IR spectra compared with of experimental data on H$_2$O ice interacting with porous silicates.

Computational codes that simulate dust grain-surface chemistry (often in tandem with gas-phase processes) are a commonly used tool that allow the chemical evolution to be studied as a function of time and physical conditions. However, there exist exceedingly few astrochemical codes that take porosity into account, either in the icy dust-grain mantles \citep[e.g.][]{Awad05,Taquet12,Garrod13,Song24} or in the grains themselves \citep{PB06,CG21}.
 
The usual computational approach taken to interstellar grain-surface chemistry is to assume a spherical grain with a single, representative radius, commonly 0.1 $\mu$m. Thus, these grains are typically intended to represent the smaller subunits of interstellar dust, rather than larger  aggregates. Upon this surface, atoms and/or molecules adsorbed from the gas phase are allowed to react, usually driven by the diffusion of relatively mobile physisorbed species such as atomic H. Under dense conditions, in which the grains are very cold, this leads to the build-up ice mantles that are chemically stable, and which are mostly composed of small molecules that include water, CO and CO$_2$. In such simulations, the surface of the grain is often implicitly assumed to be smooth, as manifested in the single-valued binding energies and diffusion barriers that are assigned individually to each chemical species on the surface. While such methods may be considered microscopically to be rather crude, they have been successful in capturing the broad behavior of grain-surface chemistry and ice-mantle compositions in a range of astrophysical environments.

In fact, multiple different approaches exist to simulate the chemistry on a technical level. These range from the kind of rate equation-based models \citep[such as the classic work by][and its successors]{Hasegawa92} that use single-valued input parameters to provide the abundances of chemical species on the average grain, to much more complex Monte Carlo kinetics treatments \citep[e.g.,][]{Chang05,Cuppen05,Garrod13} that allow microscopic surfaces to be dictated explicitly and individual atoms and molecules to be tracked on the surface. Needless to say, the latter method is far more computationally expensive than the former, and is usually limited to a very rigid set of physical conditions. Nevertheless, it is possible to take porosity into account, on some level, using either technique.

Porosity has the potential to influence the behavior of grain-surface chemical models in several important ways. Firstly, the presence of porosity increases the available surface area in comparison to the typically considered solid, spherical grain. This may slow down diffusive surface reactions, but could also result in slower (and thinner) ice-mantle build-up, leading to longer exposure of less mobile surface species (e.g.,~CO) to chemical attack by mobile reactants like atomic H. The underlying grain-material would also remain exposed for longer, leading to more rapid surface diffusion rates for some species.

Depending on the nature of the porosity, i.e., whether a more open, fractal structure, or a closed system of pocket-like enclosures, atoms and molecules residing within the pores may be restricted from desorbing effectively, and may even become entirely trapped within. This could have several effects, such as enhancing reactivity among volatile species, retaining volatile products beyond their desorption temperatures, and potentially shielding trapped product molecules from chemical or photolytic attack over long timescales (see also Sect.~\ref{sec:4.2}).

\subsubsection{Information needed to refine chemical models}

Interstellar chemical models of grain-surface chemistry on porous grains have a long way to go. While microscopically exact treatments (e.g.~{MIMICK}, above) are well suited to tackling the structural aspects of the problem, the widespread applicability of these models is severely limited by their extreme computational demands. Such models also do not, as yet, include photolysis of molecules, which could provide a clearer distinction between the molecular content of ice stored inside potentially protective grain-pore structures versus outside them. Some degree of code optimization might allow the idea of the trapping of volatiles to be tested more comprehensively.

Unfortunately, it is not clear that current treatments of porosity using simpler modeling methods are capable of accurately reproducing the formation of ices within the pores, due to their lack of microscopic detail. This would manifest in the treatment of the possible filling and enclosure of grain pores as the ices grow. Neither the porosity in the grains nor in their ice mantles is simple to simulate accurately in rate-based models. However, due to the relatively rigid, unchanging structures of dust-grain pores, as opposed to those that might be present within the ice mantles, such models might be better suited to the former than to the latter. If porosity in interstellar grains in fact manifests as a very open, fractal structure, the influence of porosity would be relatively straight forward to model, by simply adjusting the surface area available for reactions. In this scenario, trapping and shielding would likely be no more of a concern than it is with existing smooth, spherical grain models.

A clearer picture of the true degree and nature of interstellar dust-grain porosity would be a substantial motivator for the development of more detailed treatments of such features in gas-grain chemical models. Uncertain quantities include the precise degree and nature of the porosity, as discussed in this review, which is related also to the degree of connectivity within the pore structures. This connectivity, both within the pores and between the pores and the nominal outer grain surface, may be a key factor in determining the nature of the surface kinetics occurring in the pores. The models would be greatly assisted by the availability of more explicit laboratory experiments of surface chemistry on well-defined porous grain-surface analogues, which would allow simulations of both the laboratory and interstellar environments to be constructed.

\subsection{Surfaces, pores, and their impact on gas-grain chemistry}
\label{sec:4.2}

\subsubsection{Surface area}

\begin{figure}[ht]
   \centering
   \includegraphics[width=\textwidth]{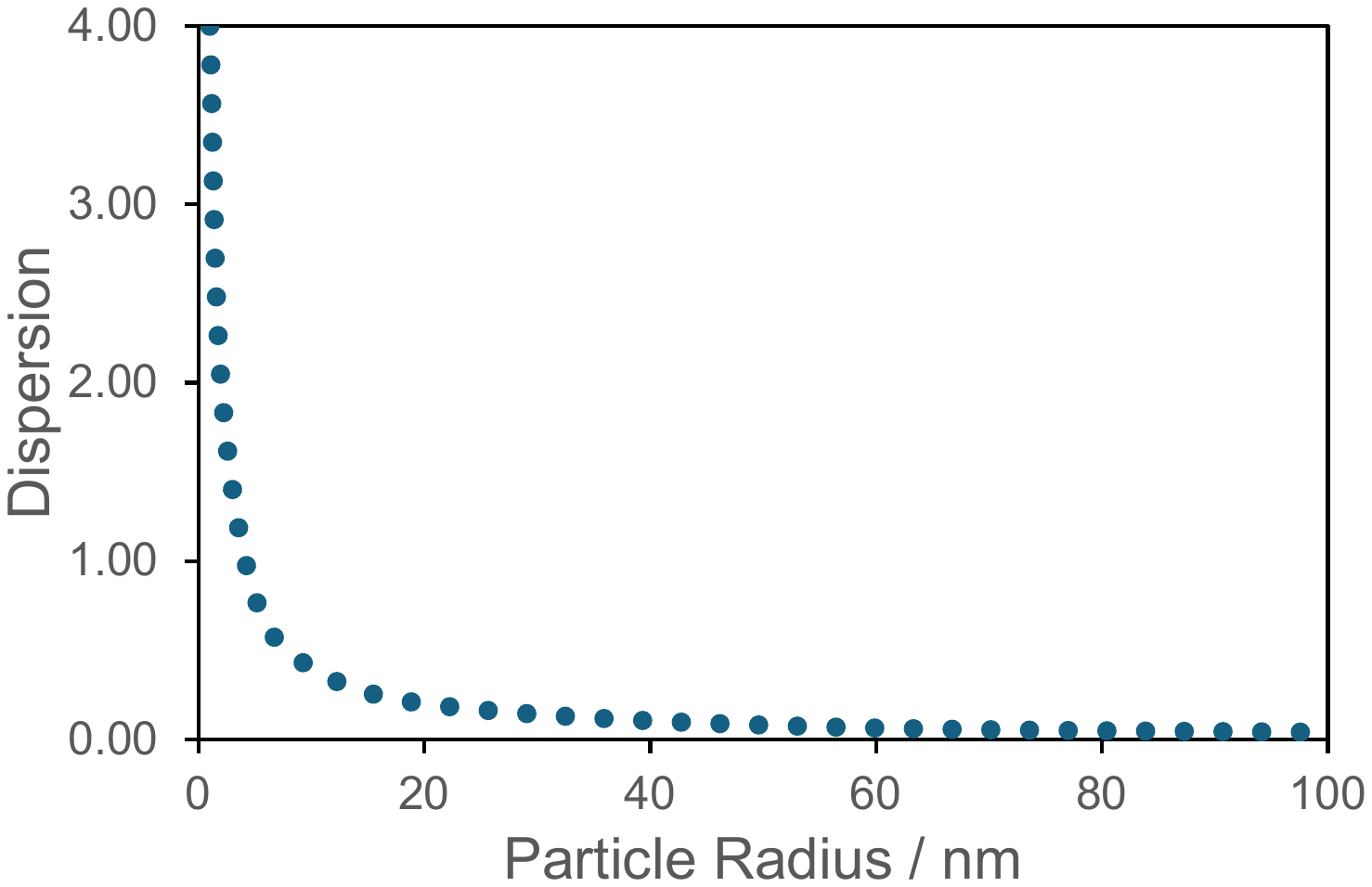}
      \caption{The effect of decreasing particle size on dispersion as given by equation 3 for a spherical particle of radius in the range given comprising of an agglomerate of 1 nm particles.}
         \label{fig:figure_D}
\end{figure}

In considering the implications of a fractal dust structure as discussed in the preceding sections, it is crucial that we understand the impact of reducing the scale of the dust particles. Figure~\ref{fig:figure_D} illustrates the issue through the concept of dispersion commonly employed in the catalysis literature \citep{RN1871}. In that literature, dispersion $D$, is defined as the ratio of the number of surface atoms ($N_S$) to the total number of atoms ($N_T$) in a particle. It is commonly used in relation to the metal particles deposited on a catalytic support:   
\begin{equation}
\label{eqn:1}
D=\frac{N_{\mathrm{S}}}{N_{\mathrm{T}}}
\end{equation}

However, we can adapt this concept to where larger particles may be agglomerates of smaller particles by considering the numbers of smaller particles comprising the larger. If we define the radius of the larger particle, r$_L$, in terms of the smaller, r$_S$, then
\begin{equation}
\label{eqn:2}
N=\frac{r_{\mathrm{L}}}{r_{\mathrm{S}}}
\end{equation}
and for spherical particles we have
\begin{equation}
\label{eqn:3}
D=\frac{4}{N} \,.
\end{equation}

In effect the dispersion gives us insight into the effective surface area of the agglomerated materials compared to a single large particle. As an illustration, for a 100 nm particle composed of 1 nm grains, we might expect the maximum total surface area of the agglomerate to be some 25 times that of a single particle, i.e., 1/D. Laboratory experiments comparing the surface areas of a compact particle and a porous fractal aggregate consisting of nm-sized monomers showed that the surface area of the aggregate is several hundred times that of a compact particle \citep{RN930}. 

There are mathematical methods allowing for estimation of a fractal dimension and a surface area of an aggregate from its geometrical parameters and interaction with light (e.g., \citealt{RN1873, RN1872}). More directly, O$_2$ adsorption can be used to characterize the surface area and porosity/morphology of grain aggregates. The technique is analogous to isothermal adsorption measurements, which are typically performed at higher temperatures \citep{RN1688, RN1689}. However, these methods have never been applied to analogues of cosmic dust particles. 

\begin{figure}[ht]
   \centering
   \includegraphics[width=0.75\textwidth]{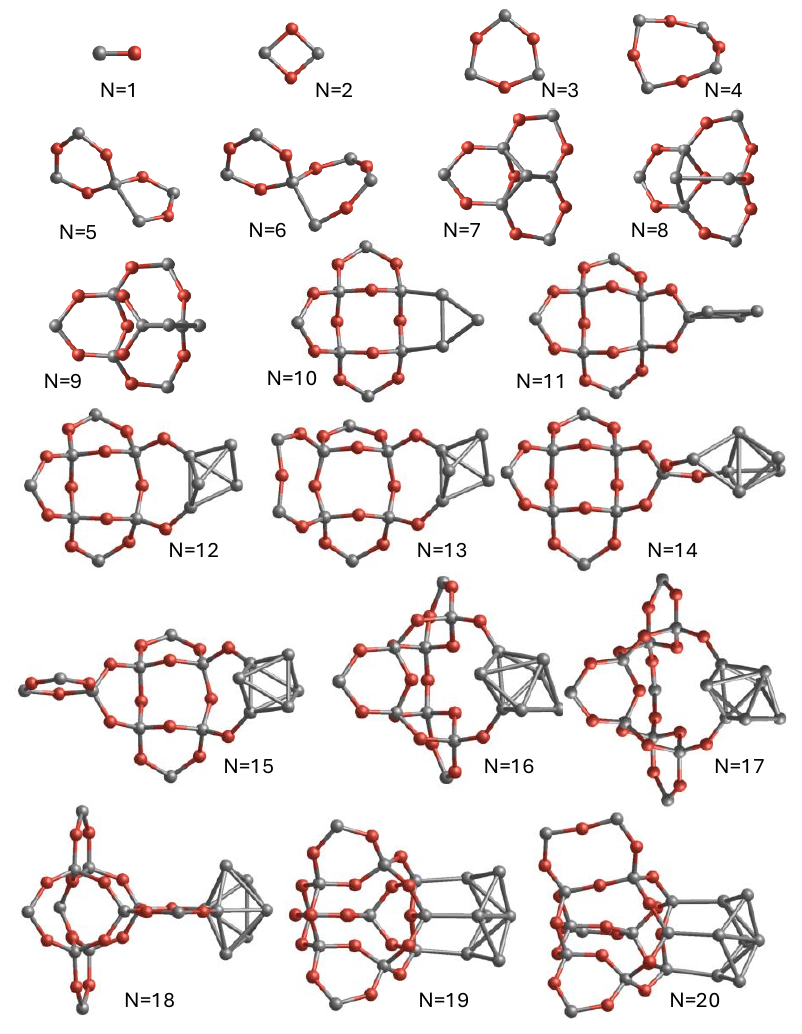}
      \caption{Small (SiO)$_N$ clusters for N = 2 -- 20. Atom color key: Si – red, O – gray. Image reproduced with permission from \citet{RN1899}, copyright by RSC.}
         \label{fig:figure_SiO_clusters}
\end{figure}

While the spherical assumption might be valid to some extent for larger particles, as grain dimensions approach the nanometre scale then particles are likely to be far from spherical as illustrated in Fig.~\ref{fig:figure_SiO_clusters} for small SiO clusters. 

This has two effects. Firstly, the sum of the small particle surface areas may be substantially larger than in the spherical limit. Additionally, and more importantly, atoms at surfaces commonly experience lower coordination than those in bulk materials and as such are naturally much more reactive. A simple comparison of the interaction of cyclohexane and benzene with an agglomerated silica cluster surface (M.R.S.\ McCoustra, unpublished data) using temperature programmed desorption methods demonstrates that the former interacts weakly with the silica surface as might be expect. While the latter is known from the literature to form a strongly bound monolayer on such surfaces \citep{RN712, RN1137}. The efficacy with which partial electron transfer occurs from $\pi$ orbitals of the benzene molecule to the low coordination number Si centres on the silica cluster mediating this process. 

In the simplest sense, increased surface area on which deposition and reaction might occur means that reactivity at the gas-grain interface might increase. This might be particularly true for surface-assisted recombination reactions (e.g., H + H $\rightarrow$ H$_2$) and reactions involving activated atoms within the grain surface (e.g., energetically-triggered formation of CO and CO$_2$ from H$_2$O on various carbon surfaces, see \citep{RN1301} for a review, where the rate of reaction scales effectively with grain surface area. However, within the Langmuir--Hinshelwood framework \rev{(considering diffusion of at least one of the reactants as a prerequisite for the reaction)} that dominates surface chemistry in space, diffusion could be slowed in an agglomerate compared to a large particle as diffusion now needs to occur between particles. As with crystallite boundaries on polycrystalline surfaces such boundaries may represent a significant barrier to diffusion and consequently slow reaction rates \citep{RN1875, RN1874, RN1887}. On the other hand, recent results demonstrated efficient diffusion of CO$_2$ and NH$_3$ molecules through (along the pores) of dust aggregates in contrast to diffusion into bulk ices of each other (Potapov, unpublished data).

\rev{When considering the additional surface area that porosity may impart to dust grains, it is necessary to consider both intrinsic porosity (associated with the material itself) and extrinsic porosity (associated with the morphology of the dust grain agglomerate). Intrinsic porosity within the materials will add to the surface area available for reactions if the pores are connected to the surface. Porosity as closed off void space will not add to the available surface area. Extrinsic porosity is related to the morphology of the agglomerate. Extended agglomerates where small particles are held together in simple chains may possess some extrinsic porosity in crevices around the contact points between the agglomerated grains, but these are unlikely to add significantly to the surface area of the agglomerate. In more compact structures, the impact of pores and voids on the surface area of the agglomerate may be significant. It is worth noting that both types of porosity may act, under the correct circumstances, to store gaseous reagents beyond their normal desorption temperatures enabling reactions on pore surfaces at higher temperatures and local pressures \citep{RN1867}.}

In relation to intrinsic porosity in the materials from which compose grains, we are considering silicates from the two main families, (FeO)$_x$(MgO)$_y$(SiO$_2$)$_z$ and (FeO)$_x$(Al$_2$O$_3$)$_y$(SiO$_2$)$_z$, and carbonaceous materials of various forms. These materials form in the outflows of large AGB stars under near thermal equilibrium conditions. The silicates may form as either crystalline or glassy solids which likely have negligible intrinsic porosity. Amorphisation by cosmic radiation may introduce defects to the lattice but generally the intrinsic porosity will remain low. In contrast, water alteration of these materials in specific environments where this is possible can produce materials that exhibit lamellar (clay) and cage (zeolite) structures that will introduce nanoscale porosity to material (e.g., \citealt{RN1903, RN1902}). Of course, these altered materials are equally subject to radiation alteration that will seek to remove the nanoscale porosity and return the material to a glassy state \citep{RN1869, RN1868}.  

In terms of carbon materials, intrinsic porosity is well-known specific systems, e.g., within fullerenes and nanotubes and the lamellar structure of graphite. Materials like diamond and hydrogenated amorphous carbon are, however, less likely to show such intrinsic porosity. The extrinsic porosity of agglomerated carbon grains will likely parallel that in silicate grains.

\subsubsection{Catalysis}
High dust grain porosity and a correspondingly large surface area as discussed above has an important consequence - the availability of active dust grain surfaces for physico-chemical processes in various astrophysical environments. There is no question that dust surfaces are available in low-density (diffuse interstellar clouds) and warm/hot (atmospheres of exoplanets, protostellar envelopes and protoplanetary disks inside the  snowline) environments due to the absence of molecular ices, which might cover dust grains and thus make the dust surface material unavailable for surface processes. However, the scientific community is still far from reaching a consensus regarding cold and dense environments, such as dense interstellar clouds, protostellar envelopes and protoplanetary disks beyond the snowline. The widely accepted view of grains in such environments presents grains as a compact refractory dust core covered by a thick (tens or hundreds of monolayers) ice layer. In such a case, processes occurring on and in the ice are independent of the properties of the dust material. However, the aforementioned study \citep{RN930} demonstrated that porous grains in cold environments can be covered by only a sub-monolayer or a few monolayers of ice, especially if that ice can agglomerate \citep{RN955}. Thus, bare dust grain surfaces are potentially available for chemical reactions in a wide range of environments. In such a case, the dust can actively participate in surface processes and, depending on the properties of the dust material, may play a role of a classical catalysis. This new view was also presented and discussed in \citep{RN1301}. 

While there remains much to explore in terms of grain surface photochemistry, our knowledge of that field is substantive in comparison to our knowledge of catalytic processes (i.e., processes in which the grain reduces the reaction and/or diffusion activation barrier accelerating the reaction) on grain surfaces. The previous research of the catalytic formation of simple molecules, such as H$_2$ and H$_2$O, and more complex species, such as NH$_4$$^+$NH$_2$COO$^-$, on dust grain analogues was recently reviewed \citep{RN1301}. Here, we highlight ideas in that area that have recently opened new doors on astrocatalysis. 

(1)	Silicate grains contain iron and other catalytically active transition metals (e.g., Ni, Co, Mn). It is well established that space weathering is known to nucleate the growth of iron (and other metal) nanoparticles in the surface regions of grains  (e.g., \citealt{RN1704, RN1703, RN1705}). Iron and its neighbours are well known to catalyse key industrial processes from the Haber--Bosch synthesis of ammonia, through Fischer--Tropsch chemistry for alkane and alcohol formation from H$_2$ and CO to chemistries tied to the bottom-up synthesis of carbon nanomaterials (PAHs and nanotubes specifically) from simple unsaturated hydrocarbons like acetylene and ethylene \citep{RN1888}.  

(2)	Carbonaceous and silicate materials can both behave as substrates for the deposition of gas transition metal atoms introduced in the gas phase by sputtering in shocked regions. Recent computational work by Rimola and co-workers nicely illustrates the potential for such single atoms to engage in catalytic chemistry \citep{RN1752, RN1753}.

(3)	Nanostructured carbons and transition metal clusters are metallic in nature. This opens a wealth of potential chemistry linked to the generation of electron-hole pairs (which dominate photochemistry on metals), and cavity-enhancement thereof, which is known to be highly efficient \citep{RN1811, RN1810}.

These factors represent opportunities for exploration by the solid state and surfaces laboratory astrophysics community and link it very strongly to the catalysis community. 

\subsubsection{Relaxation}
It is well known that surfaces effectively behave as a third body in atom and radical recombination reactions on grains by providing pathways enabling electronic and vibrational relaxation of the nascent reaction product. The formation of H$_2$ from physisorbed H atoms on a variety of surfaces nicely illustrates this effect \citep{RN911}. \rev{H$_2$ formation on graphite surfaces also nicely illustrates that unless the internal (ro-vibrational) degrees of freedom are strongly coupled into the surface then relaxation at the surface may not be as efficient as expected} \citep{RN1889, RN1890}.

\rev{Grain surfaces also provide a pathway to relaxing vibrationally and electronically excited species that either impact on surfaces from the gas phase or are already adsorbed on surfaces and excited. Of course, the more surface is available, the more effective are these processes. In both instances, however, relaxation is fundamentally governed by Fermi’s Golden Rule and the energy–time Uncertainty Principle. Under-pinned by the nature of the electronic states and the conical intersections that link them, these ideas lead us to the Jablonski Diagram, determine spectroscopic line profiles, and yield symmetric Lorentzian profiles for transitions in isolated molecules \citep{RN1877}. Efficient nonradiative mechanisms reduce electronic and vibrational excited state lifetimes and hence homogeneously, and symmetrically, broaden linewidths often resulting in Gaussian or Voigt line profiles. In the gas phase, photon-driven processes, temperature (average molecular speed – Doppler Broadening) and pressure (collision frequency – Pressure Broadening) are the principle broadening mechanisms resulting in homogeneous broadening of gas phase line profiles \citep{RN1878}. On solid surfaces, molecular translation and rotation are frustrated and weakly interacting adsorbates on simple insulating solids can therefore exhibit relatively narrow linewidths. The classic example of this is found in the work of Ewing and co-workers \citep{RN1879, RN1880}. These authors estimate the natural linewidth for the vibration of carbon monoxide (CO) on the NaCl (100) surface at 5 K to be of the order of 10$^-$$^8$ cm$^-$$^1$, and suggest that the observed linewidth, 0.07 cm$^-$$^1$, results from the residual heterogeneity of the NaCl surface. In contrast, on conductor surfaces, transition dipoles couple with the free electrons in the conductor band structure providing an efficient mechanism for relaxing excited adsorbates via electron – hole pair creation \citep{RN1881, RN1882}. The resulting line profiles are broad, for example varying from around 5 to 15 cm$^-$$^1$ for CO on conductor single crystal surfaces, and asymmetric, reflecting Fano coupling between the adsorbate vibration and the non-adiabatic electron–hole pair continuum of the substrate \citep{RN1883, RN1884}. Additionally, these systems, as with halide surfaces, are often subject to further broadening due to environmental heterogeneity  \citep{RN1885}. Environmental heterogeneity reaches its extreme with the profiles of vibrational lines of adsorbates interacting with supported metal and metal oxide catalysts \citep{RN1885} as which silicate minerals might be considered.}  

An interesting and relevant example from McCoustra and co-workers illustrates the potential for exploring the impact of surface heterogeneity and adsorbate photo-stability on surfaces in the IR through studies of line shapes for CO on amorphous silica and amorphous solid water \citep{RN1155, RN1159}. Herein, evidence is presented from the similarity of CO line profiles in these two systems, the origins of these line profiles are indeed different, with the former likely associated with an efficient IR photon-induced desorption channel and the latter with efficient vibrational energy relaxation into the amorphous solid water phonon bath.

\subsubsection{Shielding}
While the efficacy or otherwise of surface relaxation maybe crucial to defining the stability of species adsorbed on surfaces, we should recognise the simpler shielding role of the corresponding bulk beneath the surface. Both silicate and carbonaceous materials directly absorb radiation at a variety of wavelengths and especially in the short wavelength ultraviolet. As such adsorbate species are only exposed to destructive radiations from the 2$\pi$ steradians above the surface on which the adsorbate sits. Access to adsorbates by from the remaining 2$\pi$ steradians below the surface on which the adsorbate sits in attenuated by the scattering and adsorption of the substrate itself. Moreover, grain surface molecules, potentially including complex organic molecules, formed at early stages of the evolution of cosmic environments, i.e., in diffuse interstellar clouds (e.g., \citealt{RN602, RN1718, RN1324, RN1663}), can be shielded from external radiation fields due to the growth (aggregation) of the grains and, thus, survive during the travel of the grains from the ISM though protostellar envelopes and protoplanetary disks to planetary system and atmospheres of planets.   

\subsection{Planet formation, binding sites, and origin of water}
The efficiency of the formation of planets in protoplanetary disks depends on the amount of solid-state material available at locations of future planets \citep{RN1059, RN1060}. Outside of the water snowline, the density of material is enhanced due to the presence of water (the main constituent of ices in cold environments) and other molecular ices \citep{Hayashi1981}. Moreover, coagulation models \citep{RN900, RN1898, RN1221} and laboratory experiments \citep{RN1651, RN935} confirm that sticking efficiency of water-ice-coated dust grains is much higher as compared to rocky (iceless) grains. In addition, simulations demonstrated that high porosity of icy aggregates triggers significant acceleration in collisional growth \citep{RN1865, RN788, Kobayashi21}, although these models are being challenged as ALMA observations favor grains with a modest amount of porosity (see Sect.~\ref{sec:ppdobs}). Both factors are argued to have facilitated the formation of giant planets in the Solar System \citep{Stevenson1988, RN1866}, with the snowline located within the asteroid belt at about 3 AU from the Sun \citep{Morbidelli2000}.

The amount of solid mass available for planet formation beyond the snowline as well as the efficiency of dust coagulation as linked to the stickiness of ices are still being debated. Regarding the amount of solid mass, we refer the interested reader to \citep{RN1864} and references therein. Regarding the efficiency of dust coagulation, for water ice the efficiency strongly depends on the amount of water (water ice thickness) on dust grains \citep{RN1221}. The situation is worse if the CO$_2$ ice forms an outer ice layer on dust grains. It was demonstrated by laboratory experiments \citep{RN1859, RN1860} and models \citep{RN1861} that the stickiness of CO$_2$ ice is much lower as compared to H$_2$O ice. This result was favored in observational modeling of protoplanetary disks \citep{RN1858, RN1862}. We note that the case of an outer CO$_2$ ice layer is relevant to colder parts of protoplanetary disks, beyond the snowline of CO$_2$ located at about 10 AU from the Sun.  

The above discussion focuses on a general statement regarding planet formation and here porosity may also play
an important role, in particular with regards to the water content of the Earth which formed inside the nebular snow line.
In the laboratory experiments studying thermal desorption of ices, water ice desorbs completely at 160 – 180 K \citep{RN1142, RN721, RN1224}. This temperature defines the location of the snowline.  It is typically assumed that dust grains inside the snowline are dry, which motivates the need for water supply to the Earth from beyond the snowline perhaps via dynamically reshuffling of solids in the nebula due to the presence of, e.g., Jupiter \citep{Raymond2004, Rubie2015, Krijt2023}. However, more realistic laboratory experiments modeling dust grains in cold environments as dust/ice mixtures have demonstrated that a considerable part of water molecules mixed with silicate grains is trapped (strongly bound) on grains at temperatures exceeding the desorption temperature of water ice \citep{RN695, RN715, RN873, RN1715}. The amount of trapped water is directly defined by the number of binding sites, which in turn is defined by the surface area of grains which is enhanced with porous grains with variable sub-structure. The results of \citep{RN1715} demonstrated that the trapped water is physisorbed as water molecules rather than dissociatively chemisorbed (i.e. as chemically bonded -OH groups). These water molecules can persist up to about 470 K, which is well above the desorption temperature of water ice. This implies the existence of sites in porous silicates where water molecules can be very strongly bound. As mentioned in section 3.4.3, models of small molecules physisorbed inside micropores show that the molecular binding in such cases can be up to eight times larger than on the corresponding flat surface \citep{Derouane1988}. Temperature-dependent IR experiments on microporous silicate zeolites also show that water molecules can be trapped inside micropores up to at least 423 K \citep{RN1891}. These results strongly imply that micropores in porous silicate grains could strongly trap water molecules and retain them during planetary accretion processes at relatively high temperatures.

\rev{Water molecules strongly bound (trapped) in refractory dust grains may be present in the solid state in astrophysical environments, where we do not expect them, e.g., in the diffuse ISM, in protoplanetary disks inside the water snowline, and in the warm/hot atmospheres of exoplanets. Comparison of laboratory and observational spectra provided evidence of the presence of solid-state water in the diffuse ISM \citep{RN873}. Comparison of laboratory spectra and spectra of cometary particles and meteorites provided evidence for the presence of trapped water in extraterrestrial particles \citep{RN1504}. Analysis of recent JWST observations provided evidence of water in the solid state inside the water snowline in a protoplanetary disk (Potapov et al., submitted). The estimated concentration of H$_2$O is similar to that of the present-day bulk Earth and consistent with recent measurements of enstatite chondrite meteorites. This result, if generalized, would imply that most habitable zone planets are born with water budgets comparable to Earth’s.}

The microphysics of how porous grains interact with radiation may also influence the volatile composition of planet forming materials.   For instance \citep{Najita2001} noted the prevalence of energetic stellar X-ray emission that irritates dust and gas in protoplanetary disks \citep{Glassgold1997, Feigelson1999}.  They showed that when an X-ray is absorbed by an aggregate comprised of smaller particles that the energy transfer from a sub-unit to the whole grain is impeded leading to a higher local temperature for a longer period of time.  This potentially can foster local non-thermal desorption of volatile ices via ``spot heating'', which is centrally dependent on the structure of the silicate or caronaceous refractory dust.   A similar process may be present when cosmic rays, which may also be present \citep{Cleeves2013, Padovani2018}, impact volatile ice coated dust grains \citep{Sipila2021}.

Thus, the consideration of porous, fractal grains may change our understanding of the mechanisms and efficiencies of grain growth and volatile incorporation into planetesimals/pebbles and rocky planets inside disk snow lines, whether for water and organics or more volatile components such as CO or CO$_2$.   This can ultimately influence the volatile content of forming bodies.  This ranges from the water content of the Earth, a central component of its habitability, but also the materials present in other planets, asteroids, and comets.

\section{Conclusions}
\rev{Dust porosity is a crucial factor in understanding physicochemical and structural properties of dust itself as well as the evolution of dust in astrophysical environments. However, while results of laboratory experiments and dust evolution models clearly speak for high porosity of cosmic dust, results of the analysis of astronomical observations are more ambiguous. One of the problems is that models that describe observations have many degenerate parameters, e.g., porosity can be “compensated” for by a “grain size distribution”. While astronomical observations of the ISM in general do not contradict results of grain growth models and laboratory experiments, astronomical observations of planet-forming disks are more diverse.}

\rev{However, if we integrate results of the analysis of observations of different environments from the ISM though protoplanetary disks to debris disks and the Solar System, we see a logical trend, which, together with the experimental and modeling results, draw a clear picture of increasing porosity in increasingly evolved environments. The sequential growth of dust grains travelling from the diffuse ISM to the dense ISM and then to protoplanetary disks and planetary systems is accompanied by an increase of the sizes of compact grain units (monomers) and by an evolution of their porosity, which is relatively low at the beginning of the journey and very high at its end before the processes leading to the formation of planetesimals.}

\rev{There are questions immediately raised by this picture. What are the reasons for the observed diversity of astronomical observations? Is it linked to different conditions in the astrophysical environments or imperfect dust models and/or laboratory data? What is the origin of low porosity in the diffuse ISM, considering fractal growth starting from nm-sized particles? Is the degree of porosity dependent on whether grain growth in a particular environment occurs through aggregation of pre-formed subunits, or at the atomic/molecular scale (particularly as the grain core nucleates)? What are mechanisms leading to growth of compact grain units (monomers) with grain growth? Is not this result biased by the Solar System measurements?} 

\rev{There are many more open questions and considerable gaps in our knowledge. Moreover, to understand the evolution of dust and planet formation process, it is crucial to establish links between the dust particles observed in the ISM, in planet-forming disks, and in the Solar System. It is clear that observations, models, and laboratory experiments should be combined and work together to understand the complete picture of the dust evolution in space. In the following, we try to formulate a few promising research directions in observations, experiments, and modelling covering a desired science program on cosmic dust porosity.}

\rev{\emph{Observations:} To push forward our understanding of dust porosity in disks, careful studies of light scattering properties and its application to the actual disk observations are mandatory. Multiwavelength polarimetric observations of protoplanetary disks are required across a broad range of targets. Infrared observations will also be valuable for constraining grain composition, which in turn can help to solve for dust parameter degeneracy. More statistics is required for the Solar System cometary particles. For this, new sample-return missions to comets are crucial.}

\rev{\emph{Experiments:} Laboratory optical data for porous dust grains and their mixtures with molecular ices, at various temperatures from 10s (ISM) to 100s (disks, exoplanets) of K are required for reliable analysis of observations. Study of the influence of space weathering on the optical, morphological and chemical properties of analogues of cosmic dust grains would provide an important piece of puzzle of the evolution of cosmic dust grains in astrophysical environments and links between the dust particles in the ISM, planet-forming disks, the Solar System. Influence of high temperatures relevant to the exoplanet atmospheres on the dust grain growth and the porosity, chemical activity and optical properties of grains represent another important but practically uncharted territory. Considering a high porosity and a correspondingly large surface area of cosmic dust grains, the studies of physical-chemical surface processes relevant to astrophysical environments should be continued with an impact on laboratory experiments involving dust surfaces.}

\rev{\emph{Modelling:} New advanced models of porous dust grains using new laboratory data mentioned above are required for the analysis of modern multiwavelength observations of various astrophysical environments and for understanding the origin of the diversity in porosity. Radiative transfer modeling using 3D radiative transfer models coupled with a sophisticated light scattering model for porous grains/aggregates should be applied for a broad range of protoplanetary disks. Structural modeling of porous cosmic dust beginning from atomistic level is required for characterization of cosmic dust and its laboratory analogues. Considering a high porosity and a correspondingly large surface area of cosmic dust grains, the studies of physical-chemical surface processes relevant to astrophysical environments should be continued with an impact on astrochemical modeling involving dust surfaces.}

\rev{But even having possibility to realize all the ideas above, there are two big elephants in the room: (i) how to describe porosity in a model and (ii) how to combine optical constants appropriately. Optical scattering theory is solid but perhaps we need to use more Monte Carlo - based methods as we do with electrons rather than ray tracing methods to integrate transmission, absorption and scattering into a single picture. Moreover, now, in the era of quickly developing artificial intelligence, we should look for potential for a unified approach that uses machine learning to address these problems. With this review, we hope to give a start to a broad discussion on cosmic dust porosity and methods of its investigation aiming at better understanding of the properties and evolution of space.}

\backmatter
\bmhead{Acknowledgements}

We are grateful to Akimasa Kataoka, who stood at the origin of this review. We also thank Julien Milli who provided us the FITS image and the SPF data shown in Fig.~\ref{fig:milli17}. This project was supported by the Deutsche Forschungsgemeinschaft (Heisenberg Grant PO 1542/7-1, AP). MRSM acknowledges funding from UK Research and Innovation (UKRI) Engineering and Physical Sciences Research Council (EPSRC) for the project Astrocatalysis: In Operando Studies of Catalysis and Photocatalysis of Space-abundant Transition Metals (Grant Number EP-W023024-1). RT benefitted from funding from the European Research Council (ERC) under the European Union's Horizon Europe research and innovation program (grant agreement No. 101053020, project Dust2Planets, PI F. M\'enard) and was also supported by JSPS KAKENHI (Grant Number JP25K07351, JP25K01049). EAB acknowledges support from NASA's Emerging Worlds Program (Grant 80NSSC20K0333) and the Exoplanets Research Program (grant 80NSSC20K0259). STB Acknowledges support from the MICINN-funded PID2021-127957NB-I00 and TED2021-132550B-C21 project grants, and the Maria de Maeztu program for Spanish Structures of Excellence (CEX2021-001202-M) and project grant 2021SGR00354 funded by the Generalitat de Catalunya. RTG thanks the National Science Foundation for funding through the Astronomy \& Astrophysics program (Grant Number 2206516). This project has received funding within the European Union's Horizon 2020 research and innovation program from the European Research Council (ERC) for the project ``Quantum Chemistry on Interstellar Grains'' (Quantumgrain), grant agreement No. 865657 (AR). The Spanish MICINN with the projects PID2021-126427NB-I00 and CNS2023-144902 is acknowledged (AR).

\section*{Declarations}
\bmhead{Competing interests} The authors declare no competing interests.

\phantomsection
\addcontentsline{toc}{section}{References}
\bibliography{z} 

\begin{thebibliography}{512}
\providecommand{\natexlab}[1]{#1}
\providecommand{\url}[1]{{#1}}
\providecommand{\urlprefix}{URL }
\providecommand{\doi}[1]{\url{https://doi.org/#1}}
\providecommand{\eprint}[2][]{\url{#2}}
 \bibcommenthead

\bibitem[{{Acke} et~al.(2009){Acke}, {Min}, {van den Ancker}, {Bouwman}, {Ochsendorf}, {Juhasz}, and {Waters}}]{Acke09}
{Acke} B, {Min} M, {van den Ancker} ME, et~al (2009) {On the interplay between flaring and shadowing in disks around Herbig Ae/Be stars}. \aap 502(2):L17--L20. \doi{10.1051/0004-6361/200912728}, {\href{https://arxiv.org/abs/0907.2102}{{arXiv:0907.2102}}} {[astro-ph.SR]}

\bibitem[{Allamandola et~al.(1999)Allamandola, Bernstein, Sandford, and Walker}]{RN714}
Allamandola LJ, Bernstein MP, Sandford SA, et~al (1999) {Evolution of interstellar ices}. \ssr 90(1-2):219--232. \doi{10.1023/A:1005210417396}

\bibitem[{Altobelli et~al.(2016)Altobelli, Postberg, Fiege, Trieloff, Kimura, Sterken, Hsu, Hillier, Khawaja, Moragas-Klostermeyer, Blum, Burton, Srama, Kempf, and Gruen}]{RN1909}
Altobelli N, Postberg F, Fiege K, et~al (2016) {Flux and composition of interstellar dust at {Saturn} from {Cassini}'s Cosmic Dust Analyzer}. Science 352(6283):312--318. \doi{10.1126/science.aac6397}

\bibitem[{{Anderson} et~al.(1996){Anderson}, {Weitenbeck}, {Code}, {Nordsieck}, {Meade}, {Babler}, {Zellner}, {Bjorkman}, {Fox}, {Johnson}, {Sanders}, {Lupie}, and {Edgar}}]{1996AJ....112.2726A}
{Anderson} CM, {Weitenbeck} AJ, {Code} AD, et~al (1996) {Ultraviolet Interstellar Polarization of Galactic Starlight.I.Observations by the Wisconsin Ultraviolet Photo Polarimeter Experiment}. \aj 112:2726. \doi{10.1086/118217}

\bibitem[{{Andersson} et~al.(2015){Andersson}, {Lazarian}, and {Vaillancourt}}]{Andersson2015}
{Andersson} BG, {Lazarian} A, {Vaillancourt} JE (2015) {Interstellar Dust Grain Alignment}. \araa 53:501--539. \doi{10.1146/annurev-astro-082214-122414}

\bibitem[{Arakawa and Krijt(2021)}]{RN1861}
Arakawa S, Krijt S (2021) {On the stickiness of {CO} and {HO} ice particles}. \apj 910(2):130. \doi{10.3847/1538-4357/abe61d}

\bibitem[{{Arakawa} and {Nakamoto}(2016)}]{Arakawa16}
{Arakawa} S, {Nakamoto} T (2016) {Rocky Planetesimal Formation via Fluffy Aggregates of Nanograins}. \apjl 832(2):L19. \doi{10.3847/2041-8205/832/2/L19}, {\href{https://arxiv.org/abs/1611.03859}{{arXiv:1611.03859}}} {[astro-ph.EP]}

\bibitem[{{Arakawa} et~al.(2023){Arakawa}, {Okuzumi}, {Tatsuuma}, {Tanaka}, {Kokubo}, {Nishiura}, {Furuichi}, and {Nakamoto}}]{Arakawa23}
{Arakawa} S, {Okuzumi} S, {Tatsuuma} M, et~al (2023) {Size Dependence of the Bouncing Barrier in Protoplanetary Dust Growth}. \apjl 951(1):L16. \doi{10.3847/2041-8213/acdb5f}, {\href{https://arxiv.org/abs/2306.04070}{{arXiv:2306.04070}}} {[astro-ph.EP]}

\bibitem[{{Arnold} et~al.(2019){Arnold}, {Weinberger}, {Videen}, and {Zubko}}]{Arnold19}
{Arnold} JA, {Weinberger} AJ, {Videen} G, et~al (2019) {The Effect of Dust Composition and Shape on Radiation-pressure Forces and Blowout Sizes of Particles in Debris Disks}. \aj 157(4):157. \doi{10.3847/1538-3881/ab095e}, {\href{https://arxiv.org/abs/1902.10183}{{arXiv:1902.10183}}} {[astro-ph.EP]}

\bibitem[{{Arnold} et~al.(2022){Arnold}, {Weinberger}, {Videen}, and {Zubko}}]{Arnold22}
{Arnold} JA, {Weinberger} AJ, {Videen} G, et~al (2022) {Stumbling over Planetary Building Blocks: AU Microscopii as an Example of the Challenge of Retrieving Debris-disk Dust Properties}. \apj 930(2):123. \doi{10.3847/1538-4357/ac63a9}, {\href{https://arxiv.org/abs/2105.12264}{{arXiv:2105.12264}}} {[astro-ph.EP]}

\bibitem[{Arnolds(2011)}]{RN1882}
Arnolds H (2011) {Vibrational dynamics of adsorbates - {Quo} vadis?} Prog Surface Sci 86(1-2):1--40. \doi{10.1016/j.progsurf.2010.10.001}

\bibitem[{{Arriaga} et~al.(2020){Arriaga}, {Fitzgerald}, {Duch{\^e}ne}, {Kalas}, {Millar-Blanchaer}, {Perrin}, {Chen}, {Mazoyer}, {Ammons}, {Bailey}, {Barman}, {Bulger}, {Chilcote}, {Cotten}, {De Rosa}, {Doyon}, {Esposito}, {Follette}, {Gerard}, {Goodsell}, {Graham}, {Greenbaum}, {Hibon}, {Hom}, {Hung}, {Ingraham}, {Konopacky}, {Macintosh}, {Maire}, {Marchis}, {Marley}, {Marois}, {Metchev}, {Nielsen}, {Oppenheimer}, {Palmer}, {Patience}, {Poyneer}, {Pueyo}, {Rajan}, {Rameau}, {Rantakyr{\"o}}, {Ruffio}, {Savransky}, {Schneider}, {Sivaramakrishnan}, {Song}, {Soummer}, {Thomas}, {Wang}, {Ward-Duong}, and {Wolff}}]{Arriaga20}
{Arriaga} P, {Fitzgerald} MP, {Duch{\^e}ne} G, et~al (2020) {Multiband Polarimetric Imaging of HR 4796A with the Gemini Planet Imager}. \aj 160(2):79. \doi{10.3847/1538-3881/ab91b1}, {\href{https://arxiv.org/abs/2006.06818}{{arXiv:2006.06818}}} {[astro-ph.EP]}

\bibitem[{Atkins and Friedman(2005)}]{RN1877}
Atkins P, Friedman R (2005) {Molecular Quantum Mechanics}. Oxford University Press, Oxford, 4th edn

\bibitem[{{Augereau} and {Beust}(2006)}]{Augereau06}
{Augereau} JC, {Beust} H (2006) {On the AU Microscopii debris disk. Density profiles, grain properties, and dust dynamics}. \aap 455(3):987--999. \doi{10.1051/0004-6361:20054250}, {\href{https://arxiv.org/abs/astro-ph/0604313}{{arXiv:astro-ph/0604313}}} {[astro-ph]}

\bibitem[{{Augereau} et~al.(1999){Augereau}, {Lagrange}, {Mouillet}, {Papaloizou}, and {Grorod}}]{Augereau99}
{Augereau} JC, {Lagrange} AM, {Mouillet} D, et~al (1999) {On the HR 4796 A circumstellar disk}. \aap 348:557--569. \doi{10.48550/arXiv.astro-ph/9906429}, {\href{https://arxiv.org/abs/astro-ph/9906429}{{arXiv:astro-ph/9906429}}} {[astro-ph]}

\bibitem[{{Avenhaus} et~al.(2014){Avenhaus}, {Quanz}, {Meyer}, {Brittain}, {Carr}, and {Najita}}]{Avenhaus14HD100546}
{Avenhaus} H, {Quanz} SP, {Meyer} MR, et~al (2014) {HD100546 Multi-epoch Scattered Light Observations}. \apj 790(1):56. \doi{10.1088/0004-637X/790/1/56}, {\href{https://arxiv.org/abs/1405.6120}{{arXiv:1405.6120}}} {[astro-ph.SR]}

\bibitem[{{Avenhaus} et~al.(2018){Avenhaus}, {Quanz}, {Garufi}, {Perez}, {Casassus}, {Pinte}, {Bertrang}, {Caceres}, {Benisty}, and {Dominik}}]{Avenhaus18}
{Avenhaus} H, {Quanz} SP, {Garufi} A, et~al (2018) {Disks around T Tauri Stars with SPHERE (DARTTS-S). I. SPHERE/IRDIS Polarimetric Imaging of Eight Prominent T Tauri Disks}. \apj 863(1):44. \doi{10.3847/1538-4357/aab846}, {\href{https://arxiv.org/abs/1803.10882}{{arXiv:1803.10882}}} {[astro-ph.SR]}

\bibitem[{{Awad} et~al.(2005){Awad}, {Chigai}, {Kimura}, {Shalabiea}, and {Yamamoto}}]{Awad05}
{Awad} Z, {Chigai} T, {Kimura} Y, et~al (2005) {New Rate Constants of Hydrogenation of CO on H$_{2}$O-CO Ice Surfaces}. \apj 626(1):262--271. \doi{10.1086/429856}

\bibitem[{{Ballering} et~al.(2016){Ballering}, {Su}, {Rieke}, and {G{\'a}sp{\'a}r}}]{Ballering16}
{Ballering} NP, {Su} KYL, {Rieke} GH, et~al (2016) {A Comprehensive Dust Model Applied to the Resolved Beta Pictoris Debris Disk from Optical to Radio Wavelengths}. \apj 823(2):108. \doi{10.3847/0004-637X/823/2/108}, {\href{https://arxiv.org/abs/1605.01731}{{arXiv:1605.01731}}} {[astro-ph.EP]}

\bibitem[{Bannwarth et~al.(2021)Bannwarth, Caldeweyher, Ehlert, Hansen, Pracht, Seibert, Spicher, and Grimme}]{GFN-xTB}
Bannwarth C, Caldeweyher E, Ehlert S, et~al (2021) {Extended tight-binding quantum chemistry methods}. Wiley Interdiscip Rev Comput 11(2):e1493. \doi{10.1002/wcms.1493}

\bibitem[{{Bazell} and {Dwek}(1990)}]{Bazell90}
{Bazell} D, {Dwek} E (1990) {The Effects of Compositional Inhomogeneities and Fractal Dimension on the Optical Properties of Astrophysical Dust}. \apj 360:142. \doi{10.1086/169104}

\bibitem[{{Becke}(1993)}]{B3-1993}
{Becke} AD (1993) {Density-functional thermochemistry. III. The role of exact exchange}. \jcp 98(7):5648--5652. \doi{10.1063/1.464913}

\bibitem[{{Bell} et~al.(2015){Bell}, {Mamajek}, and {Naylor}}]{Bell15}
{Bell} CPM, {Mamajek} EE, {Naylor} T (2015) {A self-consistent, absolute isochronal age scale for young moving groups in the solar neighbourhood}. \mnras 454(1):593--614. \doi{10.1093/mnras/stv1981}, {\href{https://arxiv.org/abs/1508.05955}{{arXiv:1508.05955}}} {[astro-ph.SR]}

\bibitem[{Bennett et~al.(2013)Bennett, Pirim, and Orlando}]{RN1603}
Bennett CJ, Pirim C, Orlando TM (2013) {Space-Weathering of Solar System Bodies: A Laboratory Perspective}. Chem Rev 113(12):9086. \doi{10.1021/cr400153k}

\bibitem[{{Bentley} et~al.(2016){Bentley}, {Schmied}, {Mannel}, {Torkar}, {Jeszenszky}, {Romstedt}, {Levasseur-Regourd}, {Weber}, {Jessberger}, {Ehrenfreund}, {Koeberl}, and {Havnes}}]{Bentley16}
{Bentley} MS, {Schmied} R, {Mannel} T, et~al (2016) {Aggregate dust particles at comet 67P/Churyumov-Gerasimenko}. \nat 537(7618):73--75. \doi{10.1038/nature19091}, {\href{https://arxiv.org/abs/1704.00526}{{arXiv:1704.00526}}} {[astro-ph.EP]}

\bibitem[{van~den Berg et~al.(2004)van~den Berg, Bromley, Ramsahye, and Maschmeyer}]{vandenBerg-2004a}
van~den Berg AWC, Bromley ST, Ramsahye N, et~al (2004) {Diffusion of Molecular Hydrogen through Porous Materials: The Importance of Framework Flexibility}. J Phys Chem B 108(16):5088--5094. \doi{10.1021/jp037150r}

\bibitem[{Bergeld et~al.(2008)Bergeld, Kasemo, and Chakarov}]{RN1810}
Bergeld J, Kasemo B, Chakarov D (2008) {Photocatalytic reactions at the graphite/ice interface}. Phys Chem Chem Phys 10(8):1151--1155. \doi{10.1039/b714657d}

\bibitem[{{Berry} and {Percival}(1986)}]{Berry86}
{Berry} MV, {Percival} IC (1986) {Optics of Fractal Clusters Such as Smoke}. Optica Acta 33(5):577--591. \doi{10.1080/713821987}

\bibitem[{{Bertini} et~al.(2007){Bertini}, {Thomas}, and {Barbieri}}]{Bertini07}
{Bertini} I, {Thomas} N, {Barbieri} C (2007) {Modeling of the light scattering properties of cometary dust using fractal aggregates}. \aap 461(1):351--364. \doi{10.1051/0004-6361:20065461}

\bibitem[{{Bertrang} et~al.(2017){Bertrang}, {Flock}, and {Wolf}}]{Bertrang17}
{Bertrang} GHM, {Flock} M, {Wolf} S (2017) {Magnetic fields in protoplanetary discs: from MHD simulations to ALMA observations}. \mnras 464(1):L61--L64. \doi{10.1093/mnrasl/slw181}, {\href{https://arxiv.org/abs/1609.02085}{{arXiv:1609.02085}}} {[astro-ph.EP]}

\bibitem[{Besselink et~al.(2016)Besselink, Stawski, Van~Driessche, and Benning}]{RN1872}
Besselink R, Stawski TM, Van~Driessche AES, et~al (2016) {Not just fractal surfaces, but surface fractal aggregates: Derivation of the expression for the structure factor and its applications}. \jcp 145(21):211908. \doi{10.1063/1.4960953}

\bibitem[{{Birnstiel} et~al.(2018){Birnstiel}, {Dullemond}, {Zhu}, {Andrews}, {Bai}, {Wilner}, {Carpenter}, {Huang}, {Isella}, {Benisty}, {P{\'e}rez}, and {Zhang}}]{Birnstiel18}
{Birnstiel} T, {Dullemond} CP, {Zhu} Z, et~al (2018) {The Disk Substructures at High Angular Resolution Project (DSHARP). V. Interpreting ALMA Maps of Protoplanetary Disks in Terms of a Dust Model}. \apjl 869(2):L45. \doi{10.3847/2041-8213/aaf743}, {\href{https://arxiv.org/abs/1812.04043}{{arXiv:1812.04043}}} {[astro-ph.SR]}

\bibitem[{{Blum}(2018)}]{Blum18}
{Blum} J (2018) {Dust Evolution in Protoplanetary Discs and the Formation of Planetesimals. What Have We Learned from Laboratory Experiments?} \ssr 214(2):52. \doi{10.1007/s11214-018-0486-5}, {\href{https://arxiv.org/abs/1802.00221}{{arXiv:1802.00221}}} {[astro-ph.EP]}

\bibitem[{Blum and Wurm(2000)}]{RN1652}
Blum J, Wurm G (2000) {Experiments on sticking, restructuring, and fragmentation of preplanetary dust aggregates}. Icarus 143(1):138--146. \doi{10.1006/icar.1999.6234}

\bibitem[{Blum and Wurm(2008)}]{RN1604}
Blum J, Wurm G (2008) {The growth mechanisms of macroscopic bodies in protoplanetary disks}. \araa 46:21--56. \doi{10.1146/annurev.astro.46.060407.145152}

\bibitem[{Blum et~al.(2022)Blum, Bischoff, and Gundlach}]{RN1533}
Blum J, Bischoff D, Gundlach B (2022) {Formation of Comets}. Universe 8(7):381. \doi{10.3390/universe8070381}

\bibitem[{{Bohren} and {Huffman}(1983)}]{Bohren83}
{Bohren} CF, {Huffman} DR (1983) {Absorption and scattering of light by small particles}. New York: Wiley

\bibitem[{Bolina et~al.(2005)Bolina, Wolff, and Brown}]{RN1142}
Bolina AS, Wolff AJ, Brown WA (2005) {Reflection absorption infrared spectroscopy and temperature-programmed desorption studies of the adsorption and desorption of amorphous and crystalline water on a graphite surface}. J Phys Chem B 109(35):16836--16845. \doi{10.1021/jp0528111}

\bibitem[{{Botet} et~al.(1997){Botet}, {Rannou}, and {Cabane}}]{Botet97}
{Botet} R, {Rannou} P, {Cabane} M (1997) {Mean-field approximation of Mie scattering by fractal aggregates of identical spheres}. \ao 36(33):8791--8797. \doi{10.1364/AO.36.008791}

\bibitem[{Bowker(2011)}]{RN1871}
Bowker M (2011) {The Basis and Applications of Heterogeneous Catalysis}. Oxford University Press, Oxford, UK

\bibitem[{{Bradley} and {Brownlee}(1986)}]{Bradley_IDP}
{Bradley} JP, {Brownlee} DE (1986) {Cometary Particles: Thin Sectioning and Electron Beam Analysis}. Science 231(4745):1542--1544. \doi{10.1126/science.231.4745.1542}

\bibitem[{{Brisset} et~al.(2017){Brisset}, {Hei{\ss}elmann}, {Kothe}, {Weidling}, and {Blum}}]{Brisset17}
{Brisset} J, {Hei{\ss}elmann} D, {Kothe} S, et~al (2017) {Low-velocity collision behaviour of clusters composed of sub-millimetre sized dust aggregates}. \aap 603:A66. \doi{10.1051/0004-6361/201630345}, {\href{https://arxiv.org/abs/1706.07512}{{arXiv:1706.07512}}} {[astro-ph.EP]}

\bibitem[{Bromley(2024)}]{Bromley_nanosilicate_review}
Bromley ST (2024) Nanosilicates and molecular silicate dust species: properties and observational prospects. Front Astron Space Sci 11. \doi{10.3389/fspas.2024.1523977}

\bibitem[{Bromley et~al.(2016)Bromley, Martín, and Plane}]{RN1899}
Bromley ST, Martín JCG, Plane JMC (2016) Under what conditions does (sio) nucleation occur? a bottom-up kinetic modelling evaluation. Phys Chem Chem Phys 18(38):26913--26922. \doi{10.1039/c6cp03629e}

\bibitem[{{Brunngr{\"a}ber} and {Wolf}(2021)}]{Brunngraber21}
{Brunngr{\"a}ber} R, {Wolf} S (2021) {Self-scattering on large, porous grains in protoplanetary disks with dust settling}. \aap 648:A87. \doi{10.1051/0004-6361/202040033}, {\href{https://arxiv.org/abs/2103.04819}{{arXiv:2103.04819}}} {[astro-ph.SR]}

\bibitem[{Burgess and Stroud(2021)}]{RN1705}
Burgess KD, Stroud RM (2021) {Comparison of space weathering features in three particles from Itokawa}. Meteoritics \& Planetary Science 56(6):1109--1124. \doi{10.1111/maps.13692}

\bibitem[{{Burns} et~al.(1979){Burns}, {Lamy}, and {Soter}}]{Burns79}
{Burns} JA, {Lamy} PL, {Soter} S (1979) {Radiation forces on small particles in the solar system}. \icarus 40(1):1--48. \doi{10.1016/0019-1035(79)90050-2}

\bibitem[{Burris et~al.(2016)Burris, Laage, and Thompson}]{Simulated_IR_confined_water_JCP2016}
Burris PC, Laage D, Thompson WH (2016) {Simulations of the infrared, Raman, and 2D-IR photon echo spectra of water in nanoscale silica pores}. \jcp 144(19):194709. \doi{10.1063/1.4949766}

\bibitem[{Burwell(1976)}]{RN1602}
Burwell RL (1976) {Manual of symbols and terminology for physicochemical quantities and units -- Appendix II. Definitions, terminology and symbols in colloid and surface chemistry. Part II: Heterogeneous catalysis}. Pure Appl Chem 46(1):71--90

\bibitem[{{Canovas} et~al.(2013){Canovas}, {M{\'e}nard}, {Hales}, {Jord{\'a}n}, {Schreiber}, {Casassus}, {Gledhill}, and {Pinte}}]{Canovas13}
{Canovas} H, {M{\'e}nard} F, {Hales} A, et~al (2013) {Near-infrared imaging polarimetry of HD 142527}. \aap 556:A123. \doi{10.1051/0004-6361/201321924}, {\href{https://arxiv.org/abs/1306.6379}{{arXiv:1306.6379}}} {[astro-ph.SR]}

\bibitem[{Carr et~al.(1992)Carr, Misra, and Litchfield}]{RN1873}
Carr JR, Misra M, Litchfield J (1992) {Estimating Surface Area for Aggregate in the Size Range 1 mm or Larger}. Transport Res Record 1362:20. \urlprefix\url{https://onlinepubs.trb.org/Onlinepubs/trr/1992/1362/1362-003.pdf}

\bibitem[{{Carrasco-Gonz{\'a}lez} et~al.(2019){Carrasco-Gonz{\'a}lez}, {Sierra}, {Flock}, {Zhu}, {Henning}, {Chandler}, {Galv{\'a}n-Madrid}, {Mac{\'\i}as}, {Anglada}, {Linz}, {Osorio}, {Rodr{\'\i}guez}, {Testi}, {Torrelles}, {P{\'e}rez}, and {Liu}}]{CarrascoGonzalez19}
{Carrasco-Gonz{\'a}lez} C, {Sierra} A, {Flock} M, et~al (2019) {The Radial Distribution of Dust Particles in the HL Tau Disk from ALMA and VLA Observations}. \apj 883(1):71. \doi{10.3847/1538-4357/ab3d33}, {\href{https://arxiv.org/abs/1908.07140}{{arXiv:1908.07140}}} {[astro-ph.EP]}

\bibitem[{Chakarov et~al.(2001)Chakarov, Gleeson, and Kasemo}]{RN1811}
Chakarov DV, Gleeson MA, Kasemo B (2001) {Photoreactions of water and carbon at 90 K}. \jcp 115(20):9477--9483. \doi{10.1063/1.1414375}

\bibitem[{Chang et~al.(1990)Chang, Noda, and Ewing}]{RN1879}
Chang HC, Noda C, Ewing GE (1990) {Co on Nacl(100) - Model System for Investigating Vibrational-Energy Flow}. J Vacuum Sci Technol A 8(3):2644--2648. \doi{10.1116/1.576686}

\bibitem[{Chang et~al.(1995)Chang, Dai, and Ewing}]{RN1880}
Chang HC, Dai DJ, Ewing GE (1995) {Vibrational Spectroscopy as a Probe of Surface Heterogeneity - Co on Nacl(100)}. J Chinese Chem Soc 42(2):317--324. \doi{10.1002/jccs.199500043}

\bibitem[{{Chang} et~al.(2005){Chang}, {Cuppen}, and {Herbst}}]{Chang05}
{Chang} Q, {Cuppen} HM, {Herbst} E (2005) {Continuous-time random-walk simulation of H$_{2}$ formation on interstellar grains}. \aap 434(2):599--611. \doi{10.1051/0004-6361:20041842}

\bibitem[{{Chen} et~al.(2020){Chen}, {Mazoyer}, {Poteet}, {Ren}, {Duch{\^e}ne}, {Hom}, {Arriaga}, {Millar-Blanchaer}, {Arnold}, {Bailey}, {Bruzzone}, {Chilcote}, {Choquet}, {De Rosa}, {Draper}, {Esposito}, {Fitzgerald}, {Follette}, {Hibon}, {Hines}, {Kalas}, {Marchis}, {Matthews}, {Milli}, {Patience}, {Perrin}, {Pueyo}, {Rajan}, {Rantakyr{\"o}}, {Rodigas}, {Roudier}, {Schneider}, {Soummer}, {Stark}, {Wang}, {Ward-Duong}, {Weinberger}, {Wilner}, and {Wolff}}]{Chen20}
{Chen} C, {Mazoyer} J, {Poteet} CA, et~al (2020) {Multiband GPI Imaging of the HR 4796A Debris Disk}. \apj 898(1):55. \doi{10.3847/1538-4357/ab9aba}, {\href{https://arxiv.org/abs/2006.16131}{{arXiv:2006.16131}}} {[astro-ph.SR]}

\bibitem[{{Chen} et~al.(2024){Chen}, {Lawson}, {Brandt}, {Lewis}, {Uyama}, {Millar-Blanchaer}, {Tazaki}, and {Currie}}]{Chen24}
{Chen} M, {Lawson} K, {Brandt} TD, et~al (2024) {Multiband polarimetric imaging of HD 34700 with SCExAO/CHARIS}. \mnras 533(2):2473--2487. \doi{10.1093/mnras/stae1957}, {\href{https://arxiv.org/abs/2408.09038}{{arXiv:2408.09038}}} {[astro-ph.SR]}

\bibitem[{Chipera and Apps(2001)}]{RN1902}
Chipera SJ, Apps JA (2001) {Geochemical stability of natural zeolites}. Natural Zeolites: Occurrence, Properties, Applications 45:117--161. \doi{10.2138/rmg.2001.45.3}

\bibitem[{{Cho} and {Lazarian}(2007)}]{Cho07}
{Cho} J, {Lazarian} A (2007) {Grain Alignment and Polarized Emission from Magnetized T Tauri Disks}. \apj 669(2):1085--1097. \doi{10.1086/521805}, {\href{https://arxiv.org/abs/astro-ph/0611280}{{arXiv:astro-ph/0611280}}} {[astro-ph]}

\bibitem[{{Christianson} and {Garrod}(2021)}]{CG21}
{Christianson} DA, {Garrod} RT (2021) {Chemical kinetics simulations of ice chemistry on porous versus non-porous interstellar dust grains}. Front Astron Space Sci 8:21. \doi{10.3389/fspas.2021.643297}, {\href{https://arxiv.org/abs/2104.12943}{{arXiv:2104.12943}}} {[astro-ph.GA]}

\bibitem[{{Chung} et~al.(2024){Chung}, {Andrews}, {Gurwell}, {Wright}, {Long}, {Xu}, and {Liu}}]{Chung24}
{Chung} CY, {Andrews} SM, {Gurwell} MA, et~al (2024) {SMA 200{\textendash}400 GHz Survey for Dust Properties in the Icy Class II Disks in the Taurus Molecular Cloud}. \apjs 273(2):29. \doi{10.3847/1538-4365/ad528b}, {\href{https://arxiv.org/abs/2405.19867}{{arXiv:2405.19867}}} {[astro-ph.EP]}

\bibitem[{{Ch{\'y}lek} et~al.(2000){Ch{\'y}lek}, {Videen}, {Geldart}, {Dobbie}, and {Tso}}]{Chylek00}
{Ch{\'y}lek} P, {Videen} G, {Geldart} DJW, et~al (2000) {Effective Medium Approximations for Heterogeneous Particles}. San Diego : Academic Press p 274

\bibitem[{{Cleeves} et~al.(2013){Cleeves}, {Adams}, and {Bergin}}]{Cleeves2013}
{Cleeves} LI, {Adams} FC, {Bergin} EA (2013) {Exclusion of Cosmic Rays in Protoplanetary Disks: Stellar and Magnetic Effects}. \apj 772(1):5. \doi{10.1088/0004-637X/772/1/5}, {\href{https://arxiv.org/abs/1306.0902}{{arXiv:1306.0902}}} {[astro-ph.SR]}

\bibitem[{Coasne and Ugliengo(2012)}]{Coasne2012}
Coasne B, Ugliengo P (2012) {Atomistic Model of Micelle-Templated Mesoporous Silicas: Structural, Morphological, and Adsorption Properties}. Langmuir 28(30):11131--11141. \doi{10.1021/la3022529}

\bibitem[{Collings et~al.(2004)Collings, Anderson, Chen, Dever, Viti, Williams, and McCoustra}]{RN721}
Collings MP, Anderson MA, Chen R, et~al (2004) {A laboratory survey of the thermal desorption of astrophysically relevant molecules}. \mnras 354(4):1133--1140

\bibitem[{Collings et~al.(2015)Collings, Frankland, Lasne, Marchione, Rosu-Finsen, and McCoustra}]{RN1136}
Collings MP, Frankland VL, Lasne J, et~al (2015) {Probing model interstellar grain surfaces with small molecules}. \mnras 449(2):1826--1833. \doi{10.1093/mnras/stv425}

\bibitem[{Consolaro et~al.(2024)Consolaro, Rouchon, and Ersen}]{RN1868}
Consolaro VG, Rouchon V, Ersen O (2024) {Electron beam damages in zeolites: A review}. Microporous and Mesoporous Materials 364:112835. \doi{10.1016/j.micromeso.2023.112835}

\bibitem[{{Cuppen} and {Herbst}(2005)}]{Cuppen05}
{Cuppen} HM, {Herbst} E (2005) {Monte Carlo simulations of H$_{2}$ formation on grains of varying surface roughness}. \mnras 361(2):565--576. \doi{10.1111/j.1365-2966.2005.09189.x}

\bibitem[{{Cuzzi} et~al.(2014){Cuzzi}, {Estrada}, and {Davis}}]{Cuzzi14}
{Cuzzi} JN, {Estrada} PR, {Davis} SS (2014) {Utilitarian Opacity Model for Aggregate Particles in Protoplanetary Nebulae and Exoplanet Atmospheres}. \apjs 210(2):21. \doi{10.1088/0067-0049/210/2/21}, {\href{https://arxiv.org/abs/1312.1798}{{arXiv:1312.1798}}} {[astro-ph.EP]}

\bibitem[{D'Alessio et~al.(2001)D'Alessio, Calvet, and Hartmann}]{RN902}
D'Alessio P, Calvet N, Hartmann L (2001) {Accretion disks around young objects. III. Grain growth}. \apj 553(1):321--334. \doi{10.1086/320655}

\bibitem[{Dartois et~al.(2017)Dartois, Chabot, Pino, Béroff, Godard, Severin, Bender, and Trautmann}]{RN1925}
Dartois E, Chabot M, Pino T, et~al (2017) {Swift heavy ion irradiation of interstellar dust analogues Small carbonaceous species released by cosmic rays}. \aap 599:A130. \doi{10.1051/0004-6361/201629646}

\bibitem[{{Dartois} et~al.(2024){Dartois}, {Noble}, {Caselli}, {Fraser}, {Jim{\'e}nez-Serra}, {Mat{\'e}}, {McClure}, {Melnick}, {Pendleton}, {Shimonishi}, {Smith}, {Sturm}, {Taillard}, {Wakelam}, {Boogert}, {Drozdovskaya}, {Erkal}, {Harsono}, {Herrero}, {Ioppolo}, {Linnartz}, {McGuire}, {Perotti}, {Qasim}, and {Rocha}}]{Dartois2024}
{Dartois} E, {Noble} JA, {Caselli} P, et~al (2024) {Spectroscopic sizing of interstellar icy grains with JWST}. Nat Astron 8:359--367. \doi{10.1038/s41550-023-02155-x}

\bibitem[{{Davis} and {Greenstein}(1951)}]{Levertt1951}
{Davis} JLeverett, {Greenstein} JL (1951) {The Polarization of Starlight by Aligned Dust Grains.} \apj 114:206. \doi{10.1086/145464}

\bibitem[{{Debes} et~al.(2008){Debes}, {Weinberger}, and {Schneider}}]{Debes08}
{Debes} JH, {Weinberger} AJ, {Schneider} G (2008) {Complex Organic Materials in the Circumstellar Disk of HR 4796A}. \apjl 673(2):L191. \doi{10.1086/527546}, {\href{https://arxiv.org/abs/0712.3283}{{arXiv:0712.3283}}} {[astro-ph]}

\bibitem[{Delle~Piane et~al.(2013)Delle~Piane, Corno, and Ugliengo}]{DellePiane2013-JCTC}
Delle~Piane M, Corno M, Ugliengo P (2013) {Does Dispersion Dominate over H-Bonds in Drug–Surface Interactions? The Case of Silica-Based Materials As Excipients and Drug-Delivery Agents}. J Chem Theory Comput 9(5):2404--2415. \doi{10.1021/ct400073s}

\bibitem[{Delle~Piane et~al.(2014{\natexlab{a}})Delle~Piane, Corno, Pedone, Dovesi, and Ugliengo}]{DellePiane2014-JPCC}
Delle~Piane M, Corno M, Pedone A, et~al (2014{\natexlab{a}}) {Large-Scale B3LYP Simulations of Ibuprofen Adsorbed in MCM-41 Mesoporous Silica as Drug Delivery System}. J Phys Chem C 118(46):26737--26749. \doi{10.1021/jp507364h}

\bibitem[{Delle~Piane et~al.(2014{\natexlab{b}})Delle~Piane, Vaccari, Corno, and Ugliengo}]{DellePiane2014-JPCA}
Delle~Piane M, Vaccari S, Corno M, et~al (2014{\natexlab{b}}) {Silica-Based Materials as Drug Adsorbents: First Principle Investigation on the Role of Water Microsolvation on Ibuprofen Adsorption}. J Phys Chem A 118(31):5801--5807. \doi{10.1021/jp411173k}

\bibitem[{{Delle Piane} et~al.(2018){Delle Piane}, Corno, and Ugliengo}]{DellePiane2018}
{Delle Piane} M, Corno M, Ugliengo P (2018) {Chapter 9 - Ab Initio Modeling of Hydrogen Bond Interaction at Silica Surfaces With Focus on Silica/Drugs Systems}. In: Catlow CRA, {Van Speybroeck} V, {van Santen} RA (eds) {Modelling and Simulation in the Science of Micro- and Meso-Porous Materials}. Elsevier, p 297--328, \doi{10.1016/B978-0-12-805057-6.00009-0}

\bibitem[{Demtr{\"o}der(1996)}]{RN1878}
Demtr{\"o}der W (1996) {Laser Spectroscopy: Basic Concepts and Instrumentation}, 2nd edn. Springer, Berlin, Heidelberg, \doi{10.1007/978-3-662-08260-7}

\bibitem[{{Dent} et~al.(2019){Dent}, {Pinte}, {Cortes}, {M{\'e}nard}, {Hales}, {Fomalont}, and {de Gregorio-Monsalvo}}]{Dent19}
{Dent} WRF, {Pinte} C, {Cortes} PC, et~al (2019) {Submillimetre dust polarization and opacity in the HD163296 protoplanetary ring system}. \mnras 482(1):L29--L33. \doi{10.1093/mnrasl/sly181}, {\href{https://arxiv.org/abs/1809.09185}{{arXiv:1809.09185}}} {[astro-ph.EP]}

\bibitem[{Derouane et~al.(1988)Derouane, Andre, and Lucas}]{Derouane1988}
Derouane EG, Andre JM, Lucas AA (1988) {Surface curvature effects in physisorption and catalysis by microporous solids and molecular sieves}. J Catal 110(1):58--73. \doi{10.1016/0021-9517(88)90297-7}

\bibitem[{Dib et~al.(2021)Dib, Costa, Vayssilov, Aleksandrov, and Mintova}]{Dib}
Dib E, Costa IM, Vayssilov GN, et~al (2021) Complex h-bonded silanol network in zeolites revealed by ir and nmr spectroscopy combined with dft calculations. J Mater Chem A 9:27347--27352. \doi{10.1039/D1TA06908J}

\bibitem[{{Dominik} and {Tielens}(1997)}]{Dominik97}
{Dominik} C, {Tielens} AGGM (1997) {The Physics of Dust Coagulation and the Structure of Dust Aggregates in Space}. \apj 480(2):647--673. \doi{10.1086/303996}

\bibitem[{Dorschner(2010)}]{RN1696}
Dorschner J (2010) {From Dust Astrophysics Towards Dust Mineralogy -- A Historical Review}. In: Henning T (ed) Astromineralogy, Lecture Notes in Physics, vol 815, 2nd edn. Springer, Berlin, Heidelberg, p 1--60, \doi{10.1007/978-3-642-13259-9_1}

\bibitem[{{Draine}(2003)}]{Draine2003}
{Draine} BT (2003) {Interstellar Dust Grains}. \araa 41:241--289. \doi{10.1146/annurev.astro.41.011802.094840}, {\href{https://arxiv.org/abs/astro-ph/0304489}{{arXiv:astro-ph/0304489}}} {[astro-ph]}

\bibitem[{{Draine}(2006)}]{Draine06}
{Draine} BT (2006) {On the Submillimeter Opacity of Protoplanetary Disks}. \apj 636(2):1114--1120. \doi{10.1086/498130}, {\href{https://arxiv.org/abs/astro-ph/0507292}{{arXiv:astro-ph/0507292}}} {[astro-ph]}

\bibitem[{Draine(2011)}]{Draine_ismbook}
Draine BT (2011) {Physics of the Interstellar and Intergalactic Medium}. Princeton University Press

\bibitem[{{Draine}(2024{\natexlab{a}})}]{Draine2024_convex}
{Draine} BT (2024{\natexlab{a}}) {Sensitivity of Polarization to Grain Shape. I. Convex Shapes}. \apj 961(1):103. \doi{10.3847/1538-4357/ad0463}, {\href{https://arxiv.org/abs/2310.15229}{{arXiv:2310.15229}}} {[astro-ph.GA]}

\bibitem[{{Draine}(2024{\natexlab{b}})}]{Draine2024_aggregates}
{Draine} BT (2024{\natexlab{b}}) {Sensitivity of Polarization to Grain Shape. II. Aggregates}. \apj 969(2):92. \doi{10.3847/1538-4357/ad3b9a}, {\href{https://arxiv.org/abs/2404.02836}{{arXiv:2404.02836}}} {[astro-ph.GA]}

\bibitem[{{Draine}(2025)}]{Draine2025}
{Draine} BT (2025) {The Spheroidal Analog Method for Modeling Irregular Porous Aggregates}. \apj 985(1):10. \doi{10.3847/1538-4357/adc57d}, {\href{https://arxiv.org/abs/2504.02970}{{arXiv:2504.02970}}} {[astro-ph.GA]}

\bibitem[{{Draine} and {Hensley}(2021)}]{Draine2021}
{Draine} BT, {Hensley} BS (2021) {Using the Starlight Polarization Efficiency Integral to Constrain Shapes and Porosities of Interstellar Grains}. \apj 919(1):65. \doi{10.3847/1538-4357/ac0050}, {\href{https://arxiv.org/abs/2101.07277}{{arXiv:2101.07277}}} {[astro-ph.GA]}

\bibitem[{{Draine} and {Lee}(1984)}]{Draine1984}
{Draine} BT, {Lee} HM (1984) {Optical Properties of Interstellar Graphite and Silicate Grains}. \apj 285:89. \doi{10.1086/162480}

\bibitem[{{Draine} and {Tan}(2003)}]{DraineTan2003}
{Draine} BT, {Tan} JC (2003) {The Scattered X-Ray Halo around Nova Cygni 1992: Testing a Model for Interstellar Dust}. \apj 594(1):347--362. \doi{10.1086/376855}, {\href{https://arxiv.org/abs/astro-ph/0208302}{{arXiv:astro-ph/0208302}}} {[astro-ph]}

\bibitem[{{Duch{\^e}ne} et~al.(2004){Duch{\^e}ne}, {McCabe}, {Ghez}, and {Macintosh}}]{Duchene04}
{Duch{\^e}ne} G, {McCabe} C, {Ghez} AM, et~al (2004) {A Multiwavelength Scattered Light Analysis of the Dust Grain Population in the GG Tauri Circumbinary Ring}. \apj 606(2):969--982. \doi{10.1086/383126}, {\href{https://arxiv.org/abs/astro-ph/0401560}{{arXiv:astro-ph/0401560}}} {[astro-ph]}

\bibitem[{{Duch{\^e}ne} et~al.(2020){Duch{\^e}ne}, {Rice}, {Hom}, {Zalesky}, {Esposito}, {Millar-Blanchaer}, {Ren}, {Kalas}, {Fitzgerald}, {Arriaga}, {Bruzzone}, {Bulger}, {Chen}, {Chiang}, {Cotten}, {Czekala}, {De Rosa}, {Dong}, {Draper}, {Follette}, {Graham}, {Hung}, {Lopez}, {Macintosh}, {Matthews}, {Mazoyer}, {Metchev}, {Patience}, {Perrin}, {Rameau}, {Song}, {Stahl}, {Wang}, {Wolff}, {Zuckerman}, {Ammons}, {Bailey}, {Barman}, {Chilcote}, {Doyon}, {Gerard}, {Goodsell}, {Greenbaum}, {Hibon}, {Ingraham}, {Konopacky}, {Maire}, {Marchis}, {Marley}, {Marois}, {Nielsen}, {Oppenheimer}, {Palmer}, {Poyneer}, {Pueyo}, {Rajan}, {Rantakyr{\"o}}, {Ruffio}, {Savransky}, {Schneider}, {Sivaramakrishnan}, {Soummer}, {Thomas}, and {Ward-Duong}}]{Duchene20}
{Duch{\^e}ne} G, {Rice} M, {Hom} J, et~al (2020) {The Gemini Planet Imager View of the HD 32297 Debris Disk}. \aj 159(6):251. \doi{10.3847/1538-3881/ab8881}, {\href{https://arxiv.org/abs/2004.06027}{{arXiv:2004.06027}}} {[astro-ph.SR]}

\bibitem[{van Duin et~al.(2003)van Duin, Strachan, Stewman, Zhang, Xu, and Goddard}]{ReaxFF-SiO}
van Duin ACT, Strachan A, Stewman S, et~al (2003) {ReaxFFSiO Reactive Force Field for Silicon and Silicon Oxide Systems}. J Phys Chem A 107(19):3803--3811. \doi{10.1021/jp0276303}

\bibitem[{Dulieu et~al.(2019)Dulieu, Nguyen, Congiu, Baouche, and Taquet}]{RN1718}
Dulieu F, Nguyen T, Congiu E, et~al (2019) {Efficient formation route of the prebiotic molecule formamide on interstellar dust grains}. \mnras 484(1):L119--L123. \doi{10.1093/mnrasl/slz013}

\bibitem[{{Dwek}(1997)}]{Dwek1997fluffy}
{Dwek} E (1997) {Can Composite Fluffy Dust Particles Solve the Interstellar Carbon Crisis?} \apj 484(2):779--784. \doi{10.1086/304370}, {\href{https://arxiv.org/abs/astro-ph/9701109}{{arXiv:astro-ph/9701109}}} {[astro-ph]}

\bibitem[{{Dwek} et~al.(1997){Dwek}, {Arendt}, {Fixsen}, {Sodroski}, {Odegard}, {Weiland}, {Reach}, {Hauser}, {Kelsall}, {Moseley}, {Silverberg}, {Shafer}, {Ballester}, {Bazell}, and {Isaacman}}]{Dwek1997}
{Dwek} E, {Arendt} RG, {Fixsen} DJ, et~al (1997) {Detection and Characterization of Cold Interstellar Dust and Polycyclic Aromatic Hydrocarbon Emission, from COBE Observations}. \apj 475(2):565--579. \doi{10.1086/303568}, {\href{https://arxiv.org/abs/astro-ph/9610198}{{arXiv:astro-ph/9610198}}} {[astro-ph]}

\bibitem[{{Dykes} et~al.(2024){Dykes}, {Currie}, {Lawson}, {Lucas}, {Kudo}, {Chen}, {Guyon}, {Groff}, {Lozi}, {Chilcote}, {Brandt}, {Vievard}, {Skaf}, {Deo}, {El Morsy}, {Bovie}, {Uyama}, {Grady}, {Sitko}, {Hashimoto}, {Martinache}, {Jovanovic}, {Tamura}, and {Kasdin}}]{Dykes24}
{Dykes} E, {Currie} T, {Lawson} K, et~al (2024) {SCExAO/CHARIS Near-Infrared Scattered-Light Imaging and Integral Field Spectropolarimetry of the AB Aurigae Protoplanetary System}. arXiv e-prints arXiv:2410.11939. \doi{10.48550/arXiv.2410.11939}, {\href{https://arxiv.org/abs/2410.11939}{{arXiv:2410.11939}}} {[astro-ph.EP]}

\bibitem[{Eberl(1984)}]{RN1903}
Eberl DD (1984) {Clay Mineral Formation and Transformation in Rocks and Soils}. Philosophical Transactions of the Royal Society a-Mathematical Physical and Engineering Sciences 311(1517):241--257. \doi{10.1098/rsta.1984.0026}

\bibitem[{Ehrenfreund et~al.(1998)Ehrenfreund, Boogert, Gerakines, and Tielens}]{RN1601}
Ehrenfreund P, Boogert A, Gerakines P, et~al (1998) {Apolar ices}. Faraday Discussions 109:463--474. \doi{10.1039/a800608c}

\bibitem[{El-Awadi(2023)}]{RN2004}
El-Awadi GA (2023) {Review of effective techniques for surface engineering material modification for a variety of applications}. Aims Materials Science 10(4):652--692. \doi{10.3934/matersci.2023037}

\bibitem[{Ellis et~al.(2016)Ellis, Brown, Bishop, Yin, Cooke, Terry, Liu, Yin, and Palmer}]{RN1701}
Ellis PR, Brown CM, Bishop PT, et~al (2016) {The cluster beam route to model catalysts and beyond}. Faraday Discussions 188:39--56. \doi{10.1039/c5fd00178a}

\bibitem[{{Elstner} et~al.(1998){Elstner}, {Porezag}, {Jungnickel}, {Elsner}, {Haugk}, {Frauenheim}, {Suhai}, and {Seifert}}]{SCC-DFTB}
{Elstner} M, {Porezag} D, {Jungnickel} G, et~al (1998) {Self-consistent-charge density-functional tight-binding method for simulations of complex materials properties}. \prb 58(11):7260--7268. \doi{10.1103/PhysRevB.58.7260}

\bibitem[{Escatllar et~al.(2019)Escatllar, Lazaukas, Woodley, and Bromley}]{Macia_nanosilicate_2019}
Escatllar AM, Lazaukas T, Woodley SM, et~al (2019) {Structure and Properties of Nanosilicates with Olivine (Mg2SiO4)N and Pyroxene (MgSiO3)N Compositions}. ACS Earth and Space Chemistry 3(11):2390--2403. \doi{10.1021/acsearthspacechem.9b00139}

\bibitem[{{Estrada} and {Umurhan}(2023)}]{Estrada23}
{Estrada} PR, {Umurhan} OM (2023) {Formation of the First Planetesimals via the Streaming Instability in Globally Turbulent Protoplanetary Disks?} \apj 946(1):15. \doi{10.3847/1538-4357/acb7db}, {\href{https://arxiv.org/abs/2302.03163}{{arXiv:2302.03163}}} {[astro-ph.EP]}

\bibitem[{{Estrada} et~al.(2022){Estrada}, {Cuzzi}, and {Umurhan}}]{Estrada22}
{Estrada} PR, {Cuzzi} JN, {Umurhan} OM (2022) {Global Modeling of Nebulae with Particle Growth, Drift, and Evaporation Fronts. II. The Influence of Porosity on Solids Evolution}. \apj 936(1):42. \doi{10.3847/1538-4357/ac7ffd}, {\href{https://arxiv.org/abs/2207.12626}{{arXiv:2207.12626}}} {[astro-ph.EP]}

\bibitem[{{Feigelson} and {Montmerle}(1999)}]{Feigelson1999}
{Feigelson} ED, {Montmerle} T (1999) {High-Energy Processes in Young Stellar Objects}. \araa 37:363--408. \doi{10.1146/annurev.astro.37.1.363}

\bibitem[{Ferrero et~al.(2023)Ferrero, Pantaleone, Ceccarelli, Ugliengo, Sodupe, and Rimola}]{Ferrero2023}
Ferrero S, Pantaleone S, Ceccarelli C, et~al (2023) {Where Does the Energy Go during the Interstellar NH3 Formation on Water Ice? A Computational Study}. \apj 944(2):142. \doi{10.3847/1538-4357/acae8e}

\bibitem[{{Fitzgerald} et~al.(2007){Fitzgerald}, {Kalas}, {Duch{\^e}ne}, {Pinte}, and {Graham}}]{Fitzgerald07_aumic}
{Fitzgerald} MP, {Kalas} PG, {Duch{\^e}ne} G, et~al (2007) {The AU Microscopii Debris Disk: Multiwavelength Imaging and Modeling}. \apj 670(1):536--556. \doi{10.1086/521344}, {\href{https://arxiv.org/abs/0705.4196}{{arXiv:0705.4196}}} {[astro-ph]}

\bibitem[{{Fitzpatrick} and {Massa}(1986)}]{Fitzpatrick1986}
{Fitzpatrick} EL, {Massa} D (1986) {An Analysis of the Shapes of Ultraviolet Extinction Curves. I. The 2175 Angstrom Bump}. \apj 307:286. \doi{10.1086/164415}

\bibitem[{{Fitzpatrick} et~al.(2019){Fitzpatrick}, {Massa}, {Gordon}, {Bohlin}, and {Clayton}}]{Fitzpatrick2019}
{Fitzpatrick} EL, {Massa} D, {Gordon} KD, et~al (2019) {An Analysis of the Shapes of Interstellar Extinction Curves. VII. Milky Way Spectrophotometric Optical-through-ultraviolet Extinction and Its R-dependence}. \apj 886(2):108. \doi{10.3847/1538-4357/ab4c3a}, {\href{https://arxiv.org/abs/1910.08852}{{arXiv:1910.08852}}} {[astro-ph.GA]}

\bibitem[{{Fukagawa} et~al.(2010){Fukagawa}, {Tamura}, {Itoh}, {Oasa}, {Kudo}, {Hayashi}, {Kato}, {Ootsubo}, {Itoh}, {Shibai}, and {Hayashi}}]{Fukagawa10}
{Fukagawa} M, {Tamura} M, {Itoh} Y, et~al (2010) {Subaru Near-Infrared Imaging of Herbig Ae Stars}. \pasj 62:347. \doi{10.1093/pasj/62.2.347}

\bibitem[{{Fulle} and {Blum}(2017)}]{Fulle17}
{Fulle} M, {Blum} J (2017) {Fractal dust constrains the collisional history of comets}. \mnras 469:S39--S44. \doi{10.1093/mnras/stx971}

\bibitem[{{Fulle} et~al.(2015){Fulle}, {Della Corte}, {Rotundi}, {Weissman}, {Juhasz}, {Szego}, {Sordini}, {Ferrari}, {Ivanovski}, {Lucarelli}, {Accolla}, {Merouane}, {Zakharov}, {Mazzotta Epifani}, {L{\'o}pez-Moreno}, {Rodr{\'\i}guez}, {Colangeli}, {Palumbo}, {Gr{\"u}n}, {Hilchenbach}, {Bussoletti}, {Esposito}, {Green}, {Lamy}, {McDonnell}, {Mennella}, {Molina}, {Morales}, {Moreno}, {Ortiz}, {Palomba}, {Rodrigo}, {Zarnecki}, {Cosi}, {Giovane}, {Gustafson}, {Herranz}, {Jer{\'o}nimo}, {Leese}, {L{\'o}pez-Jim{\'e}nez}, and {Altobelli}}]{Fulle15}
{Fulle} M, {Della Corte} V, {Rotundi} A, et~al (2015) {Density and Charge of Pristine Fluffy Particles from Comet 67P/Churyumov-Gerasimenko}. \apjl 802(1):L12. \doi{10.1088/2041-8205/802/1/L12}

\bibitem[{{Fulle} et~al.(2016){Fulle}, {Della Corte}, {Rotundi}, {Rietmeijer}, {Green}, {Weissman}, {Accolla}, {Colangeli}, {Ferrari}, {Ivanovski}, {Lopez-Moreno}, {Epifani}, {Morales}, {Ortiz}, {Palomba}, {Palumbo}, {Rodriguez}, {Sordini}, and {Zakharov}}]{Fulle16}
{Fulle} M, {Della Corte} V, {Rotundi} A, et~al (2016) {Comet 67P/Churyumov-Gerasimenko preserved the pebbles that formed planetesimals}. \mnras 462:S132--S137. \doi{10.1093/mnras/stw2299}

\bibitem[{Fulvio et~al.(2017)Fulvio, Sandor, Jager, Akos, and Henning}]{RN982}
Fulvio D, Sandor G, Jager C, et~al (2017) {Laboratory Experiments on the Low-temperature Formation of Carbonaceous Grains in the ISM}. \apjs 233(1):14. \doi{10.3847/1538-4365/aa9224}

\bibitem[{Gadallah et~al.(2012)Gadallah, Mutschke, and Jäger}]{RN1923}
Gadallah KAK, Mutschke H, Jäger C (2012) {Mid-infrared spectroscopy of UV irradiated hydrogenated amorphous carbon materials}. \aap 544:A107. \doi{10.1051/0004-6361/201219248}

\bibitem[{Gadzuk and Luntz(1984)}]{RN1883}
Gadzuk JW, Luntz AC (1984) {On Vibrational Lineshapes of Adsorbed Molecules}. Surface Science 144(2-3):429--450. \doi{10.1016/0039-6028(84)90110-9}

\bibitem[{{Garrod}(2013)}]{Garrod13}
{Garrod} RT (2013) {Three-dimensional, Off-lattice Monte Carlo Kinetics Simulations of Interstellar Grain Chemistry and Ice Structure}. \apj 778(2):158. \doi{10.1088/0004-637X/778/2/158}, {\href{https://arxiv.org/abs/1310.2512}{{arXiv:1310.2512}}} {[astro-ph.IM]}

\bibitem[{Garrod et~al.(2022)Garrod, Jin, Matis, Jones, Willis, and Herbst}]{RN1663}
Garrod RT, Jin M, Matis KA, et~al (2022) {Formation of Complex Organic Molecules in Hot Molecular Cores through Nondiffusive Grain-surface and Ice-mantle Chemistry}. \apjs 259(1):1. \doi{10.3847/1538-4365/ac3131}

\bibitem[{{Garufi} et~al.(2022){Garufi}, {Dominik}, {Ginski}, {Benisty}, {van Holstein}, {Henning}, {Pawellek}, {Pinte}, {Avenhaus}, {Facchini}, {Galicher}, {Gratton}, {M{\'e}nard}, {Muro-Arena}, {Milli}, {Stolker}, {Vigan}, {Villenave}, {Moulin}, {Origne}, {Rigal}, {Sauvage}, and {Weber}}]{Garufi22}
{Garufi} A, {Dominik} C, {Ginski} C, et~al (2022) {A SPHERE survey of self-shadowed planet-forming disks}. \aap 658:A137. \doi{10.1051/0004-6361/202141692}, {\href{https://arxiv.org/abs/2111.07856}{{arXiv:2111.07856}}} {[astro-ph.GA]}

\bibitem[{Gavilan et~al.(2016)Gavilan, Jager, Simionovici, Lemaire, Sabri, Foy, Yagoubi, Henning, Salomon, and Martinez-Criado}]{RN353}
Gavilan L, Jager C, Simionovici A, et~al (2016) {Hard X-ray irradiation of cosmic silicate analogs: structural evolution and astrophysical implications}. \aap 587:A144

\bibitem[{Germain et~al.(2022)Germain, Tinacci, Pantaleone, Ceccarelli, and Ugliengo}]{Germain2022}
Germain A, Tinacci L, Pantaleone S, et~al (2022) {Computer Generated Realistic Interstellar Icy Grain Models: Physicochemical Properties and Interaction with NH3}. ACS Earth Space Chem 6(5):1286--1298. \doi{10.1021/acsearthspacechem.2c00004}

\bibitem[{Gierada et~al.(2016)Gierada, Petit, Handzlik, and Tielens}]{Gierada2016}
Gierada M, Petit I, Handzlik J, et~al (2016) Hydration in silica based mesoporous materials: a dft model. Phys Chem Chem Phys 18:32962--32972. \doi{10.1039/C6CP05460A}

\bibitem[{Gignone et~al.(2015)Gignone, Delle~Piane, Corno, Ugliengo, and Onida}]{DellePiane2015-JPCC}
Gignone A, Delle~Piane M, Corno M, et~al (2015) {Simulation and Experiment Reveal a Complex Scenario for the Adsorption of an Antifungal Drug in Ordered Mesoporous Silica}. J Phys Chem C 119(23):13068--13079. \doi{10.1021/acs.jpcc.5b02666}

\bibitem[{{Ginski} et~al.(2016){Ginski}, {Stolker}, {Pinilla}, {Dominik}, {Boccaletti}, {de Boer}, {Benisty}, {Biller}, {Feldt}, {Garufi}, {Keller}, {Kenworthy}, {Maire}, {M{\'e}nard}, {Mesa}, {Milli}, {Min}, {Pinte}, {Quanz}, {van Boekel}, {Bonnefoy}, {Chauvin}, {Desidera}, {Gratton}, {Girard}, {Keppler}, {Kopytova}, {Lagrange}, {Langlois}, {Rouan}, and {Vigan}}]{Ginski16}
{Ginski} C, {Stolker} T, {Pinilla} P, et~al (2016) {Direct detection of scattered light gaps in the transitional disk around HD 97048 with VLT/SPHERE}. \aap 595:A112. \doi{10.1051/0004-6361/201629265}, {\href{https://arxiv.org/abs/1609.04027}{{arXiv:1609.04027}}} {[astro-ph.EP]}

\bibitem[{{Ginski} et~al.(2023){Ginski}, {Tazaki}, {Dominik}, and {Stolker}}]{Ginski23}
{Ginski} C, {Tazaki} R, {Dominik} C, et~al (2023) {Observed Polarized Scattered Light Phase Functions of Planet-forming Disks}. \apj 953(1):92. \doi{10.3847/1538-4357/acdc97}, {\href{https://arxiv.org/abs/2301.04617}{{arXiv:2301.04617}}} {[astro-ph.EP]}

\bibitem[{{Ginski} et~al.(2024){Ginski}, {Garufi}, {Benisty}, {Tazaki}, {Dominik}, {Ribas}, {Engler}, {Birnstiel}, {Chauvin}, {Columba}, {Facchini}, {Goncharov}, {Hagelberg}, {Henning}, {Hogerheijde}, {van Holstein}, {Huang}, {Muto}, {Pinilla}, {Kanagawa}, {Kim}, {Kurtovic}, {Langlois}, {Manara}, {Milli}, {Momose}, {Orihara}, {Pawellek}, {Pinte}, {Rab}, {Schmidt}, {Snik}, {Wahhaj}, {Williams}, and {Zurlo}}]{Ginski24}
{Ginski} C, {Garufi} A, {Benisty} M, et~al (2024) {The SPHERE view of the Chamaeleon I star-forming region. The full census of planet-forming disks with GTO and DESTINYS programs}. \aap 685:A52. \doi{10.1051/0004-6361/202244005}, {\href{https://arxiv.org/abs/2403.02149}{{arXiv:2403.02149}}} {[astro-ph.GA]}

\bibitem[{{Glassgold} et~al.(1997){Glassgold}, {Najita}, and {Igea}}]{Glassgold1997}
{Glassgold} AE, {Najita} J, {Igea} J (1997) {X-Ray Ionization of Protoplanetary Disks}. \apj 480(1):344--350. \doi{10.1086/303952}

\bibitem[{Gobrecht et~al.(2016)Gobrecht, Cherchneff, Sarangi, Plane, and Bromley}]{RN820}
Gobrecht D, Cherchneff I, Sarangi A, et~al (2016) {Dust formation in the oxygen-rich AGB star IK Tauri}. \aap 585:A6

\bibitem[{Goh et~al.(2024)Goh, Yu, Zavabeti, Shi, Guo, He, Yang, Dong, Webley, Ellis, and Li}]{RN1867}
Goh JM, Yu Z, Zavabeti A, et~al (2024) {Gas storage within nanoporous material encapsulated by ice}. J Materials Chem A 12(45):31204--31213. \doi{10.1039/d4ta06629d}

\bibitem[{{Golimowski} et~al.(2006){Golimowski}, {Ardila}, {Krist}, {Clampin}, {Ford}, {Illingworth}, {Bartko}, {Ben{\'\i}tez}, {Blakeslee}, {Bouwens}, {Bradley}, {Broadhurst}, {Brown}, {Burrows}, {Cheng}, {Cross}, {Demarco}, {Feldman}, {Franx}, {Goto}, {Gronwall}, {Hartig}, {Holden}, {Homeier}, {Infante}, {Jee}, {Kimble}, {Lesser}, {Martel}, {Mei}, {Menanteau}, {Meurer}, {Miley}, {Motta}, {Postman}, {Rosati}, {Sirianni}, {Sparks}, {Tran}, {Tsvetanov}, {White}, {Zheng}, and {Zirm}}]{Golimowski06}
{Golimowski} DA, {Ardila} DR, {Krist} JE, et~al (2006) {Hubble Space Telescope ACS Multiband Coronagraphic Imaging of the Debris Disk around {\ensuremath{\beta}} Pictoris}. \aj 131(6):3109--3130. \doi{10.1086/503801}, {\href{https://arxiv.org/abs/astro-ph/0602292}{{arXiv:astro-ph/0602292}}} {[astro-ph]}

\bibitem[{{G{\'o}mez Mart{\'\i}n} et~al.(2024){G{\'o}mez Mart{\'\i}n}, {Mu{\~n}oz}, {Martikainen}, {Guirado}, {Tanarro}, {Pel{\'a}ez}, {Mat{\'e}}, {Jim{\'e}nez-Redondo}, {Herrero}, {Peiteado}, and {Jardiel}}]{Gomez24}
{G{\'o}mez Mart{\'\i}n} JC, {Mu{\~n}oz} O, {Martikainen} J, et~al (2024) {Experimental Phase Function and Degree of Linear Polarization of Light Scattered by Hydrogenated Amorphous Carbon Circumstellar Dust Analogs}. \apjs 270(1):2. \doi{10.3847/1538-4365/ad0379}

\bibitem[{{Gonz{\'a}lez} and {Abascal}(2011)}]{TIP4P2005}
{Gonz{\'a}lez} MA, {Abascal} JLF (2011) {A flexible model for water based on TIP4P/2005}. \jcp 135(22):224516--224516. \doi{10.1063/1.3663219}

\bibitem[{{Gordon} et~al.(2021){Gordon}, {Misselt}, {Bouwman}, {Clayton}, {Decleir}, {Hines}, {Pendleton}, {Rieke}, {Smith}, and {Whittet}}]{Gordon2021}
{Gordon} KD, {Misselt} KA, {Bouwman} J, et~al (2021) {Milky Way Mid-Infrared Spitzer Spectroscopic Extinction Curves: Continuum and Silicate Features}. \apj 916(1):33. \doi{10.3847/1538-4357/ac00b7}, {\href{https://arxiv.org/abs/2105.05087}{{arXiv:2105.05087}}} {[astro-ph.GA]}

\bibitem[{{Graham} et~al.(2007){Graham}, {Kalas}, and {Matthews}}]{Graham07}
{Graham} JR, {Kalas} PG, {Matthews} BC (2007) {The Signature of Primordial Grain Growth in the Polarized Light of the AU Microscopii Debris Disk}. \apj 654(1):595--605. \doi{10.1086/509318}, {\href{https://arxiv.org/abs/astro-ph/0609332}{{arXiv:astro-ph/0609332}}} {[astro-ph]}

\bibitem[{{Greenberg} and {Hage}(1990)}]{Greenberg1990}
{Greenberg} JM, {Hage} JI (1990) {From Interstellar Dust to Comets: A Unification of Observational Constraints}. \apj 361:260. \doi{10.1086/169191}

\bibitem[{{Greenberg} et~al.(1961){Greenberg}, {Pedersen}, and {Pedersen}}]{Greenberg61}
{Greenberg} JM, {Pedersen} NE, {Pedersen} JC (1961) {Microwave Analog to the Scattering of Light by Nonspherical Particles}. J Appl Phys 32(2):233--242. \doi{10.1063/1.1735984}

\bibitem[{Gregg and Sing(1982)}]{RN1688}
Gregg SJ, Sing KS (1982) {Surface Area, and Porosity, 2nd ed. }. (Academic, London, 1982)

\bibitem[{{Grynko} et~al.(2020){Grynko}, {Shkuratov}, and {F{\"o}rstner}}]{Grynko20}
{Grynko} Y, {Shkuratov} Y, {F{\"o}rstner} J (2020) {Light backscattering from large clusters of densely packed irregular particles}. \jqsrt 255:107234. \doi{10.1016/j.jqsrt.2020.107234}

\bibitem[{{Guidi} et~al.(2022){Guidi}, {Isella}, {Testi}, {Chandler}, {Liu}, {Schmid}, {Rosotti}, {Meng}, {Jennings}, {Williams}, {Carpenter}, {de Gregorio-Monsalvo}, {Li}, {Liu}, {Ortolani}, {Quanz}, {Ricci}, and {Tazzari}}]{Guidi22}
{Guidi} G, {Isella} A, {Testi} L, et~al (2022) {Distribution of solids in the rings of the HD 163296 disk: a multiwavelength study}. \aap 664:A137. \doi{10.1051/0004-6361/202142303}, {\href{https://arxiv.org/abs/2207.01496}{{arXiv:2207.01496}}} {[astro-ph.EP]}

\bibitem[{{Guillet} et~al.(2020){Guillet}, {Girart}, {Maury}, and {Alves}}]{Guillet20}
{Guillet} V, {Girart} JM, {Maury} AJ, et~al (2020) {Polarized emission by aligned grains in the Mie regime: Application to protoplanetary disks observed by ALMA}. \aap 634:L15. \doi{10.1051/0004-6361/201937314}, {\href{https://arxiv.org/abs/2001.08400}{{arXiv:2001.08400}}} {[astro-ph.GA]}

\bibitem[{Gundlach and Blum(2015)}]{RN935}
Gundlach B, Blum J (2015) {The Stickiness of Micrometer-Sized Water-Ice Particles}. \apj 798(1):34. \doi{10.1088/0004-637X/798/1/34}

\bibitem[{Gundlach et~al.(2011)Gundlach, Kilias, Beitz, and Blum}]{RN1651}
Gundlach B, Kilias S, Beitz E, et~al (2011) {Micrometer-sized ice particles for planetary-science experiments - I. Preparation, critical rolling friction force, and specific surface energy}. Icarus 214(2):717--723. \doi{10.1016/j.icarus.2011.05.005}

\bibitem[{{Gupta} et~al.(2016){Gupta}, {Vaidya}, and {Dutta}}]{Gupta16}
{Gupta} R, {Vaidya} DB, {Dutta} R (2016) {Composite circumstellar dust grains}. \mnras 462(1):867--875. \doi{10.1093/mnras/stw1710}, {\href{https://arxiv.org/abs/1607.06253}{{arXiv:1607.06253}}} {[astro-ph.GA]}

\bibitem[{{Gustafson} and {Kolokolova}(1999)}]{Gustafson99}
{Gustafson} B{\r{A}}S, {Kolokolova} L (1999) {A systematic study of light scattering by aggregate particles using the microwave analog technique: Angular and wavelength dependence of intensity and polarization}. \jgr 104(D24):31711--31720. \doi{10.1029/1999JD900327}

\bibitem[{{G{\"u}ttler} et~al.(2019){G{\"u}ttler}, {Mannel}, {Rotundi}, {Merouane}, {Fulle}, {Bockel{\'e}e-Morvan}, {Lasue}, {Levasseur-Regourd}, {Blum}, {Naletto}, {Sierks}, {Hilchenbach}, {Tubiana}, {Capaccioni}, {Paquette}, {Flandes}, {Moreno}, {Agarwal}, {Bodewits}, {Bertini}, {Tozzi}, {Hornung}, {Langevin}, {Kr{\"u}ger}, {Longobardo}, {Della Corte}, {T{\'o}th}, {Filacchione}, {Ivanovski}, {Mottola}, and {Rinaldi}}]{Guttler19}
{G{\"u}ttler} C, {Mannel} T, {Rotundi} A, et~al (2019) {Synthesis of the morphological description of cometary dust at comet 67P/Churyumov-Gerasimenko}. \aap 630:A24. \doi{10.1051/0004-6361/201834751}, {\href{https://arxiv.org/abs/1902.10634}{{arXiv:1902.10634}}} {[astro-ph.EP]}

\bibitem[{Güttler et~al.(2010)Güttler, Blum, Zsom, Ormel, and Dullemond}]{RN1653}
Güttler C, Blum J, Zsom A, et~al (2010) {The outcome of protoplanetary dust growth: pebbles, boulders, or planetesimals? I. Mapping the zoo of laboratory collision experiments}. \aap 513:A56. \doi{10.1051/0004-6361/200912852}

\bibitem[{{Hadamcik} et~al.(2002){Hadamcik}, {Renard}, {Worms}, {Levasseur-Regourd}, and {Masson}}]{Hadamcik02}
{Hadamcik} E, {Renard} JB, {Worms} JC, et~al (2002) {Polarization of Light Scattered by Fluffy Particles (PROGRA $^{2}$ Experiment)}. \icarus 155(2):497--508. \doi{10.1006/icar.2001.6732}

\bibitem[{{Hadamcik} et~al.(2006){Hadamcik}, {Renard}, {Levasseur-Regourd}, and {Lasue}}]{Hadamcik06}
{Hadamcik} E, {Renard} JB, {Levasseur-Regourd} AC, et~al (2006) {Light scattering by fluffy particles with the PROGRA\^2 experiment: Mixtures of materials}. \jqsrt 100:143--156. \doi{10.1016/j.jqsrt.2005.11.032}

\bibitem[{{Hadamcik} et~al.(2007){Hadamcik}, {Renard}, {Rietmeijer}, {Levasseur-Regourd}, {Hill}, {Karner}, and {Nuth}}]{Hadamcik07}
{Hadamcik} E, {Renard} JB, {Rietmeijer} FJM, et~al (2007) {Light scattering by fluffy Mg-Fe-SiO and C mixtures as cometary analogs (PROGRA $^{2}$ experiment)}. \icarus 190(2):660--671. \doi{10.1016/j.icarus.2007.03.010}

\bibitem[{{Hage} and {Greenberg}(1990)}]{Hage90}
{Hage} JI, {Greenberg} JM (1990) {A Model for the Optical Properties of Porous Grains}. \apj 361:251. \doi{10.1086/169190}

\bibitem[{{Halder} et~al.(2018){Halder}, {Deb Roy}, and {Das}}]{Halder18}
{Halder} P, {Deb Roy} P, {Das} HS (2018) {Dependence of light scattering properties on porosity, size and composition of dust aggregates}. \icarus 312:45--60. \doi{10.1016/j.icarus.2018.04.026}, {\href{https://arxiv.org/abs/1804.08324}{{arXiv:1804.08324}}} {[astro-ph.EP]}

\bibitem[{{Hapke}(1981)}]{Hapke81}
{Hapke} B (1981) {Bidirectional reflectance spectroscopy. I - Theory}. \jgr 86:3039--3054. \doi{10.1029/JB086iB04p03039}

\bibitem[{Hapke(2001)}]{RN1704}
Hapke B (2001) {Space weathering from Mercury to the asteroid belt}. \jgr 106(E5):10039--10073. \doi{10.1029/2000je001338}

\bibitem[{Harmon et~al.(1997)Harmon, Ostro, Benner, Rosema, Jurgens, Winkler, Yeomans, Choate, Cormier, Giorgini, Mitchell, Chodas, Rose, Kelley, Slade, and Thomas}]{RN813}
Harmon JK, Ostro SJ, Benner LAM, et~al (1997) {Radar detection of the nucleus and Coma of Comet Hyakutake (C/1996 B2)}. Science 278(5345):1921--1924

\bibitem[{{Hasegawa} and {Herbst}(1993)}]{HH93}
{Hasegawa} TI, {Herbst} E (1993) {Three-Phase Chemical Models of Dense Interstellar Clouds - Gas Dust Particle Mantles and Dust Particle Surfaces}. \mnras 263:589. \doi{10.1093/mnras/263.3.589}

\bibitem[{{Hasegawa} et~al.(1992){Hasegawa}, {Herbst}, and {Leung}}]{Hasegawa92}
{Hasegawa} TI, {Herbst} E, {Leung} CM (1992) {Models of Gas-Grain Chemistry in Dense Interstellar Clouds with Complex Organic Molecules}. \apjs 82:167. \doi{10.1086/191713}

\bibitem[{{Hasegawa} et~al.(2021){Hasegawa}, {Suzuki}, {Tanaka}, {Kobayashi}, and {Wada}}]{Hasegawa21}
{Hasegawa} Y, {Suzuki} TK, {Tanaka} H, et~al (2021) {Collisional Growth and Fragmentation of Dust Aggregates with Low Mass Ratios. I. Critical Collision Velocity for Water Ice}. \apj 915(1):22. \doi{10.3847/1538-4357/abf6cf}, {\href{https://arxiv.org/abs/2104.06711}{{arXiv:2104.06711}}} {[astro-ph.EP]}

\bibitem[{{Hayashi}(1981)}]{Hayashi1981}
{Hayashi} C (1981) {Structure of the Solar Nebula, Growth and Decay of Magnetic Fields and Effects of Magnetic and Turbulent Viscosities on the Nebula}. Progress of Theoretical Physics Supplement 70:35--53. \doi{10.1143/PTPS.70.35}

\bibitem[{Helling(2019)}]{RN852}
Helling C (2019) {Exoplanet Clouds}. Annual Review of Earth and Planetary Sciences, Vol 47 47:583--606. \doi{10.1146/annurev-earth-053018-060401}

\bibitem[{{Heng} and {Draine}(2009)}]{Heng09}
{Heng} K, {Draine} BT (2009) {Constraining the Porosities of Interstellar Dust Grains}. arXiv e-prints arXiv:0906.0773. \doi{10.48550/arXiv.0906.0773}, {\href{https://arxiv.org/abs/0906.0773}{{arXiv:0906.0773}}} {[astro-ph.GA]}

\bibitem[{Henning(2010)}]{RN1695}
Henning T (ed)  (2010) Astromineralogy, Lecture Notes in Physics, vol 815, 2nd edn. Springer, Berlin, Heidelberg, \doi{10.1007/978-3-642-13259-9}

\bibitem[{{Henning}(2010)}]{Henning2010}
{Henning} T (2010) {Cosmic silicates}. \araa 48:21--46. \doi{10.1146/annurev-astro-081309-130815}

\bibitem[{Henning and Salama(1998)}]{RN381}
Henning T, Salama F (1998) {Carbon in the Universe}. Science 282(5397):2204--2210. \doi{10.1126/science.282.5397.2204}

\bibitem[{{Henning} and {Stognienko}(1996)}]{Henning96}
{Henning} T, {Stognienko} R (1996) {Dust opacities for protoplanetary accretion disks: influence of dust aggregates.} \aap 311:291--303

\bibitem[{{Henning} et~al.(1995){Henning}, {Michel}, and {Stognienko}}]{Henning95}
{Henning} T, {Michel} B, {Stognienko} R (1995) {Dust opacities in dense regions}. \planss 43(10):1333--1343. \doi{10.1016/0032-0633(95)00003-N}

\bibitem[{Hensley and Draine(2017)}]{Hensley_2017}
Hensley BS, Draine BT (2017) {Modeling the Anomalous Microwave Emission with Spinning Nanoparticles: No PAHs Required}. \apj 836(2):179. \doi{10.3847/1538-4357/aa5c37}

\bibitem[{{Hensley} and {Draine}(2020)}]{Hensley2020}
{Hensley} BS, {Draine} BT (2020) {Detection of PAH Absorption and Determination of the Mid-infrared Diffuse Interstellar Extinction Curve from the Sight Line toward Cyg OB2-12}. \apj 895(1):38. \doi{10.3847/1538-4357/ab8cc3}, {\href{https://arxiv.org/abs/2002.02457}{{arXiv:2002.02457}}} {[astro-ph.GA]}

\bibitem[{{Hensley} and {Draine}(2021)}]{Hensley2021}
{Hensley} BS, {Draine} BT (2021) {Observational Constraints on the Physical Properties of Interstellar Dust in the Post-Planck Era}. \apj 906(2):73. \doi{10.3847/1538-4357/abc8f1}, {\href{https://arxiv.org/abs/2009.00018}{{arXiv:2009.00018}}} {[astro-ph.GA]}

\bibitem[{{Hensley} and {Draine}(2023)}]{Hensley2023}
{Hensley} BS, {Draine} BT (2023) {The Astrodust+PAH Model: A Unified Description of the Extinction, Emission, and Polarization from Dust in the Diffuse Interstellar Medium}. \apj 948(1):55. \doi{10.3847/1538-4357/acc4c2}, {\href{https://arxiv.org/abs/2208.12365}{{arXiv:2208.12365}}} {[astro-ph.GA]}

\bibitem[{{Herranen} et~al.(2021){Herranen}, {Lazarian}, and {Hoang}}]{Herranen21}
{Herranen} J, {Lazarian} A, {Hoang} T (2021) {Alignment of Irregular Grains by Radiative Torques: Efficiency Study}. \apj 913(1):63. \doi{10.3847/1538-4357/abf096}, {\href{https://arxiv.org/abs/2006.16563}{{arXiv:2006.16563}}} {[astro-ph.GA]}

\bibitem[{{Hildebrand}(1983)}]{Hildebrand1983}
{Hildebrand} RH (1983) {The determination of cloud masses and dust characteristics from submillimetre thermal emission.} \qjras 24:267--282

\bibitem[{{Hildebrand} et~al.(2000){Hildebrand}, {Davidson}, {Dotson}, {Dowell}, {Novak}, and {Vaillancourt}}]{Hildebrand2000}
{Hildebrand} RH, {Davidson} JA, {Dotson} JL, et~al (2000) {A Primer on Far-Infrared Polarimetry}. \pasp 112(775):1215--1235. \doi{10.1086/316613}

\bibitem[{Hirashita and Il'in(2022)}]{RN1910}
Hirashita H, Il'in VB (2022) {Evolution of dust grain size distribution and grain porosity in galaxies}. \mnras 509(4):5771--5789. \doi{10.1093/mnras/stab3455}

\bibitem[{{Homma} and {Nakamoto}(2018)}]{Homma18}
{Homma} K, {Nakamoto} T (2018) {Collisional Growth of Icy Dust Aggregates in the Disk Formation Stage: Difficulties for Planetesimal Formation via Direct Collisional Growth outside the Snowline}. \apj 868(2):118. \doi{10.3847/1538-4357/aae0fb}, {\href{https://arxiv.org/abs/1809.06733}{{arXiv:1809.06733}}} {[astro-ph.EP]}

\bibitem[{{Hull} et~al.(2018){Hull}, {Yang}, {Li}, {Kataoka}, {Stephens}, {Andrews}, {Bai}, {Cleeves}, {Hughes}, {Looney}, {P{\'e}rez}, and {Wilner}}]{Hull18}
{Hull} CLH, {Yang} H, {Li} ZY, et~al (2018) {ALMA Observations of Polarization from Dust Scattering in the IM Lup Protoplanetary Disk}. \apj 860(1):82. \doi{10.3847/1538-4357/aabfeb}, {\href{https://arxiv.org/abs/1804.06269}{{arXiv:1804.06269}}} {[astro-ph.SR]}

\bibitem[{{Hunziker} et~al.(2021){Hunziker}, {Schmid}, {Ma}, {Menard}, {Avenhaus}, {Boccaletti}, {Beuzit}, {Chauvin}, {Dohlen}, {Dominik}, {Engler}, {Ginski}, {Gratton}, {Henning}, {Langlois}, {Milli}, {Mouillet}, {Tschudi}, {van Holstein}, and {Vigan}}]{Hunziker21}
{Hunziker} S, {Schmid} HM, {Ma} J, et~al (2021) {HD 142527: quantitative disk polarimetry with SPHERE}. \aap 648:A110. \doi{10.1051/0004-6361/202040166}, {\href{https://arxiv.org/abs/2103.08462}{{arXiv:2103.08462}}} {[astro-ph.EP]}

\bibitem[{Hörz et~al.(2006)Hörz, Bastien, Borg, Bradley, Bridges, Brownlee, Burchell, Chi, Cintala, Dai, Djouadi, Dominguez, Economou, Fairey, Floss, Franchi, Graham, Green, Heck, Hoppe, Huth, Ishii, Kearsley, Kissel, Leitner, Leroux, Marhas, Messenger, Schwandt, See, Snead, Stadermann, Stephan, Stroud, Teslich, Trigo-Rodriguez, Tuzzolino, Troadec, Tsou, Warren, Westphal, Wozniakiewicz, Wright, and Zinner}]{RN811}
Hörz F, Bastien R, Borg J, et~al (2006) {Impact features on Stardust: Implications for comet 81P/Wild 2 dust}. Science 314(5806):1716--1719. \doi{10.1126/science.1135705}

\bibitem[{{Imai} et~al.(1996){Imai}, {Yasumori}, {Hirashima}, {Awazu}, and {Onuki}}]{Imai1996}
{Imai} H, {Yasumori} M, {Hirashima} H, et~al (1996) {Significant densification of sol-gel derived amorphous silica films by vacuum ultraviolet irradiation}. J Appl Phys 79(11):8304--8309. \doi{10.1063/1.362541}

\bibitem[{{Indebetouw} et~al.(2005){Indebetouw}, {Mathis}, {Babler}, {Meade}, {Watson}, {Whitney}, {Wolff}, {Wolfire}, {Cohen}, {Bania}, {Benjamin}, {Clemens}, {Dickey}, {Jackson}, {Kobulnicky}, {Marston}, {Mercer}, {Stauffer}, {Stolovy}, and {Churchwell}}]{Indebetouw2005}
{Indebetouw} R, {Mathis} JS, {Babler} BL, et~al (2005) {The Wavelength Dependence of Interstellar Extinction from 1.25 to 8.0 {\ensuremath{\mu}}m Using GLIMPSE Data}. \apj 619(2):931--938. \doi{10.1086/426679}, {\href{https://arxiv.org/abs/astro-ph/0406403}{{arXiv:astro-ph/0406403}}} {[astro-ph]}

\bibitem[{Ioppolo et~al.(2021)Ioppolo, Fedoseev, Chuang, Cuppen, Clements, Jin, Garrod, Qasim, Kofman, van Dishoeck, and Linnartz}]{RN1324}
Ioppolo S, Fedoseev G, Chuang KJ, et~al (2021) {A non-energetic mechanism for glycine formation in the interstellar medium}. Nat Astron 5:197. \doi{10.1038/s41550-020-01249-0}

\bibitem[{{Ishizuka} et~al.(2015){Ishizuka}, {Kimura}, and {Sakon}}]{Ishizuka2015}
{Ishizuka} S, {Kimura} Y, {Sakon} I (2015) {In Situ Infrared Measurements of Free-flying Silicate during Condensation in the Laboratory}. \apj 803(2):88. \doi{10.1088/0004-637X/803/2/88}

\bibitem[{Islam et~al.(2007)Islam, Latimer, and Price}]{RN1889}
Islam F, Latimer ER, Price SD (2007) {The formation of vibrationally excited HD from atomic recombination on cold graphite surfaces}. \jcp 127(6):064701. \doi{10.1063/1.2754684}

\bibitem[{J\"ager et~al.(1998)J\"ager, Mutschke, and Henning}]{RN1650}
J\"ager C, Mutschke H, Henning T (1998) {Optical properties of carbonaceous dust analogues}. \aap 332(1):291--299

\bibitem[{J\"ager et~al.(2008)J\"ager, Mutschke, Henning, and Huisken}]{RN407}
J\"ager C, Mutschke H, Henning T, et~al (2008) {Spectral Properties of Gas-Phase Condensed Fullerene-Like Carbon Nanoparticles from Far-Ultraviolet to Infrared Wavelengths}. \apj 689(1):249--259. \doi{10.1086/592729}

\bibitem[{{J{\"a}ger} et~al.(2009){J{\"a}ger}, {Mutschke}, {Henning}, and {Huisken}}]{RN408}
{J{\"a}ger} C, {Mutschke} H, {Henning} T, et~al (2009) {Analogs of Cosmic Dust}. In: {Henning} T, {Gr{\"u}n} E, {Steinacker} J (eds) {Cosmic Dust - Near and Far}, ASP Conference Series, vol 414. Astronomical Society of the Pacific, p 319

\bibitem[{J\"ager et~al.(2016)J\"ager, Sabri, Wendler, and Henning}]{RN1405}
J\"ager C, Sabri T, Wendler E, et~al (2016) {Ion-induced processing of cosmic silicates: a possible formation pathway to GEMs}. ApJ 831:66. \doi{10.3847/0004-637X/831/1/66}

\bibitem[{{Jang} et~al.(2024){Jang}, {Waters}, {Kaeufer}, {Tamanai}, {Perotti}, {Christiaens}, {Kamp}, {Henning}, {Min}, {Arabhavi}, {Barrado}, {van Dishoeck}, {Gasman}, {Grant}, {G{\"u}del}, {Lagage}, {Lahuis}, {Schwarz}, {Tabone}, and {Temmink}}]{Jiang24}
{Jang} H, {Waters} R, {Kaeufer} T, et~al (2024) {Dust mineralogy and variability of the inner PDS 70 disk: Insights from JWST/MIRI MRS and Spitzer IRS observations}. \aap 691:A148. \doi{10.1051/0004-6361/202451589}, {\href{https://arxiv.org/abs/2408.16367}{{arXiv:2408.16367}}} {[astro-ph.EP]}

\bibitem[{{Jessberger} et~al.(2001){Jessberger}, {Stephan}, {Rost}, {Arndt}, {Maetz}, {Stadermann}, {Brownlee}, {Bradley}, and {Kurat}}]{Jessberger2001}
{Jessberger} EK, {Stephan} T, {Rost} D, et~al (2001) {Properties of Interplanetary Dust: Information from Collected Samples}. In: {Gr{\"u}n} E, {Gustafson} BAS, {Dermott} S, et~al (eds) {Interplanetary Dust}. Springer, Berlin, Heidelberg, p 253--294, \doi{10.1007/978-3-642-56428-4_6}

\bibitem[{Jiang et~al.(2024)Jiang, Macías, Guerra-Alvarado, and Carrasco-Gonz\'alez}]{RN1862}
Jiang HC, Macías E, Guerra-Alvarado OM, et~al (2024) {Grain-size measurements in protoplanetary disks indicate fragile pebbles and low turbulence}. \aap 682:A32. \doi{10.1051/0004-6361/202348271}

\bibitem[{{Joblin} et~al.(1992){Joblin}, {Leger}, and {Martin}}]{Joblin1992}
{Joblin} C, {Leger} A, {Martin} P (1992) {Contribution of Polycyclic Aromatic Hydrocarbon Molecules to the Interstellar Extinction Curve}. \apjl 393:L79. \doi{10.1086/186456}

\bibitem[{Johansen and Lambrechts(2017)}]{RN1280}
Johansen A, Lambrechts M (2017) {Forming Planets via Pebble Accretion}. Annual Review of Earth and Planetary Sciences, Vol 45 45:359--387. \doi{10.1146/annurev-earth-063016-020226}

\bibitem[{{Johansen} et~al.(2014){Johansen}, {Blum}, {Tanaka}, {Ormel}, {Bizzarro}, and {Rickman}}]{Johansen14}
{Johansen} A, {Blum} J, {Tanaka} H, et~al (2014) {The Multifaceted Planetesimal Formation Process}. In: {Beuther} H, {Klessen} RS, {Dullemond} CP, et~al (eds) {Protostars and Planets VI}, pp 547--570, \doi{10.2458/azu_uapress_9780816531240-ch024}, {\href{https://arxiv.org/abs/1402.1344}{{arXiv:1402.1344}}}

\bibitem[{{Jones}(1988)}]{Jones1998}
{Jones} AP (1988) {Modelling interstellar extinction. I - Porous grains}. \mnras 234:209--218. \doi{10.1093/mnras/234.2.209}

\bibitem[{Jones(2016)}]{RN1508}
Jones AP (2016) {Dust evolution, a global view: III. Core/mantle grains, organic nano-globules, comets and surface chemistry}. Royal Society Open Science 3(12):160224. \doi{10.1098/rsos.160224}

\bibitem[{Jones and Nuth(2011)}]{RN1393}
Jones AP, Nuth JA (2011) {Dust destruction in the ISM: a re-evaluation of dust lifetimes}. \aap 530:A44. \doi{10.1051/0004-6361/201014440}

\bibitem[{{Jones} and {Williams}(1984)}]{Jones1984}
{Jones} AP, {Williams} DA (1984) {The 3 MU M ice band in Taurus : implications for interstellar chemistry.} \mnras 209:955--960. \doi{10.1093/mnras/209.4.955}

\bibitem[{{Jones} et~al.(1996){Jones}, {Tielens}, and {Hollenbach}}]{Jones1996}
{Jones} AP, {Tielens} AGGM, {Hollenbach} DJ (1996) {Grain Shattering in Shocks: The Interstellar Grain Size Distribution}. \apj 469:740. \doi{10.1086/177823}

\bibitem[{Jones et~al.(2013)Jones, Fanciullo, Kohler, Verstraete, Guillet, Bocchio, and Ysard}]{RN795}
Jones AP, Fanciullo L, Kohler M, et~al (2013) {The evolution of amorphous hydrocarbons in the ISM: dust modelling from a new vantage point}. \aap 558:A62

\bibitem[{{Jones} et~al.(2013){Jones}, {Fanciullo}, {K{\"o}hler}, {Verstraete}, {Guillet}, {Bocchio}, and {Ysard}}]{Jones2013}
{Jones} AP, {Fanciullo} L, {K{\"o}hler} M, et~al (2013) {The evolution of amorphous hydrocarbons in the ISM: dust modelling from a new vantage point}. \aap 558:A62. \doi{10.1051/0004-6361/201321686}, {\href{https://arxiv.org/abs/1411.6293}{{arXiv:1411.6293}}} {[astro-ph.GA]}

\bibitem[{{Karssemeijer} et~al.(2014{\natexlab{a}}){Karssemeijer}, {de Wijs}, and {Cuppen}}]{CO2-H2O_2014}
{Karssemeijer} LJ, {de Wijs} GA, {Cuppen} HM (2014{\natexlab{a}}) {Interactions of adsorbed CO2 on water ice at low temperatures}. Physical Chemistry Chemical Physics (Incorporating Faraday Transactions) 16(29):15630. \doi{10.1039/C4CP01622J}, {\href{https://arxiv.org/abs/1406.6161}{{arXiv:1406.6161}}} {[astro-ph.GA]}

\bibitem[{{Karssemeijer} et~al.(2014{\natexlab{b}}){Karssemeijer}, {Ioppolo}, {van Hemert}, {van der Avoird}, {Allodi}, {Blake}, and {Cuppen}}]{CO-H2O_2014}
{Karssemeijer} LJ, {Ioppolo} S, {van Hemert} MC, et~al (2014{\natexlab{b}}) {Dynamics of CO in Amorphous Water-ice Environments}. \apj 781(1):16. \doi{10.1088/0004-637X/781/1/16}, {\href{https://arxiv.org/abs/1311.6643}{{arXiv:1311.6643}}} {[astro-ph.GA]}

\bibitem[{Kataoka et~al.(2013)Kataoka, Tanaka, Okuzumi, and Wada}]{RN788}
Kataoka A, Tanaka H, Okuzumi S, et~al (2013) {Fluffy dust forms icy planetesimals by static compression}. \aap 557:L4. \doi{10.1051/0004-6361/201322151}

\bibitem[{{Kataoka} et~al.(2014){Kataoka}, {Okuzumi}, {Tanaka}, and {Nomura}}]{Kataoka14}
{Kataoka} A, {Okuzumi} S, {Tanaka} H, et~al (2014) {Opacity of fluffy dust aggregates}. \aap 568:A42. \doi{10.1051/0004-6361/201323199}, {\href{https://arxiv.org/abs/1312.1459}{{arXiv:1312.1459}}} {[astro-ph.EP]}

\bibitem[{{Kataoka} et~al.(2015){Kataoka}, {Muto}, {Momose}, {Tsukagoshi}, {Fukagawa}, {Shibai}, {Hanawa}, {Murakawa}, and {Dullemond}}]{Kataoka15}
{Kataoka} A, {Muto} T, {Momose} M, et~al (2015) {Millimeter-wave Polarization of Protoplanetary Disks due to Dust Scattering}. \apj 809(1):78. \doi{10.1088/0004-637X/809/1/78}, {\href{https://arxiv.org/abs/1504.04812}{{arXiv:1504.04812}}} {[astro-ph.EP]}

\bibitem[{{Kataoka} et~al.(2019){Kataoka}, {Okuzumi}, and {Tazaki}}]{Kataoka19}
{Kataoka} A, {Okuzumi} S, {Tazaki} R (2019) {Millimeter-wave Polarization Due to Grain Alignment by the Gas Flow in Protoplanetary Disks}. \apjl 874(1):L6. \doi{10.3847/2041-8213/ab0c9a}, {\href{https://arxiv.org/abs/1903.03529}{{arXiv:1903.03529}}} {[astro-ph.EP]}

\bibitem[{Keil et~al.(2000)Keil, Krishna, and Coppens}]{Diffusion_in_zeolites_Koppens2000}
Keil FJ, Krishna R, Coppens MO (2000) {Modeling of Diffusion in Zeolites}. Reviews in Chemical Engineering 16(2):71--197. \doi{doi:10.1515/REVCE.2000.16.2.71}

\bibitem[{Kempf et~al.(1999)Kempf, Pfalzner, and Henning}]{RN1901}
Kempf S, Pfalzner S, Henning TK (1999) {N-particle-simulations of dust growth - I. Growth driven by Brownian motion}. Icarus 141(2):388--398. \doi{10.1006/icar.1999.6171}

\bibitem[{{Kimura}(2001)}]{Kimura01}
{Kimura} H (2001) {Light-scattering properties of fractal aggregates: numerical calculations by a superposition technique and the discrete-dipole approximation}. \jqsrt 70:581--594. \doi{10.1016/S0022-4073(01)00031-0}

\bibitem[{{Kimura} et~al.(2006){Kimura}, {Kolokolova}, and {Mann}}]{Kimura06}
{Kimura} H, {Kolokolova} L, {Mann} I (2006) {Light scattering by cometary dust numerically simulated with aggregate particles consisting of identical spheres}. \aap 449(3):1243--1254. \doi{10.1051/0004-6361:20041783}

\bibitem[{{Kimura} et~al.(2016){Kimura}, {Kolokolova}, {Li}, and {Lebreton}}]{Kimura2016review}
{Kimura} H, {Kolokolova} L, {Li} A, et~al (2016) {Light Scattering and Thermal Emission by Primitive Dust Particles in Planetary Systems}. arXiv e-prints arXiv:1603.03123. \doi{10.48550/arXiv.1603.03123}, {\href{https://arxiv.org/abs/1603.03123}{{arXiv:1603.03123}}} {[astro-ph.EP]}

\bibitem[{Kimura et~al.(2020)Kimura, Wada, Kobayashi, Senshu, Hirai, Yoshida, Kobayashi, Hong, Arai, Ishibashi, and Yamada}]{RN1221}
Kimura H, Wada K, Kobayashi H, et~al (2020) {Is water ice an efficient facilitator for dust coagulation?} \mnras 498:1801

\bibitem[{{Kimura} et~al.(2022){Kimura}, {Tanaka}, {Inatomi}, {Ferguson}, and {Nuth}}]{Kimura2022}
{Kimura} Y, {Tanaka} KK, {Inatomi} Y, et~al (2022) {Inefficient Growth of SiOx Grains: Implications for Circumstellar Outflows}. \apjl 934(1):L10. \doi{10.3847/2041-8213/ac8002}

\bibitem[{{Kirchschlager} and {Bertrang}(2020)}]{Kirchschlager20}
{Kirchschlager} F, {Bertrang} GHM (2020) {Self-scattering of non-spherical dust grains. The limitations of perfect compact spheres}. \aap 638:A116. \doi{10.1051/0004-6361/202037943}, {\href{https://arxiv.org/abs/2004.13742}{{arXiv:2004.13742}}} {[astro-ph.SR]}

\bibitem[{{Kirchschlager} and {Wolf}(2014)}]{Kirchschlager14}
{Kirchschlager} F, {Wolf} S (2014) {Effect of dust grain porosity on the appearance of protoplanetary disks}. \aap 568:A103. \doi{10.1051/0004-6361/201323176}, {\href{https://arxiv.org/abs/1407.6575}{{arXiv:1407.6575}}} {[astro-ph.SR]}

\bibitem[{{Kirchschlager} et~al.(2019){Kirchschlager}, {Bertrang}, and {Flock}}]{Kirch19}
{Kirchschlager} F, {Bertrang} GHM, {Flock} M (2019) {Intrinsic polarization of elongated porous dust grains}. \mnras 488(1):1211--1219. \doi{10.1093/mnras/stz1763}, {\href{https://arxiv.org/abs/1906.10699}{{arXiv:1906.10699}}} {[astro-ph.EP]}

\bibitem[{Knight et~al.(2019)Knight, Kalugin, Coker, and Ilgen}]{Knight_water_confinement_2019}
Knight AW, Kalugin NG, Coker E, et~al (2019) {Water properties under nano-scale confinement}. Scientific Reports 9(1). \doi{10.1038/s41598-019-44651-z}

\bibitem[{{Kobayashi} and {Tanaka}(2021)}]{Kobayashi21}
{Kobayashi} H, {Tanaka} H (2021) {Rapid Formation of Gas-giant Planets via Collisional Coagulation from Dust Grains to Planetary Cores}. \apj 922(1):16. \doi{10.3847/1538-4357/ac289c}, {\href{https://arxiv.org/abs/2110.00919}{{arXiv:2110.00919}}} {[astro-ph.EP]}

\bibitem[{{K{\"o}hler} et~al.(2008){K{\"o}hler}, {Mann}, and {Li}}]{Kohler08}
{K{\"o}hler} M, {Mann} I, {Li} A (2008) {Complex Organic Materials in the HR 4796A Disk?} \apjl 686(2):L95. \doi{10.1086/592961}, {\href{https://arxiv.org/abs/0808.4113}{{arXiv:0808.4113}}} {[astro-ph]}

\bibitem[{{K{\"o}hler} et~al.(2011){K{\"o}hler}, {Guillet}, and {Jones}}]{Kohler11}
{K{\"o}hler} M, {Guillet} V, {Jones} A (2011) {Aggregate dust connections and emissivity enhancements}. \aap 528:A96. \doi{10.1051/0004-6361/201016379}

\bibitem[{{K{\"o}hler} et~al.(2012){K{\"o}hler}, {Stepnik}, {Jones}, {Guillet}, {Abergel}, {Ristorcelli}, and {Bernard}}]{Kohler2012}
{K{\"o}hler} M, {Stepnik} B, {Jones} AP, et~al (2012) {Dust coagulation processes as constrained by far-infrared observations}. \aap 548:A61. \doi{10.1051/0004-6361/201218975}

\bibitem[{{Kolokolova} and {Gustafson}(2001)}]{Kolokolova01}
{Kolokolova} L, {Gustafson} BAS (2001) {Scattering by inhomogeneous particles: microwave analog experiments and comparison to effective medium theories}. \jqsrt 70:611--625. \doi{10.1016/S0022-4073(01)00033-4}

\bibitem[{{Kolokolova} and {Kimura}(2010)}]{Kolokolova10}
{Kolokolova} L, {Kimura} H (2010) {Effects of electromagnetic interaction in the polarization of light scattered by cometary and other types of cosmic dust}. \aap 513:A40. \doi{10.1051/0004-6361/200913681}

\bibitem[{{Kolokolova} et~al.(2007){Kolokolova}, {Kimura}, {Kiselev}, and {Rosenbush}}]{Kolokolova07}
{Kolokolova} L, {Kimura} H, {Kiselev} N, et~al (2007) {Two different evolutionary types of comets proved by polarimetric and infrared properties of their dust}. \aap 463(3):1189--1196. \doi{10.1051/0004-6361:20065069}, {\href{https://arxiv.org/abs/astro-ph/0703220}{{arXiv:astro-ph/0703220}}} {[astro-ph]}

\bibitem[{{Kothe} et~al.(2013){Kothe}, {Blum}, {Weidling}, and {G{\"u}ttler}}]{Kothe13}
{Kothe} S, {Blum} J, {Weidling} R, et~al (2013) {Free collisions in a microgravity many-particle experiment. III. The collision behavior of sub-millimeter-sized dust aggregates}. \icarus 225(1):75--85. \doi{10.1016/j.icarus.2013.02.034}, {\href{https://arxiv.org/abs/1302.5532}{{arXiv:1302.5532}}} {[astro-ph.EP]}

\bibitem[{{Kozasa} et~al.(1992){Kozasa}, {Blum}, and {Mukai}}]{Kozasa92}
{Kozasa} T, {Blum} J, {Mukai} T (1992) {Optical properties of dust aggregates. I. Wavelength dependence}. \aap 263(1-2):423--432

\bibitem[{{Kozasa} et~al.(1993){Kozasa}, {Blum}, {Okamoto}, and {Mukai}}]{Kozasa93}
{Kozasa} T, {Blum} J, {Okamoto} H, et~al (1993) {Optical properties of dust aggregates. II. Angular dependence of scattered light}. \aap 276:278

\bibitem[{{Kramer} et~al.(2003){Kramer}, {Richer}, {Mookerjea}, {Alves}, and {Lada}}]{Kramer2003}
{Kramer} C, {Richer} J, {Mookerjea} B, et~al (2003) {Dust properties of the dark cloud IC 5146. Submillimeter and NIR imaging}. \aap 399:1073--1082. \doi{10.1051/0004-6361:20021823}, {\href{https://arxiv.org/abs/astro-ph/0212265}{{arXiv:astro-ph/0212265}}} {[astro-ph]}

\bibitem[{Krasnokutski et~al.(2014)Krasnokutski, Rouille, Jager, Huisken, Zhukovska, and Henning}]{RN825}
Krasnokutski SA, Rouille G, Jager C, et~al (2014) {Formation of Silicon Oxide Grains at Low Temperature}. \apj 782(1):15. \doi{10.1088/0004-637X/782/1/15}

\bibitem[{Krause and Blum(2004)}]{RN786}
Krause M, Blum J (2004) {Growth and form of planetary seedlings: Results from a sounding rocket microgravity aggregation experiment}. \prl 93(2):021103. \doi{10.1103/PhysRevLett.93.021103}

\bibitem[{{Krijt} et~al.(2015){Krijt}, {Ormel}, {Dominik}, and {Tielens}}]{Krijt15}
{Krijt} S, {Ormel} CW, {Dominik} C, et~al (2015) {Erosion and the limits to planetesimal growth}. \aap 574:A83. \doi{10.1051/0004-6361/201425222}, {\href{https://arxiv.org/abs/1412.3593}{{arXiv:1412.3593}}} {[astro-ph.EP]}

\bibitem[{{Krijt} et~al.(2016){Krijt}, {Ormel}, {Dominik}, and {Tielens}}]{Krijt16}
{Krijt} S, {Ormel} CW, {Dominik} C, et~al (2016) {A panoptic model for planetesimal formation and pebble delivery}. \aap 586:A20. \doi{10.1051/0004-6361/201527533}, {\href{https://arxiv.org/abs/1511.07762}{{arXiv:1511.07762}}} {[astro-ph.SR]}

\bibitem[{{Krijt} et~al.(2023){Krijt}, {Kama}, {McClure}, {Teske}, {Bergin}, {Shorttle}, {Walsh}, and {Raymond}}]{Krijt2023}
{Krijt} S, {Kama} M, {McClure} M, et~al (2023) {Chemical Habitability: Supply and Retention of Life's Essential Elements During Planet Formation}. In: {Inutsuka} S, {Aikawa} Y, {Muto} T, et~al (eds) {Protostars and Planets VII}, p 1031, \doi{10.48550/arXiv.2203.10056}, {\href{https://arxiv.org/abs/2203.10056}{{arXiv:2203.10056}}}

\bibitem[{{Krist} et~al.(2010){Krist}, {Stapelfeldt}, {Bryden}, {Rieke}, {Su}, {Chen}, {Beichman}, {Hines}, {Rebull}, {Tanner}, {Trilling}, {Clampin}, and {G{\'a}sp{\'a}r}}]{Krist10}
{Krist} JE, {Stapelfeldt} KR, {Bryden} G, et~al (2010) {HST and Spitzer Observations of the HD 207129 Debris Ring}. \aj 140(4):1051--1061. \doi{10.1088/0004-6256/140/4/1051}, {\href{https://arxiv.org/abs/1008.2793}{{arXiv:1008.2793}}} {[astro-ph.SR]}

\bibitem[{Krivov(2010)}]{RN424}
Krivov AV (2010) {Debris disks: seeing dust, thinking of planetesimals and planets}. Research in Astronomy and Astrophysics 10(5):383--414. \doi{10.1088/1674-4527/10/5/001}

\bibitem[{Kroto et~al.(1985)Kroto, Heath, Obrien, Curl, and Smalley}]{RN425}
Kroto HW, Heath JR, Obrien SC, et~al (1985) {C-60 - Buckminsterfullerene}. Nature 318(6042):162--163. Aud42 Times Cited:9763 Cited References Count:9

\bibitem[{Langreth(1985)}]{RN1884}
Langreth DC (1985) {Energy-Transfer at Surfaces - Asymmetric Line-Shapes and the Electron-Hole Pair Mechanism}. \prl 54(2):126--129. \doi{10.1103/PhysRevLett.54.126}

\bibitem[{Lasue and Levasseur-Regourd(2006)}]{RN812}
Lasue J, Levasseur-Regourd AC (2006) {Porous irregular aggregates of sub-micron sized grains to reproduce cometary dust light scattering observations}. \jqsrt 100(1-3):220--236. \doi{10.1016/j.jqsrt.2005.11.040}

\bibitem[{Latimer et~al.(2008)Latimer, Islam, and Price}]{RN1890}
Latimer ER, Islam F, Price SD (2008) {Studies of HD formed in excited vibrational states from atomic recombination on cold graphite surfaces}. Chem Phys Lett 455(4-6):174--177. \doi{10.1016/j.cplett.2008.02.105}

\bibitem[{Le~Caër et~al.(2011)Le~Caër, Pin, Esnouf, Raffy, Renault, Brubach, Creff, and Roy}]{Trapped_water_network_PCCP2011}
Le~Caër S, Pin S, Esnouf S, et~al (2011) A trapped water network in nanoporous material: the role of interfaces. Phys Chem Chem Phys 13:17658--17666. \doi{10.1039/C1CP21980D}

\bibitem[{{Lebreton} et~al.(2012){Lebreton}, {Augereau}, {Thi}, {Roberge}, {Donaldson}, {Schneider}, {Maddison}, {M{\'e}nard}, {Riviere-Marichalar}, {Mathews}, {Kamp}, {Pinte}, {Dent}, {Barrado}, {Duch{\^e}ne}, {Gonzalez}, {Grady}, {Meeus}, {Pantin}, {Williams}, and {Woitke}}]{Lebreton12}
{Lebreton} J, {Augereau} JC, {Thi} WF, et~al (2012) {An icy Kuiper belt around the young solar-type star HD 181327}. \aap 539:A17. \doi{10.1051/0004-6361/201117714}, {\href{https://arxiv.org/abs/1112.3398}{{arXiv:1112.3398}}} {[astro-ph.EP]}

\bibitem[{{Lee} et~al.(1988){Lee}, {Yang}, and {Parr}}]{LYP-1988}
{Lee} C, {Yang} W, {Parr} RG (1988) {Development of the Colle-Salvetti correlation-energy formula into a functional of the electron density}. \prb 37(2):785--789. \doi{10.1103/PhysRevB.37.785}

\bibitem[{{Li}(2005)}]{Li2005}
{Li} A (2005) {Can Fluffy Dust Alleviate the Subsolar Interstellar Abundance Problem?} \apj 622(2):965--969. \doi{10.1086/428038}, {\href{https://arxiv.org/abs/astro-ph/0503564}{{arXiv:astro-ph/0503564}}} {[astro-ph]}

\bibitem[{Li and Draine(2001)}]{Li_2001}
Li A, Draine BT (2001) {On Ultrasmall Silicate Grains in the Diffuse Interstellar Medium}. \apj 550(2):L213. \doi{10.1086/319640}

\bibitem[{{Li} and {Greenberg}(1998)}]{Li98}
{Li} A, {Greenberg} JM (1998) {A comet dust model for the beta Pictoris disk}. \aap 331:291--313

\bibitem[{{Li} and {Lunine}(2003)}]{Li03_HR4796}
{Li} A, {Lunine} JI (2003) {Modeling the Infrared Emission from the HR 4796A Disk}. \apj 590(1):368--378. \doi{10.1086/374865}, {\href{https://arxiv.org/abs/astro-ph/0311071}{{arXiv:astro-ph/0311071}}} {[astro-ph]}

\bibitem[{Li and Greenberg(1997)}]{RN1600}
Li AG, Greenberg JM (1997) {A unified model of interstellar dust}. \aap 323(2):566--584

\bibitem[{{Lin} et~al.(2023){Lin}, {Li}, {Yang}, {Mu{\~n}oz}, {Looney}, {Stephens}, {Hull}, {Fern{\'a}ndez-L{\'o}pez}, and {Harrison}}]{DanielLin22}
{Lin} ZYD, {Li} ZY, {Yang} H, et~al (2023) {(Sub)millimetre dust polarization of protoplanetary discs from scattering by large millimetre-sized irregular grains}. \mnras 520(1):1210--1223. \doi{10.1093/mnras/stad173}, {\href{https://arxiv.org/abs/2206.12357}{{arXiv:2206.12357}}} {[astro-ph.EP]}

\bibitem[{{Lin} et~al.(2024{\natexlab{a}}){Lin}, {Li}, {Stephens}, {Fern{\'a}ndez-L{\'o}pez}, {Carrasco-Gonz{\'a}lez}, {Chandler}, {Pasetto}, {Looney}, {Yang}, {Harrison}, {Sadavoy}, {Henning}, {Hughes}, {Kataoka}, {Kwon}, {Muto}, and {Segura-Cox}}]{DanielLin24HLTau}
{Lin} ZYD, {Li} ZY, {Stephens} IW, et~al (2024{\natexlab{a}}) {Panchromatic (Sub)millimeter polarization observations of HL Tau unveil aligned scattering grains}. \mnras 528(1):843--862. \doi{10.1093/mnras/stae040}, {\href{https://arxiv.org/abs/2309.10055}{{arXiv:2309.10055}}} {[astro-ph.EP]}

\bibitem[{{Lin} et~al.(2024{\natexlab{b}}){Lin}, {Li}, {Yang}, {Looney}, {Stephens}, {Fern{\'a}ndez-L{\'o}pez}, and {Harrison}}]{DanielLin24Align}
{Lin} ZYD, {Li} ZY, {Yang} H, et~al (2024{\natexlab{b}}) {Badminton birdie-like aerodynamic alignment of drifting dust grains by subsonic gaseous flows in protoplanetary discs}. \mnras 534(4):3713--3733. \doi{10.1093/mnras/stae2248}, {\href{https://arxiv.org/abs/2407.10025}{{arXiv:2407.10025}}} {[astro-ph.EP]}

\bibitem[{Lissauer and Stevenson(2007)}]{RN1866}
Lissauer JJ, Stevenson DJ (2007) {Formation of Giant Planets}. In Protostars and Planets V (eds B Reipurth, D Jewitt, and K Keil), University of Arizona Press, Tucson pp 591--606. \doi{10.1016/j.cplett.2008.02.105}

\bibitem[{{Liu}(2019)}]{Liu19}
{Liu} HB (2019) {The Anomalously Low (Sub)Millimeter Spectral Indices of Some Protoplanetary Disks May Be Explained By Dust Self-scattering}. \apjl 877(2):L22. \doi{10.3847/2041-8213/ab1f8e}, {\href{https://arxiv.org/abs/1904.00333}{{arXiv:1904.00333}}} {[astro-ph.SR]}

\bibitem[{Lodders and Fegley(1999)}]{RN1077}
Lodders K, Fegley B (1999) {Condensation chemistry of circumstellar grains}. In: Bertre TL, Lebre A, Waelkens C (eds) Asymptotic Giant Branch Stars, pp 279--289, \doi{10.1017/S0074180900203185}

\bibitem[{{Lodge} et~al.(2024{\natexlab{a}}){Lodge}, {Wakeford}, and {Leinhardt}}]{Lodge24}
{Lodge} MG, {Wakeford} HR, {Leinhardt} ZM (2024{\natexlab{a}}) {Aerosols are not spherical cows: using discrete dipole approximation to model the properties of fractal particles}. \mnras 527(4):11113--11137. \doi{10.1093/mnras/stad3743}, {\href{https://arxiv.org/abs/2312.02301}{{arXiv:2312.02301}}} {[astro-ph.EP]}

\bibitem[{{Lodge} et~al.(2024{\natexlab{b}}){Lodge}, {Wakeford}, and {Leinhardt}}]{Lodge24a}
{Lodge} MG, {Wakeford} HR, {Leinhardt} ZM (2024{\natexlab{b}}) {MANTA-Ray: Supercharging speeds for calculating the optical properties of fractal aggregates in the long-wavelength limit}. \mnras \doi{10.1093/mnras/stae2451}, {\href{https://arxiv.org/abs/2410.21400}{{arXiv:2410.21400}}} {[astro-ph.EP]}

\bibitem[{{Lomax} et~al.(2018){Lomax}, {Wisniewski}, {Roberge}, {Donaldson}, {Debes}, {Malumuth}, and {Weinberger}}]{Lomax18}
{Lomax} JR, {Wisniewski} JP, {Roberge} A, et~al (2018) {Optical Coronagraphic Spectroscopy of AU Mic: Evidence of Time Variable Colors?} \aj 155(2):62. \doi{10.3847/1538-3881/aaa1a7}, {\href{https://arxiv.org/abs/1705.09291}{{arXiv:1705.09291}}} {[astro-ph.SR]}

\bibitem[{{Lorek} et~al.(2018){Lorek}, {Lacerda}, and {Blum}}]{Lorek18}
{Lorek} S, {Lacerda} P, {Blum} J (2018) {Local growth of dust- and ice-mixed aggregates as cometary building blocks in the solar nebula}. \aap 611:A18. \doi{10.1051/0004-6361/201630175}

\bibitem[{{Lumme} et~al.(1997){Lumme}, {Rahola}, and {Hovenier}}]{Lumme97}
{Lumme} K, {Rahola} J, {Hovenier} JW (1997) {Light Scattering by Dense Clusters of Spheres}. \icarus 126(2):455--469. \doi{10.1006/icar.1996.5650}

\bibitem[{{Ma} and {Schmid}(2022)}]{Ma22}
{Ma} J, {Schmid} HM (2022) {A model grid for the reflected light from transition disks}. \aap 663:A110. \doi{10.1051/0004-6361/202142954}, {\href{https://arxiv.org/abs/2204.06370}{{arXiv:2204.06370}}} {[astro-ph.EP]}

\bibitem[{{Ma} et~al.(2024{\natexlab{a}}){Ma}, {Ginski}, {Tazaki}, {Dominik}, {Schmid}, and {M{\'e}nard}}]{Ma24b}
{Ma} J, {Ginski} C, {Tazaki} R, et~al (2024{\natexlab{a}}) {Temporal and chromatic variation of polarized scattered light in the outer disk of PDS 70}. arXiv e-prints arXiv:2411.04091. \doi{10.48550/arXiv.2411.04091}, {\href{https://arxiv.org/abs/2411.04091}{{arXiv:2411.04091}}} {[astro-ph.EP]}

\bibitem[{{Ma} et~al.(2024{\natexlab{b}}){Ma}, {Schmid}, and {Stolker}}]{Ma24a}
{Ma} J, {Schmid} HM, {Stolker} T (2024{\natexlab{b}}) {Color measurements of the polarized light scattered by the dust in protoplanetary disks}. \aap 683:A18. \doi{10.1051/0004-6361/202347782}, {\href{https://arxiv.org/abs/2312.14045}{{arXiv:2312.14045}}} {[astro-ph.EP]}

\bibitem[{{Mac{\'\i}as} et~al.(2021){Mac{\'\i}as}, {Guerra-Alvarado}, {Carrasco-Gonz{\'a}lez}, {Ribas}, {Espaillat}, {Huang}, and {Andrews}}]{Macias21}
{Mac{\'\i}as} E, {Guerra-Alvarado} O, {Carrasco-Gonz{\'a}lez} C, et~al (2021) {Characterizing the dust content of disk substructures in TW Hydrae}. \aap 648:A33. \doi{10.1051/0004-6361/202039812}, {\href{https://arxiv.org/abs/2102.04648}{{arXiv:2102.04648}}} {[astro-ph.EP]}

\bibitem[{{MacKinnon} and {Rietmeijer}(1987)}]{MacKinnon1987}
{MacKinnon} IDR, {Rietmeijer} FJM (1987) {Mineralogy of chondritic interplanetary dust particles.} Rev Geophys 25:1527--1553. \doi{10.1029/RG025i007p01527}

\bibitem[{{Mackowski}(1995)}]{Mackowski95}
{Mackowski} DW (1995) {Electrostatics analysis of radiative absorption by sphere clusters in the Rayleigh limit: application to soot particles}. \ao 34(18):3535. \doi{10.1364/AO.34.003535}

\bibitem[{{Mackowski}(2006)}]{Mackowski06}
{Mackowski} DW (2006) {A simplified model to predict the effects of aggregation on the absorption properties of soot particles}. \jqsrt 100(1-3):237--249. \doi{10.1016/j.jqsrt.2005.11.041}

\bibitem[{{Mamajek} and {Bell}(2014)}]{Mamajek14}
{Mamajek} EE, {Bell} CPM (2014) {On the age of the {\ensuremath{\beta}} Pictoris moving group}. \mnras 445(3):2169--2180. \doi{10.1093/mnras/stu1894}, {\href{https://arxiv.org/abs/1409.2737}{{arXiv:1409.2737}}} {[astro-ph.SR]}

\bibitem[{{Mannel} et~al.(2016){Mannel}, {Bentley}, {Schmied}, {Jeszenszky}, {Levasseur-Regourd}, {Romstedt}, and {Torkar}}]{Mannel16}
{Mannel} T, {Bentley} MS, {Schmied} R, et~al (2016) {Fractal cometary dust - a window into the early Solar system}. \mnras 462:S304--S311. \doi{10.1093/mnras/stw2898}

\bibitem[{{Mannel} et~al.(2019){Mannel}, {Bentley}, {Boakes}, {Jeszenszky}, {Ehrenfreund}, {Engrand}, {Koeberl}, {Levasseur-Regourd}, {Romstedt}, {Schmied}, {Torkar}, and {Weber}}]{Mannel19}
{Mannel} T, {Bentley} MS, {Boakes} PD, et~al (2019) {Dust of comet 67P/Churyumov-Gerasimenko collected by Rosetta/MIDAS: classification and extension to the nanometer scale}. \aap 630:A26. \doi{10.1051/0004-6361/201834851}

\bibitem[{Marchione et~al.(2019)Marchione, Rosu-Finsen, Taj, Lasne, Abdulgalil, Thrower, Frankland, Collings, and McCoustra}]{RN954}
Marchione D, Rosu-Finsen A, Taj S, et~al (2019) {Surface Science Investigations of Icy Mantle Growth on Interstellar Dust Grains in Cooling Environments}. ACS Earth Space Chem 3(9):1915--1931. \doi{10.1021/acsearthspacechem.9b00052}

\bibitem[{Martínez-Gonz\'alez et~al.(2018)Martínez-Gonz\'alez, Navarro-Ruiz, and Rimola}]{Minerals-Rimola2018}
Martínez-Gonz\'alez JA, Navarro-Ruiz J, Rimola A (2018) {Multiscale Computational Simulation of Amorphous Silicates’ Structural, Dielectric, and Vibrational Spectroscopic Properties}. Minerals 8(8). \doi{10.3390/min8080353}

\bibitem[{Mates-Torres and Rimola(2024)}]{PolyCleaver}
Mates-Torres E, Rimola A (2024) {Unlocking the surface chemistry of ionic minerals: a high-throughput pipeline for modeling realistic interfaces}. J Appl Crystallogr 57(2):503--508. \doi{10.1107/S1600576724001286}

\bibitem[{{Mathis}(1990)}]{Mathis1990}
{Mathis} JS (1990) {Interstellar dust and extinction}. \araa 28:37--70. \doi{10.1146/annurev.aa.28.090190.000345}

\bibitem[{{Mathis}(1996)}]{Mathis1996}
{Mathis} JS (1996) {Dust Models with Tight Abundance Constraints}. \apj 472:643. \doi{10.1086/178094}

\bibitem[{{Mathis} and {Whiffen}(1989)}]{Mathis1989}
{Mathis} JS, {Whiffen} G (1989) {Composite Interstellar Grains}. \apj 341:808. \doi{10.1086/167538}

\bibitem[{{Mathis} et~al.(1977){Mathis}, {Rumpl}, and {Nordsieck}}]{MRN1977}
{Mathis} JS, {Rumpl} W, {Nordsieck} KH (1977) {The size distribution of interstellar grains.} \apj 217:425--433. \doi{10.1086/155591}

\bibitem[{{Mathis} et~al.(1995){Mathis}, {Cohen}, {Finley}, and {Krautter}}]{Mathis1995}
{Mathis} JS, {Cohen} D, {Finley} JP, et~al (1995) {The X-Ray Halo of Nova V1974 Cygni (Nova Cygni 1992) and the Nature of Interstellar Dust}. \apj 449:320. \doi{10.1086/176057}

\bibitem[{{Matthews} et~al.(2015){Matthews}, {Kennedy}, {Sibthorpe}, {Holland}, {Booth}, {Kalas}, {MacGregor}, {Wilner}, {Vandenbussche}, {Olofsson}, {Blommaert}, {Brandeker}, {Dent}, {de Vries}, {Di Francesco}, {Fridlund}, {Graham}, {Greaves}, {Heras}, {Hogerheijde}, {Ivison}, {Pantin}, and {Pilbratt}}]{Matthews15}
{Matthews} BC, {Kennedy} G, {Sibthorpe} B, et~al (2015) {The AU Mic Debris Disk: Far-infrared and Submillimeter Resolved Imaging}. \apj 811(2):100. \doi{10.1088/0004-637X/811/2/100}, {\href{https://arxiv.org/abs/1509.06415}{{arXiv:1509.06415}}} {[astro-ph.SR]}

\bibitem[{{McCabe} et~al.(2002){McCabe}, {Duch{\^e}ne}, and {Ghez}}]{McCabe02}
{McCabe} C, {Duch{\^e}ne} G, {Ghez} AM (2002) {NICMOS Images of the GG Tauri Circumbinary Disk}. \apj 575(2):974--988. \doi{10.1086/341479}, {\href{https://arxiv.org/abs/astro-ph/0204465}{{arXiv:astro-ph/0204465}}} {[astro-ph]}

\bibitem[{{McClure}(2009)}]{McClure2009}
{McClure} M (2009) {Observational 5-20 {\ensuremath{\mu}}m Interstellar Extinction Curves Toward Star-Forming Regions Derived From Spitzer IRS Spectra}. \apjl 693(2):L81--L85. \doi{10.1088/0004-637X/693/2/L81}, {\href{https://arxiv.org/abs/0810.4561}{{arXiv:0810.4561}}} {[astro-ph]}

\bibitem[{{Meakin}(1991)}]{Meakin91}
{Meakin} P (1991) {Fractal aggregates in geophysics}. Reviews of Geophysics 29(3):317--354. \doi{10.1029/91RG00688}

\bibitem[{Mennella et~al.(2001)Mennella, Munoz~Caro, Ruiterkamp, Schutte, Greenberg, Brucato, and Colangeli}]{RN1924}
Mennella V, Munoz~Caro GM, Ruiterkamp R, et~al (2001) {UV photodestruction of CH bonds and the evolution of the 3.4 $\mu$m feature carrier II.: The case of hydrogenated carbon grains}. \aap 367(1):355--361. \doi{10.1051/0004-6361:20000340}

\bibitem[{Mennella et~al.(2020)Mennella, Ciarniello, Raponi, Capaccioni, Filacchione, Suhasaria, Popa, Kappel, Moroz, Vinogradoff, Pommerol, Rousseau, Istiqomah, Bockelee-Morvan, Carlson, and Pilorget}]{Mennella_2020}
Mennella V, Ciarniello M, Raponi A, et~al (2020) {Hydroxylated Mg-rich Amorphous Silicates: A New Component of the 3.2 $\mu$m Absorption Band of Comet 67P/Churyumov–Gerasimenko}. The Astrophysical Journal Letters 897(2):L37. \doi{10.3847/2041-8213/ab919e}

\bibitem[{{Michikoshi} and {Kokubo}(2017)}]{Michikoshi17}
{Michikoshi} S, {Kokubo} E (2017) {Dynamics of Porous Dust Aggregates and Gravitational Instability of Their Disk}. \apj 842(1):61. \doi{10.3847/1538-4357/aa7388}, {\href{https://arxiv.org/abs/1705.04520}{{arXiv:1705.04520}}} {[astro-ph.EP]}

\bibitem[{{Michoulier} et~al.(2024){Michoulier}, {Gonzalez}, and {Price}}]{Michoulier24}
{Michoulier} S, {Gonzalez} JF, {Price} DJ (2024) {Compaction during fragmentation and bouncing produces realistic dust grain porosities in protoplanetary discs}. \aap 688:A31. \doi{10.1051/0004-6361/202449719}, {\href{https://arxiv.org/abs/2406.15622}{{arXiv:2406.15622}}} {[astro-ph.EP]}

\bibitem[{{Milli} et~al.(2012){Milli}, {Mouillet}, {Lagrange}, {Boccaletti}, {Mawet}, {Chauvin}, and {Bonnefoy}}]{Milli12}
{Milli} J, {Mouillet} D, {Lagrange} AM, et~al (2012) {Impact of angular differential imaging on circumstellar disk images}. \aap 545:A111. \doi{10.1051/0004-6361/201219687}, {\href{https://arxiv.org/abs/1207.5909}{{arXiv:1207.5909}}} {[astro-ph.EP]}

\bibitem[{{Milli} et~al.(2015){Milli}, {Mawet}, {Pinte}, {Lagrange}, {Mouillet}, {Girard}, {Augereau}, {De Boer}, {Pueyo}, and {Choquet}}]{Milli15}
{Milli} J, {Mawet} D, {Pinte} C, et~al (2015) {New constraints on the dust surrounding HR 4796A}. \aap 577:A57. \doi{10.1051/0004-6361/201423950}, {\href{https://arxiv.org/abs/1407.2539}{{arXiv:1407.2539}}} {[astro-ph.EP]}

\bibitem[{{Milli} et~al.(2017){Milli}, {Vigan}, {Mouillet}, {Lagrange}, {Augereau}, {Pinte}, {Mawet}, {Schmid}, {Boccaletti}, {Matr{\`a}}, {Kral}, {Ertel}, {Chauvin}, {Bazzon}, {M{\'e}nard}, {Beuzit}, {Thalmann}, {Dominik}, {Feldt}, {Henning}, {Min}, {Girard}, {Galicher}, {Bonnefoy}, {Fusco}, {de Boer}, {Janson}, {Maire}, {Mesa}, {Schlieder}, and {SPHERE Consortium}}]{Milli17}
{Milli} J, {Vigan} A, {Mouillet} D, et~al (2017) {Near-infrared scattered light properties of the HR 4796 A dust ring. A measured scattering phase function from 13.6{\textdegree} to 166.6{\textdegree}}. \aap 599:A108. \doi{10.1051/0004-6361/201527838}, {\href{https://arxiv.org/abs/1701.00750}{{arXiv:1701.00750}}} {[astro-ph.EP]}

\bibitem[{{Milli} et~al.(2019){Milli}, {Engler}, {Schmid}, {Olofsson}, {M{\'e}nard}, {Kral}, {Boccaletti}, {Th{\'e}bault}, {Choquet}, {Mouillet}, {Lagrange}, {Augereau}, {Pinte}, {Chauvin}, {Dominik}, {Perrot}, {Zurlo}, {Henning}, {Beuzit}, {Avenhaus}, {Bazzon}, {Moulin}, {Llored}, {Moeller-Nilsson}, {Roelfsema}, and {Pragt}}]{Milli19}
{Milli} J, {Engler} N, {Schmid} HM, et~al (2019) {Optical polarised phase function of the HR 4796A dust ring}. \aap 626:A54. \doi{10.1051/0004-6361/201935363}, {\href{https://arxiv.org/abs/1905.03603}{{arXiv:1905.03603}}} {[astro-ph.EP]}

\bibitem[{{Min} et~al.(2003){Min}, {Hovenier}, and {de Koter}}]{Min03}
{Min} M, {Hovenier} JW, {de Koter} A (2003) {Shape effects in scattering and absorption by randomly oriented particles small compared to the wavelength}. \aap 404:35--46. \doi{10.1051/0004-6361:20030456}

\bibitem[{{Min} et~al.(2005){Min}, {Hovenier}, and {de Koter}}]{Min05}
{Min} M, {Hovenier} JW, {de Koter} A (2005) {Modeling optical properties of cosmic dust grains using a distribution of hollow spheres}. \aap 432(3):909--920. \doi{10.1051/0004-6361:20041920}, {\href{https://arxiv.org/abs/astro-ph/0503068}{{arXiv:astro-ph/0503068}}} {[astro-ph]}

\bibitem[{{Min} et~al.(2006{\natexlab{a}}){Min}, {Dominik}, {Hovenier}, {de Koter}, and {Waters}}]{Min06}
{Min} M, {Dominik} C, {Hovenier} JW, et~al (2006{\natexlab{a}}) {The 10 {\ensuremath{\mu}}m amorphous silicate feature of fractal aggregates and compact particles with complex shapes}. \aap 445(3):1005--1014. \doi{10.1051/0004-6361:20053212}, {\href{https://arxiv.org/abs/astro-ph/0509376}{{arXiv:astro-ph/0509376}}} {[astro-ph]}

\bibitem[{{Min} et~al.(2006{\natexlab{b}}){Min}, {Hovenier}, {Dominik}, {de Koter}, and {Yurkin}}]{Min06JQSRT}
{Min} M, {Hovenier} JW, {Dominik} C, et~al (2006{\natexlab{b}}) {Absorption and scattering properties of arbitrarily shaped particles in the Rayleigh domain}. \jqsrt 97(2):161--180. \doi{10.1016/j.jqsrt.2005.05.059}, {\href{https://arxiv.org/abs/astro-ph/0504067}{{arXiv:astro-ph/0504067}}} {[astro-ph]}

\bibitem[{{Min} et~al.(2007){Min}, {Waters}, {de Koter}, {Hovenier}, {Keller}, and {Markwick-Kemper}}]{Min07}
{Min} M, {Waters} LBFM, {de Koter} A, et~al (2007) {The shape and composition of interstellar silicate grains}. \aap 462(2):667--676. \doi{10.1051/0004-6361:20065436}, {\href{https://arxiv.org/abs/astro-ph/0611329}{{arXiv:astro-ph/0611329}}} {[astro-ph]}

\bibitem[{{Min} et~al.(2008){Min}, {Hovenier}, {Waters}, and {de Koter}}]{Min08}
{Min} M, {Hovenier} JW, {Waters} LBFM, et~al (2008) {The infrared emission spectra of compositionally inhomogeneous aggregates composed of irregularly shaped constituents}. \aap 489(1):135--141. \doi{10.1051/0004-6361:200809534}, {\href{https://arxiv.org/abs/0806.4038}{{arXiv:0806.4038}}} {[astro-ph]}

\bibitem[{{Min} et~al.(2010){Min}, {Kama}, {Dominik}, and {Waters}}]{Min10}
{Min} M, {Kama} M, {Dominik} C, et~al (2010) {The lunar phases of dust grains orbiting Fomalhaut}. \aap 509:L6. \doi{10.1051/0004-6361/200913065}, {\href{https://arxiv.org/abs/1001.0516}{{arXiv:1001.0516}}} {[astro-ph.EP]}

\bibitem[{Min et~al.(2011)Min, Dullemond, Kama, and Dominik}]{RN1059}
Min M, Dullemond CP, Kama M, et~al (2011) {The thermal structure and the location of the snow line in the protosolar nebula: Axisymmetric models with full 3-D radiative transfer}. Icarus 212(1):416--426. \doi{10.1016/j.icarus.2010.12.002}

\bibitem[{{Min} et~al.(2016){Min}, {Rab}, {Woitke}, {Dominik}, and {M{\'e}nard}}]{Min16}
{Min} M, {Rab} C, {Woitke} P, et~al (2016) {Multiwavelength optical properties of compact dust aggregates in protoplanetary disks}. \aap 585:A13. \doi{10.1051/0004-6361/201526048}, {\href{https://arxiv.org/abs/1510.05426}{{arXiv:1510.05426}}} {[astro-ph.EP]}

\bibitem[{Minissale et~al.(2022)Minissale, Aikawa, Bergin, Bertin, Brown, Cazaux, Charnley, Coutens, Cuppen, Guzman, Linnartz, McCoustra, Rimola, Schrauwen, Toubin, Ugliengo, Watanabe, Wakelam, and Dulieu}]{Minissale2022}
Minissale M, Aikawa Y, Bergin E, et~al (2022) {Thermal Desorption of Interstellar Ices: A Review on the Controlling Parameters and Their Implications from Snowlines to Chemical Complexity}. ACS Earth Space Chem 6(3):597--630. \doi{10.1021/acsearthspacechem.1c00357}

\bibitem[{{Miret-Roig} et~al.(2020){Miret-Roig}, {Galli}, {Brandner}, {Bouy}, {Barrado}, {Olivares}, {Antoja}, {Romero-G{\'o}mez}, {Figueras}, and {Lillo-Box}}]{MiretRoig20}
{Miret-Roig} N, {Galli} PAB, {Brandner} W, et~al (2020) {Dynamical traceback age of the {\ensuremath{\beta}} Pictoris moving group}. \aap 642:A179. \doi{10.1051/0004-6361/202038765}, {\href{https://arxiv.org/abs/2007.10997}{{arXiv:2007.10997}}} {[astro-ph.GA]}

\bibitem[{{Mishchenko} et~al.(2000){Mishchenko}, {Hovenier}, and {Travis}}]{Mishchenko00}
{Mishchenko} MI, {Hovenier} JW, {Travis} LD (2000) {Light scattering by nonspherical particles: theory, measurements, and applications}. San Diego, Academic Press

\bibitem[{{Miyake} and {Nakagawa}(1993)}]{Miyake93}
{Miyake} K, {Nakagawa} Y (1993) {Effects of Particle Size Distribution on Opacity Curves of Protoplanetary Disks around T Tauri Stars}. \icarus 106(1):20--41. \doi{10.1006/icar.1993.1156}

\bibitem[{{Morbidelli} et~al.(2000){Morbidelli}, {Chambers}, {Lunine}, {Petit}, {Robert}, {Valsecchi}, and {Cyr}}]{Morbidelli2000}
{Morbidelli} A, {Chambers} J, {Lunine} JI, et~al (2000) {Source regions and time scales for the delivery of water to Earth}. \maps 35(6):1309--1320. \doi{10.1111/j.1945-5100.2000.tb01518.x}

\bibitem[{Morbidelli et~al.(2015)Morbidelli, Lambrechts, Jacobson, and Bitsch}]{RN1864}
Morbidelli A, Lambrechts M, Jacobson S, et~al (2015) {The great dichotomy of the Solar System: Small terrestrial embryos and massive giant planet cores}. Icarus 258:418--429. \doi{10.1016/j.icarus.2015.06.003}

\bibitem[{{Mori} and {Kataoka}(2021)}]{Mori21}
{Mori} T, {Kataoka} A (2021) {Modeling of the ALMA HL Tau Polarization by Mixture of Grain Alignment and Self-scattering}. \apj 908(2):153. \doi{10.3847/1538-4357/abd08a}, {\href{https://arxiv.org/abs/2012.01735}{{arXiv:2012.01735}}} {[astro-ph.EP]}

\bibitem[{{Mu{\~n}oz} et~al.(2012){Mu{\~n}oz}, {Moreno}, {Guirado}, {Dabrowska}, {Volten}, and {Hovenier}}]{Munoz12}
{Mu{\~n}oz} O, {Moreno} F, {Guirado} D, et~al (2012) {The Amsterdam-Granada Light Scattering Database}. \jqsrt 113(7):565--574. \doi{10.1016/j.jqsrt.2012.01.014}

\bibitem[{{Mu{\~n}oz} et~al.(2017){Mu{\~n}oz}, {Moreno}, {Vargas-Mart{\'\i}n}, {Guirado}, {Escobar-Cerezo}, {Min}, and {Hovenier}}]{Munoz17}
{Mu{\~n}oz} O, {Moreno} F, {Vargas-Mart{\'\i}n} F, et~al (2017) {Experimental Phase Functions of Millimeter-sized Cosmic Dust Grains}. \apj 846(1):85. \doi{10.3847/1538-4357/aa7ff2}, {\href{https://arxiv.org/abs/1707.04158}{{arXiv:1707.04158}}} {[astro-ph.GA]}

\bibitem[{{Mu{\~n}oz} et~al.(2020){Mu{\~n}oz}, {Moreno}, {G{\'o}mez-Mart{\'\i}n}, {Vargas-Mart{\'\i}n}, {Guirado}, {Ramos}, {Bustamante}, {Bertini}, {Frattin}, {Markannen}, {Tubiana}, {Fulle}, {G{\"u}ttler}, {Sierks}, {Rotundi}, {Della Corte}, {Ivanovski}, {Zakharov}, {Bockel{\'e}e-Morvan}, {Blum}, {Merouane}, {Levasseur-Regourd}, {Kolokolova}, {Jardiel}, and {Caballero}}]{Munoz20}
{Mu{\~n}oz} O, {Moreno} F, {G{\'o}mez-Mart{\'\i}n} JC, et~al (2020) {Experimental Phase Function and Degree of Linear Polarization Curves of Millimeter-sized Cosmic Dust Analogs}. \apjs 247(1):19. \doi{10.3847/1538-4365/ab6851}

\bibitem[{{Mu{\~n}oz} et~al.(2021){Mu{\~n}oz}, {Frattin}, {Jardiel}, {G{\'o}mez-Mart{\'\i}n}, {Moreno}, {Ramos}, {Guirado}, {Peiteado}, {Caballero}, {Milli}, and {M{\'e}nard}}]{Munoz21}
{Mu{\~n}oz} O, {Frattin} E, {Jardiel} T, et~al (2021) {Retrieving Dust Grain Sizes from Photopolarimetry: An Experimental Approach}. \apjs 256(1):17. \doi{10.3847/1538-4365/ac0efa}, {\href{https://arxiv.org/abs/2109.05764}{{arXiv:2109.05764}}} {[astro-ph.EP]}

\bibitem[{{Mukai} et~al.(1982){Mukai}, {Mukai}, {Giese}, {Weiss}, and {Zerull}}]{Mukai82}
{Mukai} S, {Mukai} T, {Giese} RH, et~al (1982) {Scattering of Radiation by a Large Particle with a Random Rough Surface}. Moon and Planets 26(2):197--208. \doi{10.1007/BF00929281}

\bibitem[{{Mukai} et~al.(1992){Mukai}, {Ishimoto}, {Kozasa}, {Blum}, and {Greenberg}}]{Mukai92}
{Mukai} T, {Ishimoto} H, {Kozasa} T, et~al (1992) {Radiation pressure forces of fluffy porous grains}. \aap 262(1):315--320

\bibitem[{{Mulders} et~al.(2013){Mulders}, {Min}, {Dominik}, {Debes}, and {Schneider}}]{Mulders13}
{Mulders} GD, {Min} M, {Dominik} C, et~al (2013) {Why circumstellar disks are so faint in scattered light: the case of HD 100546}. \aap 549:A112. \doi{10.1051/0004-6361/201219522}, {\href{https://arxiv.org/abs/1210.4132}{{arXiv:1210.4132}}} {[astro-ph.SR]}

\bibitem[{Musat et~al.(2008)Musat, Renault, Candelaresi, Palmer, Le~Caër, Righini, and Pommeret}]{H-bond_confined_water_ACIE2008}
Musat R, Renault JP, Candelaresi M, et~al (2008) {Finite Size Effects on Hydrogen Bonds in Confined Water}. Angewandte Chemie International Edition 47(42):8033--8035. \doi{https://doi.org/10.1002/anie.200802630}

\bibitem[{Musiolik et~al.(2016{\natexlab{a}})Musiolik, Teiser, Jankowski, and Wurm}]{RN1860}
Musiolik G, Teiser J, Jankowski T, et~al (2016{\natexlab{a}}) Collisions of co ice grains in planet formation. \apj 818(1):16. \doi{10.3847/0004-637x/818/1/16}

\bibitem[{Musiolik et~al.(2016{\natexlab{b}})Musiolik, Teiser, Jankowski, and Wurm}]{RN1859}
Musiolik G, Teiser J, Jankowski T, et~al (2016{\natexlab{b}}) Ice grain collisions in comparison: Co , h o, and their mixtures. \apj 827(1):63. \doi{10.3847/0004-637x/827/1/63}

\bibitem[{{Mutschke} et~al.(2009){Mutschke}, {Min}, and {Tamanai}}]{Mutscheke09}
{Mutschke} H, {Min} M, {Tamanai} A (2009) {Laboratory-based grain-shape models for simulating dust infrared spectra}. \aap 504(3):875--882. \doi{10.1051/0004-6361/200912267}, {\href{https://arxiv.org/abs/0907.3350}{{arXiv:0907.3350}}} {[astro-ph.IM]}

\bibitem[{{Najita} et~al.(2001){Najita}, {Bergin}, and {Ullom}}]{Najita2001}
{Najita} J, {Bergin} EA, {Ullom} JN (2001) {X-Ray Desorption of Molecules from Grains in Protoplanetary Disks}. \apj 561(2):880--889. \doi{10.1086/323320}, {\href{https://arxiv.org/abs/astro-ph/0108055}{{arXiv:astro-ph/0108055}}} {[astro-ph]}

\bibitem[{{Natta} et~al.(2004){Natta}, {Testi}, {Neri}, {Shepherd}, and {Wilner}}]{Natta04}
{Natta} A, {Testi} L, {Neri} R, et~al (2004) {A search for evolved dust in Herbig Ae stars}. \aap 416:179--186. \doi{10.1051/0004-6361:20035620}, {\href{https://arxiv.org/abs/astro-ph/0311624}{{arXiv:astro-ph/0311624}}} {[astro-ph]}

\bibitem[{Navarro-Ruiz et~al.(2014)Navarro-Ruiz, Ugliengo, Rimola, and Sodupe}]{Navarro-Ruiz2014}
Navarro-Ruiz J, Ugliengo P, Rimola A, et~al (2014) {B3LYP Periodic Study of the Physicochemical Properties of the Nonpolar (010) Mg-Pure and Fe-Containing Olivine Surfaces}. J Phys Chem A 118(31):5866--5875. \doi{10.1021/jp4118198}

\bibitem[{{Nieva} and {Przybilla}(2012)}]{Nieva2012}
{Nieva} MF, {Przybilla} N (2012) {Present-day cosmic abundances. A comprehensive study of nearby early B-type stars and implications for stellar and Galactic evolution and interstellar dust models}. \aap 539:A143. \doi{10.1051/0004-6361/201118158}, {\href{https://arxiv.org/abs/1203.5787}{{arXiv:1203.5787}}} {[astro-ph.SR]}

\bibitem[{{Ohashi} et~al.(2023){Ohashi}, {Momose}, {Kataoka}, {Higuchi}, {Tsukagoshi}, {Ueda}, {Codella}, {Podio}, {Hanawa}, {Sakai}, {Kobayashi}, {Okuzumi}, and {Tanaka}}]{Ohashi23}
{Ohashi} S, {Momose} M, {Kataoka} A, et~al (2023) {Dust Enrichment and Grain Growth in a Smooth Disk around the DG Tau Protostar Revealed by ALMA Triple Bands Frequency Observations}. \apj 954(2):110. \doi{10.3847/1538-4357/ace9b9}, {\href{https://arxiv.org/abs/2307.14526}{{arXiv:2307.14526}}} {[astro-ph.EP]}

\bibitem[{Okuzumi and Tazaki(2019)}]{RN1858}
Okuzumi S, Tazaki R (2019) {Nonsticky Ice at the Origin of the Uniformly Polarized Submillimeter Emission from the HL Tau Disk}. \apj 878(2):132. \doi{10.3847/1538-4357/ab204d}

\bibitem[{Okuzumi et~al.(2009)Okuzumi, Tanaka, and Sakagami}]{RN1906}
Okuzumi S, Tanaka H, Sakagami MA (2009) {Numerical Modeling of the Coagulation and Porosity Evolution of Dust Aggregates}. \apj 707(2):1247--1263. \doi{10.1088/0004-637x/707/2/1247}

\bibitem[{Okuzumi et~al.(2012)Okuzumi, Tanaka, Kobayashi, and Wada}]{RN1865}
Okuzumi S, Tanaka H, Kobayashi H, et~al (2012) {Rapid Coagulation of Porous Dust Aggregates Outside the Snow Line: A Pathway to Successful Icy Planetesimal Formation}. \apj 752(2):106. \doi{10.1088/0004-637x/752/2/106}

\bibitem[{{Olofsson} et~al.(2020){Olofsson}, {Milli}, {Bayo}, {Henning}, and {Engler}}]{Olofsson20}
{Olofsson} J, {Milli} J, {Bayo} A, et~al (2020) {The challenge of measuring the phase function of debris discs. Application to HR 4796 A}. \aap 640:A12. \doi{10.1051/0004-6361/202038237}, {\href{https://arxiv.org/abs/2006.08595}{{arXiv:2006.08595}}} {[astro-ph.SR]}

\bibitem[{{Olofsson} et~al.(2022){Olofsson}, {Th{\'e}bault}, {Kennedy}, and {Bayo}}]{Olofsson22}
{Olofsson} J, {Th{\'e}bault} P, {Kennedy} GM, et~al (2022) {The halo around HD 32297: {\ensuremath{\mu}}m-sized cometary dust}. \aap 664:A122. \doi{10.1051/0004-6361/202243794}, {\href{https://arxiv.org/abs/2206.07068}{{arXiv:2206.07068}}} {[astro-ph.EP]}

\bibitem[{Ormel et~al.(2007)Ormel, Spaans, and Tielens}]{RN1482}
Ormel CW, Spaans M, Tielens AGGM (2007) {Dust coagulation in protoplanetary disks: porosity matters}. \aap 461(1):215--232. \doi{10.1051/0004-6361:20065949}

\bibitem[{{Ormel} et~al.(2009){Ormel}, {Paszun}, {Dominik}, and {Tielens}}]{Ormel2009}
{Ormel} CW, {Paszun} D, {Dominik} C, et~al (2009) {Dust coagulation and fragmentation in molecular clouds. I. How collisions between dust aggregates alter the dust size distribution}. \aap 502(3):845--869. \doi{10.1051/0004-6361/200811158}, {\href{https://arxiv.org/abs/0906.1770}{{arXiv:0906.1770}}} {[astro-ph.SR]}

\bibitem[{{Ormel} et~al.(2011){Ormel}, {Min}, {Tielens}, {Dominik}, and {Paszun}}]{Ormel2011}
{Ormel} CW, {Min} M, {Tielens} AGGM, et~al (2011) {Dust coagulation and fragmentation in molecular clouds. II. The opacity of the dust aggregate size distribution}. \aap 532:A43. \doi{10.1051/0004-6361/201117058}, {\href{https://arxiv.org/abs/1106.3265}{{arXiv:1106.3265}}} {[astro-ph.SR]}

\bibitem[{{Ossenkopf}(1991)}]{Ossenkopf91}
{Ossenkopf} V (1991) {Effective-medium theories for cosmic dust grains}. \aap 251(1):210--219

\bibitem[{Ossenkopf(1993)}]{RN903}
Ossenkopf V (1993) {Dust Coagulation in Dense Molecular Clouds - the Formation of Fluffy Aggregates}. \aap 280(2):617--646

\bibitem[{{Ossenkopf} and {Henning}(1994)}]{Ossenkopf1994}
{Ossenkopf} V, {Henning} T (1994) {Dust opacities for protostellar cores.} \aap 291:943--959

\bibitem[{Ott(2010)}]{RN1697}
Ott U (2010) {The Most Primitive Meteorites}. In: Henning T (ed) Astromineralogy, Lecture Notes in Physics, vol 815, 2nd edn. Springer, Berlin, Heidelberg, p 277--311, \doi{doi.org/10.1007/978-3-642-13259-9_7}

\bibitem[{{Padovani} et~al.(2018){Padovani}, {Ivlev}, {Galli}, and {Caselli}}]{Padovani2018}
{Padovani} M, {Ivlev} AV, {Galli} D, et~al (2018) {Cosmic-ray ionisation in circumstellar discs}. \aap 614:A111. \doi{10.1051/0004-6361/201732202}, {\href{https://arxiv.org/abs/1803.09348}{{arXiv:1803.09348}}} {[astro-ph.HE]}

\bibitem[{{Pagani} et~al.(2010){Pagani}, {Steinacker}, {Bacmann}, {Stutz}, and {Henning}}]{Pagani2010}
{Pagani} L, {Steinacker} J, {Bacmann} A, et~al (2010) {The Ubiquity of Micrometer-Sized Dust Grains in the Dense Interstellar Medium}. Science 329(5999):1622. \doi{10.1126/science.1193211}, {\href{https://arxiv.org/abs/1110.4180}{{arXiv:1110.4180}}} {[astro-ph.GA]}

\bibitem[{Palmer et~al.(2016)Palmer, Cao, and Yin}]{RN1700}
Palmer RE, Cao L, Yin F (2016) {Note: Proof of principle of a new type of cluster beam source with potential for scale-up}. Rev Sci Instrum 87(4):046103. \doi{10.1063/1.4947229}

\bibitem[{Pareras et~al.(2023)Pareras, Cabedo, McCoustra, and Rimola}]{RN1752}
Pareras G, Cabedo V, McCoustra M, et~al (2023) {Single-atom catalysis in space: Computational exploration of Fischer-Tropsch reactions in astrophysical environments}. \aap 680:A57. \doi{10.1051/0004-6361/202347877}

\bibitem[{Pareras et~al.(2024)Pareras, Cabedo, McCoustra, and Rimola}]{RN1753}
Pareras G, Cabedo V, McCoustra M, et~al (2024) {Single-atom catalysis in space II: ketene – acetaldehyde – ethanol and methane synthesis via Fischer-Tropsch chain growth}. \aap 687:A230. \doi{10.1051/0004-6361/202449378}

\bibitem[{{Paszun} and {Dominik}(2009)}]{Paszun09}
{Paszun} D, {Dominik} C (2009) {Collisional evolution of dust aggregates. From compaction to catastrophic destruction}. \aap 507(2):1023--1040. \doi{10.1051/0004-6361/200810682}, {\href{https://arxiv.org/abs/0909.3168}{{arXiv:0909.3168}}} {[astro-ph.SR]}

\bibitem[{{Pedone} et~al.(2022){Pedone}, {Bertani}, {Brugnoli}, and {Pallini}}]{Pedone2022}
{Pedone} A, {Bertani} M, {Brugnoli} L, et~al (2022) {Interatomic potentials for oxide glasses: Past, present, and future}. J Non-Cryst Solids: X 15:100115. \doi{10.1016/j.nocx.2022.100115}

\bibitem[{{Penttil{\"a}} et~al.(2021){Penttil{\"a}}, {Markkanen}, {V{\"a}is{\"a}nen}, {R{\"a}bin{\"a}}, {Yurkin}, and {Muinonen}}]{Penttila21}
{Penttil{\"a}} A, {Markkanen} J, {V{\"a}is{\"a}nen} T, et~al (2021) {How much is enough? The convergence of finite sample scattering properties to those of infinite media}. \jqsrt 262:107524. \doi{10.1016/j.jqsrt.2021.107524}

\bibitem[{{Perdew} et~al.(1996){Perdew}, {Burke}, and {Ernzerhof}}]{PBE1996}
{Perdew} JP, {Burke} K, {Ernzerhof} M (1996) {Generalized Gradient Approximation Made Simple}. \prl 77(18):3865--3868. \doi{10.1103/PhysRevLett.77.3865}

\bibitem[{{Perets} and {Biham}(2006)}]{PB06}
{Perets} HB, {Biham} O (2006) {Molecular hydrogen formation on porous dust grains}. \mnras 365(3):801--806. \doi{10.1111/j.1365-2966.2005.09803.x}, {\href{https://arxiv.org/abs/astro-ph/0506492}{{arXiv:astro-ph/0506492}}} {[astro-ph]}

\bibitem[{{Perrin} et~al.(2009){Perrin}, {Schneider}, {Duchene}, {Pinte}, {Grady}, {Wisniewski}, and {Hines}}]{Perrin09}
{Perrin} MD, {Schneider} G, {Duchene} G, et~al (2009) {The Case of AB Aurigae's Disk in Polarized Light: Is there Truly a Gap?} \apjl 707(2):L132--L136. \doi{10.1088/0004-637X/707/2/L132}, {\href{https://arxiv.org/abs/0911.1130}{{arXiv:0911.1130}}} {[astro-ph.EP]}

\bibitem[{{Perrin} et~al.(2015){Perrin}, {Duchene}, {Millar-Blanchaer}, {Fitzgerald}, {Graham}, {Wiktorowicz}, {Kalas}, {Macintosh}, {Bauman}, {Cardwell}, {Chilcote}, {De Rosa}, {Dillon}, {Doyon}, {Dunn}, {Erikson}, {Gavel}, {Goodsell}, {Hartung}, {Hibon}, {Ingraham}, {Kerley}, {Konapacky}, {Larkin}, {Maire}, {Marchis}, {Marois}, {Mittal}, {Morzinski}, {Oppenheimer}, {Palmer}, {Patience}, {Poyneer}, {Pueyo}, {Rantakyr{\"o}}, {Sadakuni}, {Saddlemyer}, {Savransky}, {Soummer}, {Sivaramakrishnan}, {Song}, {Thomas}, {Wallace}, {Wang}, and {Wolff}}]{Perrin15}
{Perrin} MD, {Duchene} G, {Millar-Blanchaer} M, et~al (2015) {Polarimetry with the Gemini Planet Imager: Methods, Performance at First Light, and the Circumstellar Ring around HR 4796A}. \apj 799(2):182. \doi{10.1088/0004-637X/799/2/182}, {\href{https://arxiv.org/abs/1407.2495}{{arXiv:1407.2495}}} {[astro-ph.EP]}

\bibitem[{{Petrova} et~al.(2000){Petrova}, {Jockers}, and {Kiselev}}]{Petrova00}
{Petrova} EV, {Jockers} K, {Kiselev} NN (2000) {Light Scattering by Aggregates with Sizes Comparable to the Wavelength: An Application to Cometary Dust}. \icarus 148(2):526--536. \doi{10.1006/icar.2000.6504}

\bibitem[{{Petrova} et~al.(2004){Petrova}, {Tishkovets}, and {Jockers}}]{Petrova04}
{Petrova} EV, {Tishkovets} VP, {Jockers} K (2004) {Polarization of Light Scattered by Solar System Bodies and the Aggregate Model of Dust Particles}. Solar System Research 38(4):309--324. \doi{10.1023/B:SOLS.0000037466.32514.fe}

\bibitem[{Piane et~al.(2016)Piane, Corno, and Ugliengo}]{DellePiane2016-TCA}
Piane MD, Corno M, Ugliengo P (2016) Propionic acid derivatives confined in mesoporous silica: monomers or dimers? the case of ibuprofen investigated by static and dynamic ab initio simulations. Theor Chem Acc 135:1--10. \doi{10.1007/s00214-016-1817-9}

\bibitem[{Pieters and Noble(2016)}]{RN1703}
Pieters CM, Noble SK (2016) {Space weathering on airless bodies}. \jgr 121(10):1865--1884. \doi{10.1002/2016je005128}

\bibitem[{Pino et~al.(2019)Pino, Chabot, Béroff, Godard, Fernandez-Villoria, Lei, Breuer, Herder, Wucher, Bender, Severin, Trautmann, and Dartois}]{RN1856}
Pino T, Chabot M, Béroff K, et~al (2019) {Release of large polycyclic aromatic hydrocarbons and fullerenes by cosmic rays from interstellar dust Swift heavy ion irradiations of interstellar carbonaceous dust analogue}. \aap 623:A134. \doi{10.1051/0004-6361/201834855}

\bibitem[{{Pinte} et~al.(2007){Pinte}, {Fouchet}, {M{\'e}nard}, {Gonzalez}, and {Duch{\^e}ne}}]{Pinte07}
{Pinte} C, {Fouchet} L, {M{\'e}nard} F, et~al (2007) {On the stratified dust distribution of the GG Tauri circumbinary ring}. \aap 469(3):963--971. \doi{10.1051/0004-6361:20077137}, {\href{https://arxiv.org/abs/0704.2747}{{arXiv:0704.2747}}} {[astro-ph]}

\bibitem[{{Pinte} et~al.(2008){Pinte}, {Padgett}, {M{\'e}nard}, {Stapelfeldt}, {Schneider}, {Olofsson}, {Pani{\'c}}, {Augereau}, {Duch{\^e}ne}, {Krist}, {Pontoppidan}, {Perrin}, {Grady}, {Kessler-Silacci}, {van Dishoeck}, {Lommen}, {Silverstone}, {Hines}, {Wolf}, {Blake}, {Henning}, and {Stecklum}}]{Pinte08}
{Pinte} C, {Padgett} DL, {M{\'e}nard} F, et~al (2008) {Probing dust grain evolution in IM Lupi's circumstellar disc. Multi-wavelength observations and modelling of the dust disc}. \aap 489(2):633--650. \doi{10.1051/0004-6361:200810121}, {\href{https://arxiv.org/abs/0808.0619}{{arXiv:0808.0619}}} {[astro-ph]}

\bibitem[{Pitman and van Duin(2012)}]{ReaxFF-Zeolite}
Pitman MC, van Duin ACT (2012) {Dynamics of Confined Reactive Water in Smectite Clay–Zeolite Composites}. J Am Chem Soc 134(6):3042--3053. \doi{10.1021/ja208894m}

\bibitem[{Potapov and Bouwman(2022)}]{RN1534}
Potapov A, Bouwman J (2022) {Importance of laboratory experimental studies of silicate grains for exoplanet atmosphere characterization}. Front Astron Space Sci 9:912302. \doi{10.3389/fspas.2022.912302}

\bibitem[{Potapov and McCoustra(2021)}]{RN1301}
Potapov A, McCoustra MRS (2021) {Physics and chemistry on the surface of cosmic dust grains: a laboratory view}. Int Rev Phys Chem 40:299--364. \doi{10.1080/0144235X.2021.1918498}

\bibitem[{Potapov et~al.(2017)Potapov, J\"ager, Henning, Jonusas, and Krim}]{RN602}
Potapov A, J\"ager C, Henning T, et~al (2017) {The Formation of Formaldehyde on Interstellar Carbonaceous Grain Analogs by O/H Atom Addition}. \apj 846(2):131. \doi{10.3847/1538-4357/aa85e8}

\bibitem[{Potapov et~al.(2018a)Potapov, Mutschke, Seeber, Henning, and J\"ager}]{RN695}
Potapov A, Mutschke H, Seeber P, et~al (2018a) {Low-temperature Optical Properties of Interstellar and Circumstellar Icy Silicate Grain Analogs in the Mid-infrared Spectral Region}. \apj 861:84. \doi{10.3847/1538-4357/aac6d3}

\bibitem[{Potapov et~al.(2018b)Potapov, J\"ager, and Henning}]{RN715}
Potapov A, J\"ager C, Henning T (2018b) {Temperature Programmed Desorption of Water Ice from the Surface of Amorphous Carbon and Silicate Grains as Related to Planet-forming Disks}. \apj 865(1):58. \doi{10.3847/1538-4357/aad803}

\bibitem[{Potapov et~al.(2019)Potapov, Theule, J\"ager, and Henning}]{RN767}
Potapov A, Theule P, J\"ager C, et~al (2019) {Evidence of surface catalytic effect on cosmic dust grain analogues: the ammonia and carbon dioxide surface reaction}. ApJL 878:L20. \doi{10.3847/2041-8213/ab2538}

\bibitem[{Potapov et~al.(2020)Potapov, J\"ager, and Henning}]{RN930}
Potapov A, J\"ager C, Henning T (2020) {Ice coverage of dust grains in cold astrophysical environments}. \prl 124:221103. \doi{10.1103/PhysRevLett.124.221103}

\bibitem[{Potapov et~al.(2021)Potapov, Bouwman, J\"ager, and Henning}]{RN873}
Potapov A, Bouwman J, J\"ager C, et~al (2021) {Dust/ice mixing in cold regions and solid-state water in the diffuse interstellar medium}. Nat Astron 5:78. \doi{10.1038/s41550-020-01214-x}

\bibitem[{Potapov et~al.(2022)Potapov, Palumbo, Dionnet, Longobardo, Jäger, Baratta, Rotundi, and Henning}]{RN1504}
Potapov A, Palumbo M, Dionnet Z, et~al (2022) {Exploring refractory organics in extraterrestrial particles}. ApJ 935:158. \doi{10.3847/1538-4357/ac7f32}

\bibitem[{Potapov et~al.(2024)Potapov, J\"ager, Mutschke, and Henning}]{RN1715}
Potapov A, J\"ager C, Mutschke H, et~al (2024) {Trapped Water on Silicates in the Laboratory and in Astrophysical Environments}. \apj 965(1):48. \doi{10.3847/1538-4357/ad2c07}

\bibitem[{Pratontep et~al.(2005)Pratontep, Carroll, Xirouchaki, Streun, and Palmer}]{RN1699}
Pratontep S, Carroll SJ, Xirouchaki C, et~al (2005) {Size-selected cluster beam source based on radio frequency magnetron plasma sputtering and gas condensation}. Review of Scientific Instruments 76(4):045103. \doi{10.1063/1.1869332}

\bibitem[{Rabkin et~al.(1999)Rabkin, Semenov, Bischoff, and Gust}]{RN1875}
Rabkin E, Semenov V, Bischoff E, et~al (1999) {Evidence for slow self-diffusion along special CSL grain boundaries from the kinetics of discontinuous ordering in Fe-50 at.\% Co}. Intergranular and Interphase Boundaries in Materials, Iib98 294-2:601--604. \doi{10.4028/www.scientific.net/MSF.294-296.601}

\bibitem[{{Rannou} et~al.(1997){Rannou}, {Cabane}, {Botet}, and {Chassefi{\`e}re}}]{Rannou97}
{Rannou} P, {Cabane} M, {Botet} R, et~al (1997) {A new interpretation of scattered light measurements at Titan's limb}. \jgr 102(E5):10997--11014. \doi{10.1029/97JE00719}

\bibitem[{{Rannou} et~al.(2024){Rannou}, {Botet}, and {Tazaki}}]{Rannou24}
{Rannou} P, {Botet} R, {Tazaki} R (2024) {Improvement of the Mean Field T-Matrix method for scattering by fractal aggregates of identical spheres in astrophysical environments}. \icarus 424:116247. \doi{10.1016/j.icarus.2024.116247}

\bibitem[{{Raymond} et~al.(2004){Raymond}, {Quinn}, and {Lunine}}]{Raymond2004}
{Raymond} SN, {Quinn} T, {Lunine} JI (2004) {Making other earths: dynamical simulations of terrestrial planet formation and water delivery}. \icarus 168(1):1--17. \doi{10.1016/j.icarus.2003.11.019}, {\href{https://arxiv.org/abs/astro-ph/0308159}{{arXiv:astro-ph/0308159}}} {[astro-ph]}

\bibitem[{{Rebollido} et~al.(2024){Rebollido}, {Stark}, {Kammerer}, {Perrin}, {Lawson}, {Pueyo}, {Chen}, {Hines}, {Girard}, {Worthen}, {Ingerbretsen}, {Betti}, {Clampin}, {Golimowski}, {Hoch}, {Lewis}, {Lu}, {van der Marel}, {Rickman}, {Seager}, {Soummer}, {Valenti}, {Ward-Duong}, and {Mountain}}]{Rebollido24}
{Rebollido} I, {Stark} CC, {Kammerer} J, et~al (2024) {JWST-TST High Contrast: Asymmetries, Dust Populations, and Hints of a Collision in the {\ensuremath{\beta}} Pictoris Disk with NIRCam and MIRI}. \aj 167(2):69. \doi{10.3847/1538-3881/ad1759}, {\href{https://arxiv.org/abs/2401.05271}{{arXiv:2401.05271}}} {[astro-ph.EP]}

\bibitem[{{Ren} et~al.(2020){Ren}, {Pueyo}, {Chen}, {Choquet}, {Debes}, {Duch{\^e}ne}, {M{\'e}nard}, and {Perrin}}]{Ren20}
{Ren} B, {Pueyo} L, {Chen} C, et~al (2020) {Using Data Imputation for Signal Separation in High-contrast Imaging}. \apj 892(2):74. \doi{10.3847/1538-4357/ab7024}, {\href{https://arxiv.org/abs/2001.00563}{{arXiv:2001.00563}}} {[astro-ph.IM]}

\bibitem[{{Ren} et~al.(2023{\natexlab{a}}){Ren}, {Benisty}, {Ginski}, {Tazaki}, {Wallack}, {Milli}, {Garufi}, {Bae}, {Facchini}, {M{\'e}nard}, {Pinilla}, {Swastik}, {Teague}, and {Wahhaj}}]{Ren23}
{Ren} BB, {Benisty} M, {Ginski} C, et~al (2023{\natexlab{a}}) {Protoplanetary disks in K$_{s}$-band total intensity and polarized light}. \aap 680:A114. \doi{10.1051/0004-6361/202347353}, {\href{https://arxiv.org/abs/2310.08589}{{arXiv:2310.08589}}} {[astro-ph.EP]}

\bibitem[{{Ren} et~al.(2023{\natexlab{b}}){Ren}, {Rebollido}, {Choquet}, {Zhou}, {Perrin}, {Schneider}, {Milli}, {Wolff}, {Chen}, {Debes}, {Hagan}, {Hines}, {Millar-Blanchaer}, {Pueyo}, {Roberge}, {Serabyn}, and {Soummer}}]{Ren23debriscolor}
{Ren} BB, {Rebollido} I, {Choquet} {\'E}, et~al (2023{\natexlab{b}}) {Debris disk color with the Hubble Space Telescope}. \aap 672:A114. \doi{10.1051/0004-6361/202245458}, {\href{https://arxiv.org/abs/2302.04273}{{arXiv:2302.04273}}} {[astro-ph.EP]}

\bibitem[{{Renard} et~al.(2024){Renard}, {Hadamcik}, and {Worms}}]{Renard24}
{Renard} JB, {Hadamcik} E, {Worms} JC (2024) {The laboratory PROGRA2 database to interpret the linear polarization and brightness phase curves of light scattered by solid particles in clouds and layers}. \jqsrt 320:108980. \doi{10.1016/j.jqsrt.2024.108980}

\bibitem[{{Rich} et~al.(2019){Rich}, {Wisniewski}, {Currie}, {Fukagawa}, {Grady}, {Sitko}, {Pikhartova}, {Hashimoto}, {Abe}, {Brandner}, {Brandt}, {Carson}, {Chilcote}, {Dong}, {Feldt}, {Goto}, {Groff}, {Guyon}, {Hayano}, {Hayashi}, {Hayashi}, {Henning}, {Hodapp}, {Ishii}, {Iye}, {Janson}, {Jovanovic}, {Kandori}, {Kasdin}, {Knapp}, {Kudo}, {Kusakabe}, {Kuzuhara}, {Kwon}, {Lozi}, {Martinache}, {Matsuo}, {Mayama}, {McElwain}, {Miyama}, {Morino}, {Moro-Martin}, {Nakagawa}, {Nishimura}, {Pyo}, {Serabyn}, {Suto}, {Russel}, {Suzuki}, {Takami}, {Takato}, {Terada}, {Thalmann}, {Turner}, {Uyama}, {Wagner}, {Watanabe}, {Yamada}, {Takami}, {Usuda}, and {Tamura}}]{Rich19}
{Rich} EA, {Wisniewski} JP, {Currie} T, et~al (2019) {Multi-epoch Direct Imaging and Time-variable Scattered Light Morphology of the HD 163296 Protoplanetary Disk}. \apj 875(1):38. \doi{10.3847/1538-4357/ab0f3b}, {\href{https://arxiv.org/abs/1811.07785}{{arXiv:1811.07785}}} {[astro-ph.SR]}

\bibitem[{Rimola and Bromley(2021)}]{Rimola-Bromley2021}
Rimola A, Bromley ST (2021) Formation of interstellar silicate dust via nanocluster aggregation: Insights from quantum chemistry simulations. Front Astron Space Sci 8. \doi{10.3389/fspas.2021.659494}

\bibitem[{Rimola et~al.(2013)Rimola, Costa, Sodupe, Lambert, and Ugliengo}]{Rimola2013}
Rimola A, Costa D, Sodupe M, et~al (2013) {Silica Surface Features and Their Role in the Adsorption of Biomolecules: Computational Modeling and Experiments}. Chem Rev 113(6):4216--4313. \doi{10.1021/cr3003054}

\bibitem[{Rimola et~al.(2021)Rimola, Ferrero, Germain, Corno, and Ugliengo}]{Minerals-Rimola2021}
Rimola A, Ferrero S, Germain A, et~al (2021) {Computational Surface Modelling of Ices and Minerals of Interstellar Interest—Insights and Perspectives}. Minerals 11(1). \doi{10.3390/min11010026}

\bibitem[{{Rodigas} et~al.(2014){Rodigas}, {Debes}, {Hinz}, {Mamajek}, {Pecaut}, {Currie}, {Bailey}, {Defrere}, {De Rosa}, {Hill}, {Leisenring}, {Schneider}, {Skemer}, {Skrutskie}, {Vaitheeswaran}, and {Ward-Duong}}]{Rodigas14}
{Rodigas} TJ, {Debes} JH, {Hinz} PM, et~al (2014) {Does the Debris Disk around HD 32297 Contain Cometary Grains?} \apj 783(1):21. \doi{10.1088/0004-637X/783/1/21}, {\href{https://arxiv.org/abs/1401.3343}{{arXiv:1401.3343}}} {[astro-ph.EP]}

\bibitem[{{Rodigas} et~al.(2015){Rodigas}, {Stark}, {Weinberger}, {Debes}, {Hinz}, {Close}, {Chen}, {Smith}, {Males}, {Skemer}, {Puglisi}, {Follette}, {Morzinski}, {Wu}, {Briguglio}, {Esposito}, {Pinna}, {Riccardi}, {Schneider}, and {Xompero}}]{Rodigas15}
{Rodigas} TJ, {Stark} CC, {Weinberger} A, et~al (2015) {On the Morphology and Chemical Composition of the HR 4796A Debris Disk}. \apj 798(2):96. \doi{10.1088/0004-637X/798/2/96}, {\href{https://arxiv.org/abs/1410.7753}{{arXiv:1410.7753}}} {[astro-ph.SR]}

\bibitem[{Rosu-Finsen et~al.(2016)Rosu-Finsen, Marchione, Salter, Stubbing, Brown, and McCoustra}]{RN955}
Rosu-Finsen A, Marchione D, Salter TL, et~al (2016) {Peeling the astronomical onion}. Phys Chem Chem Phys 18(46):31930--31935. \doi{10.1039/c6cp05751a}

\bibitem[{Rouillé et~al.(2014)Rouillé, Jager, Krasnokutski, Krebsz, and Henning}]{RN824}
Rouillé G, Jager C, Krasnokutski SA, et~al (2014) {Cold condensation of dust in the ISM}. Faraday Discussions 168:449--460. \doi{10.1039/C4FD00010B}

\bibitem[{{Roumesy} et~al.(2025){Roumesy}, {M{\'e}nard}, {Tazaki}, {Duch{\^e}ne}, {Martinien}, and {Zerna}}]{Roumesy25}
{Roumesy} M, {M{\'e}nard} F, {Tazaki} R, et~al (2025) {DRAGyS -- A comprehensive tool to extract scattering phase functions in protoplanetary disks}. arXiv e-prints arXiv:2505.20070. \doi{10.48550/arXiv.2505.20070}, {\href{https://arxiv.org/abs/2505.20070}{{arXiv:2505.20070}}} {[astro-ph.IM]}

\bibitem[{Rouquerol et~al.(1999)Rouquerol, Rouquerol, and Sing}]{RN1689}
Rouquerol F, Rouquerol J, Sing K (1999) {Adsorption by Powders \& Porous Solids: Principles, Methodology and Applications}. Academic, San Diego

\bibitem[{{Rubie} et~al.(2015){Rubie}, {Jacobson}, {Morbidelli}, {O'Brien}, {Young}, {de Vries}, {Nimmo}, {Palme}, and {Frost}}]{Rubie2015}
{Rubie} DC, {Jacobson} SA, {Morbidelli} A, et~al (2015) {Accretion and differentiation of the terrestrial planets with implications for the compositions of early-formed Solar System bodies and accretion of water}. \icarus 248:89--108. \doi{10.1016/j.icarus.2014.10.015}, {\href{https://arxiv.org/abs/1410.3509}{{arXiv:1410.3509}}} {[astro-ph.EP]}

\bibitem[{Rubin and Ma(2017)}]{RN1694}
Rubin AE, Ma C (2017) {Meteoritic minerals and their origins}. Chemie Der Erde-Geochemistry 77(3):325--385. \doi{10.1016/j.chemer.2017.01.005}

\bibitem[{Ryberg(1985)}]{RN1885}
Ryberg R (1985) {Vibrational Line-Shape of Chemisorbed {CO}}. Phys Rev B 32(4):2671--2673. \doi{10.1103/PhysRevB.32.2671}

\bibitem[{Sabri et~al.(2014)Sabri, Gavilan, Jager, Lemaire, Vidali, Mutschke, and Henning}]{RN784}
Sabri T, Gavilan L, Jager C, et~al (2014) {Interstellar Silicate Analogs for Grain-Surface Reaction Experiments: Gas-Phase Condensation and Characterization of the Silicate Dust Grains}. \apj 780(2):180. \doi{10.1088/0004-637X/780/2/180}

\bibitem[{Saliba et~al.(2016)Saliba, Ruch, Volksen, Magbitang, Dubois, and Michel}]{Pore_size_vs_hydroxylation_MMM2016}
Saliba S, Ruch P, Volksen W, et~al (2016) {Combined influence of pore size distribution and surface hydrophilicity on the water adsorption characteristics of micro- and mesoporous silica}. Microporous and Mesoporous Materials 226:221--228. \doi{10.1016/j.micromeso.2015.12.029}

\bibitem[{Samra et~al.(2020)Samra, Helling, and Min}]{RN1433}
Samra D, Helling C, Min M (2020) {Mineral snowflakes on exoplanets and brown dwarfs: Effects of micro-porosity, size distributions, and particle shape}. \aap 639:A107. \doi{10.1051/0004-6361/202037553}

\bibitem[{Sanzone et~al.(2021)Sanzone, Yin, and Sun}]{RN1702}
Sanzone G, Yin JL, Sun HL (2021) {Scaling up of cluster beam deposition technology for catalysis application}. Front Chem Sci Eng 15(6):1360--1379. \doi{10.1007/s11705-021-2101-7}

\bibitem[{{Sargent} et~al.(2006){Sargent}, {Forrest}, {D'Alessio}, {Li}, {Najita}, {Watson}, {Calvet}, {Furlan}, {Green}, {Kim}, {Sloan}, {Chen}, {Hartmann}, and {Houck}}]{Sargent06}
{Sargent} B, {Forrest} WJ, {D'Alessio} P, et~al (2006) {Dust Processing in Disks around T Tauri Stars}. \apj 645(1):395--415. \doi{10.1086/504283}, {\href{https://arxiv.org/abs/astro-ph/0605415}{{arXiv:astro-ph/0605415}}} {[astro-ph]}

\bibitem[{{Savage} and {Sembach}(1996)}]{Savage1996}
{Savage} BD, {Sembach} KR (1996) {Interstellar Abundances from Absorption-Line Observations with the Hubble Space Telescope}. \araa 34:279--330. \doi{10.1146/annurev.astro.34.1.279}

\bibitem[{{Schegerer} et~al.(2006){Schegerer}, {Wolf}, {Voshchinnikov}, {Przygodda}, and {Kessler-Silacci}}]{Schegerer06}
{Schegerer} A, {Wolf} S, {Voshchinnikov} NV, et~al (2006) {Analysis of the dust evolution in the circumstellar disks of T Tauri stars}. \aap 456(2):535--548. \doi{10.1051/0004-6361:20054560}, {\href{https://arxiv.org/abs/astro-ph/0604059}{{arXiv:astro-ph/0604059}}} {[astro-ph]}

\bibitem[{{Schlafly} et~al.(2016){Schlafly}, {Meisner}, {Stutz}, {Kainulainen}, {Peek}, {Tchernyshyov}, {Rix}, {Finkbeiner}, {Covey}, {Green}, {Bell}, {Burgett}, {Chambers}, {Draper}, {Flewelling}, {Hodapp}, {Kaiser}, {Magnier}, {Martin}, {Metcalfe}, {Wainscoat}, and {Waters}}]{Schlafly2016}
{Schlafly} EF, {Meisner} AM, {Stutz} AM, et~al (2016) {The Optical-infrared Extinction Curve and Its Variation in the Milky Way}. \apj 821(2):78. \doi{10.3847/0004-637X/821/2/78}, {\href{https://arxiv.org/abs/1602.03928}{{arXiv:1602.03928}}} {[astro-ph.GA]}

\bibitem[{{Schlafly} et~al.(2017){Schlafly}, {Peek}, {Finkbeiner}, and {Green}}]{Schlafly2017}
{Schlafly} EF, {Peek} JEG, {Finkbeiner} DP, et~al (2017) {Mapping the Extinction Curve in 3D: Structure on Kiloparsec Scales}. \apj 838(1):36. \doi{10.3847/1538-4357/aa619d}, {\href{https://arxiv.org/abs/1612.02818}{{arXiv:1612.02818}}} {[astro-ph.GA]}

\bibitem[{{Schmid}(2021)}]{Schmid21}
{Schmid} HM (2021) {Quadrant polarization parameters for the scattered light of circumstellar disks. Analysis of debris disk models and observations of HR 4796A}. \aap 655:A83. \doi{10.1051/0004-6361/202140405}, {\href{https://arxiv.org/abs/2109.10099}{{arXiv:2109.10099}}} {[astro-ph.SR]}

\bibitem[{{Schneider} et~al.(2009){Schneider}, {Weinberger}, {Becklin}, {Debes}, and {Smith}}]{Schneider09}
{Schneider} G, {Weinberger} AJ, {Becklin} EE, et~al (2009) {STIS Imaging of the HR 4796A Circumstellar Debris Ring}. \aj 137(1):53--61. \doi{10.1088/0004-6256/137/1/53}, {\href{https://arxiv.org/abs/0810.0286}{{arXiv:0810.0286}}} {[astro-ph]}

\bibitem[{{Seizinger} and {Kley}(2013)}]{Seizinger13}
{Seizinger} A, {Kley} W (2013) {Bouncing behavior of microscopic dust aggregates}. \aap 551:A65. \doi{10.1051/0004-6361/201220946}, {\href{https://arxiv.org/abs/1301.3629}{{arXiv:1301.3629}}} {[astro-ph.EP]}

\bibitem[{Senanayake et~al.(2021)Senanayake, Greathouse, Ilgen, and Thompson}]{Simulated_IR_confined_water_JCP2021}
Senanayake HS, Greathouse JA, Ilgen AG, et~al (2021) {Simulations of the IR and Raman spectra of water confined in amorphous silica slit pores}. \jcp 154(10):104503. \doi{10.1063/5.0040739}

\bibitem[{{Serkowski} et~al.(1975){Serkowski}, {Mathewson}, and {Ford}}]{Serkowski1975}
{Serkowski} K, {Mathewson} DS, {Ford} VL (1975) {Wavelength dependence of interstellar polarization and ratio of total to selective extinction.} \apj 196:261--290. \doi{10.1086/153410}

\bibitem[{{Shen} et~al.(2008){Shen}, {Draine}, and {Johnson}}]{Shen08}
{Shen} Y, {Draine} BT, {Johnson} ET (2008) {Modeling Porous Dust Grains with Ballistic Aggregates. I. Geometry and Optical Properties}. \apj 689(1):260--275. \doi{10.1086/592765}, {\href{https://arxiv.org/abs/0801.1996}{{arXiv:0801.1996}}} {[astro-ph]}

\bibitem[{{Shen} et~al.(2009){Shen}, {Draine}, and {Johnson}}]{Shen09}
{Shen} Y, {Draine} BT, {Johnson} ET (2009) {Modeling Porous Dust Grains with Ballistic Aggregates. II. Light Scattering Properties}. \apj 696(2):2126--2137. \doi{10.1088/0004-637X/696/2/2126}, {\href{https://arxiv.org/abs/0901.2177}{{arXiv:0901.2177}}} {[astro-ph.EP]}

\bibitem[{{Sicilia-Aguilar} et~al.(2007){Sicilia-Aguilar}, {Hartmann}, {Watson}, {Bohac}, {Henning}, and {Bouwman}}]{SiciliaAguilar07}
{Sicilia-Aguilar} A, {Hartmann} LW, {Watson} D, et~al (2007) {Silicate Dust in Evolved Protoplanetary Disks: Growth, Sedimentation, and Accretion}. \apj 659(2):1637--1660. \doi{10.1086/512121}, {\href{https://arxiv.org/abs/astro-ph/0701321}{{arXiv:astro-ph/0701321}}} {[astro-ph]}

\bibitem[{{Sierra} and {Lizano}(2020)}]{Sierra20}
{Sierra} A, {Lizano} S (2020) {Effects of Scattering, Temperature Gradients, and Settling on the Derived Dust Properties of Observed Protoplanetary Disks}. \apj 892(2):136. \doi{10.3847/1538-4357/ab7d32}, {\href{https://arxiv.org/abs/2003.02982}{{arXiv:2003.02982}}} {[astro-ph.EP]}

\bibitem[{{Sierra} et~al.(2021){Sierra}, {P{\'e}rez}, {Zhang}, {Law}, {Guzm{\'a}n}, {Qi}, {Bosman}, {{\"O}berg}, {Andrews}, {Long}, {Teague}, {Booth}, {Walsh}, {Wilner}, {M{\'e}nard}, {Cataldi}, {Czekala}, {Bae}, {Huang}, {Bergner}, {Ilee}, {Benisty}, {Le Gal}, {Loomis}, {Tsukagoshi}, {Liu}, {Yamato}, and {Aikawa}}]{Sierra21}
{Sierra} A, {P{\'e}rez} LM, {Zhang} K, et~al (2021) {Molecules with ALMA at Planet-forming Scales (MAPS). XIV. Revealing Disk Substructures in Multiwavelength Continuum Emission}. \apjs 257(1):14. \doi{10.3847/1538-4365/ac1431}, {\href{https://arxiv.org/abs/2109.06433}{{arXiv:2109.06433}}} {[astro-ph.EP]}

\bibitem[{{Silsbee} and {Draine}(2016)}]{2016ApJ...818..133S}
{Silsbee} K, {Draine} BT (2016) {Radiation Pressure on Fluffy Submicron-sized Grains}. \apj 818(2):133. \doi{10.3847/0004-637X/818/2/133}, {\href{https://arxiv.org/abs/1508.00646}{{arXiv:1508.00646}}} {[astro-ph.EP]}

\bibitem[{Singh and White(2018)}]{RN1891}
Singh J, White RL (2018) {A variable temperature infrared spectroscopy study of NaA zeolite dehydration}. Vibrational Spectroscopy 94:37--42. \doi{10.1016/j.vibspec.2017.11.003}

\bibitem[{{Sipil{\"a}} et~al.(2021){Sipil{\"a}}, {Silsbee}, and {Caselli}}]{Sipila2021}
{Sipil{\"a}} O, {Silsbee} K, {Caselli} P (2021) {A Revised Description of the Cosmic Ray Induced Desorption of Interstellar Ices}. \apj 922(2):126. \doi{10.3847/1538-4357/ac23ce}, {\href{https://arxiv.org/abs/2106.04593}{{arXiv:2106.04593}}} {[astro-ph.GA]}

\bibitem[{Smit(2008)}]{Smit2008}
Smit B (2008) {Molecular Simulations of Zeolites: Adsorption, Diffusion, and Shape Selectivity}. Chem Rev 108(10):4125--4184. \doi{10.1021/cr8002642}

\bibitem[{{Smith} and {Dwek}(1998)}]{Smith1998}
{Smith} RK, {Dwek} E (1998) {Soft X-Ray Scattering and Halos from Dust}. \apj 503(2):831--842. \doi{10.1086/306018}, {\href{https://arxiv.org/abs/astro-ph/9710232}{{arXiv:astro-ph/9710232}}} {[astro-ph]}

\bibitem[{Smith et~al.(2011)Smith, Matthiesen, Knox, and Kay}]{RN1224}
Smith RS, Matthiesen J, Knox J, et~al (2011) {Crystallization Kinetics and Excess Free Energy of H$_2$O and D$_2$O Nanoscale Films of Amorphous Solid Water}. J Phys Chem A 115(23):5908--5917. \doi{10.1021/jp110297q}

\bibitem[{Sobrado et~al.(2023)Sobrado, Santoro, Mart{\'i}nez, Merino, Joblin, Cernicharo, and Gago}]{RN1698}
Sobrado J, Santoro G, Mart{\'i}nez L, et~al (2023) The stardust machine project. In: Mennella V, Joblin C (eds) European Conference on Laboratory Astrophysics ECLA2020, Astrophysics and Space Science Proceedings, vol~59. Springer, Cham, pp 101--110, \doi{10.1007/978-3-031-29003-9_12}

\bibitem[{{Song} et~al.(2024){Song}, {Chang}, {Meng}, and {Zhang}}]{Song24}
{Song} Z, {Chang} Q, {Meng} Q, et~al (2024) {Modeling long carbon-chain species formation with porous multiphase models}. \aap 691:A40. \doi{10.1051/0004-6361/202450647}

\bibitem[{{Sorensen}(2001)}]{Sorensen01}
{Sorensen} CM (2001) {Light Scattering by Fractal Aggregates: A Review}. Aerosol Science Technology 35(2):648--687. \doi{10.1080/02786820117868}

\bibitem[{{Stammler} and {Birnstiel}(2022)}]{Stammler22}
{Stammler} SM, {Birnstiel} T (2022) {DustPy: A Python Package for Dust Evolution in Protoplanetary Disks}. \apj 935(1):35. \doi{10.3847/1538-4357/ac7d58}, {\href{https://arxiv.org/abs/2207.00322}{{arXiv:2207.00322}}} {[astro-ph.EP]}

\bibitem[{{Stecher} and {Donn}(1965)}]{Strecher1965}
{Stecher} TP, {Donn} B (1965) {On Graphite and Interstellar Extinction}. \apj 142:1681. \doi{10.1086/148461}

\bibitem[{{Stephens} et~al.(2017){Stephens}, {Yang}, {Li}, {Looney}, {Kataoka}, {Kwon}, {Fern{\'a}ndez-L{\'o}pez}, {Hull}, {Hughes}, {Segura-Cox}, {Mundy}, {Crutcher}, and {Rao}}]{Stephens17}
{Stephens} IW, {Yang} H, {Li} ZY, et~al (2017) {ALMA Reveals Transition of Polarization Pattern with Wavelength in HL Tau{\textquoteright}s Disk}. \apj 851(1):55. \doi{10.3847/1538-4357/aa998b}, {\href{https://arxiv.org/abs/1710.04670}{{arXiv:1710.04670}}} {[astro-ph.SR]}

\bibitem[{{Stephens} et~al.(2023){Stephens}, {Lin}, {Fern{\'a}ndez-L{\'o}pez}, {Li}, {Looney}, {Yang}, {Harrison}, {Kataoka}, {Carrasco-Gonzalez}, {Okuzumi}, and {Tazaki}}]{Stephens23}
{Stephens} IW, {Lin} ZYD, {Fern{\'a}ndez-L{\'o}pez} M, et~al (2023) {Aligned grains and scattered light found in gaps of planet-forming disk}. \nat 623(7988):705--708. \doi{10.1038/s41586-023-06648-7}, {\href{https://arxiv.org/abs/2311.08452}{{arXiv:2311.08452}}} {[astro-ph.GA]}

\bibitem[{Sterken et~al.(2015)Sterken, Strub, Krüger, von Steiger, and Frisch}]{RN1678}
Sterken VJ, Strub P, Krüger H, et~al (2015) {Sixteen Years of {Ulysses} Interstellar Dust Measurements in the Solar System. III. Simulations and Data Unveil New Insights into Local Interstellar Dust}. \apj 812(2):141. \doi{10.1088/0004-637x/812/2/141}

\bibitem[{{Stevenson} and {Lunine}(1988)}]{Stevenson1988}
{Stevenson} DJ, {Lunine} JI (1988) {Rapid formation of Jupiter by diffusive redistribution of water vapor in the solar nebula}. \icarus 75(1):146--155. \doi{10.1016/0019-1035(88)90133-9}

\bibitem[{{Stognienko} et~al.(1995){Stognienko}, {Henning}, and {Ossenkopf}}]{Stognienko95}
{Stognienko} R, {Henning} T, {Ossenkopf} V (1995) {Optical properties of coagulated particles.} \aap 296:797

\bibitem[{{Stolker} et~al.(2016){Stolker}, {Dominik}, {Avenhaus}, {Min}, {de Boer}, {Ginski}, {Schmid}, {Juhasz}, {Bazzon}, {Waters}, {Garufi}, {Augereau}, {Benisty}, {Boccaletti}, {Henning}, {Langlois}, {Maire}, {M{\'e}nard}, {Meyer}, {Pinte}, {Quanz}, {Thalmann}, {Beuzit}, {Carbillet}, {Costille}, {Dohlen}, {Feldt}, {Gisler}, {Mouillet}, {Pavlov}, {Perret}, {Petit}, {Pragt}, {Rochat}, {Roelfsema}, {Salasnich}, {Soenke}, and {Wildi}}]{Stolker16}
{Stolker} T, {Dominik} C, {Avenhaus} H, et~al (2016) {Shadows cast on the transition disk of HD 135344B. Multiwavelength VLT/SPHERE polarimetric differential imaging}. \aap 595:A113. \doi{10.1051/0004-6361/201528039}, {\href{https://arxiv.org/abs/1603.00481}{{arXiv:1603.00481}}} {[astro-ph.EP]}

\bibitem[{Suyama et~al.(2008)Suyama, Wada, and Tanaka}]{RN901}
Suyama T, Wada K, Tanaka H (2008) {Numerical simulation of density evolution of dust aggregates in protoplanetary disks. I. Head-on collisions}. \apj 684(2):1310--1322. \doi{10.1086/590143}

\bibitem[{Taj et~al.(2017)Taj, Baird, Rosu-Finsen, and McCoustra}]{RN1155}
Taj S, Baird D, Rosu-Finsen A, et~al (2017) {Surface heterogeneity and inhomogeneous broadening of vibrational line profiles}. Phys Chem Chem Phys 19(11):7990--7995. \doi{10.1039/c6cp07530d}

\bibitem[{Taj et~al.(2021)Taj, Rosu-Finsen, and McCoustra}]{RN1159}
Taj S, Rosu-Finsen A, McCoustra MRS (2021) {Impact of Surface Heterogeneity on IR Line Profiles of Adsorbed Carbon Monoxide on Models of Interstellar Grain Surfaces}. \mnras 504:5806–5812. \doi{10.1093/mnras/stab1174}

\bibitem[{{Takami} et~al.(2013){Takami}, {Karr}, {Hashimoto}, {Kim}, {Wisniewski}, {Henning}, {Grady}, {Kandori}, {Hodapp}, {Kudo}, {Kusakabe}, {Chou}, {Itoh}, {Momose}, {Mayama}, {Currie}, {Follette}, {Kwon}, {Abe}, {Brandner}, {Brandt}, {Carson}, {Egner}, {Feldt}, {Guyon}, {Hayano}, {Hayashi}, {Hayashi}, {Ishii}, {Iye}, {Janson}, {Knapp}, {Kuzuhara}, {McElwain}, {Matsuo}, {Miyama}, {Morino}, {Moro-Martin}, {Nishimura}, {Pyo}, {Serabyn}, {Suto}, {Suzuki}, {Takato}, {Terada}, {Thalmann}, {Tomono}, {Turner}, {Watanabe}, {Yamada}, {Takami}, {Usuda}, and {Tamura}}]{Takami13}
{Takami} M, {Karr} JL, {Hashimoto} J, et~al (2013) {High-contrast Near-infrared Imaging Polarimetry of the Protoplanetary Disk around RY TAU}. \apj 772(2):145. \doi{10.1088/0004-637X/772/2/145}, {\href{https://arxiv.org/abs/1306.1887}{{arXiv:1306.1887}}} {[astro-ph.SR]}

\bibitem[{{Takami} et~al.(2014){Takami}, {Hasegawa}, {Muto}, {Gu}, {Dong}, {Karr}, {Hashimoto}, {Kusakabe}, {Chapillon}, {Tang}, {Itoh}, {Carson}, {Follette}, {Mayama}, {Sitko}, {Janson}, {Grady}, {Kudo}, {Akiyama}, {Kwon}, {Takahashi}, {Suenaga}, {Abe}, {Brandner}, {Brandt}, {Currie}, {Egner}, {Feldt}, {Guyon}, {Hayano}, {Hayashi}, {Hayashi}, {Henning}, {Hodapp}, {Honda}, {Ishii}, {Iye}, {Kandori}, {Knapp}, {Kuzuhara}, {McElwain}, {Matsuo}, {Miyama}, {Morino}, {Moro-Martin}, {Nishimura}, {Pyo}, {Serabyn}, {Suto}, {Suzuki}, {Takato}, {Terada}, {Thalmann}, {Tomono}, {Turner}, {Wisniewski}, {Watanabe}, {Yamada}, {Takami}, {Usuda}, and {Tamura}}]{Takami14}
{Takami} M, {Hasegawa} Y, {Muto} T, et~al (2014) {Surface Geometry of Protoplanetary Disks Inferred From Near-Infrared Imaging Polarimetry}. \apj 795(1):71. \doi{10.1088/0004-637X/795/1/71}, {\href{https://arxiv.org/abs/1409.1390}{{arXiv:1409.1390}}} {[astro-ph.SR]}

\bibitem[{{Tamanai} et~al.(2006a){Tamanai}, {Mutschke}, {Blum}, and {Neuh{\"a}user}}]{Tamanai06a}
{Tamanai} A, {Mutschke} H, {Blum} J, et~al (2006a) {Experimental infrared spectroscopic measurement of light extinction for agglomerate dust grains}. \jqsrt 100(1-3):373--381. \doi{10.1016/j.jqsrt.2005.11.051}

\bibitem[{{Tamanai} et~al.(2009){Tamanai}, {Mutschke}, and {Blum}}]{Tamanai09}
{Tamanai} A, {Mutschke} H, {Blum} J (2009) {IR Spectroscopic Measurements of Free-Flying Silicate Dust Grains}. In: {Henning} T, {Gr{\"u}n} E, {Steinacker} J (eds) {Cosmic Dust - Near and Far}, ASP Conference Series, vol 414. Astronomical Society of the Pacific, p 438

\bibitem[{{Tamura} et~al.(2006){Tamura}, {Fukagawa}, {Kimura}, {Yamamoto}, {Suto}, and {Abe}}]{Tamura06}
{Tamura} M, {Fukagawa} M, {Kimura} H, et~al (2006) {First Two-Micron Imaging Polarimetry of {\ensuremath{\beta}} Pictoris}. \apj 641(2):1172--1177. \doi{10.1086/500575}, {\href{https://arxiv.org/abs/astro-ph/0603344}{{arXiv:astro-ph/0603344}}} {[astro-ph]}

\bibitem[{{Tanaka} et~al.(2023){Tanaka}, {Anayama}, and {Tazaki}}]{Tanaka23}
{Tanaka} H, {Anayama} R, {Tazaki} R (2023) {Compression of Dust Aggregates via Sequential Collisions with High Mass Ratios}. \apj 945(1):68. \doi{10.3847/1538-4357/acb92b}

\bibitem[{{Tang} et~al.(2023){Tang}, {Bromley}, and {Hammer}}]{Tang2023}
{Tang} Z, {Bromley} ST, {Hammer} B (2023) {A machine learning potential for simulating infrared spectra of nanosilicate clusters}. \jcp 158(22):224108. \doi{10.1063/5.0150379}, {\href{https://arxiv.org/abs/2305.15846}{{arXiv:2305.15846}}} {[physics.chem-ph]}

\bibitem[{{Taquet} et~al.(2012){Taquet}, {Ceccarelli}, and {Kahane}}]{Taquet12}
{Taquet} V, {Ceccarelli} C, {Kahane} C (2012) {Multilayer modeling of porous grain surface chemistry. I. The GRAINOBLE model}. \aap 538:A42. \doi{10.1051/0004-6361/201117802}, {\href{https://arxiv.org/abs/1111.4165}{{arXiv:1111.4165}}} {[astro-ph.GA]}

\bibitem[{{Tatsuuma} et~al.(2024){Tatsuuma}, {Kataoka}, {Tanaka}, and {Guillot}}]{Tatsuuma24}
{Tatsuuma} M, {Kataoka} A, {Tanaka} H, et~al (2024) {The Bulk Densities of Small Solar System Bodies as a Probe of Planetesimal Formation}. \apj 974(1):9. \doi{10.3847/1538-4357/ad6a5d}, {\href{https://arxiv.org/abs/2407.21386}{{arXiv:2407.21386}}} {[astro-ph.EP]}

\bibitem[{{Tazaki}(2021)}]{Tazaki21}
{Tazaki} R (2021) {Analytic expressions for geometric cross-sections of fractal dust aggregates}. \mnras 504(2):2811--2821. \doi{10.1093/mnras/stab1069}, {\href{https://arxiv.org/abs/2104.06804}{{arXiv:2104.06804}}} {[astro-ph.EP]}

\bibitem[{{Tazaki} and {Dominik}(2022)}]{Tazaki22}
{Tazaki} R, {Dominik} C (2022) {The size of monomers of dust aggregates in planet-forming disks. Insights from quantitative optical and near-infrared polarimetry}. \aap 663:A57. \doi{10.1051/0004-6361/202243485}, {\href{https://arxiv.org/abs/2204.08506}{{arXiv:2204.08506}}} {[astro-ph.EP]}

\bibitem[{{Tazaki} and {Tanaka}(2018)}]{Tazaki18}
{Tazaki} R, {Tanaka} H (2018) {Light Scattering by Fractal Dust Aggregates. II. Opacity and Asymmetry Parameter}. \apj 860(1):79. \doi{10.3847/1538-4357/aac32d}, {\href{https://arxiv.org/abs/1803.03775}{{arXiv:1803.03775}}} {[astro-ph.EP]}

\bibitem[{{Tazaki} et~al.(2016){Tazaki}, {Tanaka}, {Okuzumi}, {Kataoka}, and {Nomura}}]{Tazaki16}
{Tazaki} R, {Tanaka} H, {Okuzumi} S, et~al (2016) {Light Scattering by Fractal Dust Aggregates. I. Angular Dependence of Scattering}. \apj 823(2):70. \doi{10.3847/0004-637X/823/2/70}, {\href{https://arxiv.org/abs/1603.07492}{{arXiv:1603.07492}}} {[astro-ph.EP]}

\bibitem[{{Tazaki} et~al.(2017){Tazaki}, {Lazarian}, and {Nomura}}]{Tazaki17}
{Tazaki} R, {Lazarian} A, {Nomura} H (2017) {Radiative Grain Alignment In Protoplanetary Disks: Implications for Polarimetric Observations}. \apj 839(1):56. \doi{10.3847/1538-4357/839/1/56}, {\href{https://arxiv.org/abs/1701.02063}{{arXiv:1701.02063}}} {[astro-ph.EP]}

\bibitem[{{Tazaki} et~al.(2019a){Tazaki}, {Tanaka}, {Muto}, {Kataoka}, and {Okuzumi}}]{Tazaki2019a}
{Tazaki} R, {Tanaka} H, {Muto} T, et~al (2019a) {Effect of dust size and structure on scattered-light images of protoplanetary discs}. \mnras 485(4):4951--4966. \doi{10.1093/mnras/stz662}, {\href{https://arxiv.org/abs/1903.01890}{{arXiv:1903.01890}}} {[astro-ph.EP]}

\bibitem[{{Tazaki} et~al.(2019b){Tazaki}, {Tanaka}, {Kataoka}, {Okuzumi}, and {Muto}}]{Tazaki2019b}
{Tazaki} R, {Tanaka} H, {Kataoka} A, et~al (2019b) {Unveiling Dust Aggregate Structure in Protoplanetary Disks by Millimeter-wave Scattering Polarization}. \apj 885(1):52. \doi{10.3847/1538-4357/ab45f0}, {\href{https://arxiv.org/abs/1907.00189}{{arXiv:1907.00189}}} {[astro-ph.EP]}

\bibitem[{{Tazaki} et~al.(2021){Tazaki}, {Murakawa}, {Muto}, {Honda}, and {Inoue}}]{Tazaki21ice}
{Tazaki} R, {Murakawa} K, {Muto} T, et~al (2021) {Scattering Polarization of 3 {\ensuremath{\mu}}m Water-ice Feature by Large Icy Grains}. \apj 910(1):26. \doi{10.3847/1538-4357/abdd3d}, {\href{https://arxiv.org/abs/2101.07635}{{arXiv:2101.07635}}} {[astro-ph.GA]}

\bibitem[{{Tazaki} et~al.(2023{\natexlab{a}}){Tazaki}, {Ginski}, and {Dominik}}]{Tazaki23-AggScat}
{Tazaki} R, {Ginski} C, {Dominik} C (2023{\natexlab{a}}) {AggScatVIR}. \doi{10.5281/zenodo.7547601}

\bibitem[{{Tazaki} et~al.(2023{\natexlab{b}}){Tazaki}, {Ginski}, and {Dominik}}]{Tazaki23}
{Tazaki} R, {Ginski} C, {Dominik} C (2023{\natexlab{b}}) {Fractal Aggregates of Submicron-sized Grains in the Young Planet-forming Disk around IM Lup}. \apjl 944(2):L43. \doi{10.3847/2041-8213/acb824}, {\href{https://arxiv.org/abs/2302.01119}{{arXiv:2302.01119}}} {[astro-ph.EP]}

\bibitem[{{Testi} et~al.(2014){Testi}, {Birnstiel}, {Ricci}, {Andrews}, {Blum}, {Carpenter}, {Dominik}, {Isella}, {Natta}, {Williams}, and {Wilner}}]{Testi14}
{Testi} L, {Birnstiel} T, {Ricci} L, et~al (2014) {Dust Evolution in Protoplanetary Disks}. In: {Beuther} H, {Klessen} RS, {Dullemond} CP, et~al (eds) {Protostars and Planets VI}, pp 339--361, \doi{10.2458/azu_uapress_9780816531240-ch015}, {\href{https://arxiv.org/abs/1402.1354}{{arXiv:1402.1354}}}

\bibitem[{{Thang} et~al.(2024){Thang}, {Diep}, {Hoang}, {Tram}, {Ngoc}, {Phuong}, and {Truong}}]{Thang24}
{Thang} NT, {Diep} PN, {Hoang} T, et~al (2024) {Evidence of Grain Alignment by Magnetically Enhanced Radiative Torques from Multiwavelength Dust Polarization Modeling of HL Tau}. \apj 970(2):114. \doi{10.3847/1538-4357/ad4f74}, {\href{https://arxiv.org/abs/2401.00220}{{arXiv:2401.00220}}} {[astro-ph.SR]}

\bibitem[{Thommes and Duncan(2006)}]{RN1060}
Thommes EW, Duncan M (2006) {The accretion of giant planet cores}. In: Klahr H, Brander W (eds) {Planet Formation: Theory, Observations, and Experiments}. Cambridge University Press, Cambridge, p 129--146

\bibitem[{Thrower et~al.(2009{\natexlab{a}})Thrower, Collings, Rutten, and McCoustra}]{RN1137}
Thrower JD, Collings MP, Rutten FJM, et~al (2009{\natexlab{a}}) {Laboratory investigations of the interaction between benzene and bare silicate grain surfaces}. \mnras 394(3):1510--1518. \doi{10.1111/j.1365-2966.2009.14420.x}

\bibitem[{Thrower et~al.(2009{\natexlab{b}})Thrower, Collings, Rutten, and McCoustra}]{RN712}
Thrower JD, Collings MP, Rutten FJM, et~al (2009{\natexlab{b}}) {Thermal desorption of C6H6 from surfaces of astrophysical relevance}. \jcp 131(24):244711. \doi{10.1063/1.3267634}

\bibitem[{{Tielens}(1989)}]{Tielens1989}
{Tielens} A (1989) {Dust in Dense Clouds}. In: {Allamandola} LJ, {Tielens} AGGM (eds) {Interstellar Dust}, IAU Symposium, vol 135. Kluwer Academic, Dordrecht, p 239

\bibitem[{{Tobon Valencia} et~al.(2022){Tobon Valencia}, {Geffrin}, {M{\'e}nard}, {Milli}, {Renard}, {Tortel}, {Eyraud}, {Litman}, {Rannou}, {Maalouf}, and {Laur}}]{TobonValencia22}
{Tobon Valencia} V, {Geffrin} JM, {M{\'e}nard} F, et~al (2022) {Scattering properties of protoplanetary dust analogs with microwave analogy: Aggregates of fractal dimensions from 1.5 to 2.8}. \aap 666:A68. \doi{10.1051/0004-6361/202142656}

\bibitem[{{Tobon Valencia} et~al.(2024){Tobon Valencia}, {Geffrin}, {M{\'e}nard}, {Milli}, {Renard}, {Tortel}, and {Litman}}]{TobonValencia24}
{Tobon Valencia} V, {Geffrin} JM, {M{\'e}nard} F, et~al (2024) {Scattering properties of protoplanetary dust analogs with microwave analogy: Rough compact grains}. \aap 688:A70. \doi{10.1051/0004-6361/202347529}

\bibitem[{{Trumpler}(1930)}]{Trumpler1930}
{Trumpler} RJ (1930) {Preliminary results on the distances, dimensions and space distribution of open star clusters}. Lick Obs Bull 420:154--188. \doi{10.5479/ADS/bib/1930LicOB.14.154T}

\bibitem[{{Tschudi} and {Schmid}(2021)}]{Tschudi21}
{Tschudi} C, {Schmid} HM (2021) {Quantitative polarimetry of the disk around HD 169142}. \aap 655:A37. \doi{10.1051/0004-6361/202141028}, {\href{https://arxiv.org/abs/2108.03102}{{arXiv:2108.03102}}} {[astro-ph.EP]}

\bibitem[{Twigg(1989)}]{RN1888}
Twigg MV (1989) {Catalyst Handbook}, 2nd edn. Routledge, New York

\bibitem[{Ueba(1997)}]{RN1881}
Ueba H (1997) {Vibrational relaxation and pump-probe spectroscopies of adsorbates on solid surfaces}. Progress in Surface Science 55(2):115--179. \doi{10.1016/S0079-6816(97)00021-X}

\bibitem[{{Ueda} et~al.(2020){Ueda}, {Kataoka}, and {Tsukagoshi}}]{Ueda20}
{Ueda} T, {Kataoka} A, {Tsukagoshi} T (2020) {Scattering-induced Intensity Reduction: Large Mass Content with Small Grains in the Inner Region of the TW Hya disk}. \apj 893(2):125. \doi{10.3847/1538-4357/ab8223}, {\href{https://arxiv.org/abs/2003.09353}{{arXiv:2003.09353}}} {[astro-ph.EP]}

\bibitem[{{Ueda} et~al.(2022){Ueda}, {Kataoka}, and {Tsukagoshi}}]{Ueda22}
{Ueda} T, {Kataoka} A, {Tsukagoshi} T (2022) {Massive Compact Dust Disk with a Gap around CW Tau Revealed by ALMA Multiband Observations}. \apj 930(1):56. \doi{10.3847/1538-4357/ac634d}, {\href{https://arxiv.org/abs/2203.16236}{{arXiv:2203.16236}}} {[astro-ph.EP]}

\bibitem[{{Ueda} et~al.(2024){Ueda}, {Tazaki}, {Okuzumi}, {Flock}, and {Sudarshan}}]{Ueda24}
{Ueda} T, {Tazaki} R, {Okuzumi} S, et~al (2024) {Support for fragile porous dust in a gravitationally self-regulated disk around IM Lup}. Nat Astron \doi{10.1038/s41550-024-02308-6}, {\href{https://arxiv.org/abs/2406.07427}{{arXiv:2406.07427}}} {[astro-ph.EP]}

\bibitem[{Ugliengo et~al.(2008)Ugliengo, Sodupe, Musso, Bush, Orlando, and Dovesi}]{Ugliengo2008}
Ugliengo P, Sodupe M, Musso F, et~al (2008) {Realistic models of hydroxylated amorphous silica surfaces and MCM-41 mesoporous material simulated by large-scale periodic B3LYP calculations}. Adv Mater 20(23):4579--4583. \doi{10.1002/adma.200801489}

\bibitem[{{Vaidya} and {Gupta}(2011)}]{Vaidya11}
{Vaidya} DB, {Gupta} R (2011) {Infrared emission from the composite grains: effects of inclusions and porosities on the 10 and 18 {\ensuremath{\mu}}m features}. \aap 528:A57. \doi{10.1051/0004-6361/201014899}, {\href{https://arxiv.org/abs/1101.1782}{{arXiv:1101.1782}}} {[astro-ph.SR]}

\bibitem[{{van Boekel} et~al.(2005){van Boekel}, {Min}, {Waters}, {de Koter}, {Dominik}, {van den Ancker}, and {Bouwman}}]{vanBoekel05}
{van Boekel} R, {Min} M, {Waters} LBFM, et~al (2005) {A 10 {\ensuremath{\mu}}m spectroscopic survey of Herbig Ae star disks: Grain growth and crystallization}. \aap 437(1):189--208. \doi{10.1051/0004-6361:20042339}, {\href{https://arxiv.org/abs/astro-ph/0503507}{{arXiv:astro-ph/0503507}}} {[astro-ph]}

\bibitem[{{van de Hulst}(1957)}]{VdH57}
{van de Hulst} HC (1957) {Light Scattering by Small Particles}. New York: John Wiley \& Sons

\bibitem[{{van den Berg} et~al.(2004){van den Berg}, {Bromley}, {Flikkema}, {Wojdel}, {Maschmeyer}, and {Jansen}}]{vandenBerg-2004b}
{van den Berg} AWC, {Bromley} ST, {Flikkema} E, et~al (2004) {Molecular-dynamics analysis of the diffusion of molecular hydrogen in all-silica sodalite}. \jcp 120(21):10285--10289. \doi{10.1063/1.1737368}

\bibitem[{Ventura et~al.(2012)Ventura, Di~Criscienzo, Schneider, Carini, Valiante, D'Antona, Gallerani, Maiolino, and Tornambe}]{RN823}
Ventura P, Di~Criscienzo M, Schneider R, et~al (2012) {Dust formation around AGB and SAGB stars: a trend with metallicity?} \mnras 424(3):2345--2357. \doi{10.1111/j.1365-2966.2012.21403.x}

\bibitem[{{Volten} et~al.(2007){Volten}, {Mu{\~n}oz}, {Hovenier}, {Rietmeijer}, {Nuth}, {Waters}, and {van der Zande}}]{Volten07}
{Volten} H, {Mu{\~n}oz} O, {Hovenier} JW, et~al (2007) {Experimental light scattering by fluffy aggregates of magnesiosilica, ferrosilica, and alumina cosmic dust analogs}. \aap 470(1):377--386. \doi{10.1051/0004-6361:20066744}

\bibitem[{{Voshchinnikov} and {Henning}(2008)}]{Vosh08}
{Voshchinnikov} NV, {Henning} T (2008) {Is the silicate emission feature only influenced by grain size?} \aap 483(1):L9--L12. \doi{10.1051/0004-6361:200809697}, {\href{https://arxiv.org/abs/0803.3227}{{arXiv:0803.3227}}} {[astro-ph]}

\bibitem[{{Voshchinnikov} and {Mathis}(1999)}]{Vosh99}
{Voshchinnikov} NV, {Mathis} JS (1999) {Calculating Cross Sections of Composite Interstellar Grains}. \apj 526(1):257--264. \doi{10.1086/307997}, {\href{https://arxiv.org/abs/astro-ph/9908240}{{arXiv:astro-ph/9908240}}} {[astro-ph]}

\bibitem[{{Voshchinnikov} et~al.(2005){Voshchinnikov}, {Il'in}, and {Henning}}]{Voschchinnikov05}
{Voshchinnikov} NV, {Il'in} VB, {Henning} T (2005) {Modelling the optical properties of composite and porous interstellar grains}. \aap 429:371--381. \doi{10.1051/0004-6361:200400081}, {\href{https://arxiv.org/abs/astro-ph/0409457}{{arXiv:astro-ph/0409457}}} {[astro-ph]}

\bibitem[{{Voshchinnikov} et~al.(2006){Voshchinnikov}, {Il'in}, {Henning}, and {Dubkova}}]{Voshchinnikov2006}
{Voshchinnikov} NV, {Il'in} VB, {Henning} T, et~al (2006) {Dust extinction and absorption: the challenge of porous grains}. \aap 445(1):167--177. \doi{10.1051/0004-6361:20053371}, {\href{https://arxiv.org/abs/astro-ph/0509277}{{arXiv:astro-ph/0509277}}} {[astro-ph]}

\bibitem[{{Voshchinnikov} et~al.(2007){Voshchinnikov}, {Videen}, and {Henning}}]{Vosh07}
{Voshchinnikov} NV, {Videen} G, {Henning} T (2007) {Effective medium theories for irregular fluffy structures: aggregation of small particles}. \ao 46(19):4065--4072. \doi{10.1364/AO.46.004065}, {\href{https://arxiv.org/abs/astro-ph/0703023}{{arXiv:astro-ph/0703023}}} {[astro-ph]}

\bibitem[{Wada et~al.(2007)Wada, Tanaka, Suyama, Kimura, and Yamamoto}]{RN1907}
Wada K, Tanaka H, Suyama T, et~al (2007) {Numerical simulation of dust aggregate collisions. I. Compression and disruption of two-dimensional aggregates}. \apj 661(1):320--333. \doi{10.1086/514332}

\bibitem[{Wada et~al.(2008)Wada, Tanaka, Suyama, Kimura, and Yamamoto}]{RN1900}
Wada K, Tanaka H, Suyama T, et~al (2008) {Numerical simulation of dust aggregate collisions. II. Compression and disruption of three-dimensional aggregates in head-on collisions}. \apj 677(2):1296--1308. \doi{10.1086/529511}

\bibitem[{Wada et~al.(2009)Wada, Tanaka, Suyama, Kimura, and Yamamoto}]{RN900}
Wada K, Tanaka H, Suyama T, et~al (2009) {Collisional Growth Conditions for Dust Aggregates}. \apj 702(2):1490--1501. \doi{10.1088/0004-637x/702/2/1490}

\bibitem[{Wada et~al.(2011)Wada, Tanaka, Suyama, Kimura, and Yamamoto}]{RN1863}
Wada K, Tanaka H, Suyama T, et~al (2011) {The Rebound Condition of Dust Aggregates Revealed by Numerical Simulation of Their Collisions}. \apj 737(1):36. \doi{10.1088/0004-637x/737/1/36}

\bibitem[{Wada et~al.(2013)Wada, Tanaka, Okuzumi, Kobayashi, Suyama, Kimura, and Yamamoto}]{RN1898}
Wada K, Tanaka H, Okuzumi S, et~al (2013) {Growth efficiency of dust aggregates through collisions with high mass ratios}. \aap 559:A62. \doi{10.1051/0004-6361/201322259}

\bibitem[{Wakelam et~al.(2017)Wakelam, Bron, Cazaux, Dulieu, Gry, Guillard, Habart, Hornekaer, Morisset, Nyman, Pirronello, Price, Valdivia, Vidali, and Watanabe}]{RN911}
Wakelam V, Bron E, Cazaux S, et~al (2017) {H$_2$ formation on interstellar dust grains: The viewpoints of theory, experiments, models and observations}. Molecular Astrophysics 9:1--36. \doi{10.1016/j.molap.2017.11.001}

\bibitem[{{Wang} and {Chen}(2019)}]{Wang2019}
{Wang} S, {Chen} X (2019) {The Optical to Mid-infrared Extinction Law Based on the APOGEE, Gaia DR2, Pan-STARRS1, SDSS, APASS, 2MASS, and WISE Surveys}. \apj 877(2):116. \doi{10.3847/1538-4357/ab1c61}, {\href{https://arxiv.org/abs/1904.04575}{{arXiv:1904.04575}}} {[astro-ph.GA]}

\bibitem[{Wang et~al.(2000)Wang, Wang, and Ewing}]{RN1869}
Wang SX, Wang LM, Ewing RC (2000) {Electron and ion irradiation of zeolites}. J Nucl Materials 278(2-3):233--241. \doi{10.1016/S0022-3115(99)00246-9}

\bibitem[{Wang et~al.(2017)Wang, Niu, and Wang}]{RN1874}
Wang XX, Niu LL, Wang SQ (2017) {Strong trapping and slow diffusion of helium in a tungsten grain boundary}. J Nucl Materials 487:158--166. \doi{10.1016/j.jnucmat.2017.02.010}

\bibitem[{{Warren} et~al.(1997){Warren}, {Watts}, {Thomas-Keprta}, {Wentworth}, {Dodson}, and {Zolensky}}]{RN1225}
{Warren} J, {Watts} L, {Thomas-Keprta} K, et~al (1997) {Cosmic Dust Catalog. Volum 15: Particles from Collectors L2036 and L2021}. Tech. Rep. NASA/CR-97-112971; JSC-27897; NAS 1.26:112971, NASA

\bibitem[{{Watson} et~al.(2007){Watson}, {Stapelfeldt}, {Wood}, and {M{\'e}nard}}]{Watson07}
{Watson} AM, {Stapelfeldt} KR, {Wood} K, et~al (2007) {Multiwavelength Imaging of Young Stellar Object Disks: Toward an Understanding of Disk Structure and Dust Evolution}. In: {Reipurth} B, {Jewitt} D, {Keil} K (eds) {Protostars and Planets V}, p 523, \doi{10.48550/arXiv.0707.2608}, {\href{https://arxiv.org/abs/0707.2608}{{arXiv:0707.2608}}}

\bibitem[{{Weidling} et~al.(2009){Weidling}, {G{\"u}ttler}, {Blum}, and {Brauer}}]{Weidling09}
{Weidling} R, {G{\"u}ttler} C, {Blum} J, et~al (2009) {The Physics of Protoplanetesimal Dust Agglomerates. III. Compaction in Multiple Collisions}. \apj 696(2):2036--2043. \doi{10.1088/0004-637X/696/2/2036}, {\href{https://arxiv.org/abs/0902.3082}{{arXiv:0902.3082}}} {[astro-ph.EP]}

\bibitem[{{Weingartner} and {Draine}(2001)}]{Weingarnter2001}
{Weingartner} JC, {Draine} BT (2001) {Dust Grain-Size Distributions and Extinction in the Milky Way, Large Magellanic Cloud, and Small Magellanic Cloud}. \apj 548(1):296--309. \doi{10.1086/318651}, {\href{https://arxiv.org/abs/astro-ph/0008146}{{arXiv:astro-ph/0008146}}} {[astro-ph]}

\bibitem[{{West}(1991)}]{West91}
{West} RA (1991) {Optical properties of aggregate particles whose outer diameter is comparable to the wavelength}. \ao 30(36):5316--5324. \doi{10.1364/AO.30.005316}

\bibitem[{Westphal et~al.(2014)Westphal, Stroud, Bechtel, Brenker, Butterworth, Flynn, Frank, Gainsforth, Hillier, Postberg, Simionovici, Sterken, Nittler, Allen, Anderson, Ansari, Bajt, Bastien, Bassim, Bridges, Brownlee, Burchell, Burghammer, Changela, Cloetens, Davis, Doll, Floss, Gruen, Heck, Hoppe, Hudson, Huth, Kearsley, King, Lai, Leitner, Lemelle, Leonard, Leroux, Lettieri, Marchant, Ogliore, Ong, Price, Sandford, Tresseras, Schmitz, Schoonjans, Schreiber, Silversmit, Sole, Srama, Stadermann, Stephan, Stodolna, Sutton, Trieloff, Tsou, Tyliszczak, Vekemans, Vincze, Von~Korff, Wordsworth, Zevin, Zolensky, and Dusters}]{RN1677}
Westphal AJ, Stroud RM, Bechtel HA, et~al (2014) {Evidence for interstellar origin of seven dust particles collected by the Stardust spacecraft}. Science 345(6198):786--791. \doi{10.1126/science.1252496}

\bibitem[{{Whittet}(2022)}]{Whittet_dustbook}
{Whittet} DCB (2022) {Dust in the Galactic Environment}, 3rd edn. IOP Publishing, \doi{10.1088/2514-3433/ac7204}

\bibitem[{Williams et~al.(2016)Williams, Travis, Burton, and Harding}]{Williams2016}
Williams CD, Travis KP, Burton NA, et~al (2016) {A new method for the generation of realistic atomistic models of siliceous MCM-41}. Microporous Mesoporous Mater 228:215--223. \doi{10.1016/j.micromeso.2016.03.034}

\bibitem[{{Wolff} et~al.(1998){Wolff}, {Clayton}, and {Gibson}}]{Wolff98}
{Wolff} MJ, {Clayton} GC, {Gibson} SJ (1998) {Modeling Composite and Fluffy Grains. II. Porosity and Phase Functions}. \apj 503(2):815--830. \doi{10.1086/306029}

\bibitem[{{Wright}(1987)}]{Wright87}
{Wright} EL (1987) {Long-Wavelength Absorption by Fractal Dust Grains}. \apj 320:818. \doi{10.1086/165597}

\bibitem[{Wurm and Blum(1998)}]{RN785}
Wurm G, Blum J (1998) {Experiments on preplanetary dust aggregation}. Icarus 132(1):125--136. \doi{10.1006/icar.1998.5891}

\bibitem[{{Yang} et~al.(2016){Yang}, {Li}, {Looney}, and {Stephens}}]{HaifengYang16}
{Yang} H, {Li} ZY, {Looney} L, et~al (2016) {Inclination-induced polarization of scattered millimetre radiation from protoplanetary discs: the case of HL Tau}. \mnras 456(3):2794--2805. \doi{10.1093/mnras/stv2633}, {\href{https://arxiv.org/abs/1507.08353}{{arXiv:1507.08353}}} {[astro-ph.SR]}

\bibitem[{{Yang} et~al.(2019){Yang}, {Li}, {Stephens}, {Kataoka}, and {Looney}}]{HaifengHang19}
{Yang} H, {Li} ZY, {Stephens} IW, et~al (2019) {Does HL Tau disc polarization in ALMA band 3 come from radiatively aligned grains?} \mnras 483(2):2371--2381. \doi{10.1093/mnras/sty3263}, {\href{https://arxiv.org/abs/1811.11897}{{arXiv:1811.11897}}} {[astro-ph.SR]}

\bibitem[{Yi et~al.(2021)Yi, Liu, Chen, Yang, Wei, Liu, and Wei}]{RN2003}
Yi KY, Liu DH, Chen XS, et~al (2021) {Plasma-Enhanced Chemical Vapor Deposition of Two-Dimensional Materials for Applications}. Accounts of Chemical Research 54(4):1011--1022. \doi{10.1021/acs.accounts.0c00757}

\bibitem[{{Yoshida} et~al.(2025){Yoshida}, {Nomura}, {Tsukagoshi}, {Doi}, {Furuya}, and {Kataoka}}]{Yoshida25}
{Yoshida} TC, {Nomura} H, {Tsukagoshi} T, et~al (2025) {Dust Scattering Albedo at Millimeter Wavelengths in the TW Hya Disk}. \apj 980(1):50. \doi{10.3847/1538-4357/ad9f31}, {\href{https://arxiv.org/abs/2412.10731}{{arXiv:2412.10731}}} {[astro-ph.EP]}

\bibitem[{{Ysard} et~al.(2019){Ysard}, {Koehler}, {Jimenez-Serra}, {Jones}, and {Verstraete}}]{Ysard2019}
{Ysard} N, {Koehler} M, {Jimenez-Serra} I, et~al (2019) {From grains to pebbles: the influence of size distribution and chemical composition on dust emission properties}. \aap 631:A88. \doi{10.1051/0004-6361/201936089}, {\href{https://arxiv.org/abs/1909.05015}{{arXiv:1909.05015}}} {[astro-ph.GA]}

\bibitem[{Zamirri et~al.(2018)Zamirri, Casassa, Rimola, Segado-Centellas, Ceccarelli, and Ugliengo}]{Zamirri2018}
Zamirri L, Casassa S, Rimola A, et~al (2018) {IR spectral fingerprint of carbon monoxide in interstellar water–ice models}. \mnras 480(2):1427--1444. \doi{10.1093/mnras/sty1927}

\bibitem[{Zamirri et~al.(2019)Zamirri, Macià~Escatllar, Mari{\~n}oso~Guiu, Ugliengo, and Bromley}]{Zamirri2019}
Zamirri L, Macià~Escatllar A, Mari{\~n}oso~Guiu J, et~al (2019) {What Can Infrared Spectra Tell Us about the Crystallinity of Nanosized Interstellar Silicate Dust Grains?} ACS Earth Space Chem 3(10):2323--2338. \doi{10.1021/acsearthspacechem.9b00157}

\bibitem[{Zeegers et~al.(2023)Zeegers, Guiu, Kemper, Marshall, and Bromley}]{Zeegers_2023}
Zeegers ST, Guiu JM, Kemper F, et~al (2023) Predicting observable infrared signatures of nanosilicates in the diffuse interstellar medium. Faraday Discuss 245:609--619. \doi{10.1039/D3FD00055A}

\bibitem[{{Zeller} et~al.(1966){Zeller}, {Ronca}, and {Levy}}]{Zeller1966}
{Zeller} EJ, {Ronca} LB, {Levy} PW (1966) {Proton-Induced Hydroxyl Formation on the Lunar Surface}. \jgr 71:4855. \doi{10.1029/JZ071i020p04855}

\bibitem[{{Zerull} et~al.(1993){Zerull}, {Gustafson}, {Schulz}, and {Thiele-Corbach}}]{Zerull93}
{Zerull} RH, {Gustafson} BAS, {Schulz} K, et~al (1993) {Scattering by aggregates with and without an absorbing mantle: microwave analog experiments}. \ao 32(21):4088--4100. \doi{10.1364/AO.32.004088}

\bibitem[{{Zhang} et~al.(2023){Zhang}, {Zhu}, {Ueda}, {Kataoka}, {Sierra}, {Carrasco-Gonz{\'a}lez}, and {Mac{\'\i}as}}]{SZhang23}
{Zhang} S, {Zhu} Z, {Ueda} T, et~al (2023) {Porous Dust Particles in Protoplanetary Disks: Application to the HL Tau Disk}. \apj 953(1):96. \doi{10.3847/1538-4357/acdb4e}, {\href{https://arxiv.org/abs/2306.00158}{{arXiv:2306.00158}}} {[astro-ph.EP]}

\bibitem[{{Zhong} et~al.(2024){Zhong}, {Ren}, {Ma}, {Xie}, {Ma}, {Wallack}, {Mawet}, and {Ruane}}]{Zhong24}
{Zhong} H, {Ren} BB, {Ma} B, et~al (2024) {Multiband reflectance and shadowing of the protoplanetary disk RX J1604.3-2130 in scattered light}. \aap 684:A168. \doi{10.1051/0004-6361/202348874}, {\href{https://arxiv.org/abs/2402.16698}{{arXiv:2402.16698}}} {[astro-ph.EP]}

\bibitem[{Zhou et~al.(2019)Zhou, Mousseau, and Song}]{RN1887}
Zhou X, Mousseau N, Song J (2019) {Is Hydrogen Diffusion along Grain Boundaries Fast or Slow? Atomistic Origin and Mechanistic Modeling}. \prl 122(21):215501. \doi{10.1103/PhysRevLett.122.215501}

\bibitem[{{Zhu} et~al.(2019){Zhu}, {Zhang}, {Jiang}, {Kataoka}, {Birnstiel}, {Dullemond}, {Andrews}, {Huang}, {P{\'e}rez}, {Carpenter}, {Bai}, {Wilner}, and {Ricci}}]{Zhu19}
{Zhu} Z, {Zhang} S, {Jiang} YF, et~al (2019) {One Solution to the Mass Budget Problem for Planet Formation: Optically Thick Disks with Dust Scattering}. \apjl 877(2):L18. \doi{10.3847/2041-8213/ab1f8c}, {\href{https://arxiv.org/abs/1904.02127}{{arXiv:1904.02127}}} {[astro-ph.EP]}

\bibitem[{Zhukovska and Henning(2013)}]{RN821}
Zhukovska S, Henning T (2013) {Dust input from AGB stars in the Large Magellanic Cloud}. \aap 555:A99. \doi{10.1051/0004-6361/201321368}

\bibitem[{{Zsom} et~al.(2010){Zsom}, {Ormel}, {G{\"u}ttler}, {Blum}, and {Dullemond}}]{Zsom10}
{Zsom} A, {Ormel} CW, {G{\"u}ttler} C, et~al (2010) {The outcome of protoplanetary dust growth: pebbles, boulders, or planetesimals? II. Introducing the bouncing barrier}. \aap 513:A57. \doi{10.1051/0004-6361/200912976}, {\href{https://arxiv.org/abs/1001.0488}{{arXiv:1001.0488}}} {[astro-ph.EP]}

\bibitem[{{Zubko} et~al.(2015){Zubko}, {Shkuratov}, and {Videen}}]{Zubko15}
{Zubko} E, {Shkuratov} Y, {Videen} G (2015) {Effect of morphology on light scattering by agglomerates}. \jqsrt 150:42--54. \doi{10.1016/j.jqsrt.2014.06.023}

\bibitem[{Öberg and Bergin(2021)}]{RN1361}
Öberg KI, Bergin EA (2021) {Astrochemistry and compositions of planetary systems}. Phys Rep 893:1--48. \doi{10.1016/j.physrep.2020.09.004}

\end{thebibliography}

\end{document}